\PassOptionsToPackage{table}{xcolor}

\documentclass[10pt,twocolumn,letterpaper]{article}

\usepackage[pagenumbers]{cvpr} 

\newcommand{\red}[1]{{\color{red}#1}}

\usepackage[acronym]{glossaries}
\usepackage[super]{nth}
\usepackage{multirow}
\usepackage{relsize}
\usepackage{comment}
\usepackage{caption}
\usepackage{alphalph}
\usepackage[export]{adjustbox}
\newacronym{nerf}{NeRF}{Neural Radiance Fields}
\newacronym{sdf}{SDF}{Signed Distance Field}
\newacronym{mvs}{MVS}{Multi-View Stereo}
\newacronym{nir}{NIR}{Near-Infrared}
\newacronym{ms}{MS}{Multispectral}
\newacronym{pol}{Pol}{Polarization}
\newacronym{tof}{ToF}{Time-of-Flight}
\newacronym{mlp}{MLP}{Multi-Layer Perceptron}
\newacronym{roi}{RoI}{Region of Interest}
\newacronym{multimodalstudio}{MMS}{MultimodalStudio}
\newacronym{nxdc}{NXDC}{Normalized Cross-Device Coordinates}
\newacronym{sfm}{SfM}{Structure-from-Motion}
\newacronym{gs}{GS}{Gaussian Splatting}
\newacronym{sh}{SH}{Spherical Harmonics}
\newacronym{brdf}{BRDF}{Bidirectional Reflectance Distribution Function}
\newacronym{aop}{AoLP}{Angle of Linear Polarization}
\newacronym{dop}{DoLP}{Degree of Linear Polarization}

\usepackage[accsupp]{axessibility}

\def\dname{\textit{\gls{multimodalstudio}-DATA}}
\def\fname{\textit{\gls{multimodalstudio}-FW}}

\newcommand{\g}[1]{{\leavevmode\smaller[2]\color{gray}#1}}
\definecolor{green}{rgb}{0.0, 0.68, 0.33}
\definecolor{blue}{rgb}{0.0, 0.44, 0.75}
\definecolor{orange}{rgb}{0.97, 0.58, 0.15}
\definecolor{purple}{rgb}{0.45, 0.19, 0.61}
\definecolor{grey}{rgb}{0.5, 0.5, 0.5}
\newcommand{\green}[1]{{\leavevmode\color{green}#1}}
\newcommand{\blue}[1]{{\leavevmode\color{blue}#1}}
\newcommand{\orange}[1]{{\leavevmode\color{orange}#1}}
\newcommand{\purple}[1]{{\leavevmode\color{purple}#1}}
\newcommand{\grey}[1]{{\leavevmode\color{grey}#1}}

\makeatletter
\let\@oldmaketitle\@maketitle
\renewcommand{\@maketitle}{\@oldmaketitle
  \centering
  \includegraphics[width=\linewidth]{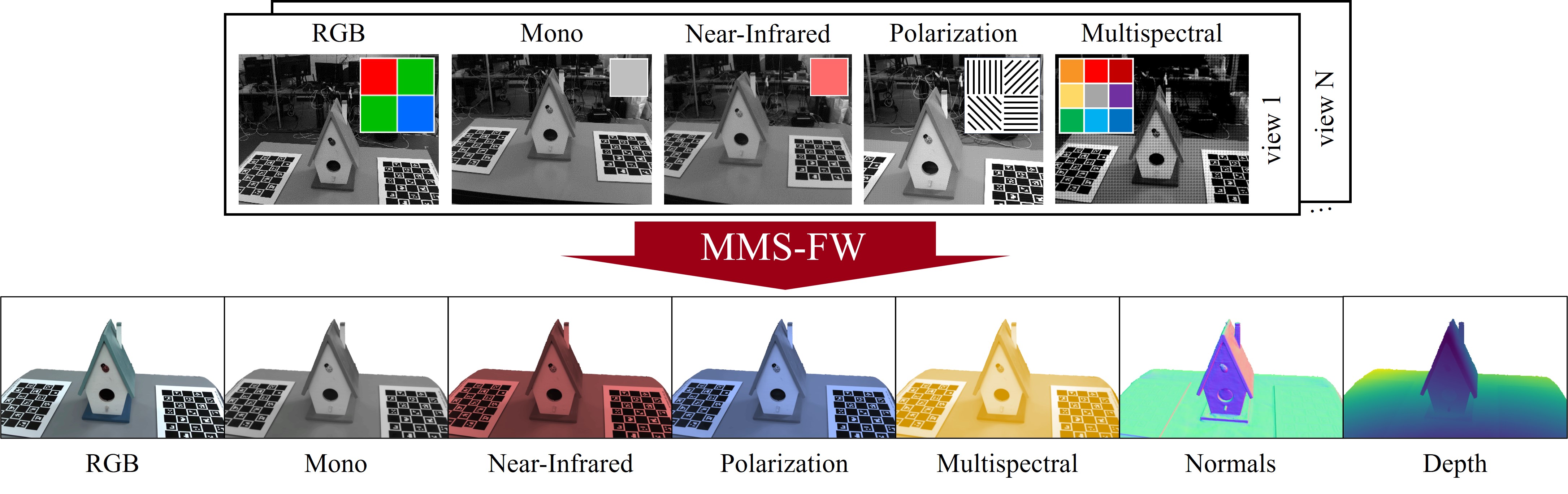}
  \vspace{-8pt}
  \captionof{figure}{Overview of the proposed framework. MMS-FW exploits unaligned multimodal frames acquired by sensors with different intrinsics to render perfectly aligned novel views for each modality. The mosaick pattern for each modality is shown in the top corners.}
  \label{fig:teaser}
  \bigskip}
\makeatother

\definecolor{cvprblue}{rgb}{0.21,0.49,0.74}
\usepackage[a-1b]{pdfx}
\hypersetup{
    pagebackref,breaklinks,colorlinks,allcolors=cvprblue,
}

\def\paperID{15741} 
\def\confName{CVPR}
\def\confYear{2025}

\title{MultimodalStudio: A Heterogeneous Sensor Dataset and Framework for Neural Rendering across Multiple Imaging Modalities}

\author{Federico Lincetto\textsuperscript{1}
\and
Gianluca Agresti\textsuperscript{2}
\and
Mattia Rossi\textsuperscript{2}
\and
Pietro Zanuttigh\textsuperscript{1}
\and
\textsuperscript{1}University of Padova \and \textsuperscript{2}Sony Europe B.V.
}

\begin{document}

\maketitle
\begin{abstract}
\gls{nerf} have shown impressive performances in the rendering of 3D scenes from arbitrary viewpoints. While RGB images are widely preferred for training volume rendering models, the interest in other radiance modalities is also growing. However, the capability of the underlying implicit neural models to learn and transfer information across heterogeneous imaging modalities has seldom been explored, mostly due to the limited training data availability. For this purpose, we present \gls{multimodalstudio}: it encompasses \dname\ and \fname. \dname\ is a multimodal multi-view dataset containing 32 scenes acquired with 5 different imaging modalities: RGB, monochrome, near-infrared, polarization and multispectral. \fname\  is a novel modular multimodal NeRF framework designed to handle multimodal raw data and able to support an arbitrary number of multi-channel devices. Through extensive experiments, we demonstrate that \fname\ trained on \dname\ can transfer information between different imaging modalities and produce higher quality renderings than using single modalities alone. We publicly release the dataset and the framework, to promote the research on multimodal volume rendering and beyond. 
\end{abstract}
\section{Introduction}
\label{sec:intro}

In recent years, \gls{nerf}~\cite{mildenhall2020nerf} and \gls{gs}~\cite{kerbl3Dgaussians} greatly advanced the field of volume rendering by enabling novel view synthesis with unprecedented photorealistic appearance. Their growing popularity has driven research to employ these approaches for related tasks. This is the case of multi-view 3D reconstruction, which received a strong boost first by the introduction of \gls{sdf} estimation methods~\cite{wang2021neus,yariv2021volume}, then thanks to the adoption of the multi-resolution feature structures~\cite{mueller2022instant,li2023neuralangelo,wang2023petneus,Chen2022ECCV,Chen2023TOG}, and finally with the exploitation of the \gls{gs} formulation~\cite{guedon2024sugar,huang20242d}. 
Other works explored the possibility to perform the material properties estimation along with a \gls{nerf} or \gls{gs} training, thus enabling the possibility to understand, to replicate and to edit the interaction of the incident light with the scene surfaces~\cite{verbin2022ref,liu2023nero,jin2023tensoir,liang2024gs}. \\
\indent Typically neural rendering methods rely exclusively on RGB images, as these are convenient to capture and relatively straightforward to process. However, we argue that they may not always be the optimal choice to address the aforementioned tasks: there exist other imaging modalities capable of measuring either the information carried by different spectral bands, for instance \gls{nir} and \gls{ms} data, or different properties of the light, such as \gls{pol}. 
Only a few works explore the usage of other imaging modalities, but they either couple RGB data with a single additional modality, or they consider a modality other than RGB, in isolation~\cite{li2024spec,li2024spectralnerf,lin2024thermalnerf,Chen2024ECCV,Xu2024ECCV,ma2024hyperspectral}.
One reason for this is the limited availability of multi-view multimodal training data. We argue that exploiting the additional information carried by multiple modalities can offer a valuable advantage in addressing the mentioned  
tasks. Moreover, a multimodal radiance field has multiple applications: it enables the generation of pseudo-real datasets for the training of learning-based methods that require perfectly aligned multimodal data; it permits to carry out generalized modality-to-modality conversion; it enables the modeling of a sensor digital twin to reproduce its properties and to investigate its behavior. For these reasons, we present \textit{MultimodalStudio}. Our contribution comprises (1)~\textit{\dname}, a novel dataset containing 32 scenes acquired from 50 viewpoints with 5 different imaging devices: these include RGB, monochrome, near-infrared, polarization and multispectral cameras and are precisely geometrically calibrated; (2)~\textit{\fname}, a novel modular NeRF framework that handles  multimodal data: it includes support for an arbitrary number of multi-channel representations, for raw data and for the specific requirements of polarized images. \\
\indent We employ \dname\ and \fname\ to conduct extensive investigations that show the benefits of using combinations of modalities for neural rendering and how the underlying implicit representation is able to transfer information across modalities, even in scenarios where some of them are available only for a limited number of viewpoints. We expect that the release of \textit{MultimodalStudio} will spread the relevance of using multiple modality combinations in volume rendering related tasks.

\section{Related Works}
\label{sec:related}

\paragraph{Novel View Synthesis}
In recent years, the advent of neural implicit representations~\cite{sitzmann2019scene} has promoted a huge acceleration in the field of novel view synthesis. Previously, this task was addressed employing enhanced \gls{mvs} algorithms~\cite{cooke2006multi,rosu2022neuralmvs,woodford2007new} or by training deep CNNs~\cite{zhou2016view,riegler2020free,choi2019extreme}. Today, interest in solutions employing diffusion models is increasing~\cite{chan2023generative,tseng2023consistent,yu2023long}, but the proposed methods are still limited in terms of view consistency and rendering resolution and require computationally expensive training procedures.
In this scenario, NeRFs~\cite{mildenhall2020nerf} and GS~\cite{kerbl3Dgaussians} are the most established solutions for neural rendering, thanks to their relatively short training time and their capability to render consistent high-resolution novel views quickly. A crucial contribution was the introduction of the multi-resolution feature structures~\cite{mueller2022instant,Chen2022ECCV,Chen2023TOG}, since they allow considerably speeding up the training and capturing fine details. Finally, the development of NeRFStudio~\cite{nerfstudio} and SDFStudio~\cite{Yu2022SDFStudio} strongly supported research in the field by offering large and comprehensive frameworks to test, compare, and develop different NeRF and GS methods. In releasing \textit{MultimodalStudio}, we take inspiration from NeRFStudio and SDFStudio to offer a novel modular framework for multimodal neural rendering.

\vspace{-13pt}
\paragraph{Multimodal Neural Rendering}
The introduction of unconventional imaging modalities in neural rendering pipelines is experiencing an increasing interest. Some methods propose to include depth information, from external sensors or estimated from color images, 
to guide the density estimation~\cite{Chang_2023_ICCV,attal2021torf,liu2023multi,lincetto2023emp,zhu2023multimodal}. 
Other methods employ additional imaging modalities to render novel views of scenes acquired by thermal or \gls{ms} sensors~\cite{lin2024thermalnerf,Xu2024ECCV,li2024spec,li2024spectralnerf,ma2024hyperspectral,chen2024thermal3d,li2024implicit}. Another field focuses on the exploitation of \gls{pol} frames to perform accurate surface reconstruction and to enable material property estimation~\cite{dave2022pandora,li2024neisf,han2024nersp}. However, the most close contributions to our work are X-NeRF~\cite{poggi2022xnerf} and NeSpoF~\cite{kim2023neural}, and both are more limited than our method in multiple aspects: 
\begin{itemize}
\item The \gls{nxdc} of X-NeRF are designed for forward-facing scenes, and lead to worse results on object-centric scenes. Instead, MMS-FW supports forward-facing, object-centric, bounded, or unbounded scenes and focuses the reconstruction on the foreground, separating it from the background.
\item X-NeRF only supports RGB, IR, and MS frames and does not permit to work with other modalities. NeSpoF aims to produce polarization renderings of novel views for each multispectral channel, thus it is extremely specific and supports only polarized-multispectral images. On the contrary, MMS-FW supports arbitrary sets of modalities captured by heterogeneous sensors, including RGB, monochrome, infrared, multispectral, and polarization. 
\item X-NeRF and NeSpoF require the same amount of frames for each modality, while MMS-FW can manage an unbalanced amount of frames for each of them.
\item X-NeRF and NeSpoF require undistorted and demosaicked images, as they do not support RAW data, while MMS-FW accepts distorted and mosaicked frames, thus it can avoid the use of costly preprocessing procedures.
\end{itemize}
The experiments on \fname\ show that different combinations of modalities can improve the novel view synthesis quality, and that the underlying implicit representation can transfer information across heterogeneous modalities.

\vspace{-13pt}
\paragraph{Multimodal Multi-view Datasets} 
The availability of multimodal multi-view dataset is scarce both in terms of scenes and in terms of involved modalities, as having access to a wide variety of heterogeneous sensors is not common, and their joint calibration challenging. A multimodal multi-view dataset that includes RGB, \gls{nir} and \gls{ms} frames has been presented along with X-NeRF~\cite{poggi2022xnerf}, but it contains just 16 forward-facing scenes and 30 viewpoints per modality, with only the RGB camera poses available. 
Also NeSpoF~\cite{kim2023neural} provides the dataset, but it is limited to only 4 real and 4 synthetic forward-facing scenes with 25 polarized-multispectral frames each. Both datasets are not suitable for a robust investigation on the combining of different modalities due to (1) the limited amount of different modalities involved, (2) the too restricted set of materials with different properties, (3) the lack of an adequate number of viewpoints, and (4) the absence of object-centric scenes. 
On the other side, \dname\ matches all the mentioned requirements. Therefore, it permits to investigate the benefits of combining multiple modalities not only in more mainstream tasks such as neural rendering and 3D reconstruction, but also for radiance spectrum estimation, modality-to-modality conversion and material property estimation.

\section{Dataset}
\label{sec:dataset}

\begin{figure}
    \includegraphics[width=0.242\linewidth]{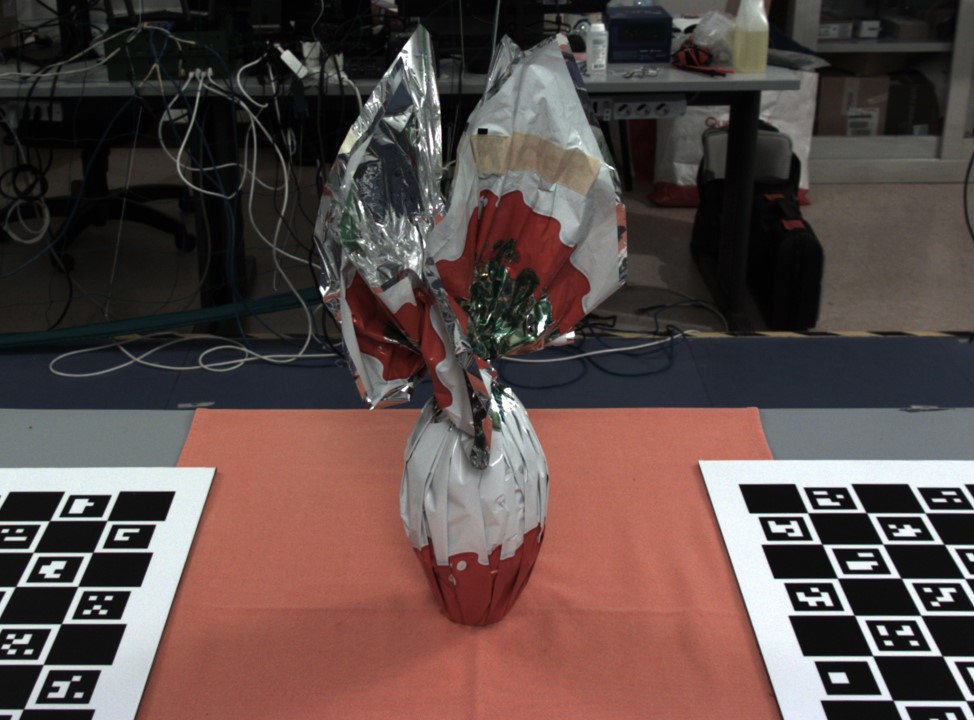}
    \hfill
    \includegraphics[width=0.242\linewidth]{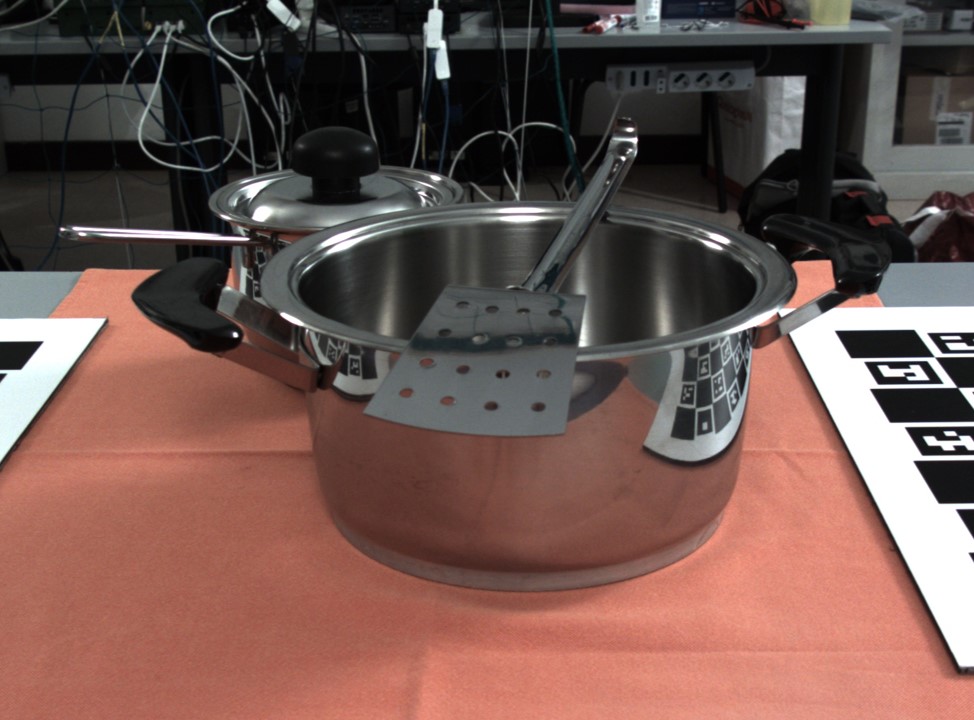}
    \hfill
    \includegraphics[width=0.242\linewidth]{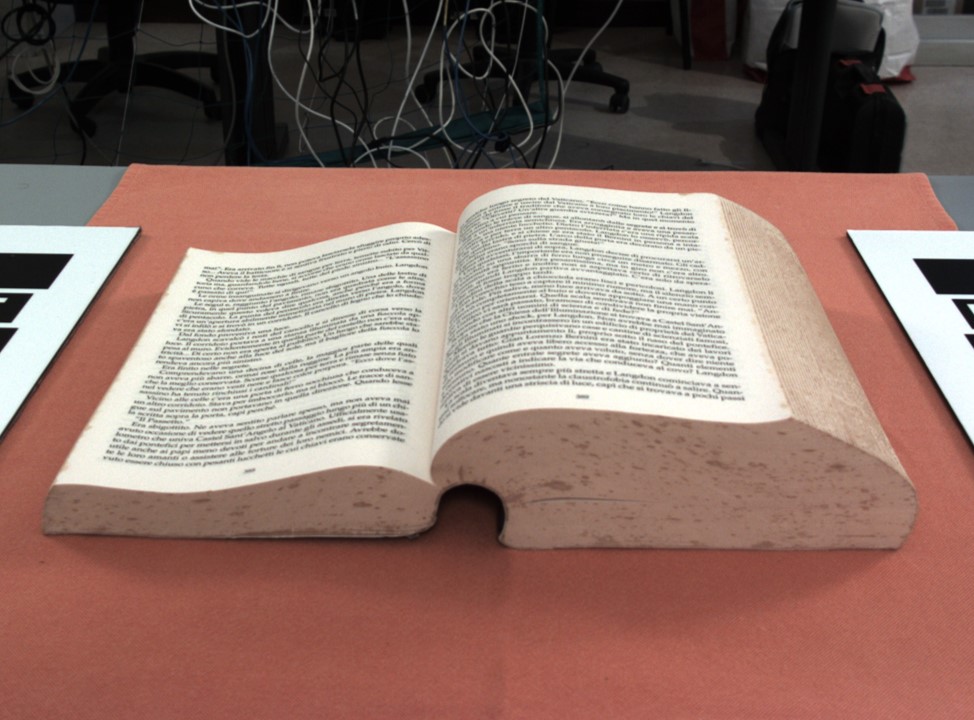}
    \hfill
    \includegraphics[width=0.242\linewidth]{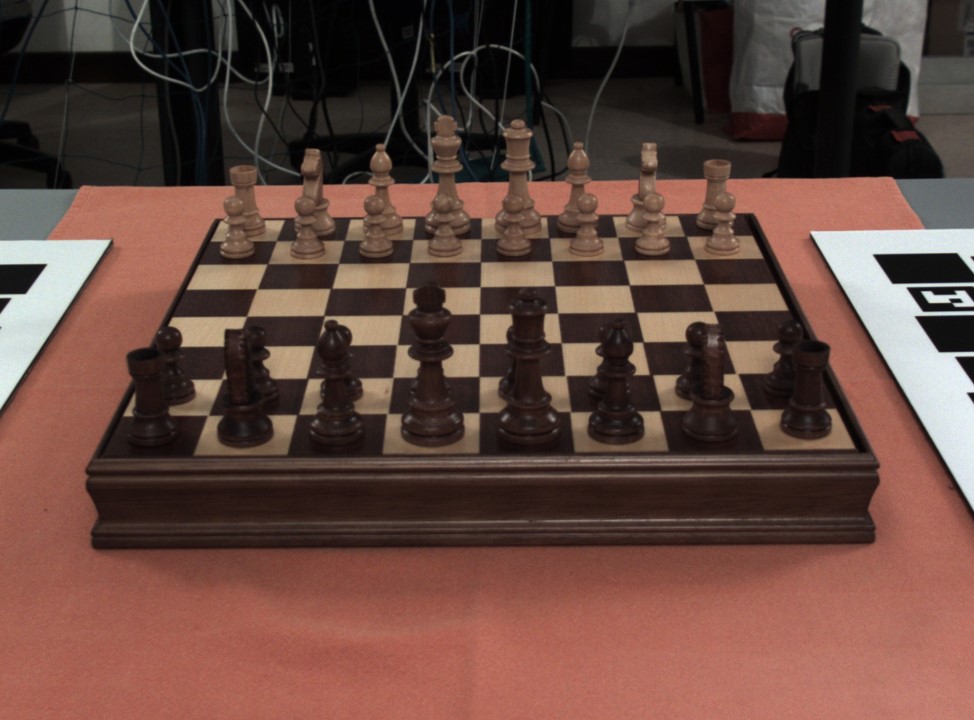}
    \hfill
    \includegraphics[width=0.242\linewidth]{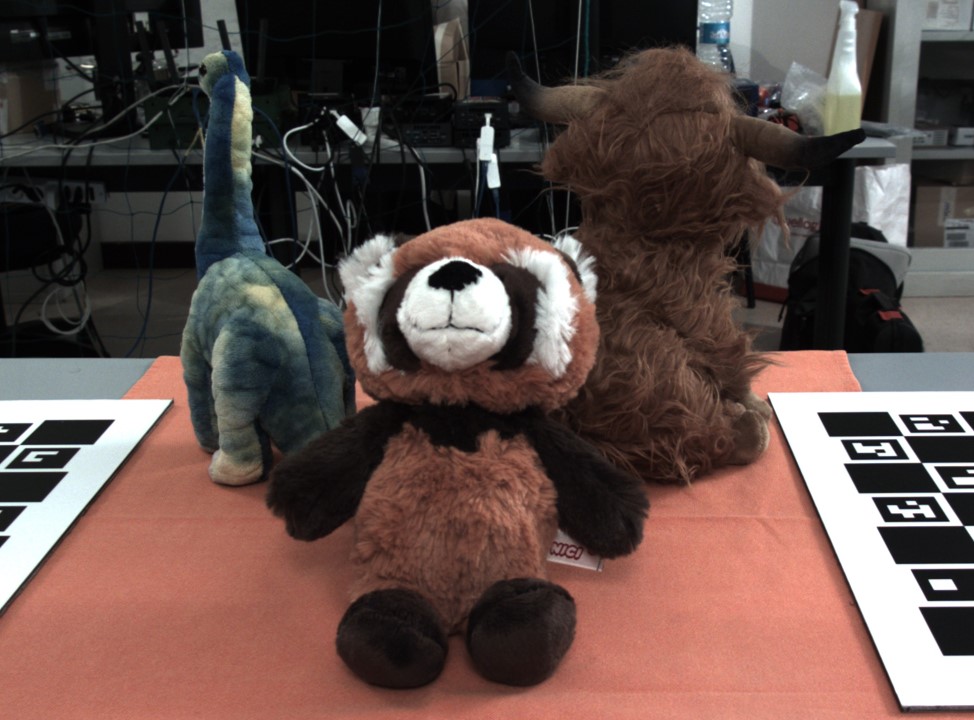}
    \hfill
    \includegraphics[width=0.242\linewidth]{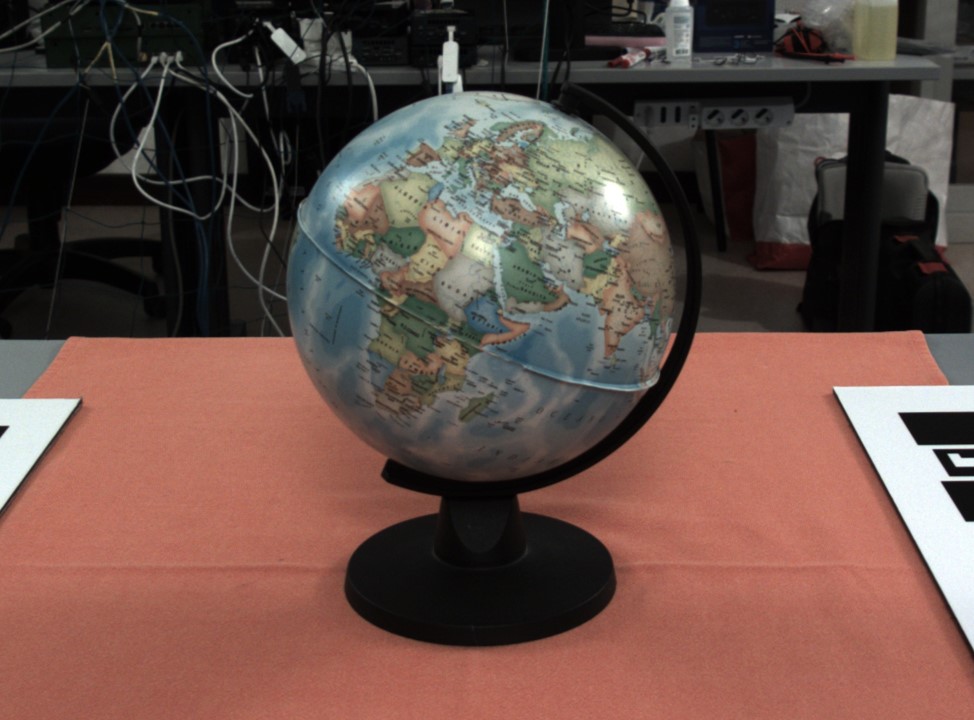}
    \hfill
    \includegraphics[width=0.242\linewidth]{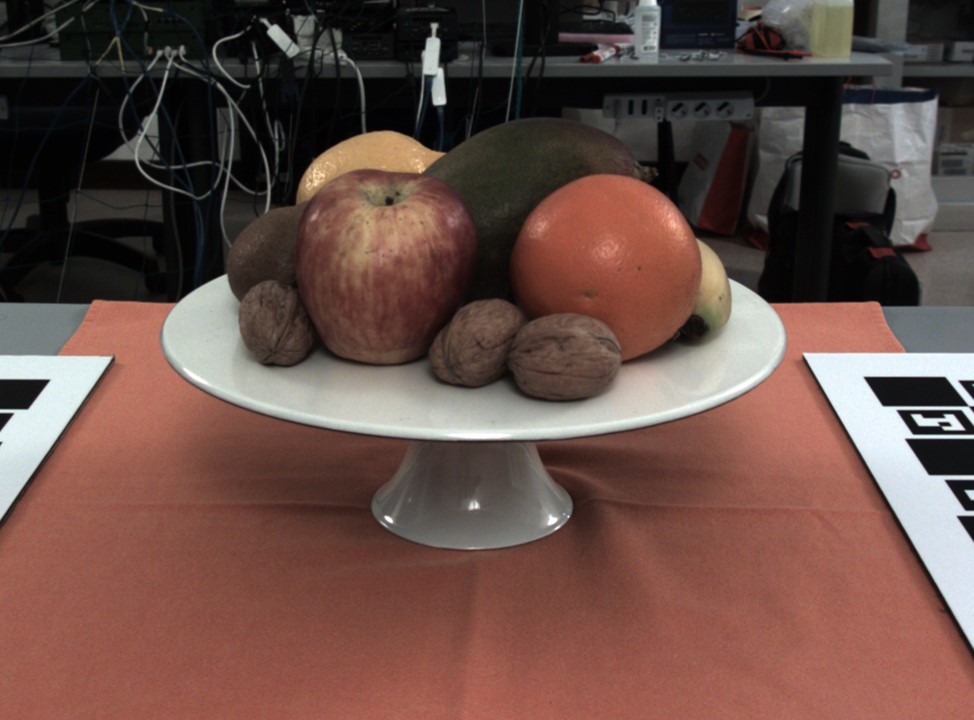}
    \hfill
    \includegraphics[width=0.242\linewidth]{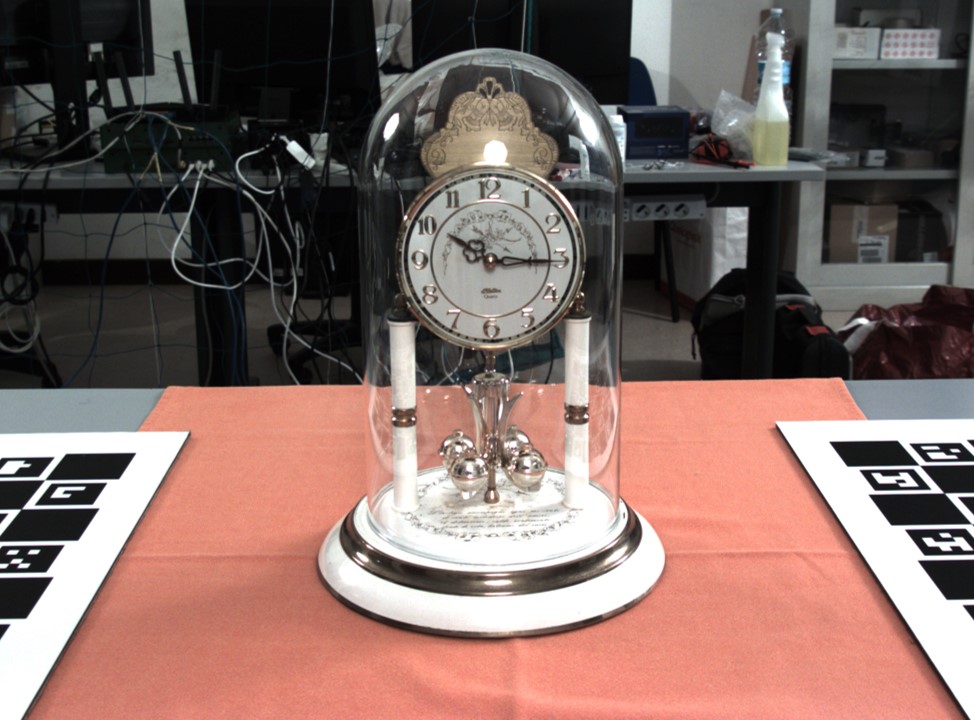}
    \hfill
    \includegraphics[width=0.242\linewidth]{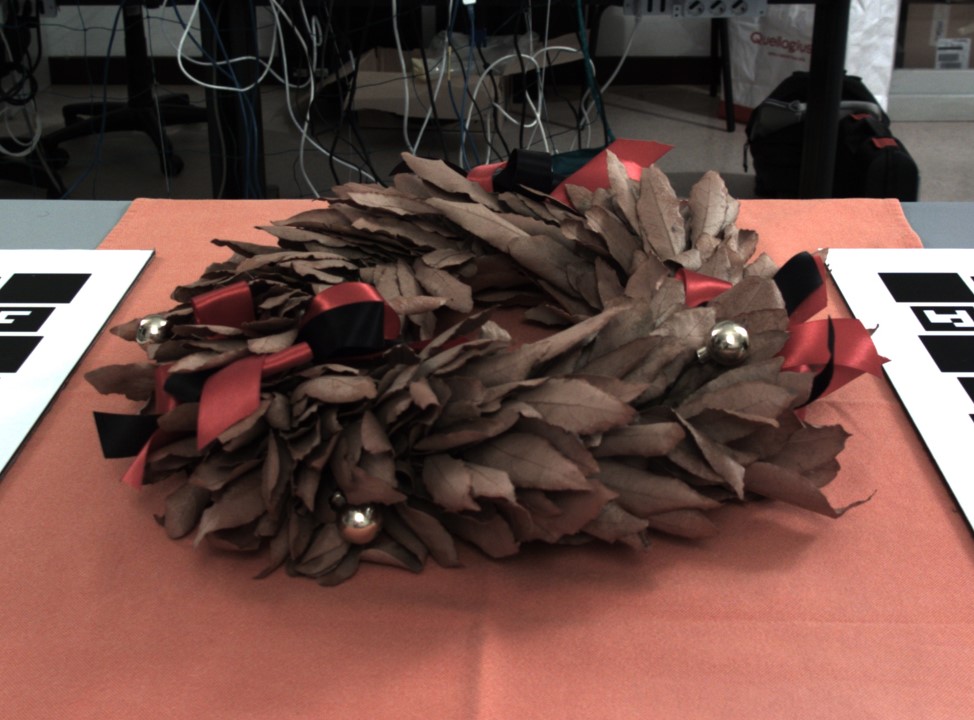}
    \hfill
    \includegraphics[width=0.242\linewidth]{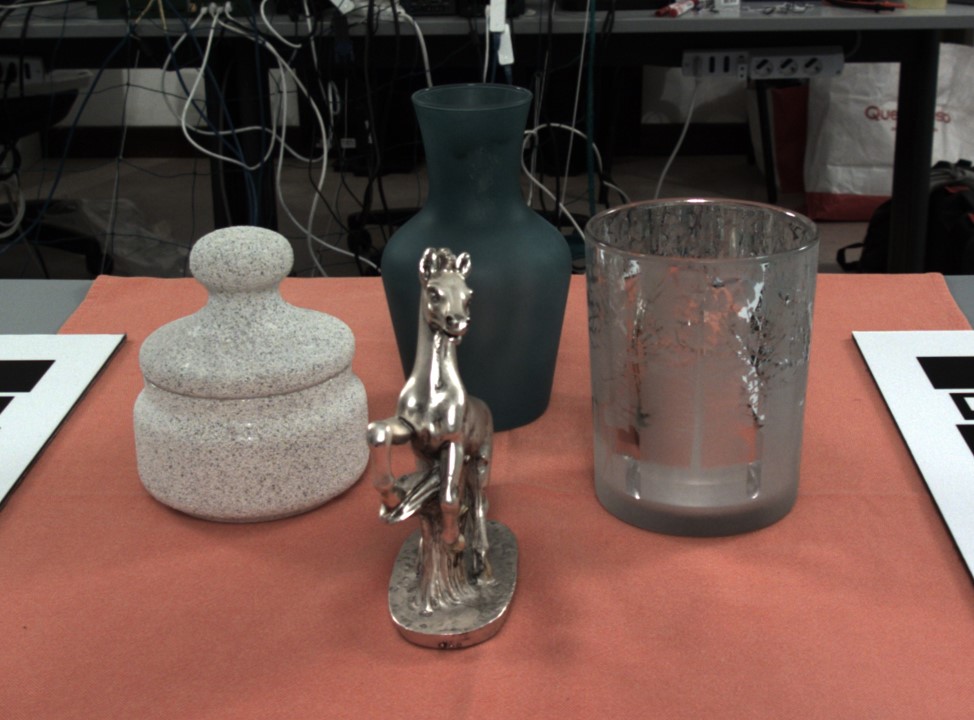}
    \hfill
    \includegraphics[width=0.242\linewidth]{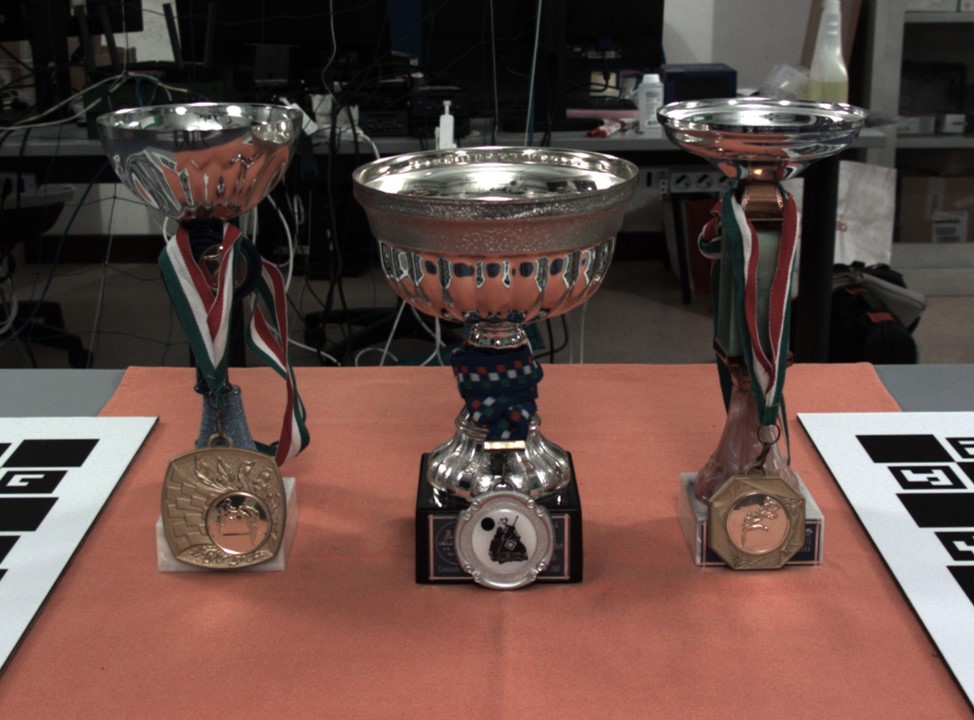}
    \hfill
    \includegraphics[width=0.242\linewidth]{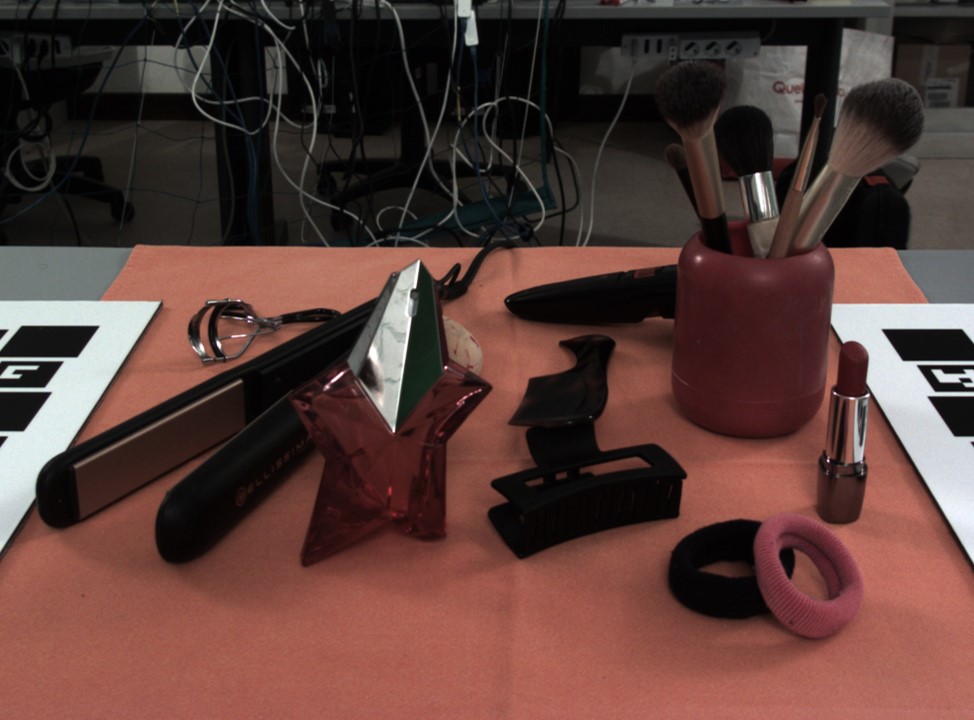}
    \caption{Sample scenes from \dname.
    The complete set is shown in the Supplementary Material.}
    \label{fig:scenes}
\end{figure}
\begin{figure}
  \centering
  \includegraphics[width=\linewidth]{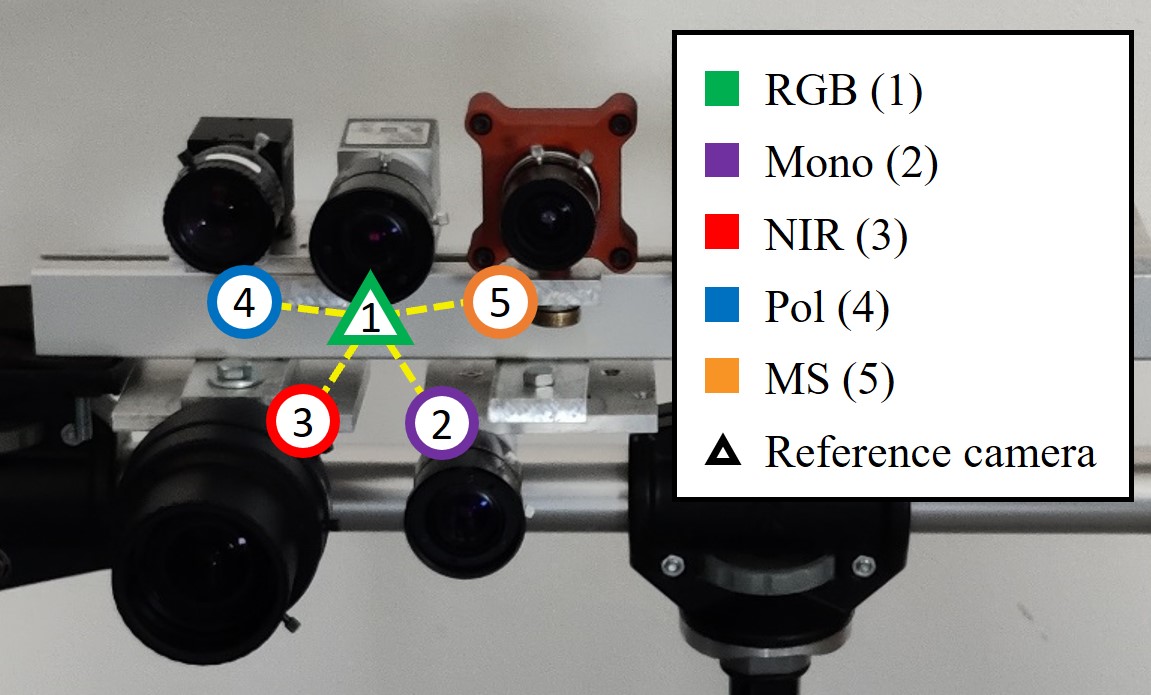}
  \caption{The imaging sensors arranged with a star topology.}
  \label{fig:rig}
  \vspace{-10pt}
\end{figure}

In this section, we present \textit{\dname}, a new multimodal multi-view dataset. It comprises a set of object-centric scenes with everyday items made of a variety of materials, including plastic, metal, wood, organic, paper, glass, and cloth (see \cref{fig:scenes} for some samples).
It was captured with different sensors, to support research on many applications like \gls{mvs}, neural rendering and 3D reconstruction.

\subsection{Overview of the Dataset}
\label{sec:overview}

Our dataset consists of 32 object-centric scenes acquired with 5 different imaging modalities. The modalities include RGB, Mono, \gls{nir}, \gls{pol} and \gls{ms} images, captured by a Basler acA2500-14gc (1), a Basler acA2500-14gm (2), a Basler acA1300-60gmNIR (3), a FLIR BFS-U3-51S5P-C (4) and a Silios CMS-C1 (5), respectively. The \gls{nir} sensor integrates both the entire visible and the near-infrared spectrum. The cameras are mounted on a metallic rig, as shown in \cref{fig:rig}, and equipped with optics providing $\sim50^o$ FOV to each.\\
\indent For each scene, we collected 50 views per modality, which results into 250 frames per scene. The acquisitions cover the object at 360° at two different heights. The dataset includes a wide variety of materials. We chose objects with diffusive, glossy, reflective, and transparent surfaces to enable the evaluation of how the different modalities capture different light-material interactions. In addition, we selected a wide range of materials with varying compositions, such as plastic, metal, wood, organic, cloth, paper, and glass materials. Such a variety of materials is characterized by a set of very different spectral responses, and this is well captured by our heterogeneous set of sensors. Refer to the supplementary material for the complete scene classification and for additional details on the setup description.

\subsection{Acquisition}
\label{sec:acquistion}
The acquisitions were performed by moving the rig all around the target object placed on the table and 
by capturing data from all modalities at each viewpoint.
Hereafter, we use the term \textit{macroframe} to refer to a complete multimodal acquisition taken from a single rig position: in total there are 50 macroframes per scene, each encompassing 5 modalities, one for each camera. The rig was moved manually, therefore the sensor positions are different from scene to scene. During each acquisition, the rig was static to avoid blur artifacts. Fixed exposure, fixed white balance and fixed black level were adopted. Finally, we chose to record and save mosaicked frames exploiting the maximal available bit depth of every sensor, namely 12 bit data for RGB, mono, \gls{ms} and \gls{nir} cameras, and 16 bit data for the \gls{pol} camera.

\subsection{Geometrical Calibration}
\label{sec:calibration}
All the sensors are accurately geometrically calibrated, assuming the Brown–Conrady distortion model~\cite{brown1996decentering}. The calibration process involves two steps: the first is the intrinsics calibration, the second is the joint pose calibration of the different sensors. We calibrated the camera extrinsics assuming a star topology, as shown in \cref{fig:scenes}. The RGB sensor is selected as the reference camera, and all the other sensors are stereo calibrated with respect to it.  We arranged a set of five ChAruco boards in a single column and we acquired $\sim$30 macroframes, at different height and distance from the boards. The choice of ChAruco boards is justified by their unique patterns, so they can be properly recognized and matched even if the board is not completely in the field of view. This commonly happens in our setup since all the sensors have a different collocation on the rig. 
In the intrinsics estimation, we kept the skew, and the radial and tangential distortion parameters higher than the \nth{2} order fixed. The reprojection errors after the intrinsic and extrinsic parameter estimation are reported in \cref{tab:reprojection_error}.

\begin{table}
    \centering
    \begin{tabular}{@{}lccc@{}}
        \toprule
        \multirow{2}{*}{Modality} & Pixel size & Intrinsics repr. & Stereo repr. \\
        & [$\mu m$] & RMSE [px] & RMSE [px] \\
        \midrule
        RGB & 2.20 & 0.18 & - \\
        Mono & 2.20 & 0.17 & 0.38 \\
        Infrared & 5.30 & 0.11 & 0.32 \\
        Polarization & 3.45 & 0.20 & 0.24 \\
        Multispectral & 5.30 & 0.36 & 0.35 \\
        \bottomrule
    \end{tabular}
    \caption{Calibration reprojection error for the different cameras.}
    \label{tab:reprojection_error}
\end{table}

\section{Framework}
\label{sec:framework}

\begin{figure}
    \centering
    \includegraphics[width=\linewidth]{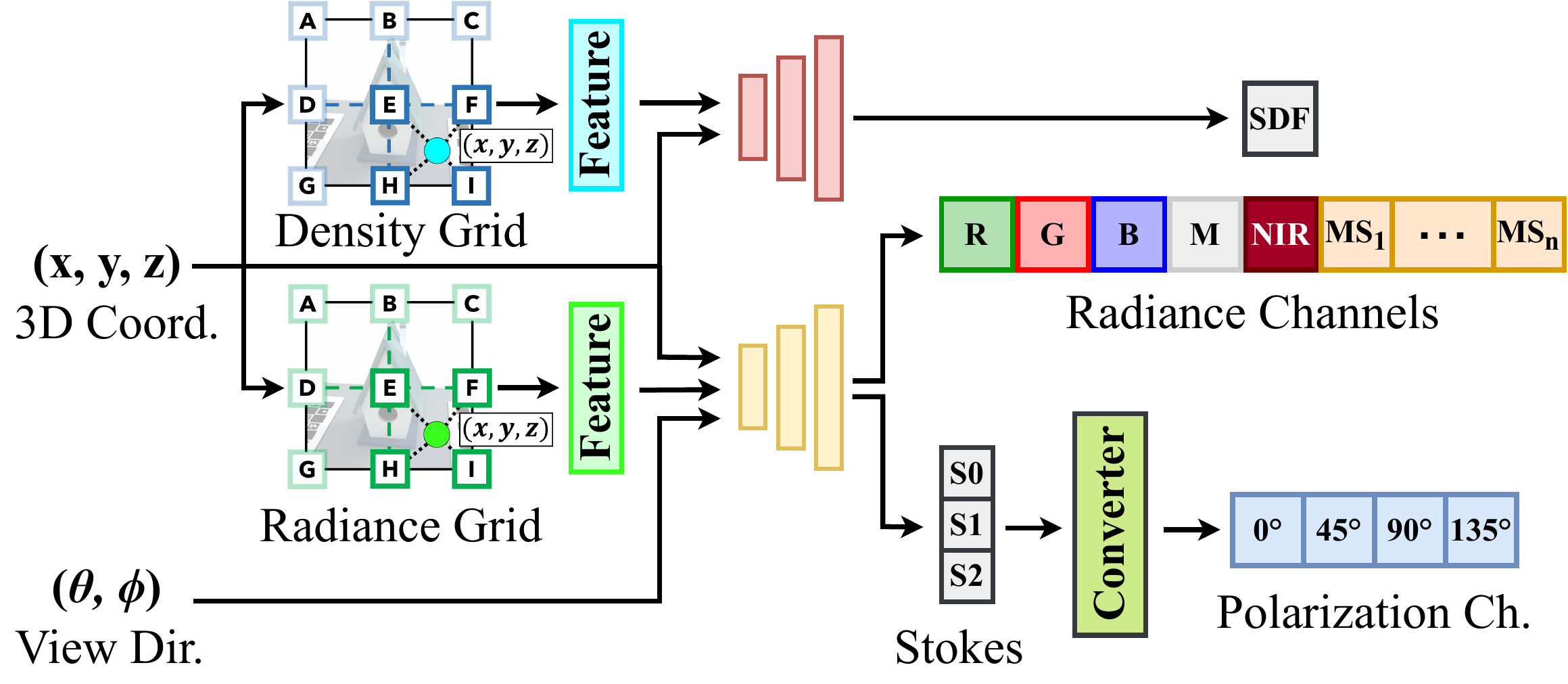}
    \caption{Schema of the pipeline used in the experiments.}
    \label{fig:schema}
\end{figure}

Along with the dataset, we introduce \textit{\fname}, a framework to enable \gls{nerf} training with multimodal data. The implementation takes inspiration from the structure of NeRFStudio~\cite{nerfstudio} and SDFStudio~\cite{Yu2022SDFStudio}, but it introduces several changes in order to deal with multiple imaging modalities. 
Moreover, we want \fname\ to be versatile and able to handle sensors of different nature and employ various NeRF methods: for this reason, one of its most important features is its modularity. The framework architecture is organized in a hierarchical manner; indeed, it is possible to easily change the configuration of the \gls{nerf} pipeline and test different implicit representation strategies by switching modules. In this section, we present the main features of \fname\ and the potentialities offered by its use.

\subsection{Multimodality}
\label{multimodality}
As anticipated, the core of \fname\ is the capability of handling multiple modalities at the same time, to employ them for neural rendering and 3D reconstruction applications. Therefore, each sensor is defined by an independent set of intrinsics and extrinsics geometrical calibration parameters and by a modality. Currently, \fname\ supports modalities of two natures: radiance modalities with a variable number of channels (\eg, RGB, \gls{ms}, \gls{nir}) and modalities involving polarization data. With ``variable number of channels" it is intended that there are no limitations, on the implementation side, on the number of sensors and on the number of channels per modality that \fname\ can handle. 
The polarized data must be handled separately due to their dependency on the camera orientation. In example, when the camera rotates around its viewing direction, the sensor polarization filters also rotate, thus measuring varying polarized intensities. 
For this reason, we directly estimate the Stokes vectors~\cite{stokes1851composition} with respect to a fixed reference system, and then convert them to the polarized intensities by considering the actual camera orientation, as proposed in~\cite{dave2022pandora,li2024neisf}.\\
\indent The development of these features required several modifications with respect to the standard \gls{nerf} pipeline. 
Specifically, at each iteration, a batch of rays is cast by randomly selecting an equal amount of pixels from each modality. We decoupled the density from the radiance estimation by initializing two separate modules. 
The implicit representation architecture used by these modules can be easily switched among the available ones: for our experiments we employ the multi-resolution hash-grid proposed in Neuralangelo~\cite{li2023neuralangelo}. It uses a hash-map that associates each 3D coordinate to a set of learnable features, then passed to a shallow MLP as additional input.
The density model relies on the \gls{sdf} estimation~\cite{park2019deepsdf} and it is shared by all the modalities, since we assume the same geometry for all the considered modalities. We employ a shared multi-resolution hash-grid also for the radiance estimation, because all the modalities share a relevant part of the information since they capture overlapping spectral bands. The idea is that a shared model promotes a better information transfer between modalities.
The hash-grid features are concatenated to the 3D coordinates and the viewing direction and passed to a shallow multi-layer perceptron that has as many output channels as the total amount of channels of all the available modalities. A schema is shown in \cref{fig:schema}. 
For each training ray of a specific modality, only the channels associated to that modality are supervised. This architecture permits that, at evaluation time, the model can estimate any channel of any training modality from whatever viewpoint, thus producing perfectly aligned multimodal renderings.\\

\vspace{-0.5cm}
\subsection{Raw Data Support}
\label{raw_frame_support}
An additional feature of \fname\ is the possibility to use mosaicked and lens distorted frames for the training. In the scenario of mosaicked images, also the multi-channel modalities can rely only on a single channel per pixel as supervision signal. The model is able to load this data and the respective information about the mosaick pattern, in order to know at any time the channel associated to each rendered pixel. However, \fname\ is agnostic about the spectral content of every pixel and of every modality: it is possible to use radiance frames acquired by sensors not included in our default set and, once defined the mosaick pattern, the framework can handle them out of the box. The only exception is represented by \gls{pol} frames, for which a custom procedure is needed for the reasons previously discussed.
Dealing with mosaick frames allows the network to learn the demosaicking interpolation process, a feature specially useful for non-standard imaging modalities (\eg, \gls{ms} and \gls{pol}), 
for which the demosaicking algorithms are not deeply investigated as for RGB.
A further analysis on this capability is presented in \cref{mosaicked_vs_demosaicked}. In addition, the possibility to cast rays from frames affected by camera distortion is essential in the case of mosaicked images, as it is not trivial to perform frame undistortion while keeping the mosaick pattern unaltered. To handle distorted frames, we employ the procedure proposed in~\cite{nerfstudio}.\\

\vspace{-0.4cm}
\subsection{Modularity}
\label{modularity}
Modularity is a key feature of \fname : it makes our framework suitable and convenient to support different applications and research works. Modularity refers to the organization of the different steps of the rendering pipeline as independent modules. Each module covers a specific functionality, has a standardized input and a standardized output. This is enforced for every step of the pipeline, in a hierarchical fashion. 
For instance, replacing the radiance module with a different radiance formulation module would not affect the functioning of the pipeline, as long as the input and output are consistent. Recursively, the radiance module itself contains modules to determine which implicit representation and what encoding are employed.
\fname\ modularity opens several possibilities by permitting to arbitrarily switch between: different pixel sampling and ray sampling strategies; different implicit representations for the radiance and density field estimation; different camera optimization options; different optimizer configurations controlling which parameters are optimized by distinct optimizers; different losses and loss scheduling for each modality; modality-specific rendering functions. We believe that this will provide a flexible tool for the community researching on multimodal neural rendering.

\section{Multimodal Rendering Experiments}
\label{sec:experiments}

\begin{figure}
    \centering
    \includegraphics[width=\linewidth]{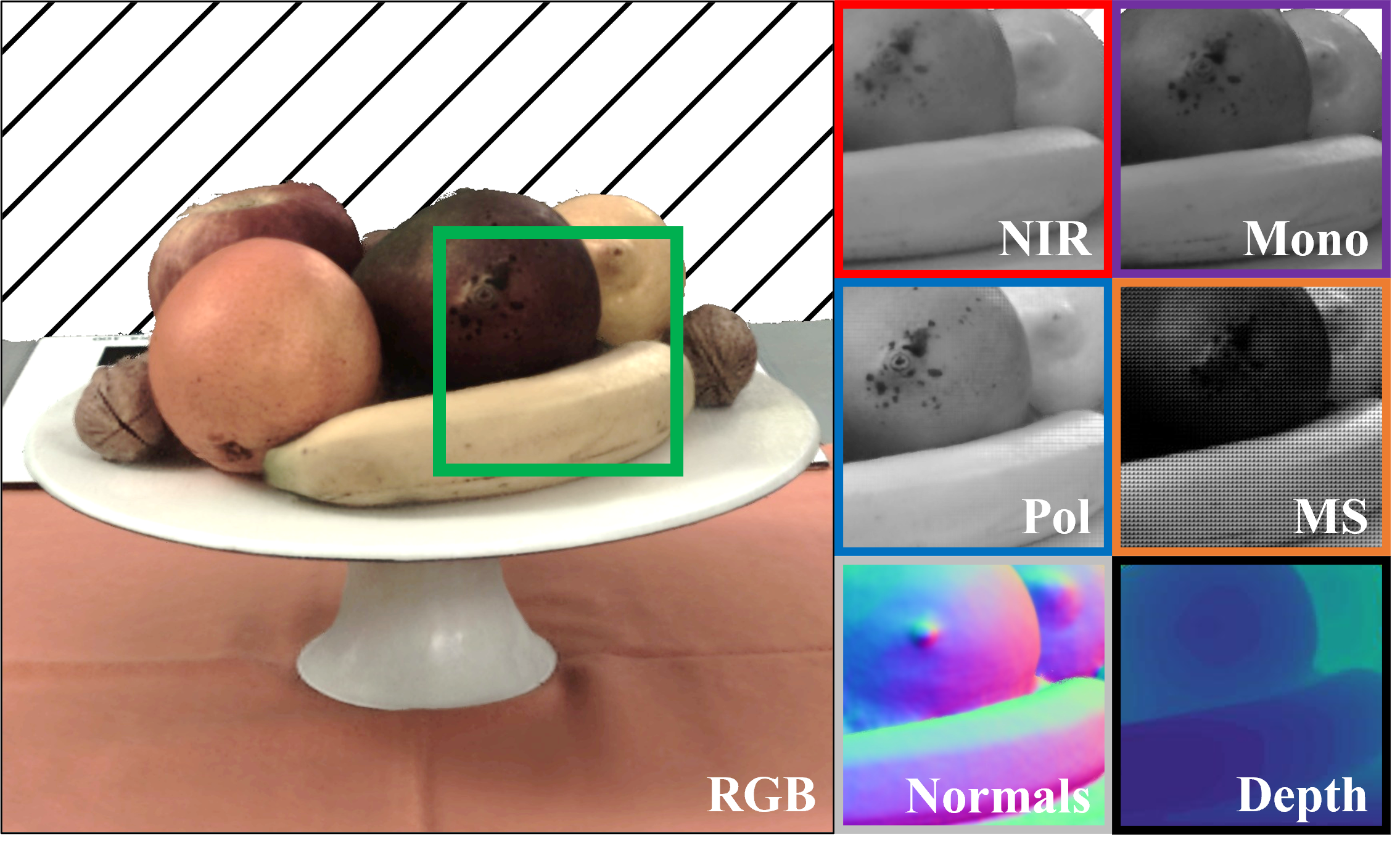}
    \caption{Qualitative results of 5-modality training on Fruits scene. Rendering of \green{RGB}, \red{NIR}, \purple{Mono}, \blue{Pol}, \orange{MS}, \grey{Normals} and \textbf{Depth}. All modalities except RGB are rendered mosaicked for visualization purposes. Crops of not aligned renderings.}
    \label{fig:qualitative_cropped}
    \vspace{-0.5cm}
\end{figure}

\begin{table*}
    \begin{minipage}{0.2\linewidth}
    \centering
    \renewcommand{\arraystretch}{1.3}
    \begin{tabular}{l@{\extracolsep{16pt}}c@{\extracolsep{16pt}}c}
        \toprule
        Mod. & PSNR$\uparrow$ & SSIM$\uparrow$ \\
        \midrule
        \green{RGB} & 27.76 & - \\
        \purple{Mono} & 28.97 & 0.92 \\
        \red{NIR} & 29.75 & 0.90 \\
        \blue{Pol} & 27.58 & - \\
        \orange{MS} & 27.31 & - \\
        \bottomrule
    \end{tabular}
    \caption{Single-modality training PSNR and SSIM on mosaicked images.}
    \label{tab:single_modality_training}
    \end{minipage}
    \hfill
    \begin{minipage}{0.72\linewidth}
        \centering
        \renewcommand{\arraystretch}{0.7}
        \begin{tabular}{@{}lccccc@{}}
            \toprule
            \multirow{2}{*}{Training Mod.} & \multirow{2}{*}{Test Mod.} & ALL views & \underline{HALF} views & ALL views & \underline{HALF} views \\
            & & ALL rays & ALL rays & \underline{HALF} rays & \underline{HALF} rays\\
            \midrule
            \multirow{2}{*}{\green{RGB} - \purple{Mono}} & \green{RGB} & 31.09 \g{(+3.33)} & 29.59 \g{(+1.83)} & 31.01 \g{(+3.25)} & 29.69 \g{(+1.93)} \\
            \arrayrulecolor{black!30}\cmidrule{2-6}
            & \purple{Mono} & 31.24 \g{(+2.27)} & 29.47 \g{(+0.50)} & 31.35 \g{(+2.38)} & 29.74 \g{(+0.77)} \\
            \arrayrulecolor{black}\midrule
            \multirow{2}{*}{\green{RGB} - \red{NIR}} & \green{RGB} & 31.00 \g{(+3.24)} & 29.37 \g{(+1.61)} & 30.93 \g{(+3.17)} & 29.65 \g{(+1.89)} \\
            \arrayrulecolor{black!30}\cmidrule{2-6}
            & \red{NIR} & 32.42 \g{(+2.67)} & 30.42 \g{(+0.67)} & 32.59 \g{(+2.84)} & 30.89 \g{(+1.14)} \\
            \arrayrulecolor{black}\midrule
            \multirow{2}{*}{\green{RGB} - \blue{Pol}} & \green{RGB} & 30.20 \g{(+2.44)} & 28.59 \g{(+0.83)} & 29.97 \g{(+2.21)} & 28.87 \g{(+1.11)} \\
            \arrayrulecolor{black!30}\cmidrule{2-6}
            & \blue{Pol} & 28.83 \g{(+1.25)} & 27.76 \g{(+0.18)} & 28.71 \g{(+1.13)} & 27.87 \g{(+0.29)} \\
            \arrayrulecolor{black}\midrule
            \multirow{2}{*}{\green{RGB} - \orange{MS}} & \green{RGB} & 30.99 \g{(+3.23)} & 28.98 \g{(+1.22)} & 30.89 \g{(+3.13)} & 29.38 \g{(+1.62)} \\
            \arrayrulecolor{black!30}\cmidrule{2-6}
            & \orange{MS} & 29.64 \g{(+2.33)} & 27.55 \g{(+0.24)} & 29.62 \g{(+2.31)} & 28.08 \g{(+0.77)} \\
            \arrayrulecolor{black}\bottomrule
        \end{tabular}
        \caption{Comparative tests in terms of PSNR (dB) of two-modality trainings with different combinations of views per modality and cast rays per iteration. Between brackets, the $\Delta$ in PSNR with respect to the single-modality case is shown. Tests on mosaicked images.}
        \vspace{-0.2cm}
        \label{tab:two_modality_training_ablation}
    \end{minipage}
\end{table*}

In this section, we present a set of experiments to show the impact of using multiple modalities to train \gls{nerf}-based models. 
We evaluate the novel view rendering quality across multiple modalities. 
Our method also supports geometry estimation, but currently the lack of a ground-truth geometry does not permit its numerical evaluation. \\
\indent We believe that different modalities encode complementary information, thus combining them should be beneficial in terms of reconstruction quality. However, it is not clear which combination is the most convenient or effective. Moreover, always considering the same number of images for each modality may be too restrictive. In a real-world scenario, due to the cost or the limited mobility of certain cameras, acquiring an even amount of frames per modality may not be possible, thus the rendering of novel viewpoints for scarcely available modalities is a key requirement.\\
\indent For these reasons, we present several experiments aiming at: (1) understanding to which extent the \gls{nerf} model can exploit different combinations of modalities, (2) investigating whether it can leverage commodity modalities (\eg, RGB) to accurately estimate data for less available modalities (\eg, \gls{ms} or \gls{pol}), and (3) exploring whether the use of unprocessed mosaicked data offers any advantage with respect to preprocessed data. Finally, we present some comparisons between \fname\ and X-NeRF~\cite{poggi2022xnerf}, the current state-of-the-art multimodal \gls{nerf} method.\\
\indent We evaluate the rendering quality in terms of PSNR, SSIM, and LPIPS. However, most of the tests are performed on mosaicked (RAW) images. The mosaick pattern interferes with the luminance, structure, and contrast evaluation of SSIM. Similarly, LPIPS employs neural networks trained on demosaicked RGB data, thus it is unclear what it measures in the presence of mosaicked images and on non-RGB modalities. Therefore, we include SSIM and LPIPS metrics only for demosaicked tests, or when applicable (e.g., LPIPS for RGB, SSIM for RAW single-channel modalities).

\subsection{Implementations Details}
\label{implementation_details}
We used \fname\ for all the tests. The model is trained on a single Nvidia A6000 GPU for 100k iterations. At each iteration, 2048 rays per modality are cast, and 64 points per ray are sampled. 
An additional vanilla \gls{nerf} model to estimate the background is employed for convergence purposes~\cite{wang2021neus,martin2021nerf}. However, the background is never considered in the evaluation metrics, as we are only interested in the rendering of the volume inside the \gls{roi}. For this reason, we precomputed a set of masks for every modality and every scene. For each scene, we trained an analogous but independent model using all the available views, without discerning between train and test sets. Then, for each pixel in a view, we render and mark it as background if all its radiance contributions come from the background model, as foreground otherwise. Note that we used this independent model exclusively for the mask creation. These same masks are used to compute the metrics in all the tests here presented and only for evaluation purposes. We used mosaicked distorted frames in all these tests, and all the metrics are averaged on the test views of all the 32 scenes of \dname\ (see the Supplementary Material for details of the train and test splits). Finally, for all the modules, we employ an AdamW optimizer and a specific learning rate schedule: we adopt a rising trend until 10\% of the training and then it is consecutively reduced when the training reaches the 50\%, 75\% and 90\%.

\subsection{Balanced Combinations of Modalities}
\label{balanced_combinations}

Here we present the results obtained by training the NeRF model with different combinations of modalities. As anticipated, we believe that complementary modalities can carry additional information in the form of constraints for the radiance estimation, thus leading to improved rendering results. We propose keeping RGB data always available: this is what usually happens in a real scenario since nowadays RGB cameras are cheap and widely diffused.
Therefore, we couple RGB frames with an additional imaging modality, train a multimodal \gls{nerf} model, and then evaluate the rendering of novel views. The evaluation is performed by comparing the metrics of the test views generated by the multimodal \gls{nerf} against the same test views generated by the \gls{nerf} models trained independently on the single modalities. 
In \cref{tab:single_modality_training} and in the first column of \cref{tab:two_modality_training_ablation} it is possible to see the results achieved with the single-modality and the two-modality trainings, respectively.
It is clear that the additional modality is effective in enabling the model to always produce better results. The RGB rendering PSNR gains $\sim$2.5-3~dB consistently, while the second modality around $\sim$2.4~dB with the exception of the polarization data that have a smaller gain.
This is explained considering that RGB shares with \gls{pol} less information than with \gls{nir} or \gls{ms}. \\
\indent The reader could argue that, with an additional modality, the framework has access to double the number of training views and, therefore, benefits from a number of ray casts per iteration that is two times bigger. To investigate whether the improvement in the results is to be attributed to this or actually to the cross-modality information transfer, we performed a set of additional comparative experiments.
(1) The training of a two-modality model with half the number of available views per modality, by selecting only one modality per \textit{macroframe}. (2) The training of a two-modality model keeping all the training views of all modalities while halving the amount of rays cast per iteration. (3) Training after halving both the number of available frames per modality and the number of rays per iteration. The results are presented in \cref{tab:two_modality_training_ablation}. 
It is possible to observe that in every configuration, even when training the model with just one view every two and casting half of the rays per iteration, the metrics achieved are higher than the single modality case (\cref{tab:single_modality_training}). These results show that the additional modality always helps in producing higher quality novel view renderings for both the modalities. \\
\indent We show in \cref{tab:five_modality_training} two additional tests, involving three- and five-modality trainings. In both cases, the additional modalities allow improving the metrics gain with respect to the single-modality case. Considering all these tests, we can conclude that including frames of other modalities in the training provides complementary information that helps the \gls{nerf} to better estimate the multimodal radiance fields. Some qualitative results are shown in \cref{fig:qualitative_cropped}.

\begin{table}
    \centering
    \renewcommand{\arraystretch}{0.7}\begin{tabular}{cccc}
        \toprule
        Train Mod. & Test Mod. & PSNR$\uparrow$ & SSIM$\uparrow$ \\
        \midrule
        \multirow{3}{*}{\begin{minipage}{2.25cm}\vspace{0.3cm}\centering\green{RGB} \\ \blue{Pol} - \orange{MS}\end{minipage}} & \green{RGB} & 31.27 \g{(+3.51)} & - \\
        \arrayrulecolor{black!30}\cmidrule{2-4}
        & \blue{Pol} & 29.51 \g{(+1.92)} & - \\
        \arrayrulecolor{black!30}\cmidrule{2-4}
        & \orange{MS} & 30.04 \g{(+2.73)} & - \\
        \arrayrulecolor{black}
        \midrule
        \multirow{5}{*}{\begin{minipage}{2.25cm}\vspace{0.7cm}\centering\green{RGB} - \purple{Mono}\ \red{NIR} - \blue{Pol} - \orange{MS}\end{minipage}} & \green{RGB} & 32.00 \g{(+4.23)} & - \\
        \arrayrulecolor{black!30}\cmidrule{2-4}
        & \purple{Mono} & 32.31 \g{(+3.34)} & 0.94 \\
        \arrayrulecolor{black!30}\cmidrule{2-4}
        & \red{NIR} & 33.63 \g{(+3.88)} & 0.93 \\
        \arrayrulecolor{black!30}\cmidrule{2-4}
        & \blue{Pol} & 29.80 \g{(+2.22)} & - \\
        \arrayrulecolor{black!30}\cmidrule{2-4}
        & \orange{MS} & 30.69 \g{(+3.39)} & - \\
        \arrayrulecolor{black}\bottomrule
    \end{tabular}
    \caption{Training results with $3$ and $5$ modalities on mosaicked images. Between brackets, the $\Delta$ in PSNR with respect to the single-modality case.}
    \label{tab:five_modality_training}
  \vspace{-5pt}
\end{table}

\subsection{Unbalanced Combinations of Modalities}
\label{unbalanced_combinations}

\begin{figure}
  \centering
  \begin{subfigure}{0.9\linewidth}
    \centering
    \includegraphics[width=\linewidth]{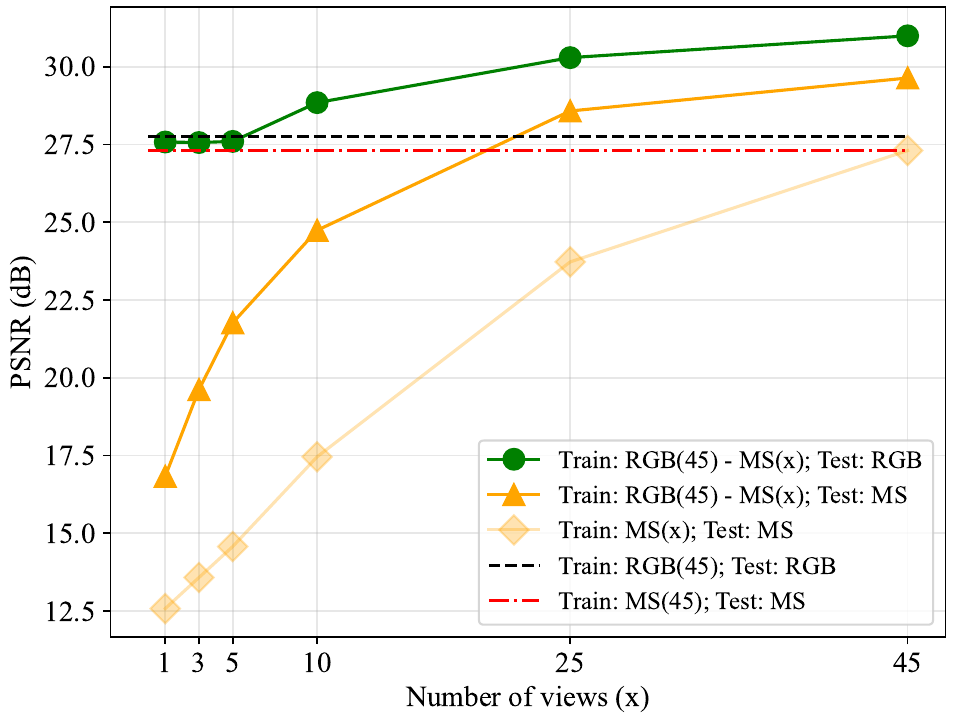}
    \caption{Test with an unbalanced combination of RGB and MS frames.}
    \label{fig:unbalanced_views_ms}
  \end{subfigure}
  \\
  \begin{subfigure}{0.9\linewidth}
    \centering
    \includegraphics[width=\linewidth]{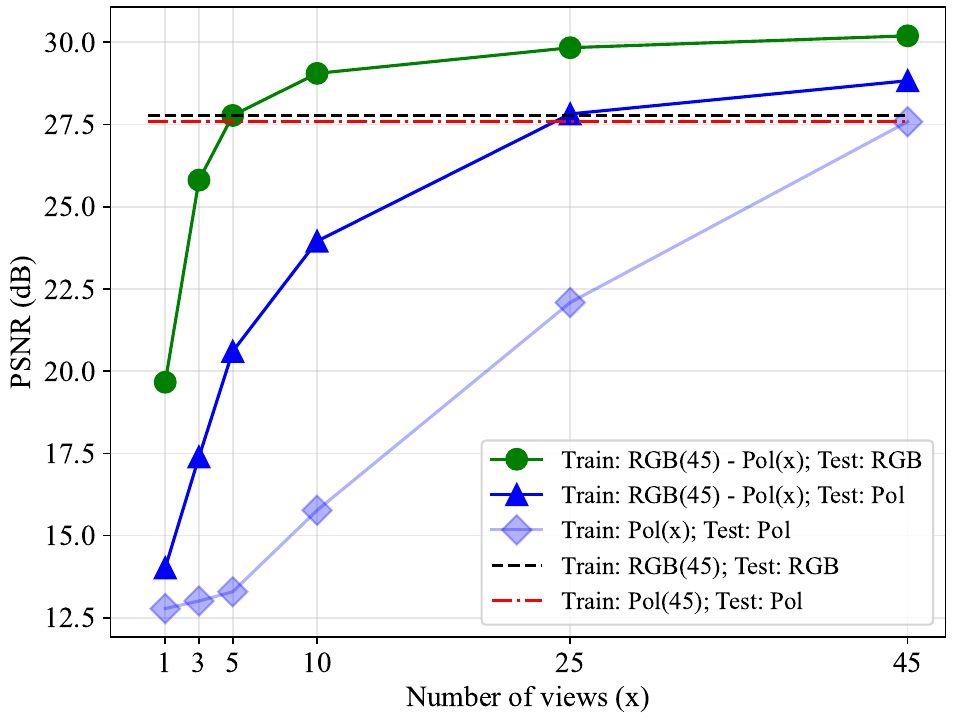}
    \caption{Test with an unbalanced combination of RGB and Pol frames.}
    \label{fig:unbalanced_views_pol}
  \end{subfigure}
  \caption{Plots showing the PSNR achieved by training a NeRF with unbalanced combinations of training frames per modality.}
  \label{fig:unbalanced_views}
  \vspace{-0.5cm}
\end{figure}

The next experiments involve the scenario with an unbalanced number of available frames per modality. 
We perform the evaluation with 2 modalities, i.e., the complete set of RGB frames and a subset of frames from a second modality. We pair RGB with \gls{ms} and \gls{pol} data and run separate tests using 1, 3, 5, 10, 25 and 45 frames respectively, in addition to the 45 RGB frames. In \cref{fig:unbalanced_views} it is possible to observe the obtained PSNR as a function of the number of available additional modality viewpoints. It is interesting to note that when the model is trained with the support of RGB frames, then the additional modality renderings are always more accurate than the ones obtained by training the model with the other modality alone, regardless of the number of training views used. Moreover, the results show that it is sufficient to have more than 5 frames of a second modality to measure an improvement also in the RGB rendering quality. Analogously, in a two-modality training with all the RGB frames and at least half of the additional modality frames, the rendering quality of the second modality outperforms the quality of the renderings produced by the single-modality model trained on all the available frames (dashed lines). These results show that the model can efficiently transfer information from one modality to another. Indeed, in a scenario where the number of available second-modality frames is limited, the multimodal model manages to exploit common information to produce more accurate second-modality renderings than when using only that single modality. \\
\indent Finally, it is possible to observe that, in the RGB-Pol case, using less than 5 Pol frames has the effect of reducing the RGB rendering quality with respect to the single-modality case (black dashed line). As said in \cref{raw_frame_support}, the additional constraints introduced by the polarization estimation tend to limit the other modality optimizations. In the case of few available Pol views, the model overfits on those frames, thus limiting the density estimation accuracy. Since the density plays a crucial role in the volume rendering, its poor estimation affects all the modalities involved, leading to poor rendering quality also for RGB. This is exacerbated by the fact that the additional modality is present since the first iteration: we expect that introducing a pretraining on RGB frames only to consolidate the density estimation, and then fine tuning on all the modalities, may reduce the problem. This possibility is left for future work. 

\subsection{Mosaicked VS Demosaicked}
\label{mosaicked_vs_demosaicked}

In this section, we compare the results achieved by using either demosaicked or raw frames to train the model. The demosaicking is a standard processing step applied immediately after a frame is captured. It is convenient, because it makes the information of every channel available for every pixel in the image. However, it relies on interpolation, which may introduce small artifacts that do not reflect the actual scene content. Moreover, while demosaicking algorithms are well established for the standard RGB Bayer pattern, for modalities with unconventional mosaick patterns the task is less trivial. They would require an ad-hoc demosaicking algorithm but this leads to a time consuming development, it is often camera-dependent, and it also requires a stronger interpolation than standard RGB frames.\\
\indent Since in \gls{nerf}s each pixel is treated independently, we investigate whether the direct use of mosaicked frames during training leads to obtain comparable or even better results than when using demosaicked frames. We compare a NeRF training with raw frames with the trainings with demosaicked frames.
Furthermore, since we noticed that in the demosaicked case each ray carries more information (i.e., multiple channels instead of one), we considered also two more fair comparisons with the raw version. In one version we employed a reduced ray batch size per iteration; in the last instead we supervised only one random channel per pixel per iteration. This ensures that the two modified versions match the amount of supervision received by the raw frame NeRF, thus offering more comparable results. In \cref{tab:mosaiked_vs_raw} we report the results obtained in terms of PSNR. It is possible to observe that using raw frames is better for some modalities, while employing demosaicked images produces slightly better results for others. However, overall the performances are very close; indeed, the model trained with raw frames can achieve comparable quality with respect to a \gls{nerf} trained, as usual, with demosaicked frames. 
As explained before, using raw mosaicked frames has the advantage of avoiding the demosaicking step, which may be problematic for unconventional modalities. 
Therefore, the raw frame support is a very valuable feature of \fname.

\subsection{Comparisons with X-NeRF}
\label{comparisons}

\begin{table}
    \centering
    \renewcommand{\arraystretch}{0.7}\begin{tabular}{cc|ccc|c}
        \toprule
        \multirow{2}{*}{Train} &  \multirow{2}{*}{Test} & \multicolumn{3}{c|}{Demosaicked} & \\
        \multirow{2}{*}{Mod.} & \multirow{2}{*}{Mod.} & \multirow{2}{*}{standard} & smaller & single & Raw \\
        & & & batch & channel \\
        \midrule
        \multirow{5}{*}{\rotatebox{90}{\begin{minipage}{2.25cm}\centering\green{RGB} - \purple{Mono}\ \red{NIR} - \blue{Pol} - \orange{MS}\end{minipage}}}& \green{RGB} & \textbf{32.34} & 32.10 & 32.28 & 32.00 \\
        \arrayrulecolor{black!30}\cmidrule{2-6}
         & \purple{Mono} & \textbf{32.43} & 32.24 & 32.21 & 32.31 \\
        \arrayrulecolor{black!30}\cmidrule{2-6}
         & \red{NIR} & \textbf{33.81} & 33.66 & 33.63 & 33.63 \\
        \arrayrulecolor{black!30}\cmidrule{2-6}
        & \blue{Pol} & 29.79 & 29.62 & 29.65 & \textbf{29.80} \\
        \arrayrulecolor{black!30}\cmidrule{2-6}
        & \orange{MS} & 30.05 & 29.90 & 29.87 & \textbf{30.69} \\
        \arrayrulecolor{black}\bottomrule
    \end{tabular}
    \caption{Comparison between PSNR (dB) obtained by training with demosaicked or raw frames.}
    \label{tab:mosaiked_vs_raw}
    \vspace{-0.5cm}
\end{table}

\begin{table}
    \centering
    \renewcommand{\arraystretch}{1}
    \begin{tabular}{@{}c@{\extracolsep{0pt}}|c@{\extracolsep{4pt}}c@{\extracolsep{8pt}}c@{\extracolsep{8pt}}c@{}c@{}c@{}}
        \toprule
        & \multirow{2}{*}{Method} & Train & Test & \multirow{2}{*}{PSNR$\uparrow$} & \multirow{2}{*}{SSIM$\uparrow$} & \multirow{2}{*}{LPIPS$\downarrow$} \\
        & & Mod. & Mod. & & & \\
         \midrule
        \multirow{6}{0.5cm}{\centering\rotatebox[origin=c]{90}{X-NeRF Dataset}} & \multirow{3}{*}{X-NeRF} & \multirow{3}{1cm}{\centering\green{RGB}\\\red{NIR}\\\orange{MS}} &\green{RGB} & 31.77 & \textbf{0.89} & NA \\
        \arrayrulecolor{black!30}\cline{4-7}
        & & & \red{NIR} & 31.60 & \textbf{0.92} & - \\
        \arrayrulecolor{black}
        \arrayrulecolor{black!30}\cline{4-7}
        & & & \orange{MS} & 33.22 & 0.91 & - \\
        \arrayrulecolor{black}
        \cmidrule{2-7}
        & \multirow{3}{*}{MMS-FW} & \multirow{3}{1cm}{\centering\green{RGB}\\\red{NIR}\\\orange{MS}} &\green{RGB} & \textbf{33.75} & 0.88 & 0.35 \\
        \arrayrulecolor{black!30}\cline{4-7}
        & & & \red{NIR} & \textbf{33.33} & 0.90 & - \\
        \arrayrulecolor{black}
        \arrayrulecolor{black!30}\cline{4-7}
        & & & \orange{MS} & \textbf{36.11} & \textbf{0.94} & - \\
        \arrayrulecolor{black}
        \midrule
        \multirow{6}{0.5cm}{\centering\rotatebox[origin=c]{90}{MMS-DATA}} & \multirow{3}{*}{X-NeRF} & \multirow{3}{1cm}{\centering\green{RGB}\\\red{NIR}\\\orange{MS}} &\green{RGB} & 25.73 & 0.84 & 0.35 \\
        \arrayrulecolor{black!30}\cline{4-7}
        & & & \red{NIR} & 26.97 & 0.84 & - \\
        \arrayrulecolor{black}
        \arrayrulecolor{black!30}\cline{4-7}
        & & & \orange{MS} & 23.63 & 0.73 & - \\
        \arrayrulecolor{black}
        \cmidrule{2-7}
        & \multirow{3}{*}{MMS-FW} & \multirow{3}{1cm}{\centering\green{RGB}\\\red{NIR}\\\orange{MS}} &\green{RGB} & \textbf{32.42} & \textbf{0.94} & \textbf{0.06} \\
        \arrayrulecolor{black!30}\cline{4-7}
        & & & \red{NIR} & \textbf{32.97} & \textbf{0.93} & - \\
        \arrayrulecolor{black}
        \arrayrulecolor{black!30}\cline{4-7}
        & & & \orange{MS} & \textbf{29.89} & \textbf{0.92} & - \\
        \arrayrulecolor{black}
        \bottomrule
    \end{tabular}
    \caption{Comparison between X-NeRF and \fname\ (Ours) on X-NeRF data and \dname. Demosaicked undistorted frames.}
    \label{tab:x-nerf_vs_ours}
    \vspace{-0.5cm}
\end{table}

Finally, we compare \fname\ with X-NeRF \cite{poggi2022xnerf}, the current state-of-the-art NeRF-based multimodal method. First, we trained our model on the X-NeRF dataset and compared the results with the ones in \cite{poggi2022xnerf}; then, we trained the two models also on \dname. Since X-NeRF trains only on demosaicked frames, we also trained \fname\ on demosaicked images. 
The results are presented in \cref{tab:x-nerf_vs_ours}: our method consistently outperforms X-NeRF in all metrics. \fname\ achieves a $\sim$2-3~dB and $\sim$6~dB higher PSNR on X-NeRF dataset and \dname, respectively. The unmatched difference in the results of the two methods when testing on X-NeRF dataset or on \dname\ can be explained by considering that the first dataset contains only forward-facing scenes, while ours contains object-centric scenes. X-NeRF proposes the \gls{nxdc}, specifically designed for forward-facing scenes. When running on other types of scenes, the \gls{nxdc} must be disabled, which leads to poorer quality results. Instead, \fname\ is more flexible in terms of scene types: it can manage either forward-facing or object-centric scenes while achieving better quality.
\section{Conclusions}
\label{sec:conclusions}

In this paper we introduce a novel multimodal dataset and \gls{nerf}-based framework. The large number of scenes, views, and different modalities included in the dataset allows for the experimentation of novel multimodal learning schemes previously not possible due to the lack of data. Furthermore, we provide a modular framework that extends the well-known \gls{nerf} pipeline to the multimodal data scenario, easing the switch between different models as well as the extension to more challenging imaging modalities. Experiments proved the capability of the proposed approach to transfer information between different modalities. \\
\indent Future work will tackle the inclusion in the dataset of 3D information data, such as from Time-of-Flight and stereo vision systems, in order to extend and evaluate also the geometry reconstruction capabilities of the proposed approach.

\vspace{-5pt}
\paragraph{Acknowledgment}
\label{sec:acknowledgment}
This collaborative work was funded by Sony Europe Limited. We warmly thank Piergiorgio Sartor for his brilliant supervision, Francesco Michielin for his help with sensor calibration, and Oliver Erdler, Yalcin Incesu, and Alexander Gatto for their support.

{
    \small
    \bibliographystyle{ieeenat_fullname}

}

\renewcommand{\maketitlesupplementary}
   {
   \newpage
       \onecolumn
       \begin{minipage}{\linewidth}
        \centering
        \hspace{-28pt}
        \Large
        \textbf{\thetitle}\\
        \vspace{0.5em}Supplementary Material \\
        \vspace{1.0em}
       \end{minipage}
   }
\clearpage
\maketitlesupplementary
\renewcommand\thesection{\Alph{section}}
\setcounter{section}{0}

\noindent In this document, we provide some additional details on \textit{MultimodalStudio}. Firstly, we show all the scenes included in \dname\ and we add a further description of them (\cref{sup_sec:dataset}). Then, we present more in-depth results of the geometrical calibration for the multimodal setup (\cref{sup_sec:calibration}). In \cref{sup_sec:renderings}, we show an extensive set of comparisons between the renderings produced by \fname\ and the ground truth for all the involved modalities, and we provide an exhaustive overview of the results achieved on every scene. Finally, we present some results and considerations on the few shot scenario (\cref{sup_sec:few_shot}).
\section{Dataset}
\label{sup_sec:dataset}

\begin{figure}[b]
    \centering
    \includegraphics[width=0.48\linewidth]{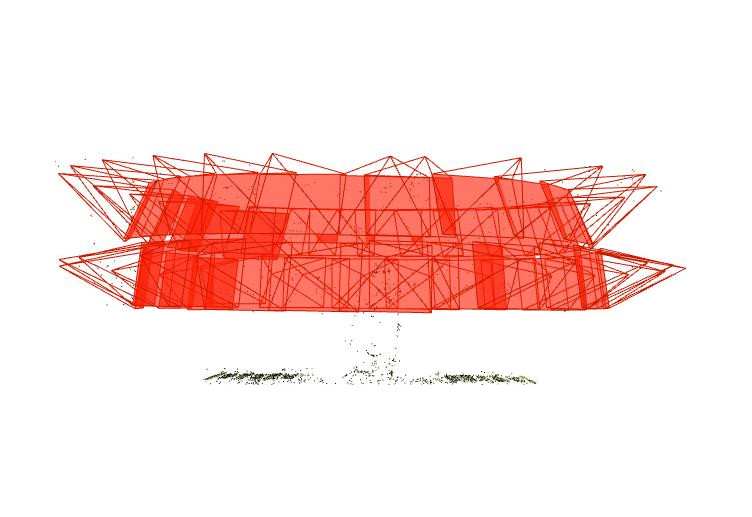}
    \includegraphics[width=0.48\linewidth]{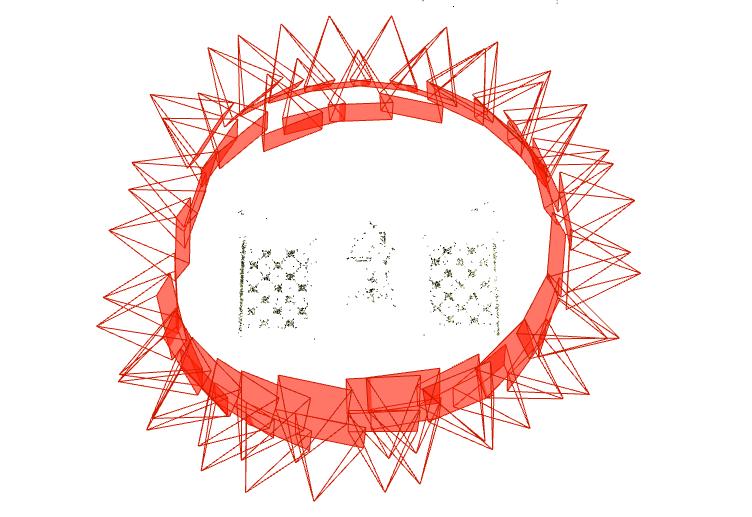}
    \caption{Overview of the RGB camera poses in a sample scene.}
    \label{sup_fig:acquisition_path}
\end{figure}

We acquired 32 scenes from 50 different viewpoints. The viewpoints are arranged as shown in \cref{sup_fig:acquisition_path}. The scene subject was placed on a table and enlightened by 3 halogen light bulbs equipped with diffusers. The sunlight could not penetrate the acquisition environment. Halogen lights were chosen because their emission spectrum is more suited to multispectral acquisition than that of LED light bulbs. LEDs are generally optimized to emit only in the visible band, but we are also interested in infrared emission given that the setup has a \gls{nir} camera. Instead, halogen light bulbs have an emission spectrum that also covers bands beyond the visible range. The three lights were placed as the three vertices of an equilateral triangle centered at the object, but higher than it. On the table, a couple of ChAruco boards were placed to ease the rig pose estimation when running the \gls{sfm} algorithm, and to estimate the arbitrary scale factor that the \gls{sfm} introduces. In \cref{sup_fig:all_scenes} a complete overview of the acquired scenes is shown. It is possible to see that they capture a wide variety of common objects made of materials of different nature. In \cref{sup_tab:scene_materials} it is possible to see which type of materials is present in each scene, for better classification. Hereunder, we report some additional details concerning specific scenes:
\begin{itemize}
    \item \textbf{Clock} and \textbf{Glass Clock:} the clock in these scenes~(\cref{fig:clock,fig:glassclock}) is almost entirely made of plastic.
    \item \textbf{Laurel Wreath:} the wreath is made of real laurel but the leaves have been dried out by time~(\cref{fig:laurelwreath}).
    \item \textbf{Orchid:} the orchid in this scene is not a real organic flower and it is entirely made of plastic~(\cref{fig:orchid}). It is also included in the scene Bouquet~(\cref{fig:bouquet}) along with real living plants.
    \item \textbf{Teddy Bear:} behind the teddy bear~(\cref{fig:teddybear}), a X-Rite colorchecker is placed for color calibration purposes.
    \item \textbf{Trophies:} the trophies upper part is metal, as well as the medals, while the middle and lower parts are made of plastic~(\cref{fig:trophies}).
    \item \textbf{Vases:} in the vases scene~(\cref{fig:vases}), the horse statue is plastic, the vase with lid is ceramic, and the remaining two vases are made of glass.
\end{itemize}
In all the other scenes, the material appearance reflects the actual material nature.\\
\indent We defined a fixed test and training split for the viewpoints (it is the same for all the scenes): we used viewpoints number 9, 19, 29, 39, and 49 as test views, while all the remaining ones are used as training views. 
Consider that the pictures have been acquired by moving the camera rig around the object in 2 circular patterns, a lower one (views 0-24) and an upper one (views 25-49) as shown in \cref{sup_fig:acquisition_path}.
However, recall that the camera rig is moved manually, as explained in Sec. 3.2
of the main paper, thus the actual viewpoint for each view can slightly change across different scenes.

\begin{figure*}
    \centering
    \begin{subfigure}{0.24\linewidth}
        \includegraphics[width=\linewidth]{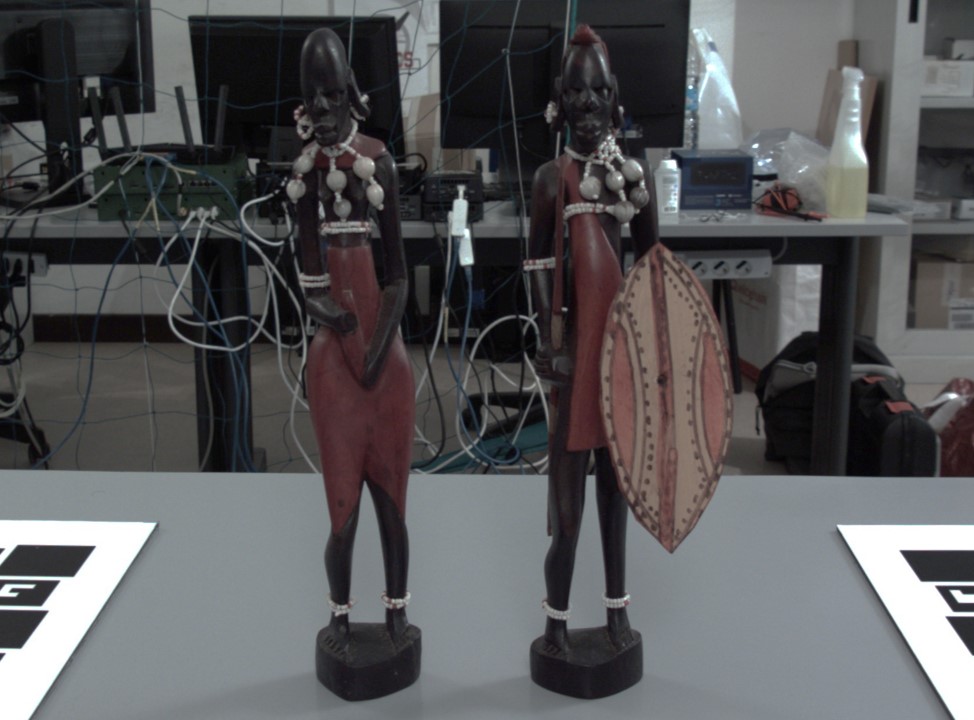}
        \caption{African Art}
        \label{fig:africanart}
    \end{subfigure}
    \hfill
    \begin{subfigure}{0.24\linewidth}
        \includegraphics[width=\linewidth]{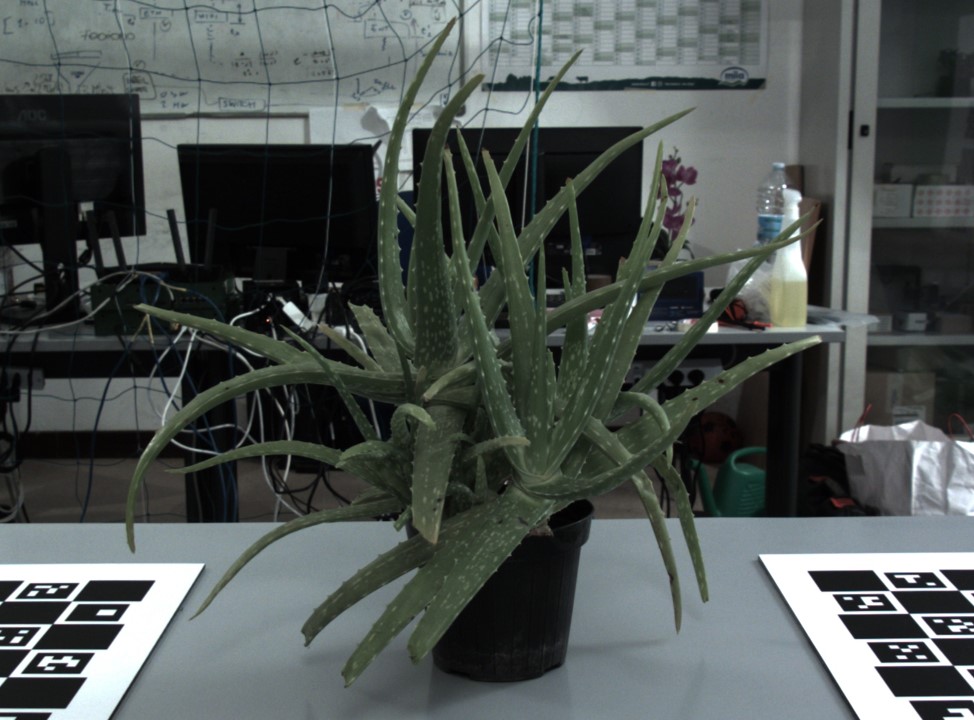}
        \caption{Aloe}
        \label{fig:aloe}
    \end{subfigure}
    \hfill
    \begin{subfigure}{0.24\linewidth}
        \includegraphics[width=\linewidth]{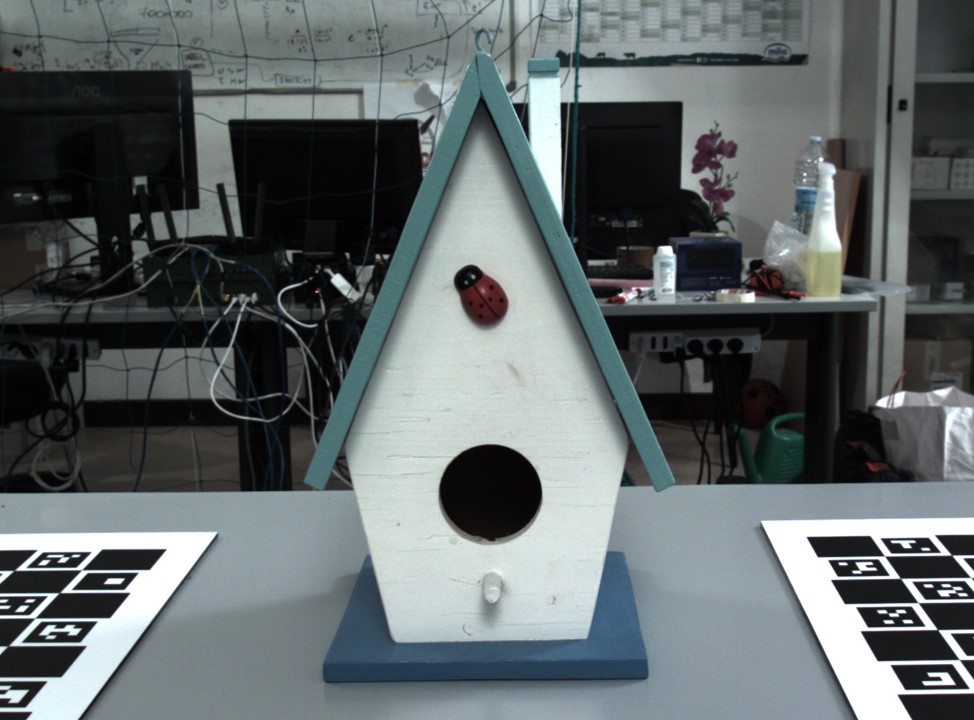}
        \caption{Bird House}
        \label{fig:birdhouse}
    \end{subfigure}
    \hfill
        \begin{subfigure}{0.24\linewidth}
        \includegraphics[width=\linewidth]{figures/book_cc.jpg}
        \caption{Book}
        \label{fig:book}
    \end{subfigure}
    \hfill
    \begin{subfigure}{0.24\linewidth}
        \includegraphics[width=\linewidth]{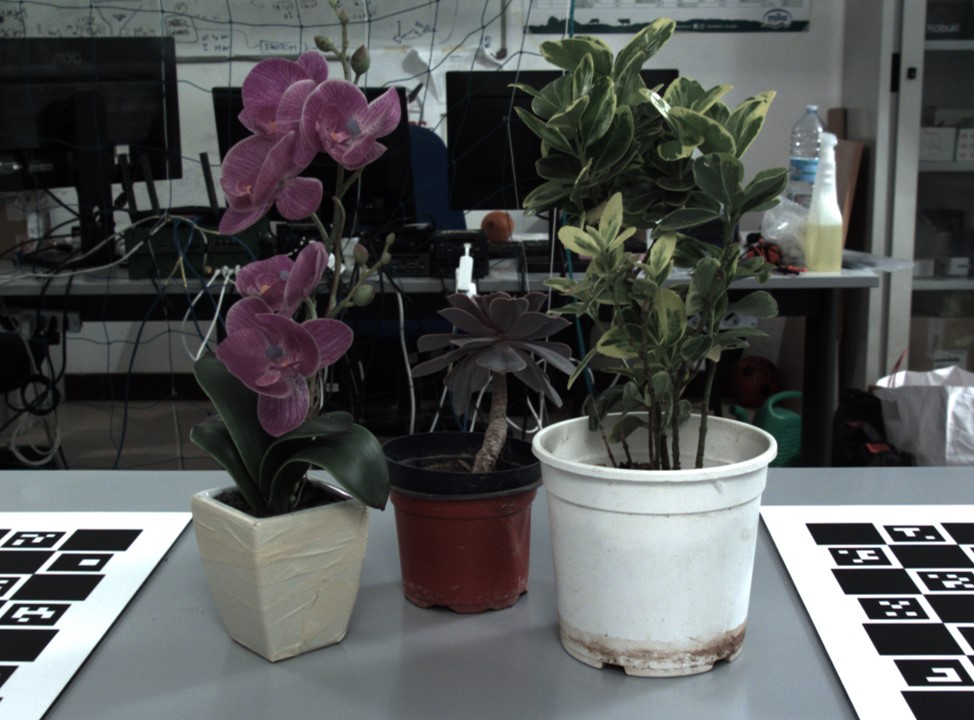}
        \caption{Bouquet}
        \label{fig:bouquet}
    \end{subfigure}
    \hfill
    \begin{subfigure}{0.24\linewidth}
        \includegraphics[width=\linewidth]{figures/chess_cc.jpg}
        \caption{Chess}
        \label{fig:chess}
    \end{subfigure}
    \hfill
    \begin{subfigure}{0.24\linewidth}
        \includegraphics[width=\linewidth]{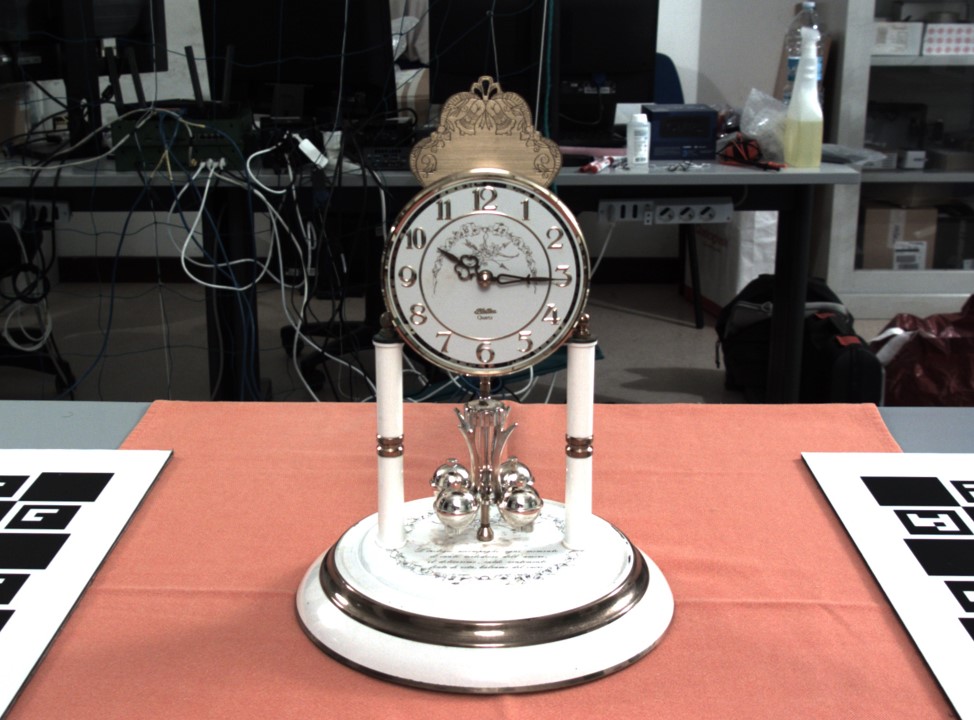}
        \caption{Clock}
        \label{fig:clock}
    \end{subfigure}
    \hfill
    \begin{subfigure}{0.24\linewidth}
        \includegraphics[width=\linewidth]{figures/easteregg2_cc.jpg}
        \caption{Easter Egg}
        \label{fig:easteregg}
    \end{subfigure}
    \hfill
    \begin{subfigure}{0.24\linewidth}
        \includegraphics[width=\linewidth]{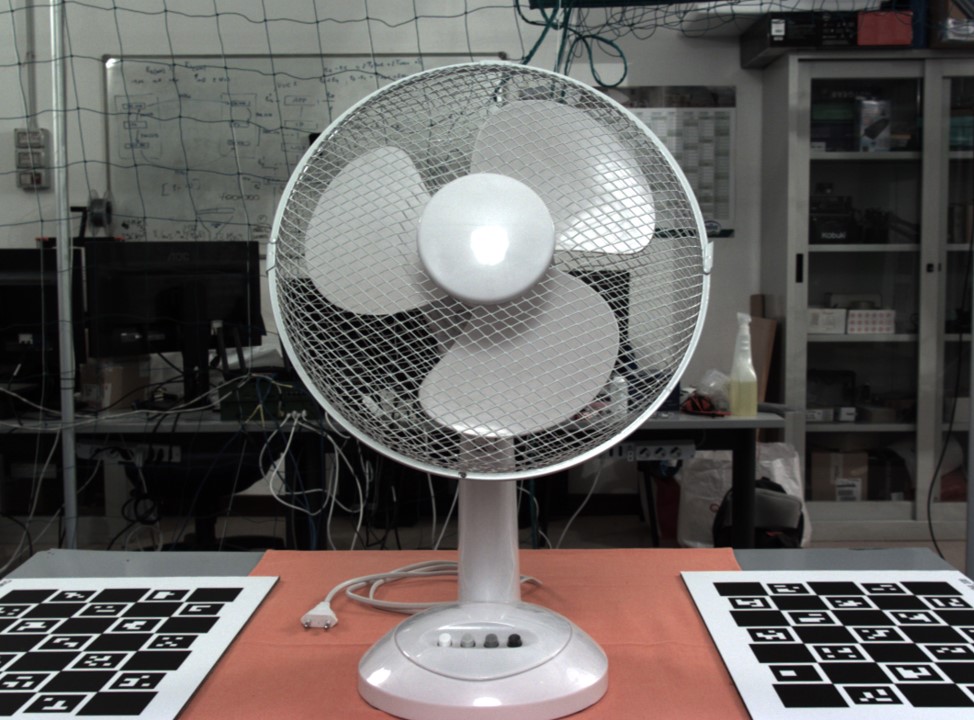}
        \caption{Fan}
        \label{fig:fan}
    \end{subfigure}
    \hfill
    \begin{subfigure}{0.24\linewidth}
        \includegraphics[width=\linewidth]{figures/forestgang1_cc.jpg}
        \caption{Forest Gang 1}
        \label{fig:forestgang1}
    \end{subfigure}
    \hfill
    \begin{subfigure}{0.24\linewidth}
        \includegraphics[width=\linewidth]{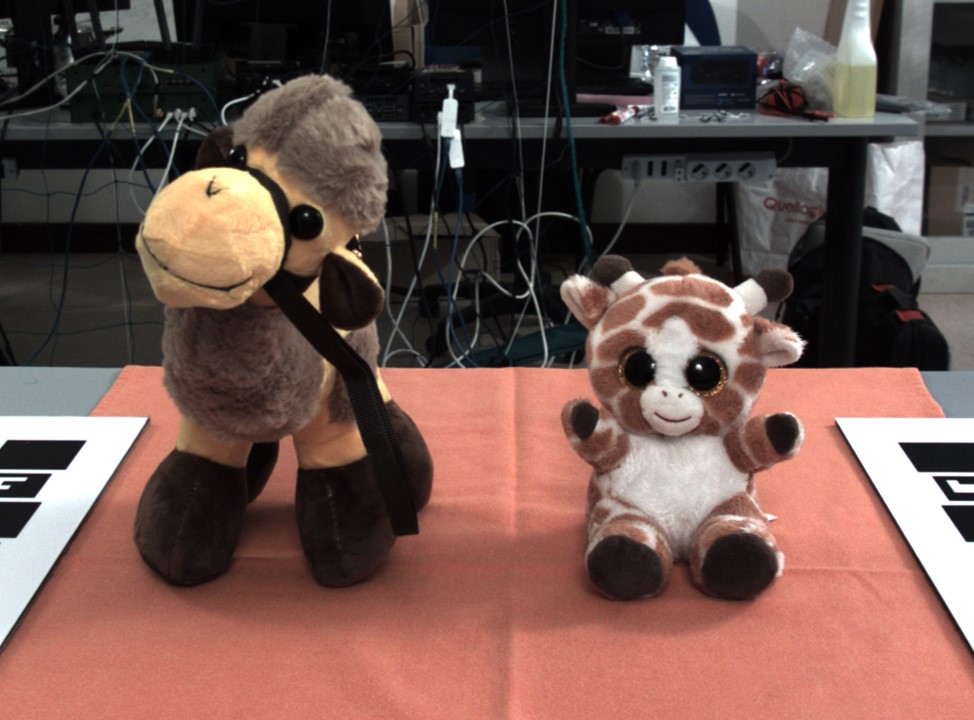}
        \caption{Forest Gang 2}
        \label{fig:forestgang2}
    \end{subfigure}
    \hfill
        \begin{subfigure}{0.24\linewidth}
        \includegraphics[width=\linewidth]{figures/fruits_cc.jpg}
        \caption{Fruits}
        \label{fig:fruits}
    \end{subfigure}
    \hfill
    \begin{subfigure}{0.24\linewidth}
        \includegraphics[width=\linewidth]{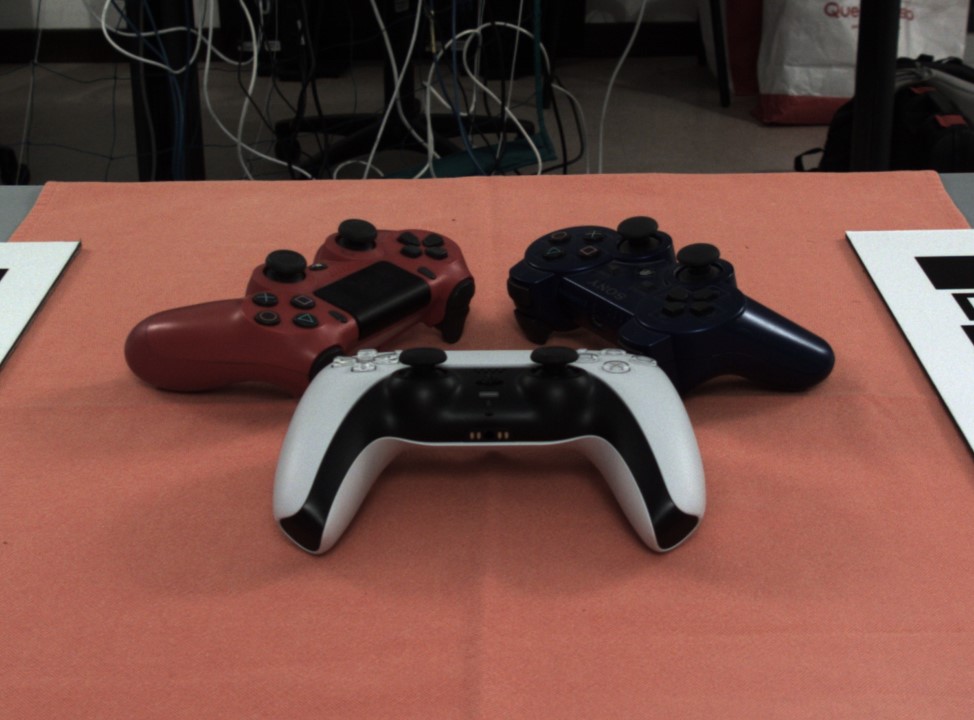}
        \caption{Gamepads}
        \label{fig:gamepads}
    \end{subfigure}
    \hfill
    \begin{subfigure}{0.24\linewidth}
        \includegraphics[width=\linewidth]{figures/glassclock_cc.jpg}
        \caption{Glass Clock}
        \label{fig:glassclock}
    \end{subfigure}
    \hfill
    \begin{subfigure}{0.24\linewidth}
        \includegraphics[width=\linewidth]{figures/globe_cc.jpg}
        \caption{Globe}
        \label{fig:globe}
    \end{subfigure}
    \hfill
        \begin{subfigure}{0.24\linewidth}
        \includegraphics[width=\linewidth]{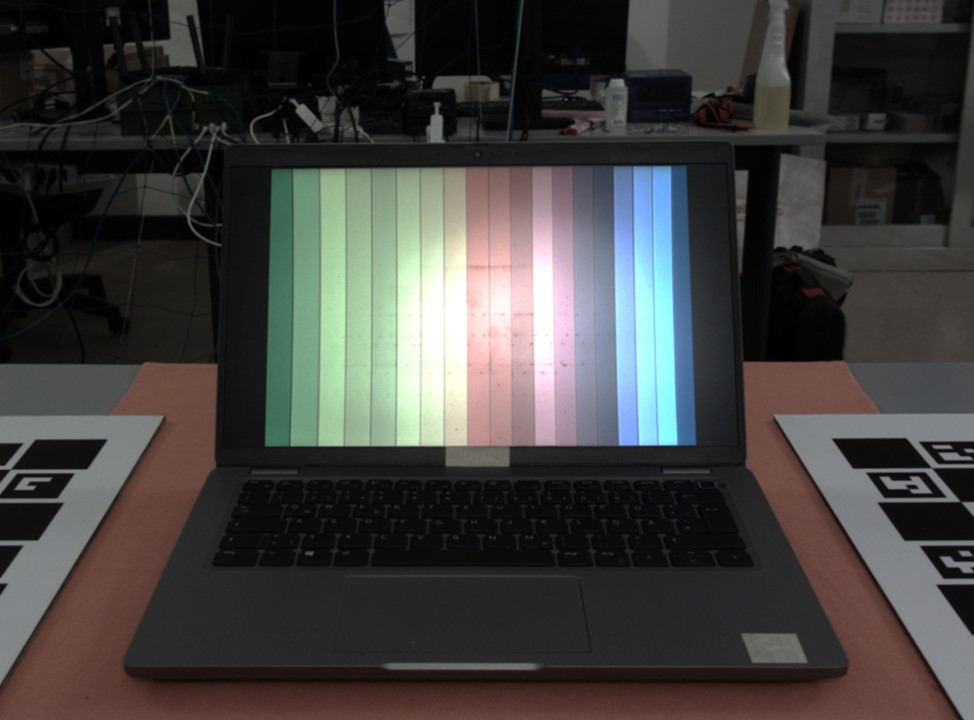}
        \caption{Laptop}
        \label{fig:laptop}
    \end{subfigure}
    \hfill
    \begin{subfigure}{0.24\linewidth}
        \includegraphics[width=\linewidth]{figures/laurelwreath_cc.jpg}
        \caption{Laurel Wreath}
        \label{fig:laurelwreath}
    \end{subfigure}
    \hfill
    \begin{subfigure}{0.24\linewidth}
        \includegraphics[width=\linewidth]{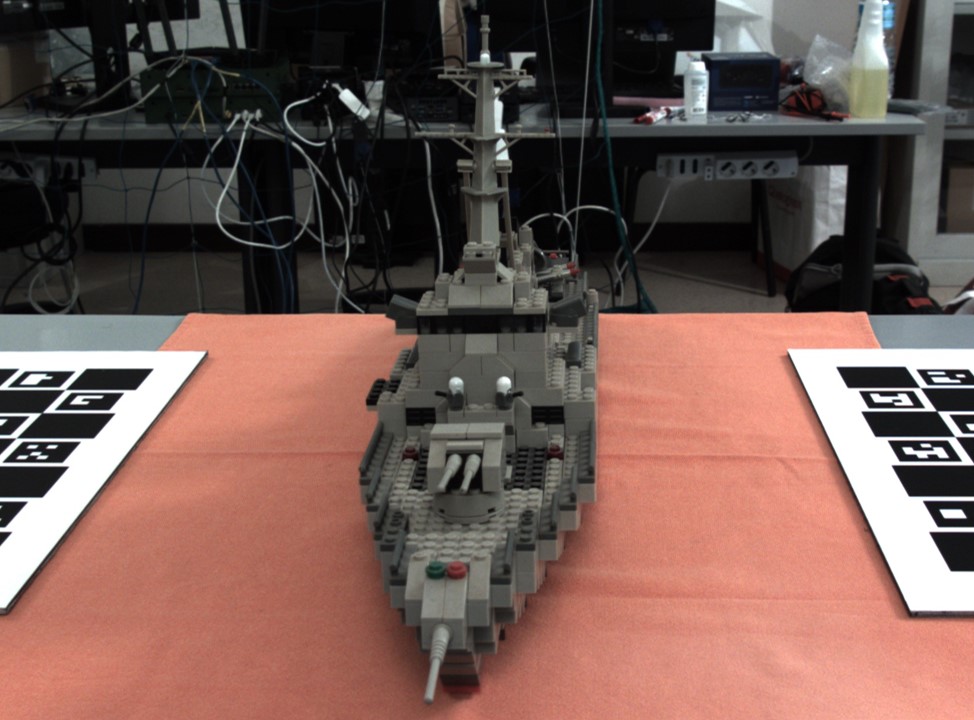}
        \caption{Lego Ship}
        \label{fig:legoship}
    \end{subfigure}
    \hfill
    \begin{subfigure}{0.24\linewidth}
        \includegraphics[width=\linewidth]{figures/makeup_cc.jpg}
        \caption{Makeup}
        \label{fig:makeup}
    \end{subfigure}
    \hfill
        \begin{subfigure}{0.24\linewidth}
        \includegraphics[width=\linewidth]{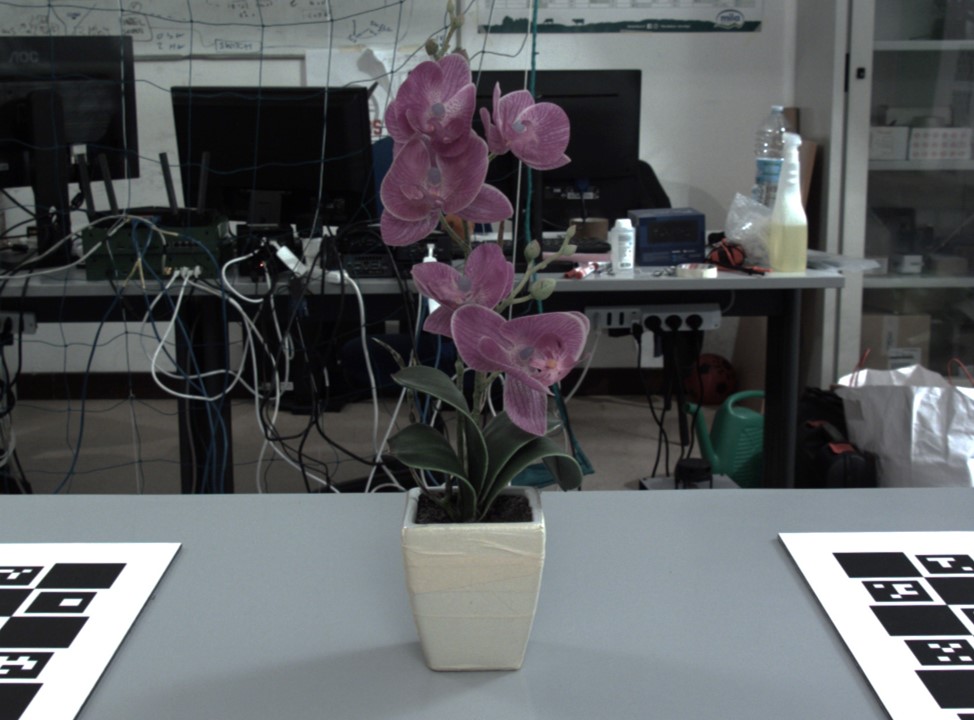}
        \caption{Orchid}
        \label{fig:orchid}
    \end{subfigure}
    \hfill
    \begin{subfigure}{0.24\linewidth}
        \includegraphics[width=\linewidth]{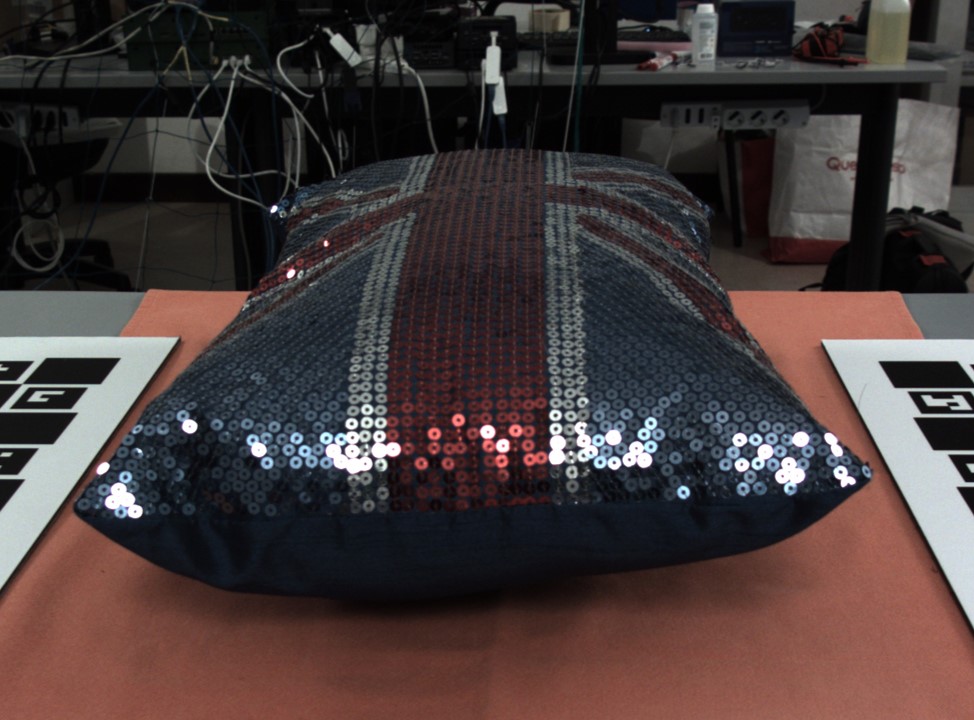}
        \caption{Pillow}
        \label{fig:pillow}
    \end{subfigure}
    \hfill
    \begin{subfigure}{0.24\linewidth}
        \includegraphics[width=\linewidth]{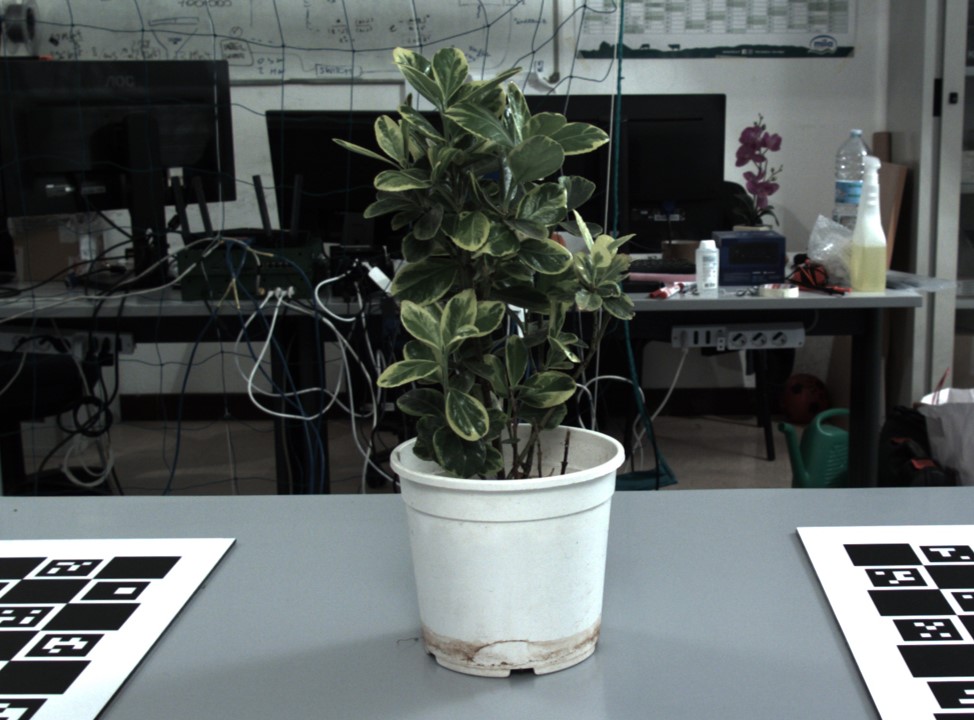}
        \caption{Plant}
        \label{fig:plant}
    \end{subfigure}
    \hfill
    \begin{subfigure}{0.24\linewidth}
        \includegraphics[width=\linewidth]{figures/steelpot_cc.jpg}
        \caption{Steel Pot}
        \label{fig:steelpot}
    \end{subfigure}
    \hfill
    \begin{subfigure}{0.24\linewidth}
        \includegraphics[width=\linewidth]{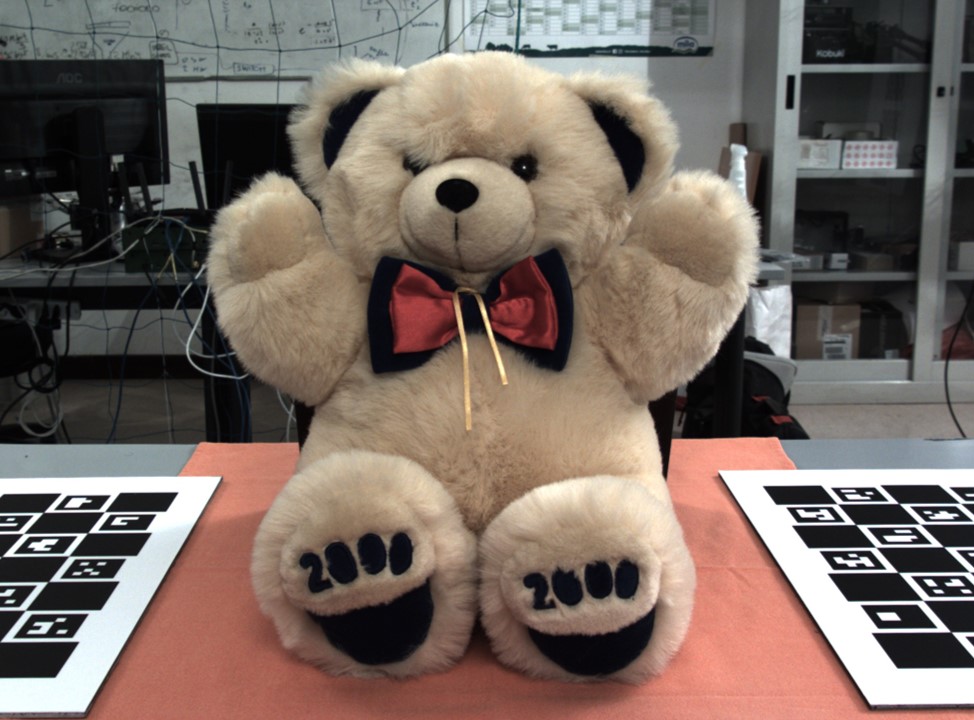}
        \caption{Teddy Bear}
        \label{fig:teddybear}
    \end{subfigure}
\end{figure*}
\begin{figure*}\ContinuedFloat
    \centering
    \begin{subfigure}{0.24\linewidth}
        \includegraphics[width=\linewidth]{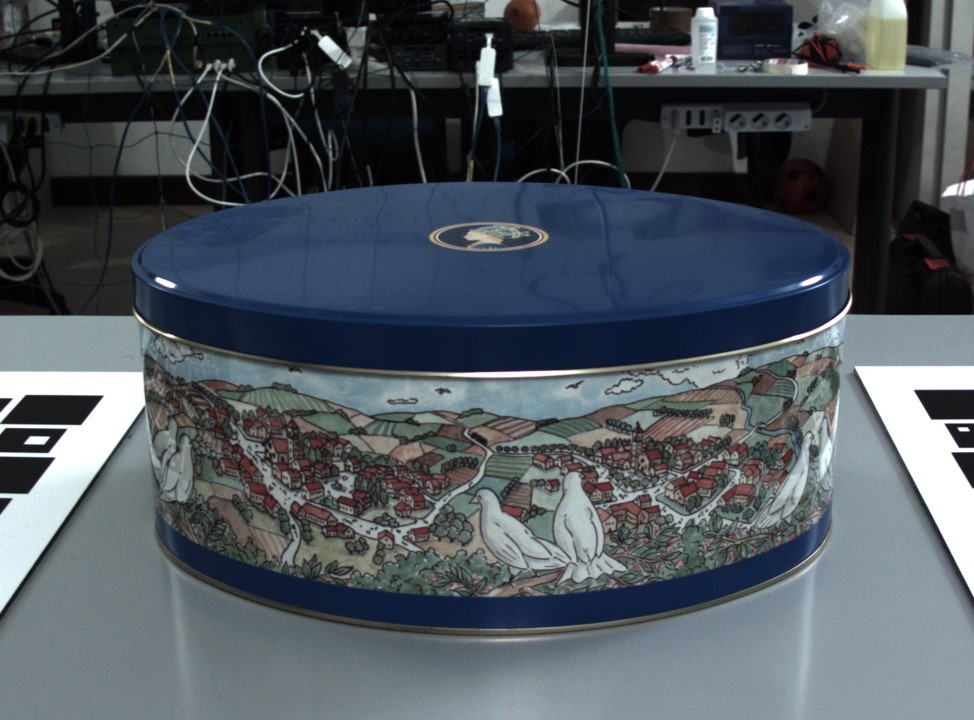}
        \caption{Tin box 1}
        \label{fig:tinbox1}
    \end{subfigure}
    \hfill
    \begin{subfigure}{0.24\linewidth}
        \includegraphics[width=\linewidth]{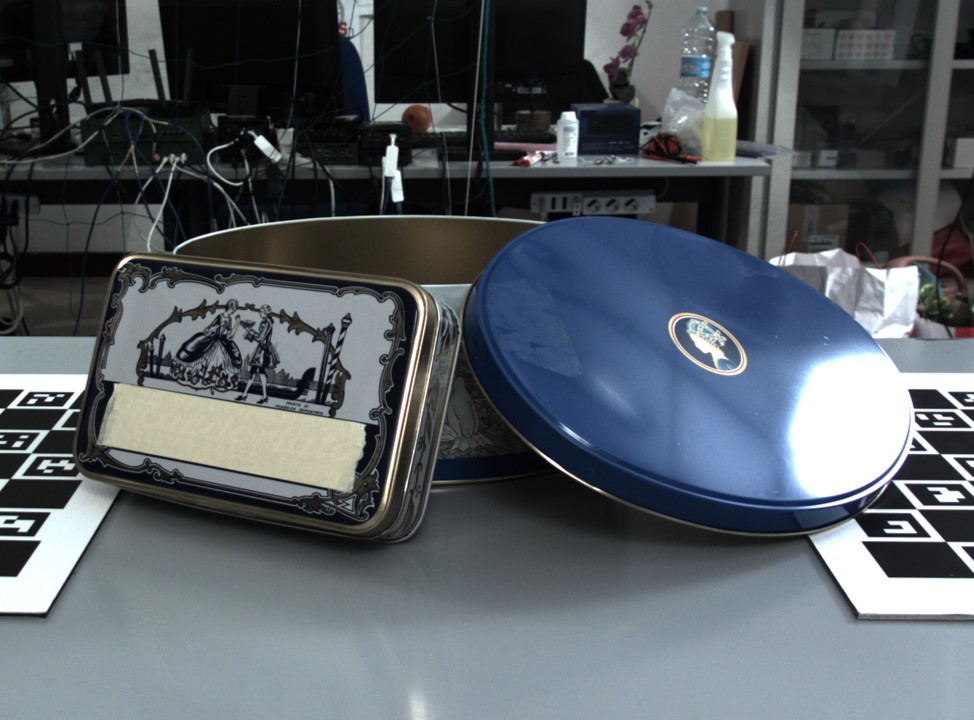}
        \caption{Tin Box 2}
        \label{fig:tinbox2}
    \end{subfigure}
    \hfill
    \begin{subfigure}{0.24\linewidth}
        \includegraphics[width=\linewidth]{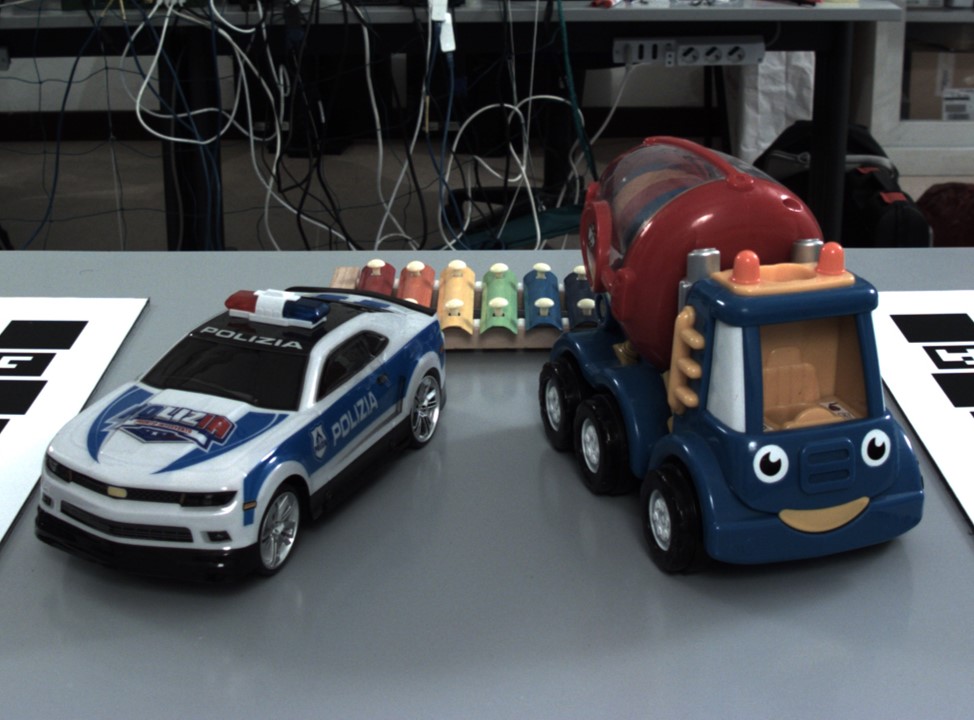}
        \caption{Toys}
        \label{fig:toys}
    \end{subfigure}
    \hfill
        \begin{subfigure}{0.24\linewidth}
        \includegraphics[width=\linewidth]{figures/trophies_cc.jpg}
        \caption{Trophies}
        \label{fig:trophies}
    \end{subfigure}
    \hfill
    \begin{subfigure}{0.24\linewidth}
        \includegraphics[width=\linewidth]{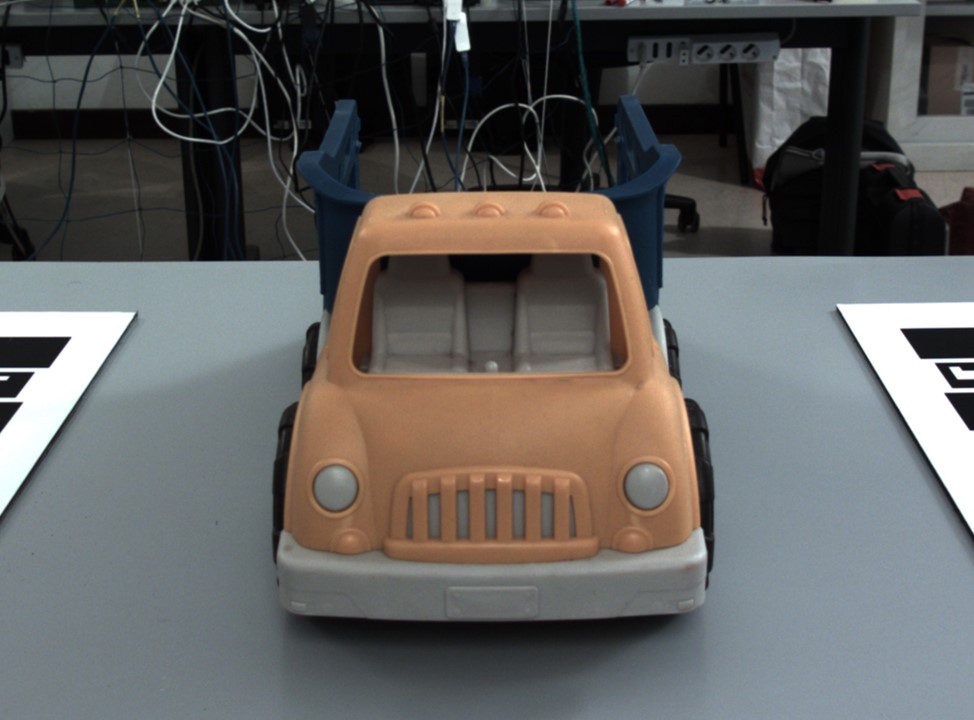}
        \caption{Truck}
        \label{fig:truck}
    \end{subfigure}
    \hfill
    \begin{subfigure}{0.24\linewidth}
        \includegraphics[width=\linewidth]{figures/vases_cc.jpg}
        \caption{Vases}
        \label{fig:vases}
    \end{subfigure}
    \hfill
    \begin{subfigure}{0.24\linewidth}
        \includegraphics[width=\linewidth]{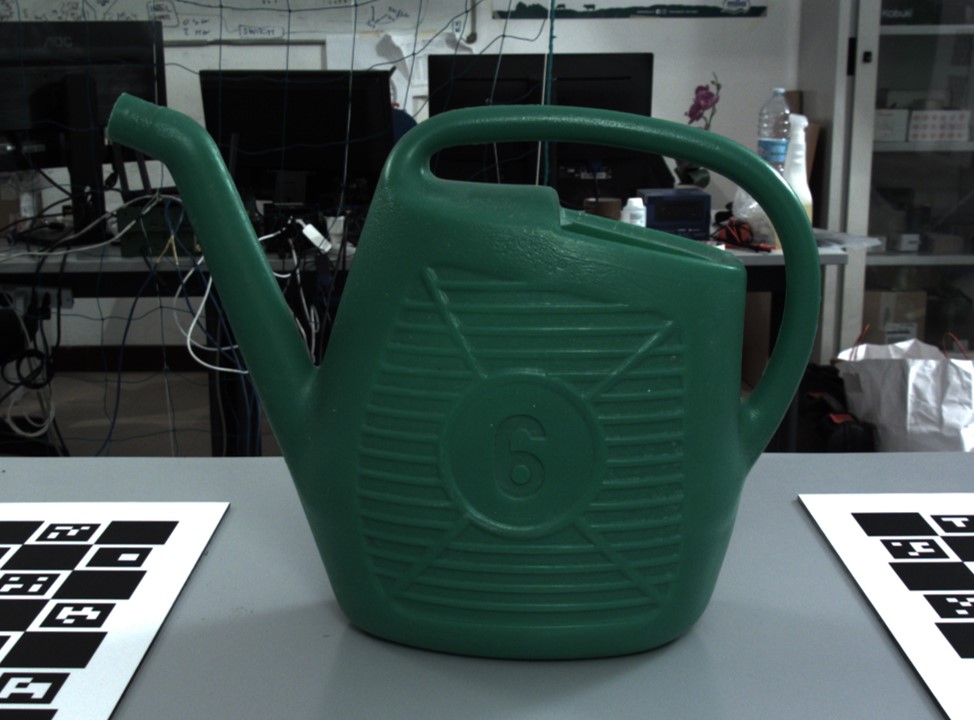}
        \caption{Watering Can 1}
        \label{fig:wateringcan1}
    \end{subfigure}
    \hfill
        \begin{subfigure}{0.24\linewidth}
        \includegraphics[width=\linewidth]{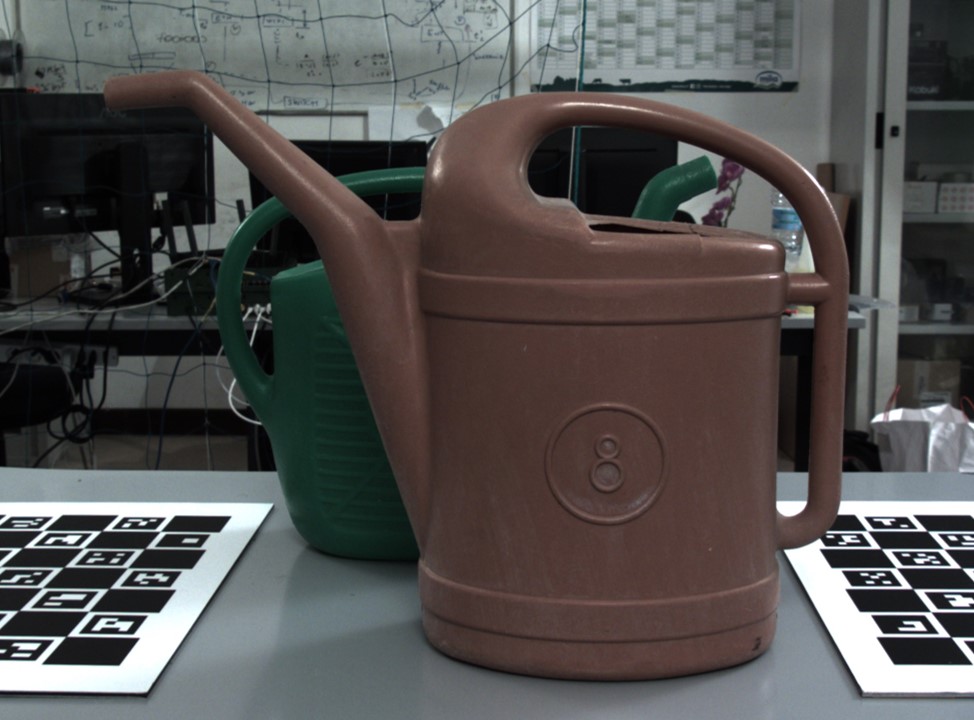}
        \caption{Watering Can 2}
        \label{fig:wateringcan2}
    \end{subfigure}
    \hfill
    \caption{The complete set of scenes of \dname.}
    \label{sup_fig:all_scenes}
\end{figure*}

\begin{table*}
    \centering
    \renewcommand{\arraystretch}{0.95}
    \setlength{\tabcolsep}{5.3pt}
    \begin{tabular}{l|cccc|ccccccc}
        \toprule
        \textbf{Scene} & \textbf{Diff.} & \textbf{Glossy} & \textbf{Reflect.} & \textbf{Transp.} & \textbf{Plastic} & \textbf{Metal} & \textbf{Wood} & \textbf{Organic} & \textbf{Paper} & \textbf{Cloth} & \textbf{Glass} \\
        \midrule
        African Art & x & x &  &  & x &  & x &  &  &  & \\
        \rowcolor{black!3} Aloe & x & x &  &  & x &  &  & x &  &  & \\
        Bird House      &  & x &  &  &  &  & x &  &  &  & \\
        \rowcolor{black!3} Book            & x &  &  &  &  &  &  &  & x &  & \\
        Bouquet         &  & x &  &  & x &  &  & x &  &  & \\
        \rowcolor{black!3} Chess           &  & x &  &  &  &  & x &  &  &  & \\
        Clock           &  & x & x &  & x &  &  &  &  &  & \\
        \rowcolor{black!3} Easter Egg      &  &  & x &  & x &  &  &  &  &  & \\
        Fan             &  & x &  &  & x & x &  &  &  &  & \\
        \rowcolor{black!3} Forest Gang 1   & x &  &  &  &  &  &  &  &  & x & \\
        Forest Gang 2   & x &  &  &  &  &  &  &  &  & x & \\
        \rowcolor{black!3} Fruits          & x & x &  &  & x &  &  & x &  &  & \\
        Gamepads        & x & x &  &  & x &  &  &  &  &  & \\
        \rowcolor{black!3} Glass Clock     &  &  & x & x & x &  &  &  &  &  & x \\
        Globe           &  & x &  &  & x &  &  &  &  &  & \\
        \rowcolor{black!3} Laptop          &  & x &  &  & x &  &  &  &  &  & \\
        Laurel Wreath   & x &  & x &  & x &  &  & x &  & x \\
        \rowcolor{black!3} Lego Ship       &  & x &  &  & x &  &  &  &  &  & \\
        Makeup          & x & x & x & x & x & x &  &  &  & x & x \\
        \rowcolor{black!3} Orchid          &  & x &  &  & x &  &  &  &  &  & \\
        Pillow          & x &  & x &  & x &  &  &  &  & x \\
        \rowcolor{black!3} Plant           &  & x &  &  & x &  &  & x &  &  & \\
        Steel Pot       &  &  & x &  &  & x &  &  &  &  & \\
        \rowcolor{black!3} Teddy Bear      & x &  &  &  & x &  &  &  &  & x & \\
        Tin Box 1       &  &  & x &  &  & x &  &  &  &  & \\
        \rowcolor{black!3} Tin Box 2       &  &  & x &  &  & x &  &  & x &  & \\
        Toys            &  & x &  & x & x &  &  &  &  &  & \\
        \rowcolor{black!3} Trophies        &  & x & x &  & x & x &  &  &  & x & \\
        Truck           & x &  &  &  & x &  &  &  &  &  & \\
        \rowcolor{black!3} Vases           & x & x &  & x & x &  &  &  &  &  & x \\
        Watering Can 1  & x &  &  &  & x &  &  &  &  &  & \\
        \rowcolor{black!3} Watering Can 2  & x &  &  &  & x &  &  &  &  &  & \\
        \bottomrule
    \end{tabular}
    \caption{Table showing which types of material are present in each scene of \dname.}
    \label{sup_tab:scene_materials}
\end{table*}
\section{Geometrical Calibration}
\label{sup_sec:calibration}

\begin{table*}
    \centering
    \begin{tabular}{cccc}
        Modality & Sensor & Error Plot & RMSE (px) \\
        \midrule
        \green{RGB} & Basler acA2500-14gm & \includegraphics[width=0.3\linewidth, valign=m]{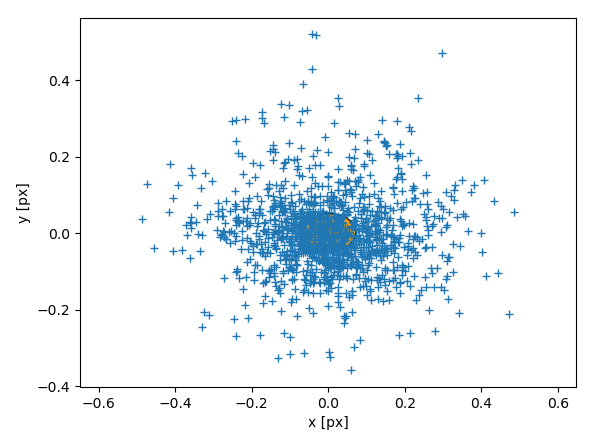} & 0.18 \\
        \purple{Mono} & Basler acA2500-14gc & \includegraphics[width=0.3\linewidth, valign=m]{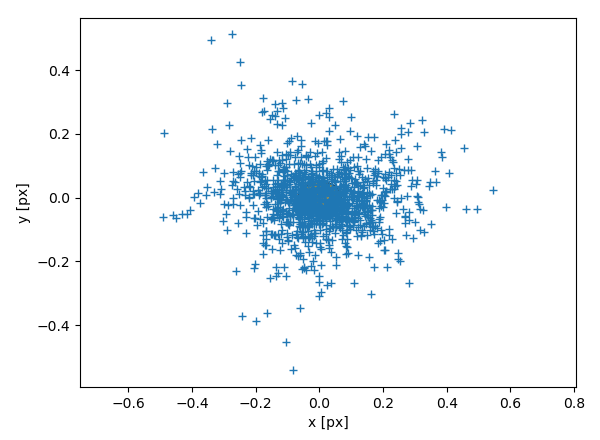} & 0.17 \\
        \red{NIR} & Basler acA1300-60gmNIR & \includegraphics[width=0.3\linewidth, valign=m]{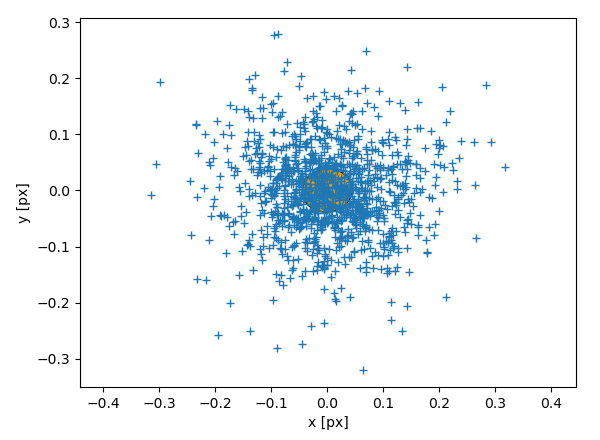} & 0.11 \\
        \blue{Pol} & FLIR Blackfly S BFS-U3-51S5P & \includegraphics[width=0.3\linewidth, valign=m]{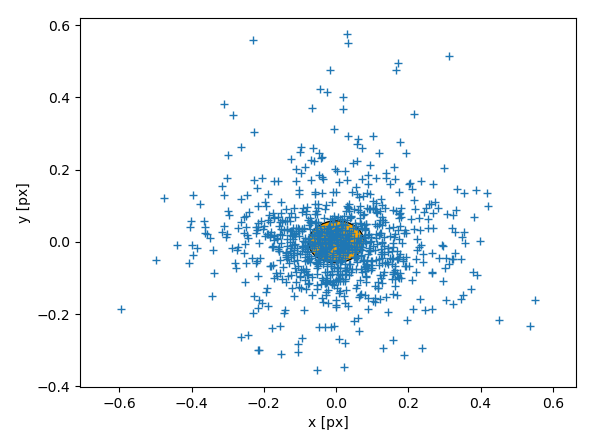} & 0.20 \\
        \orange{MS} & Silios CMS-C1 & \includegraphics[width=0.3\linewidth, valign=m]{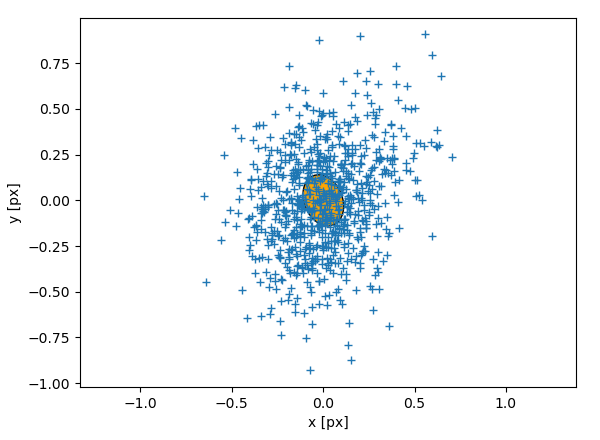} & 0.36 \\
        \midrule
        Modality & Sensor & Error Plot & RMSE (px) \\
    \end{tabular}
    \caption{Sensors used for the dataset acquisition and corresponding calibration accuracies. The error plots show the reprojection error distribution in terms of pixels. The yellow blob represents the error standard deviation.}
    \label{sup_tab:calibration_results}
\end{table*}

\begin{figure*}
    \centering
    \begin{subfigure}{0.43\linewidth}
        \centering
        \includegraphics[width=\linewidth]{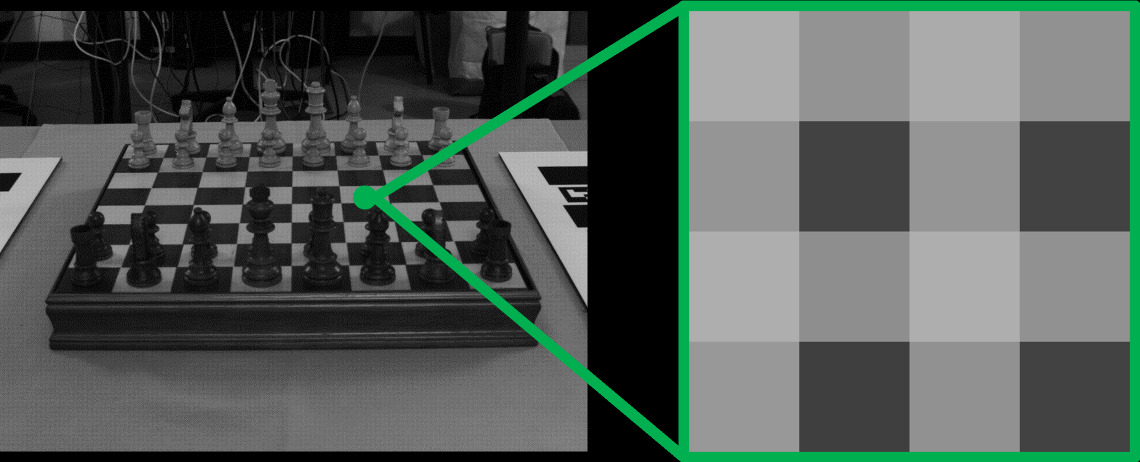}
        \caption{RGB mosaick pattern on the Chess scene.}
    \end{subfigure}
    \hspace{20pt}
    \begin{subfigure}{0.43\linewidth}
        \centering
        \includegraphics[width=\linewidth]{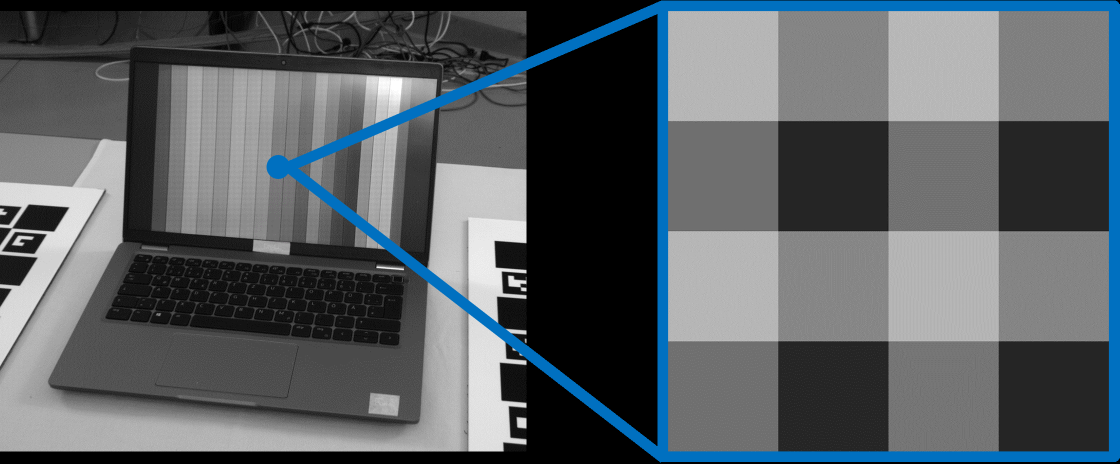}
        \caption{Polarization mosaick pattern on the Laptop scene.}
    \end{subfigure}
    \\
    \vspace{10pt}
    \begin{subfigure}{0.43\linewidth}
        \centering
        \includegraphics[width=\linewidth]{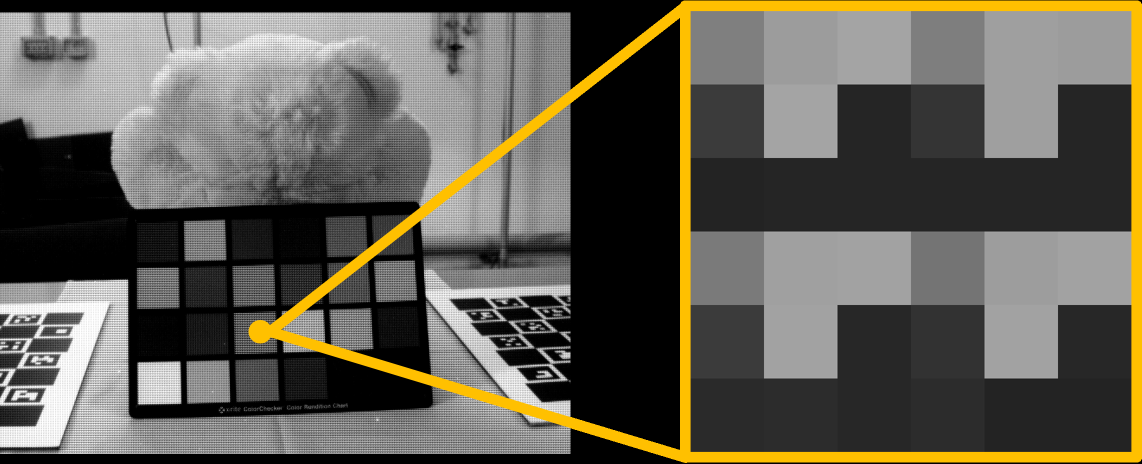}
        \caption{Multispectral mosaick pattern on the Teddy Bear scene.}
    \end{subfigure}
    \caption{RGB, Polarization and Multispectral mosaick patterns. The RGB and Pol patterns are composed by 4 pixel arranged in a $2\times2$ square. The MS pattern includes 9 pixels arranged in a $3\times3$ square. In the figures, four patterns per modality are shown.}
    \label{sup_fig:mosaick_patterns}
    \vspace{-10pt}
\end{figure*}

As anticipated in Sec. 3.3, 
the sensor calibration is performed employing five different ChAruco boards. We displaced the ChAruco boards in a column to ensure that the ChAruco patterns always span the whole vertical field-of-view of every sensor. Capturing the calibration patterns on every region of the image plane enables a better intrinsics and distortion parameter estimation, thus it is important that the matched features lay close to the frame border too. Due to this displacement, we only needed to shift the rig horizontally while capturing images to achieve the complete vertical and horizontal field-of-view coverage. Moreover, the patterns were acquired also at different distances from the cameras, in order to achieve a more robust geometrical calibration. In \cref{sup_tab:calibration_results}, the reprojection error plots for each sensor are shown. It is possible to observe that the Silios CMS-C1 multispectral camera calibration is the least accurate: this is explainable considering that the calibration was performed using the demosaicked frames. Indeed, demosaicking a $3\times3$ multispectral pattern (\cref{sup_fig:mosaick_patterns}) is not trivial: the employed bilinear interpolation introduces some block artifacts that may reduce the calibration pattern localization accuracy.

\clearpage
\section{Additional Results}
\label{sup_sec:renderings}

\begin{table}
    \centering
    \begin{tabular}{cccccc}
        \toprule
        Macroframes & Training Mod. & Test Mod. & PSNR$\uparrow$ & SSIM$\uparrow$ & LPIPS$\downarrow$ \\
        \midrule
         & \green{RGB} & \green{RGB} & 15.73 & 0.63 & 0.30 \\
        \cmidrule{2-6}
        \multirow{1}{1cm}{\centering5} & \green{RGB}-\red{NIR} & \green{RGB} & 16.75 & 0.66 & 0.27 \\
        \cmidrule{2-6}
        & \multirow{1}{2.2cm}{\centering\green{RGB}-\red{NIR}} & \multirow{2}{1.5cm}{\centering\green{RGB}} & \multirow{2}{1cm}{\centering19.46} & \multirow{2}{1cm}{\centering0.73} & \multirow{2}{1cm}{\centering0.20} \\
         & \multirow{1}{2.2cm}{\centering\purple{Mono}-\blue{Pol}-\orange{MS}} & & & & \\
        \midrule
         & \green{RGB} & \green{RGB} & 16.97 & 0.68 & 0.25 \\
        \cmidrule{2-6}
        \multirow{1}{1cm}{\centering10} & \green{RGB}-\red{NIR} & \green{RGB} & 22.17 & 0.81 & 0.15 \\
        \cmidrule{2-6}
        & \multirow{1}{2.2cm}{\centering\green{RGB}-\red{NIR}} & \multirow{2}{1.5cm}{\centering\green{RGB}} & \multirow{2}{1cm}{\centering25.84} & \multirow{2}{1cm}{\centering0.89} & \multirow{2}{1cm}{\centering0.09} \\
         & \multirow{1}{2.2cm}{\centering\purple{Mono}-\blue{Pol}-\orange{MS}} &  &  &  & \\   
        \bottomrule
    \end{tabular}
    \caption{Few-shot results averaged on \dname. Metrics computed on the demosaicked frames rendered by the model.}
    \label{sup_tab:few-shot}
\end{table}

\begin{figure*}
    \centering
    \begin{tabular}{@{\extracolsep{-6pt}}cccc}
        \multicolumn{2}{l}{\hspace{-6pt}RGB} \\
        & & & \\
        View & Rendering & Ground Truth & Error \\
        \midrule
        9 & \includegraphics[width=0.3\linewidth, valign=m]{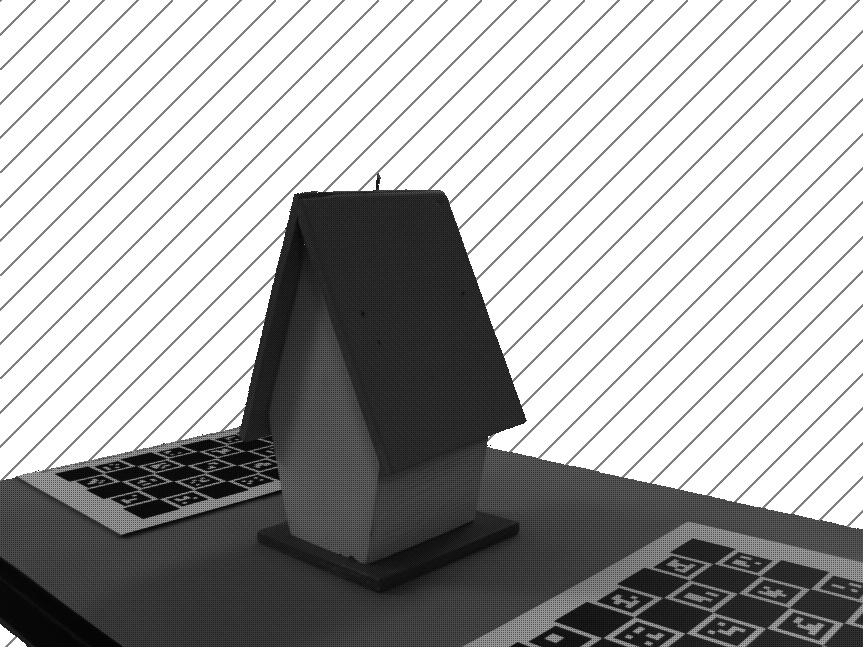} & \includegraphics[width=0.3\linewidth, valign=m]{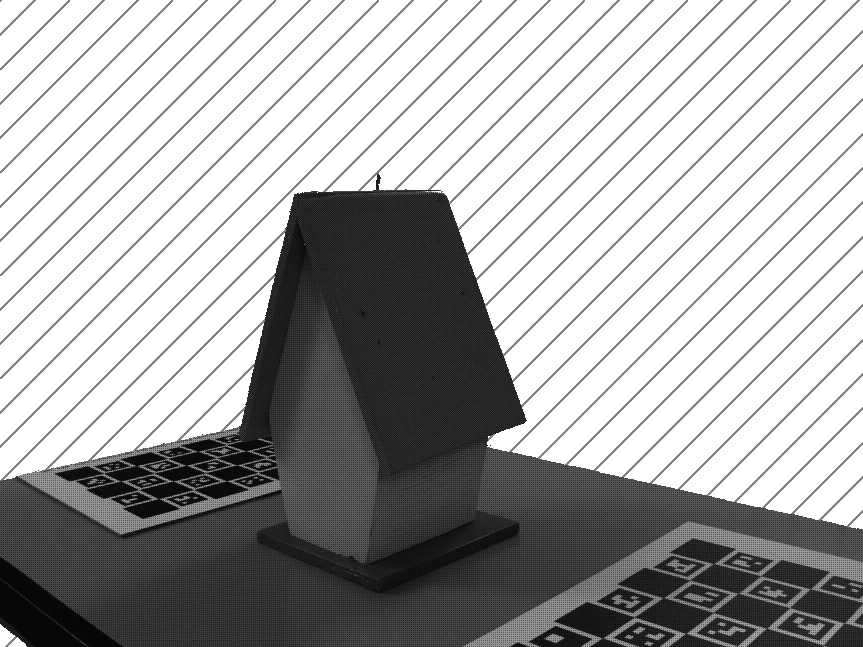} & \includegraphics[width=0.3\linewidth, valign=m]{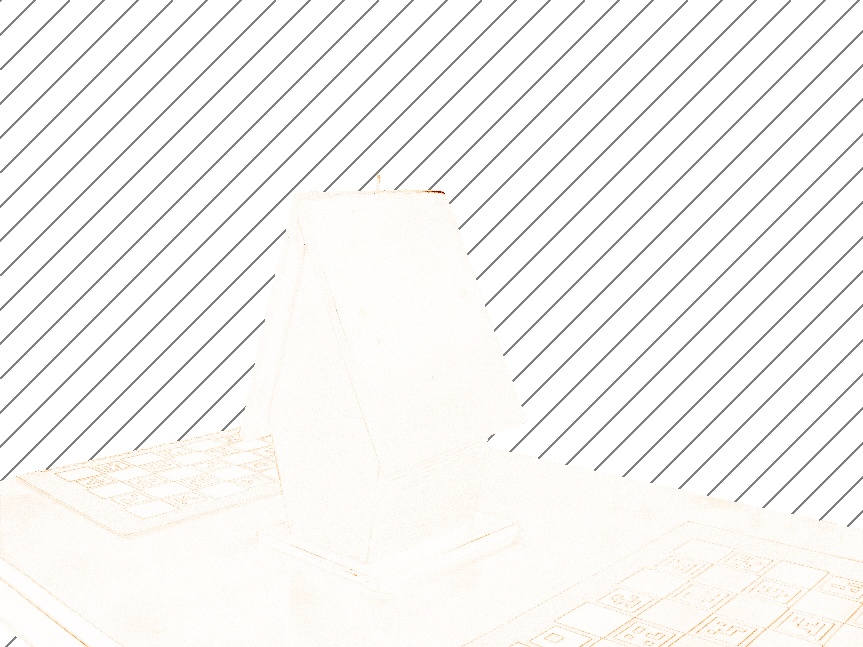} \\
        19 & \includegraphics[width=0.3\linewidth, valign=m]{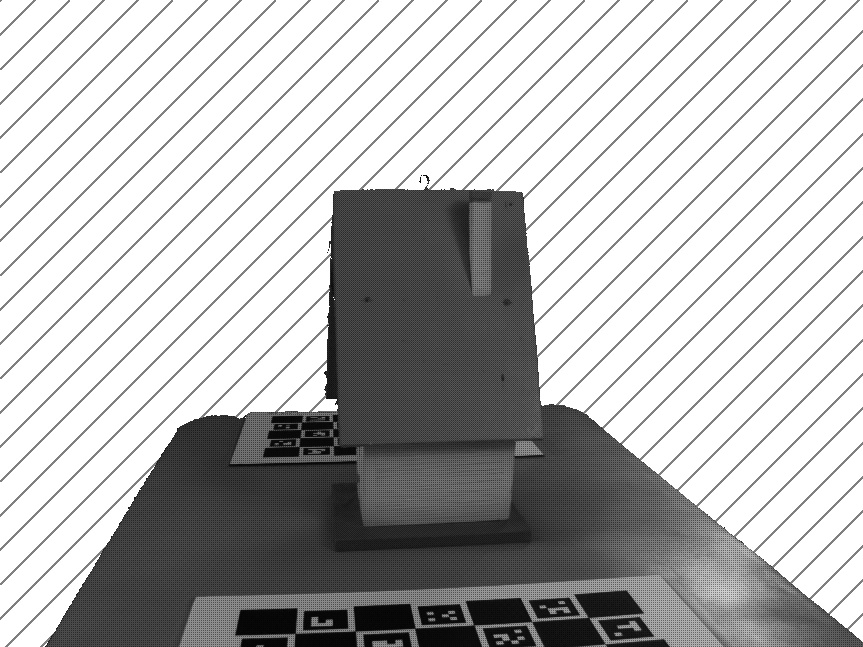} & \includegraphics[width=0.3\linewidth, valign=m]{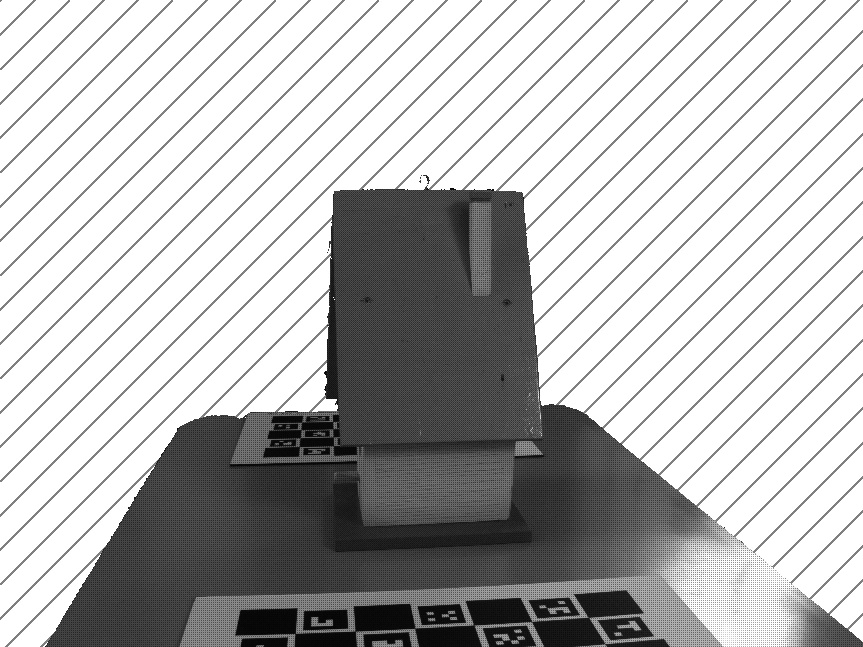} & \includegraphics[width=0.3\linewidth, valign=m]{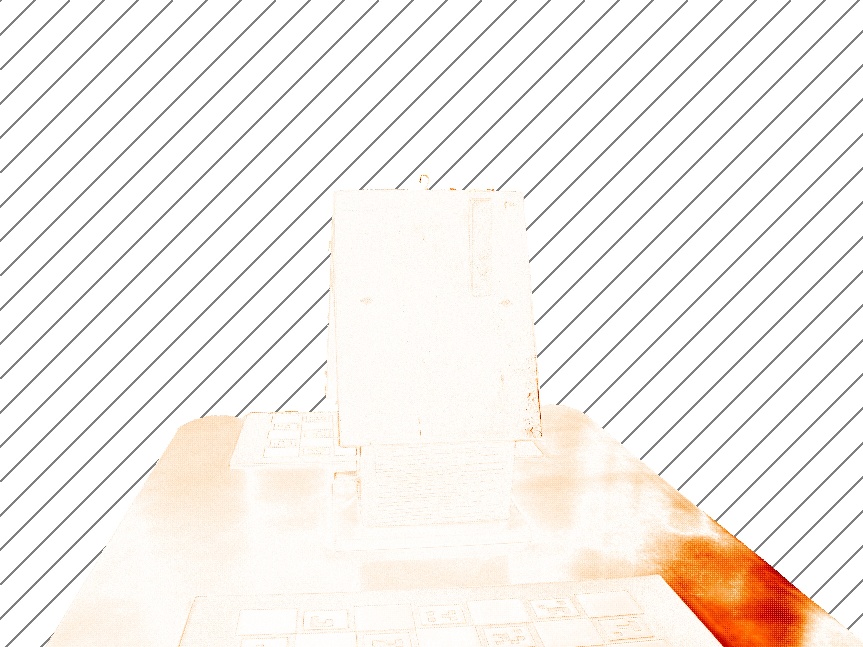} \\
        29 & \includegraphics[width=0.3\linewidth, valign=m]{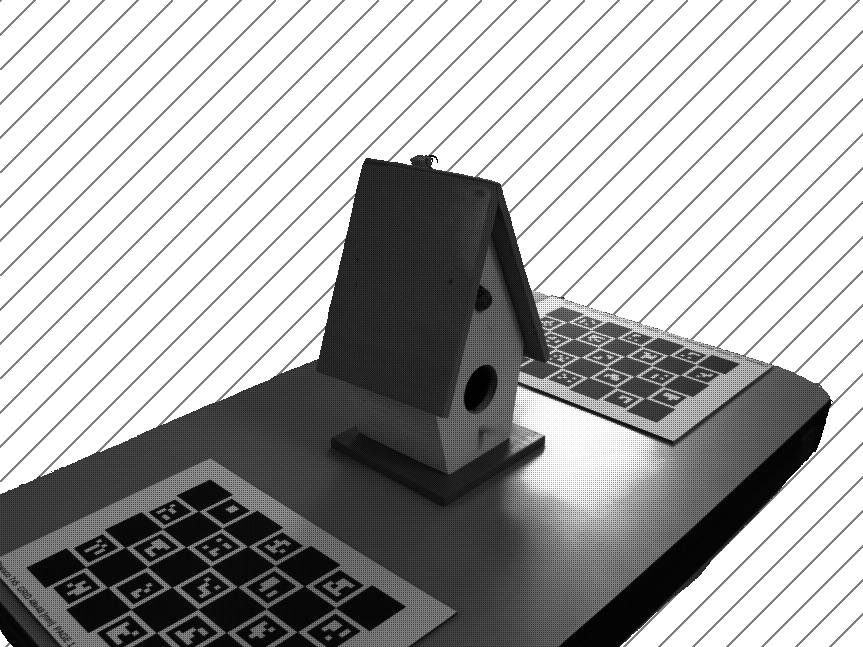} & \includegraphics[width=0.3\linewidth, valign=m]{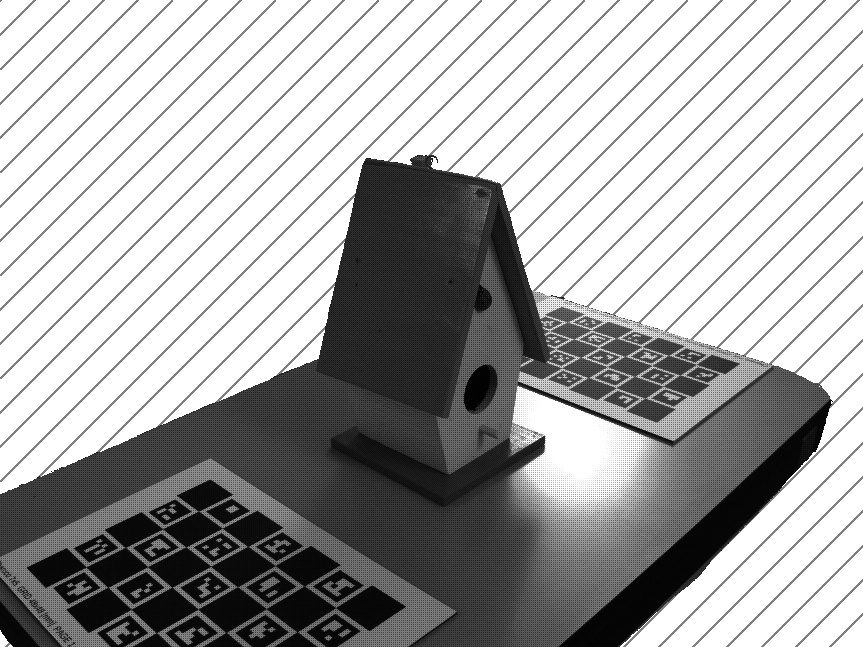} & \includegraphics[width=0.3\linewidth, valign=m]{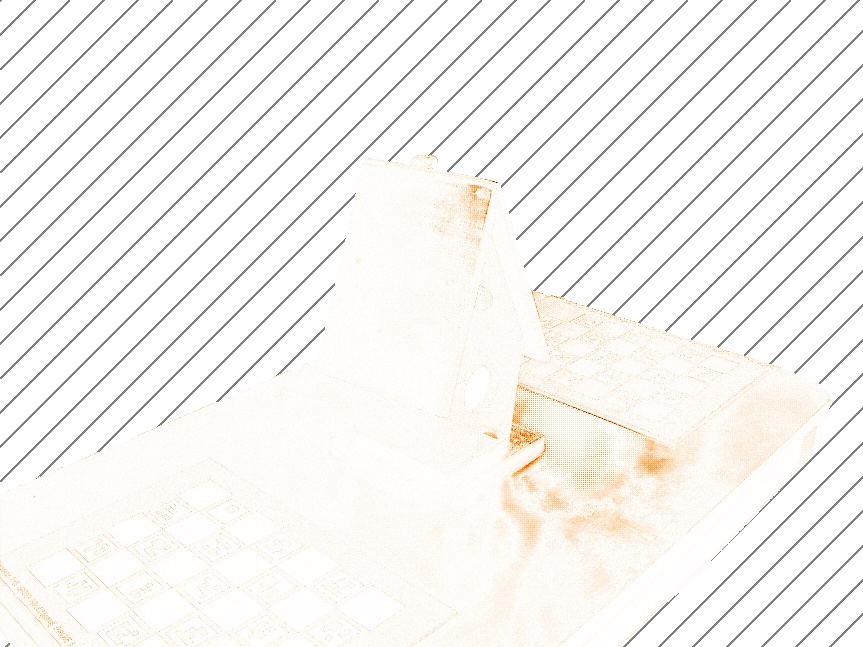} \\
        39 & \includegraphics[width=0.3\linewidth, valign=m]{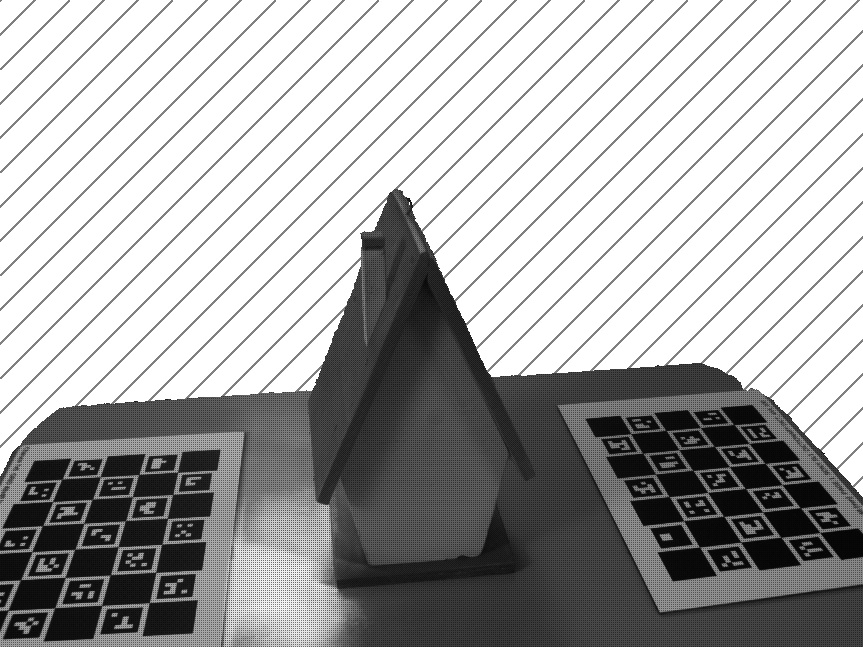} & \includegraphics[width=0.3\linewidth, valign=m]{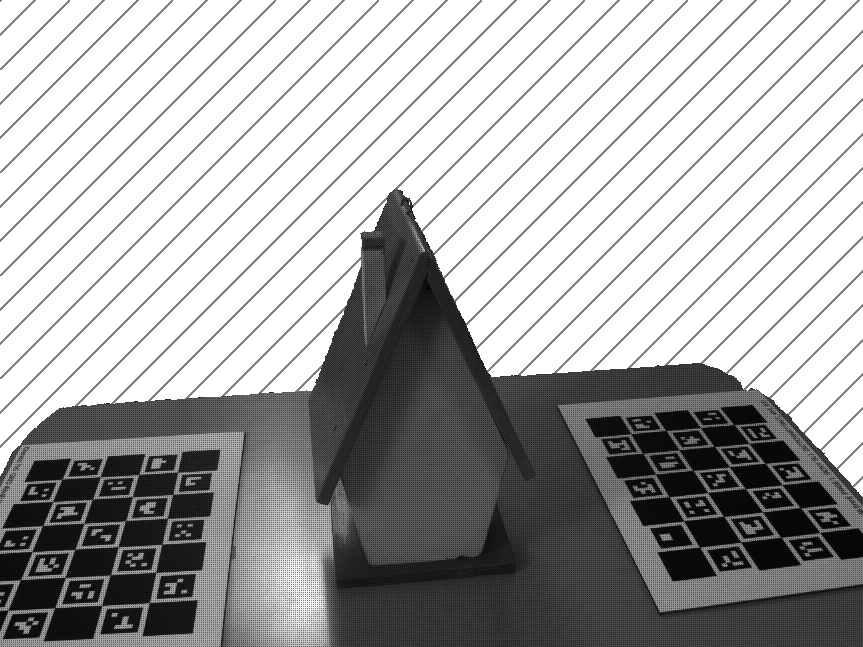} & \includegraphics[width=0.3\linewidth, valign=m]{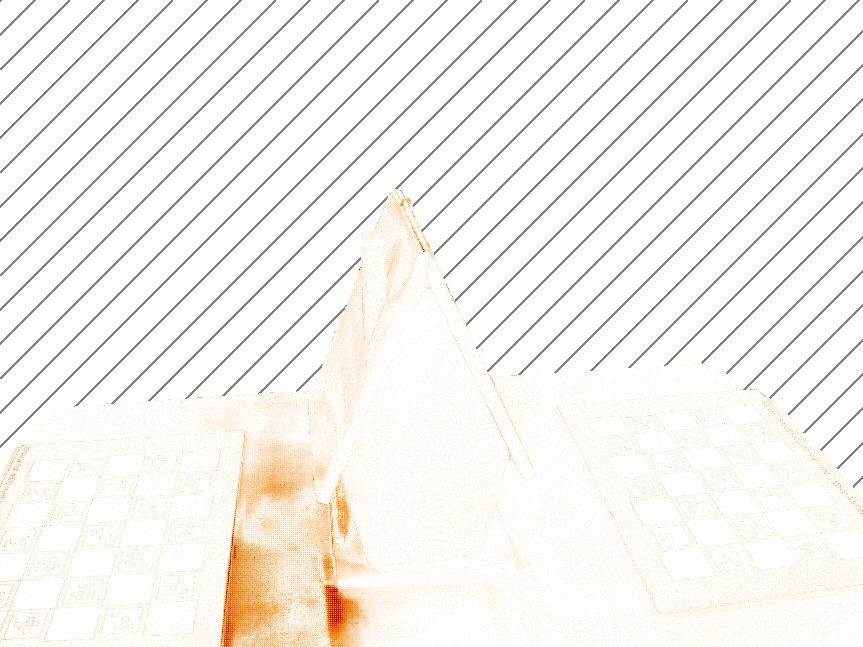} \\
        49 & \includegraphics[width=0.3\linewidth, valign=m]{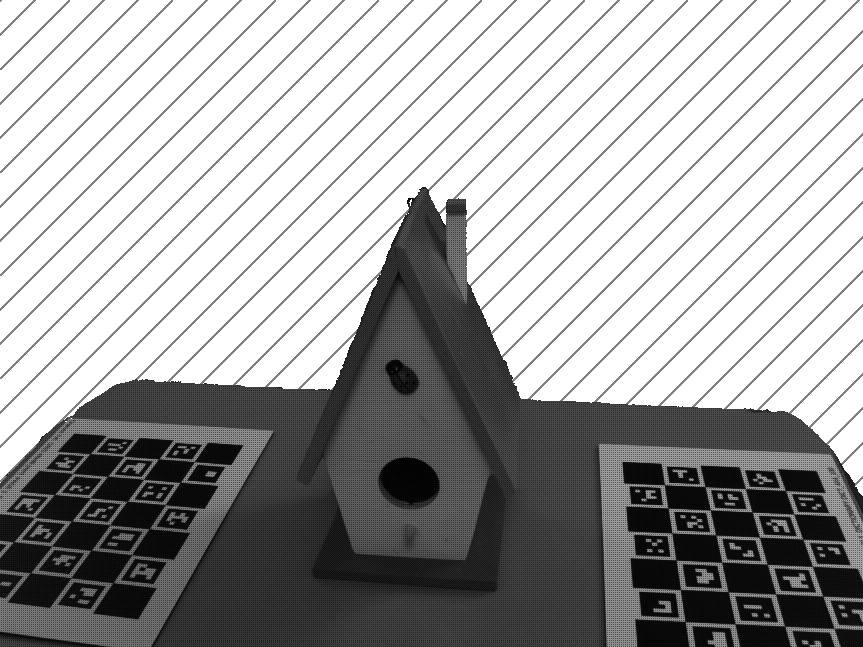} & \includegraphics[width=0.3\linewidth, valign=m]{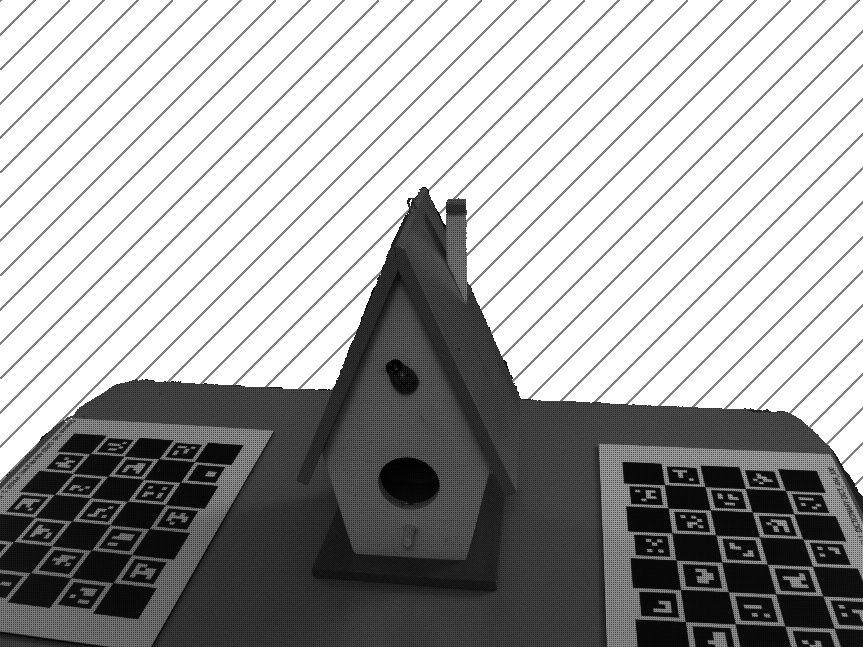} & \includegraphics[width=0.3\linewidth, valign=m]{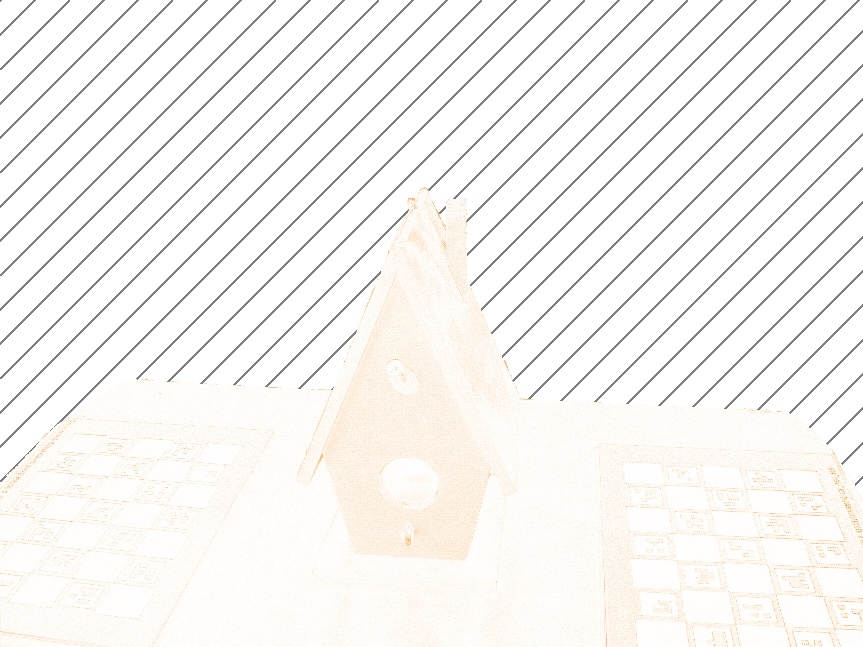} \\
        \midrule
        View & Rendering & Ground Truth & Error \\
    \end{tabular}
    \caption{Mosaicked RGB renderings, ground truth and error maps of the Bird House scene from the five different test views.}
    \label{sup_fig:rgb_raw_renedrings}
\end{figure*}

\begin{figure*}
    \centering
    \begin{tabular}{@{\extracolsep{-6pt}}cccc}
        \multicolumn{2}{l}{\hspace{-6pt}Mono} \\
        & & & \\
        View & Rendering & Ground Truth & Error \\
        \midrule
        9 & \includegraphics[width=0.3\linewidth, valign=m]{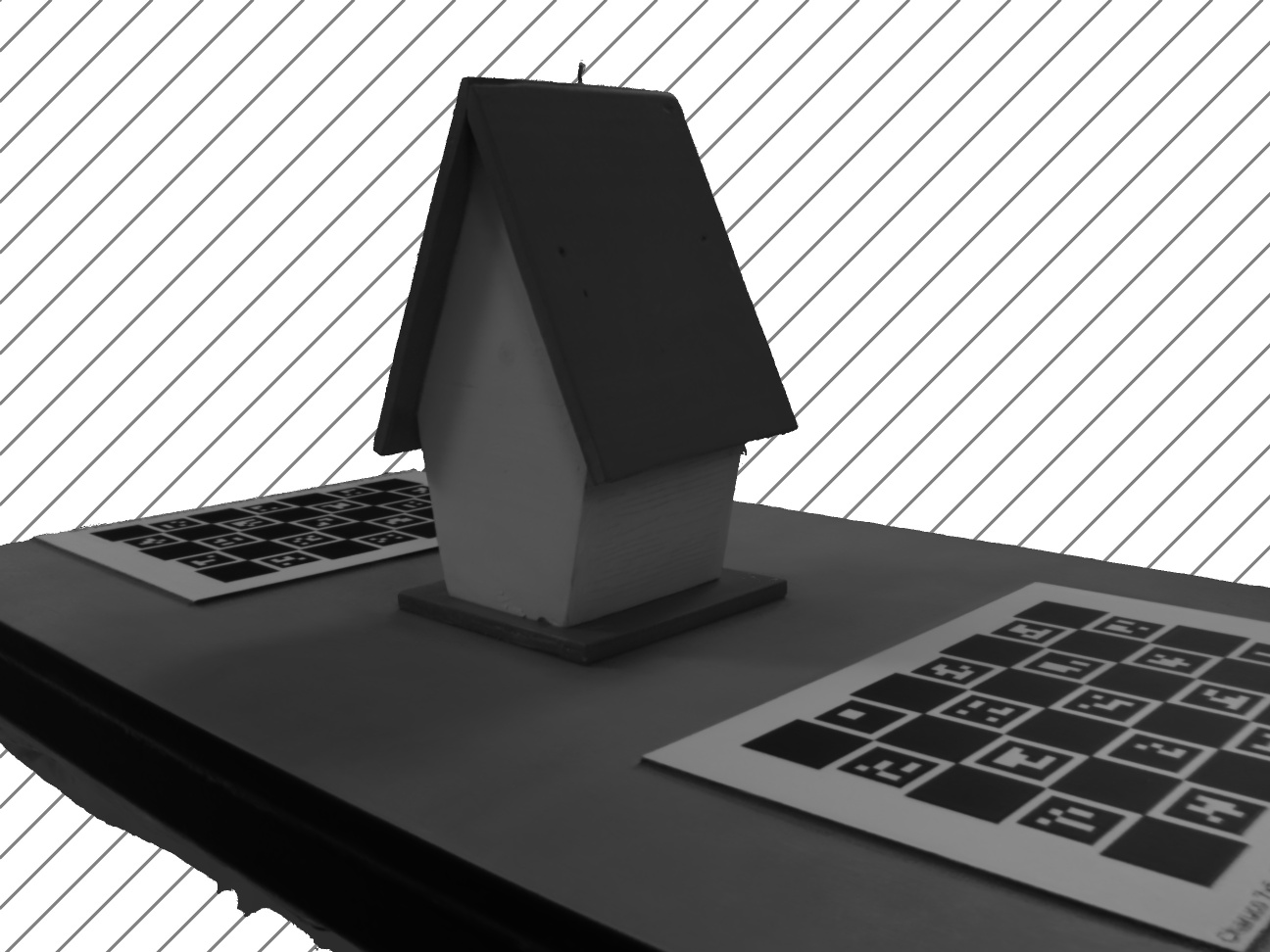} & \includegraphics[width=0.3\linewidth, valign=m]{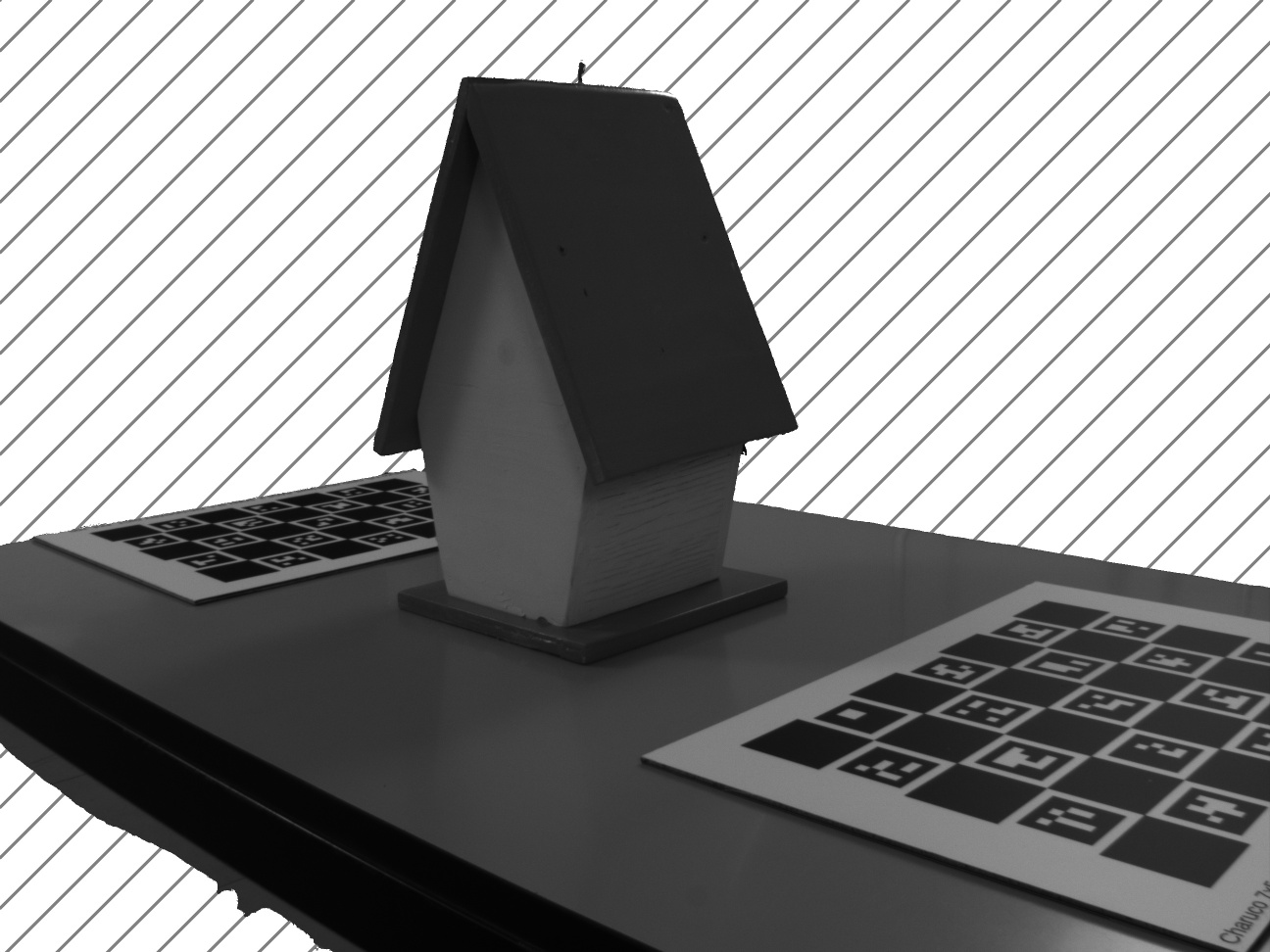} & \includegraphics[width=0.3\linewidth, valign=m]{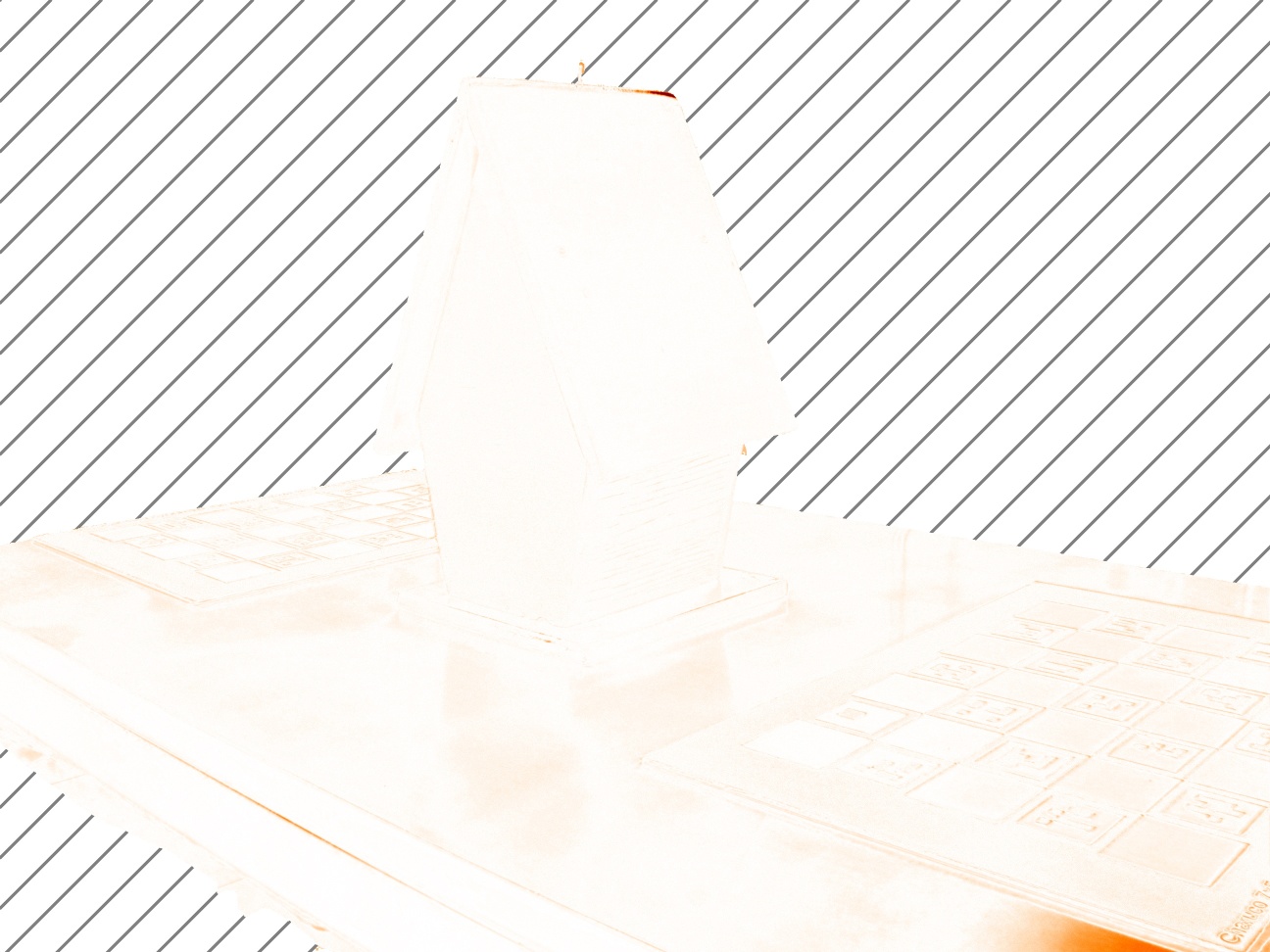} \\
        19 & \includegraphics[width=0.3\linewidth, valign=m]{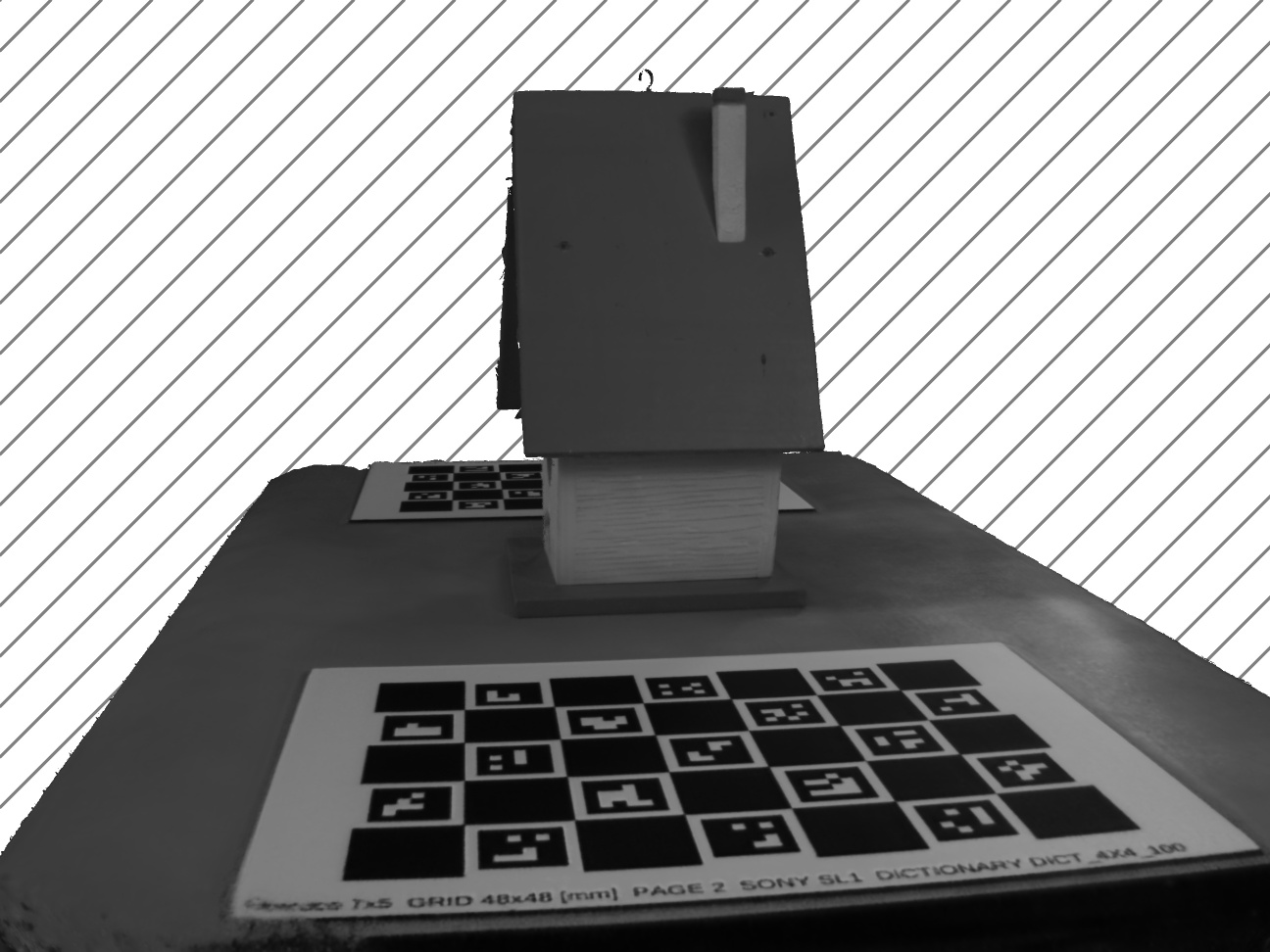} & \includegraphics[width=0.3\linewidth, valign=m]{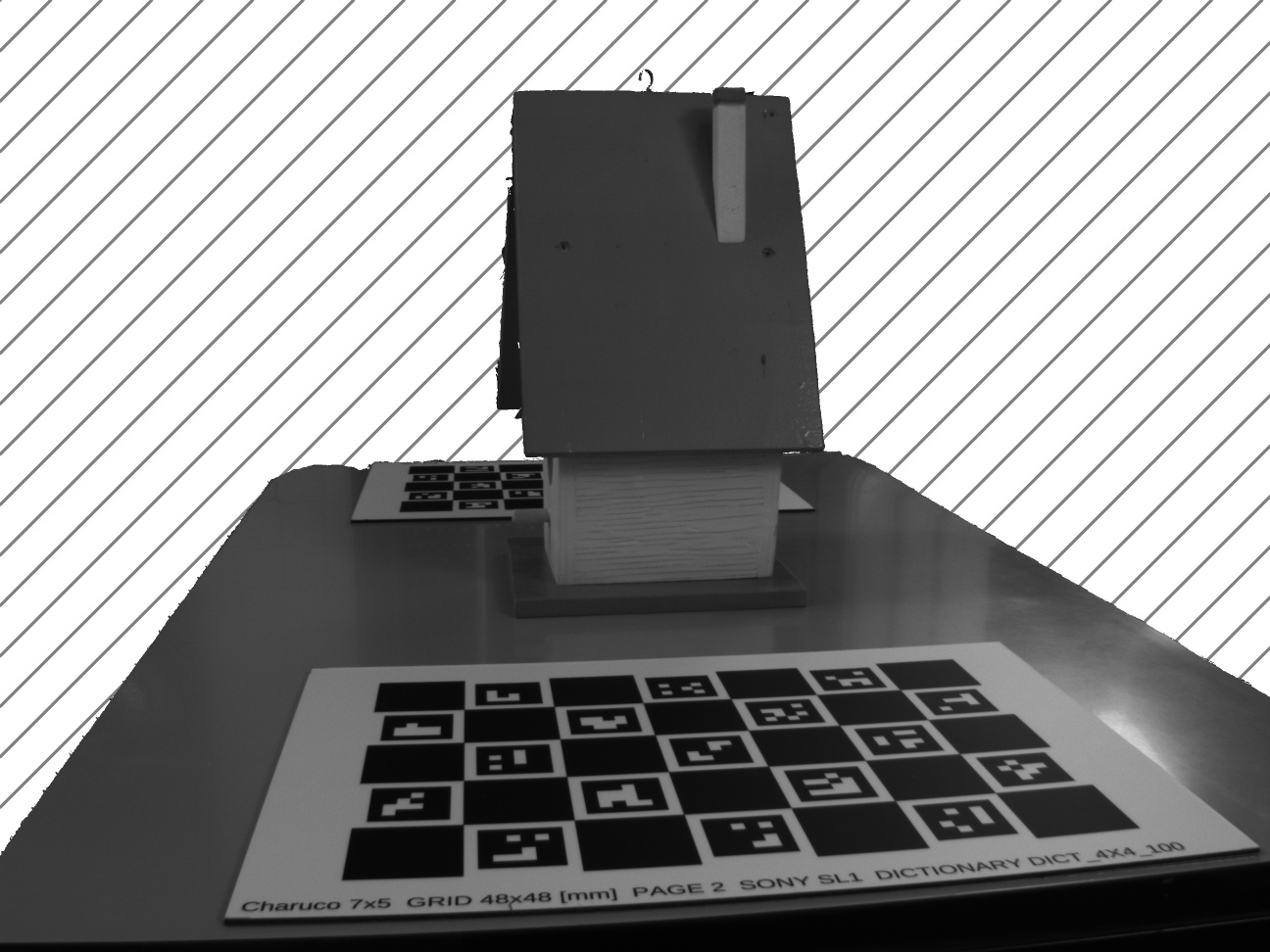} & \includegraphics[width=0.3\linewidth, valign=m]{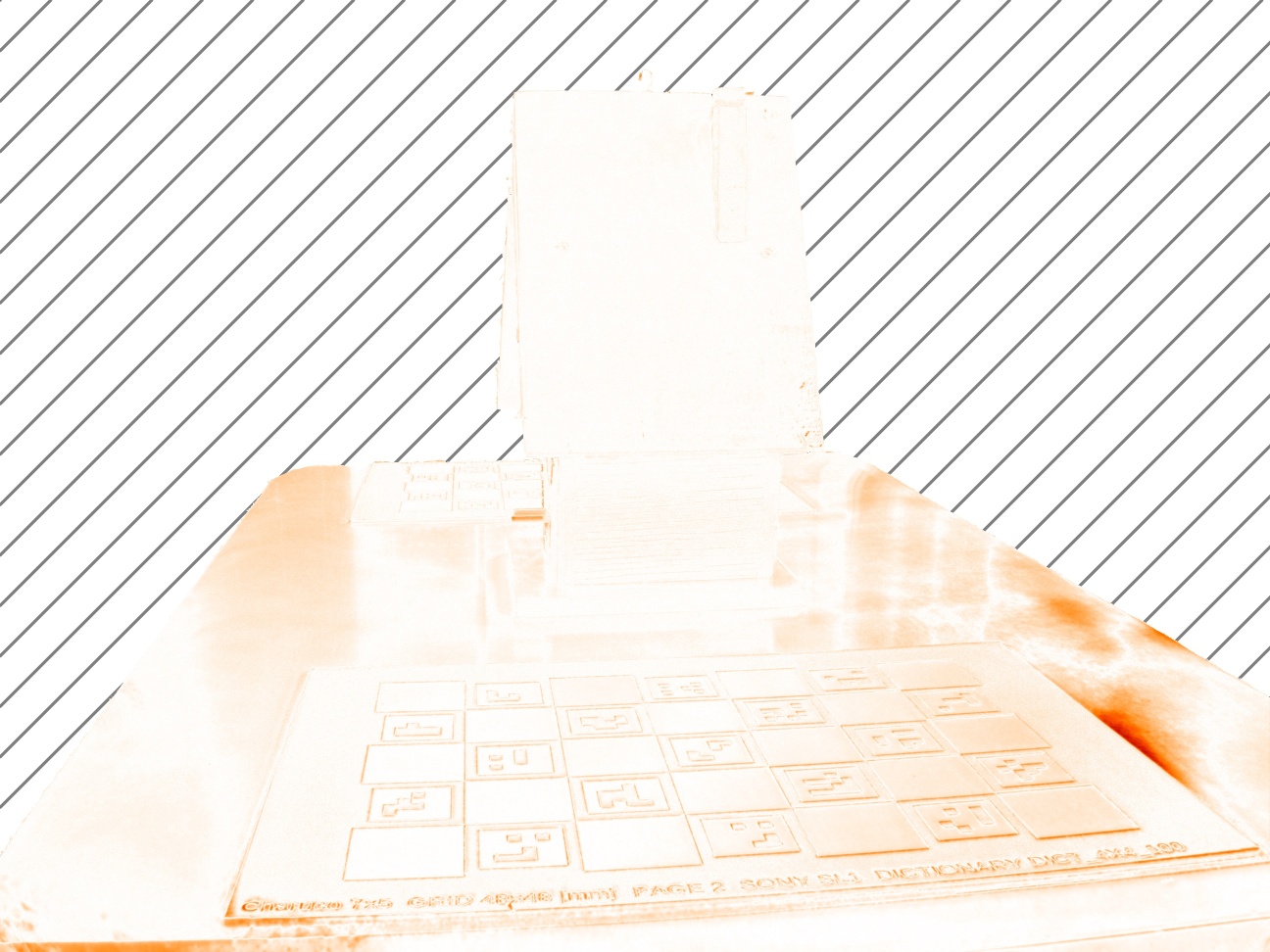} \\
        29 & \includegraphics[width=0.3\linewidth, valign=m]{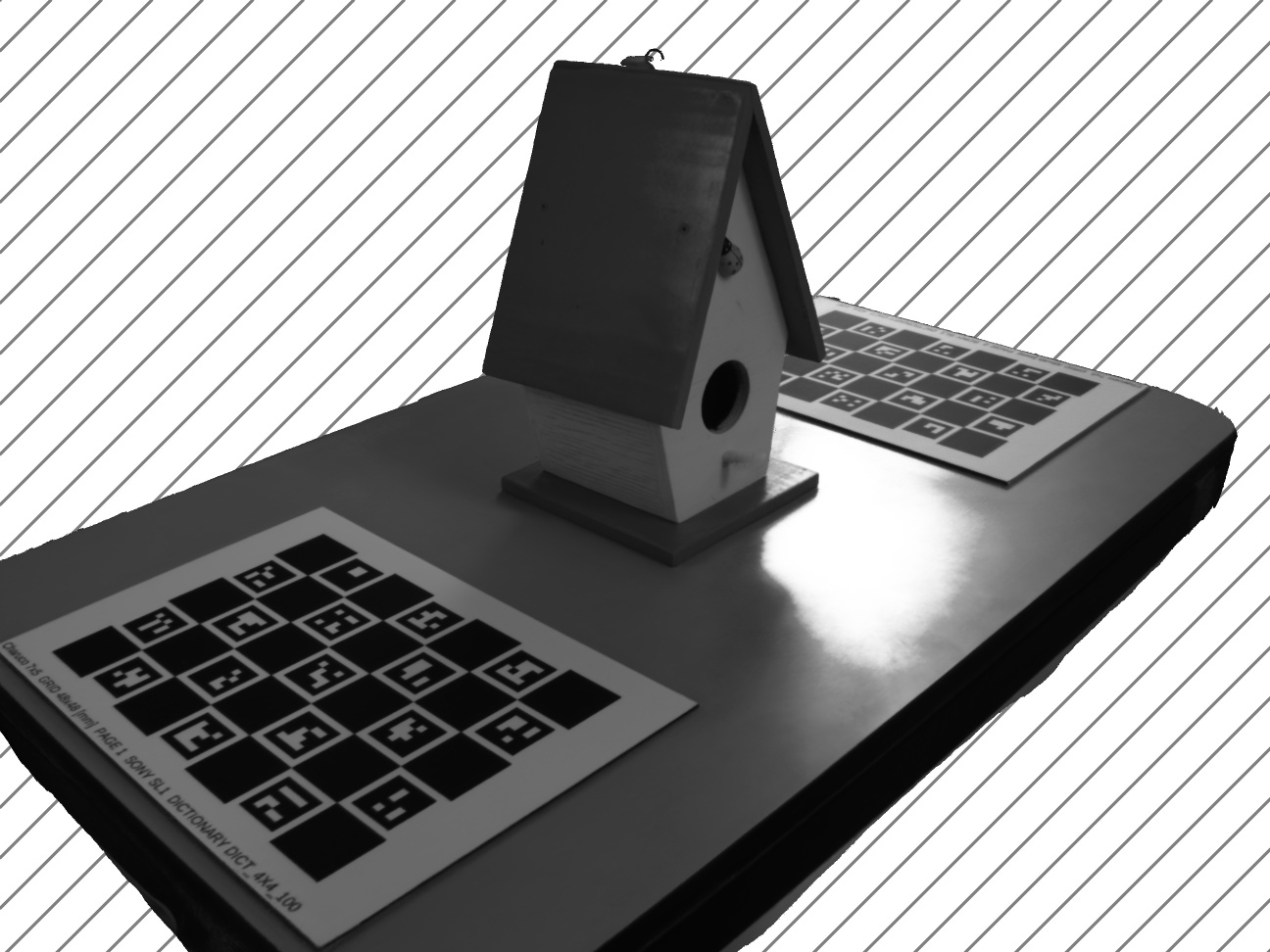} & \includegraphics[width=0.3\linewidth, valign=m]{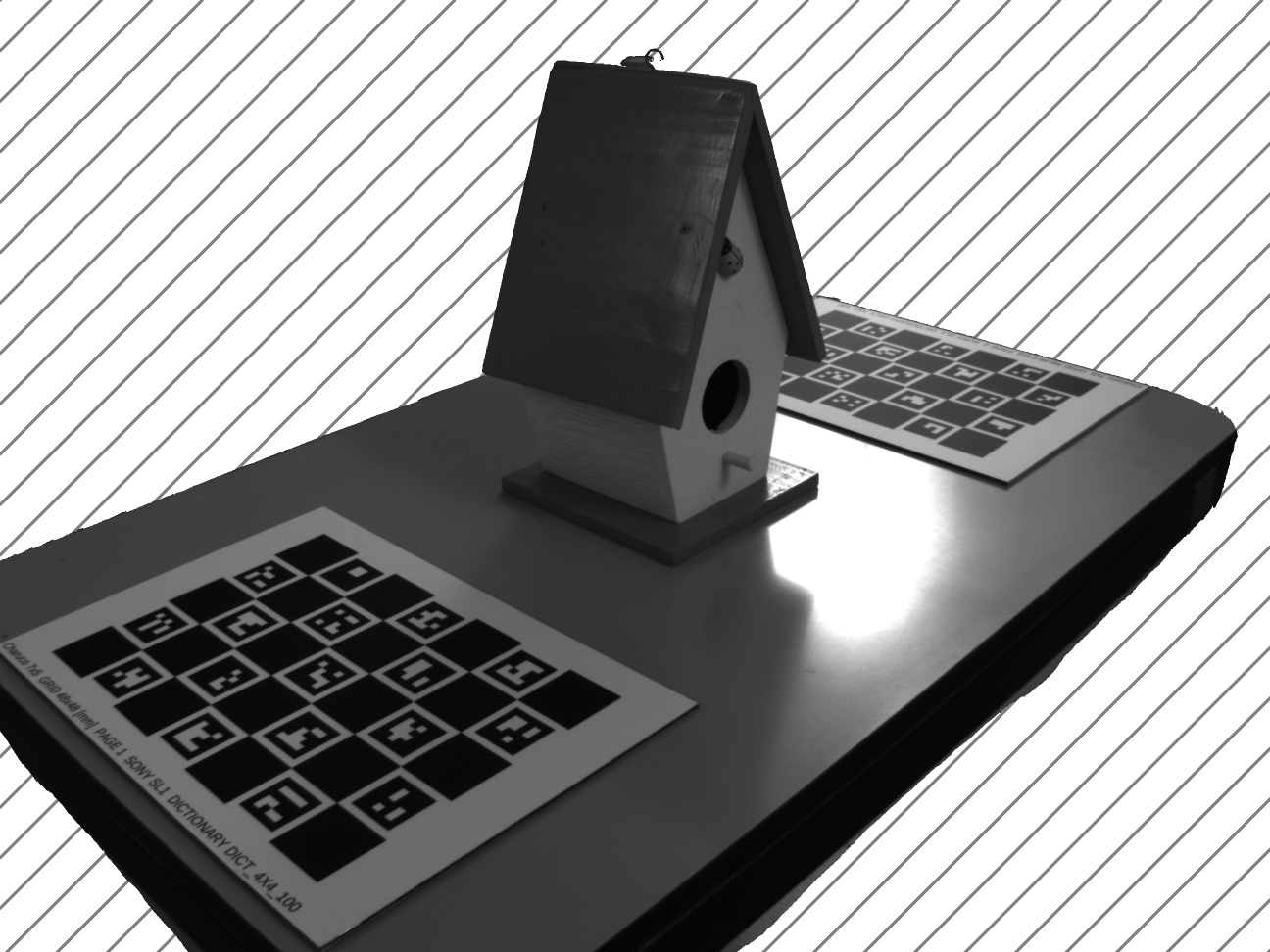} & \includegraphics[width=0.3\linewidth, valign=m]{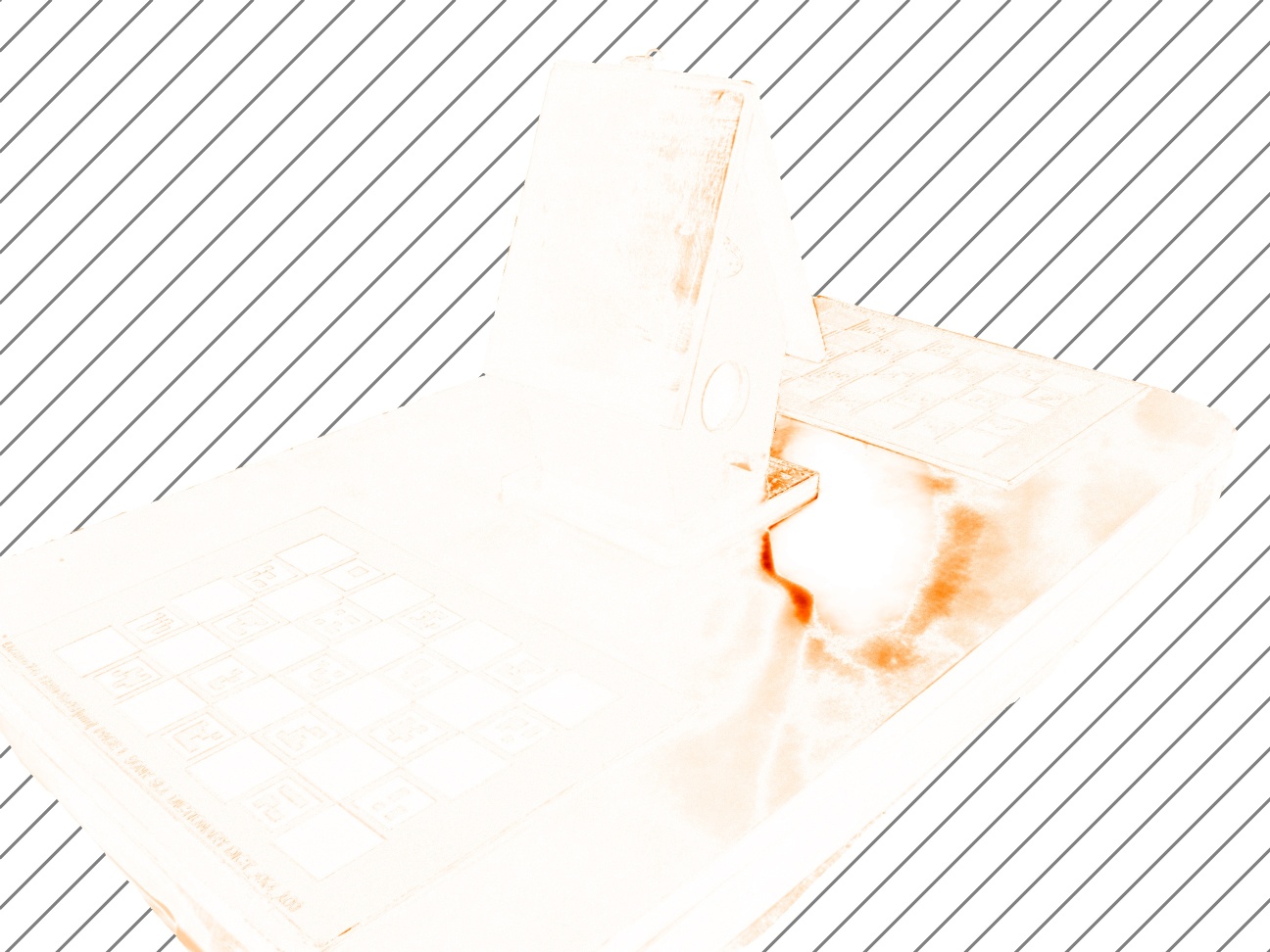} \\
        39 & \includegraphics[width=0.3\linewidth, valign=m]{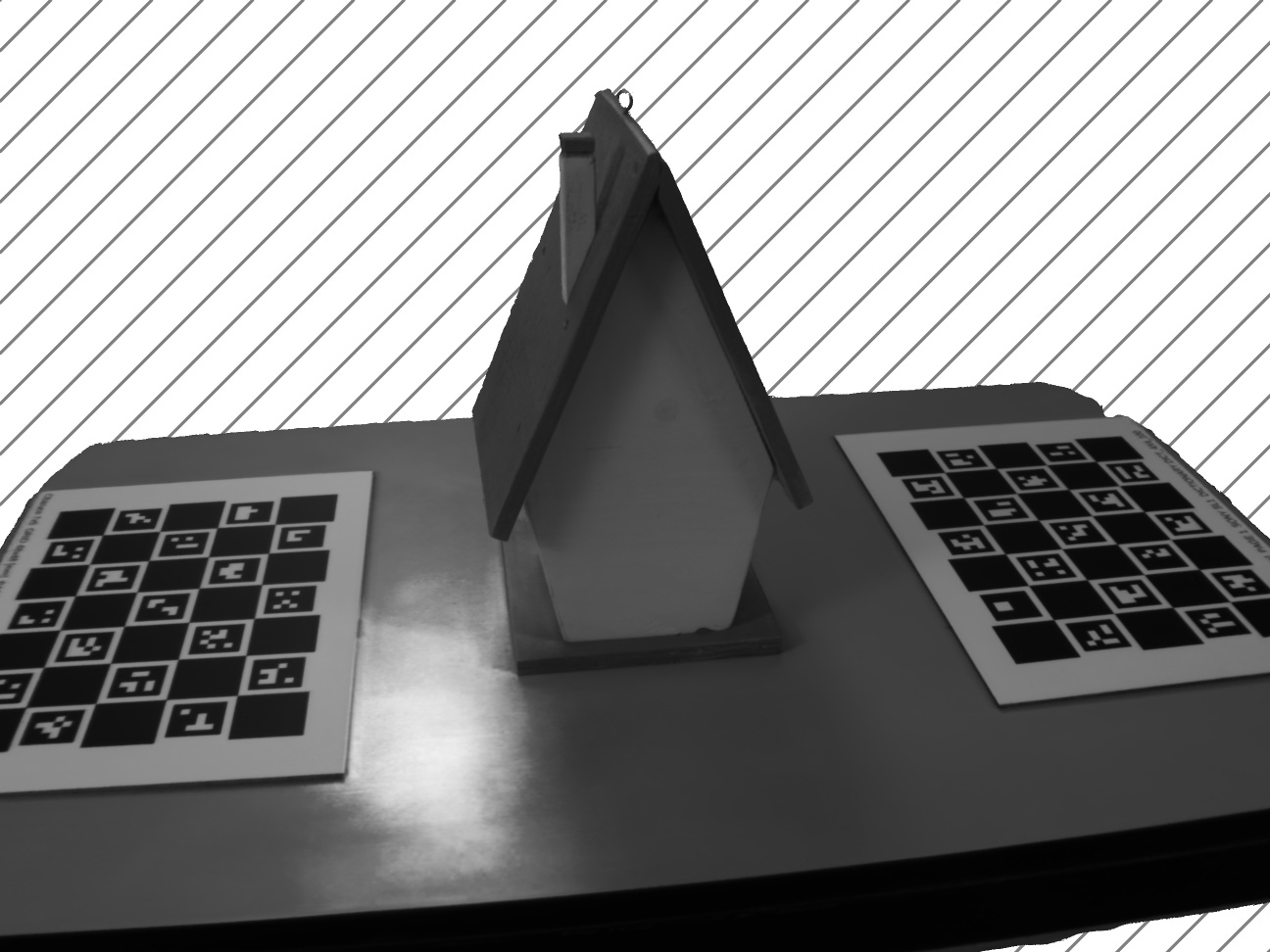} & \includegraphics[width=0.3\linewidth, valign=m]{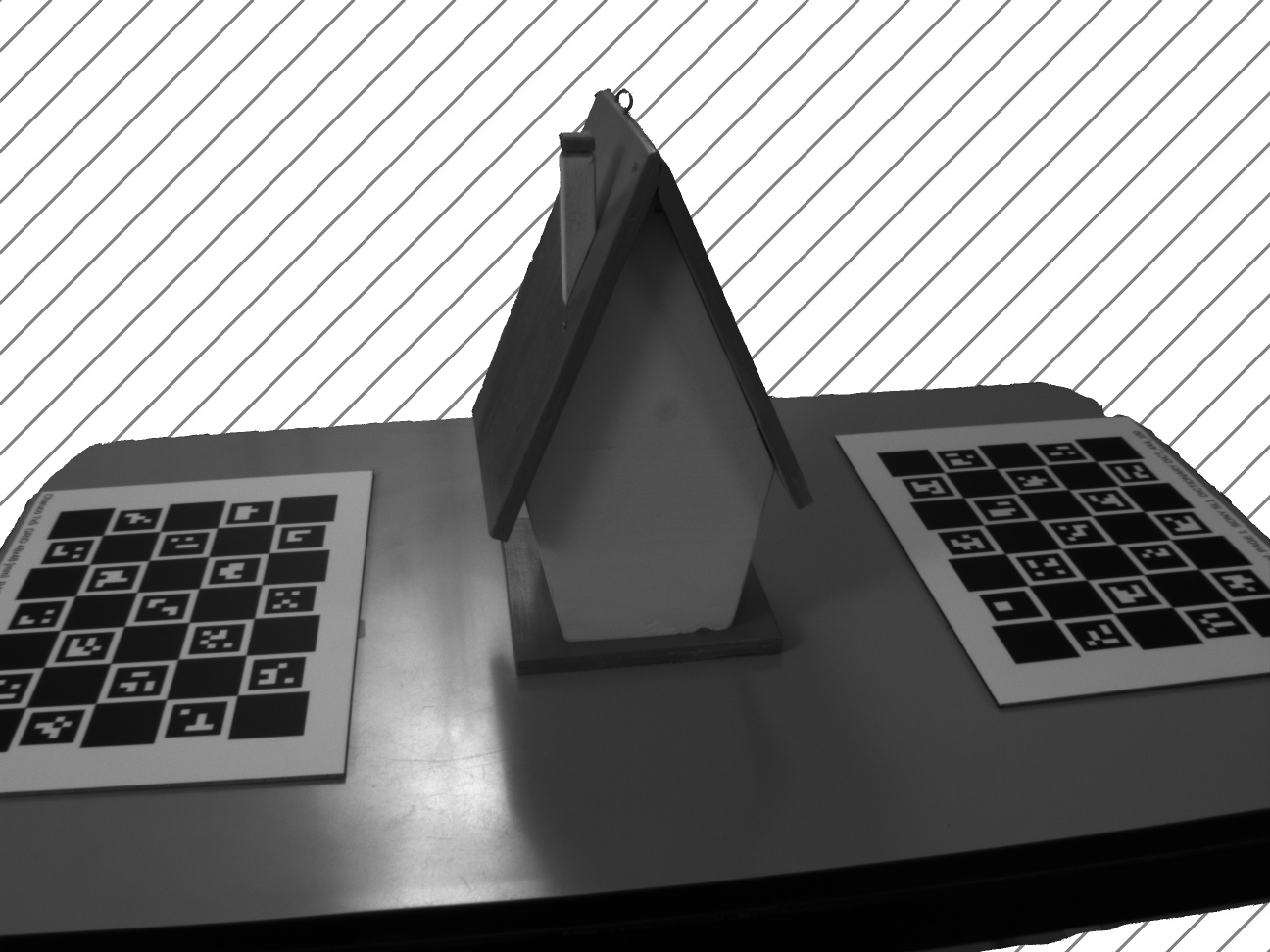} & \includegraphics[width=0.3\linewidth, valign=m]{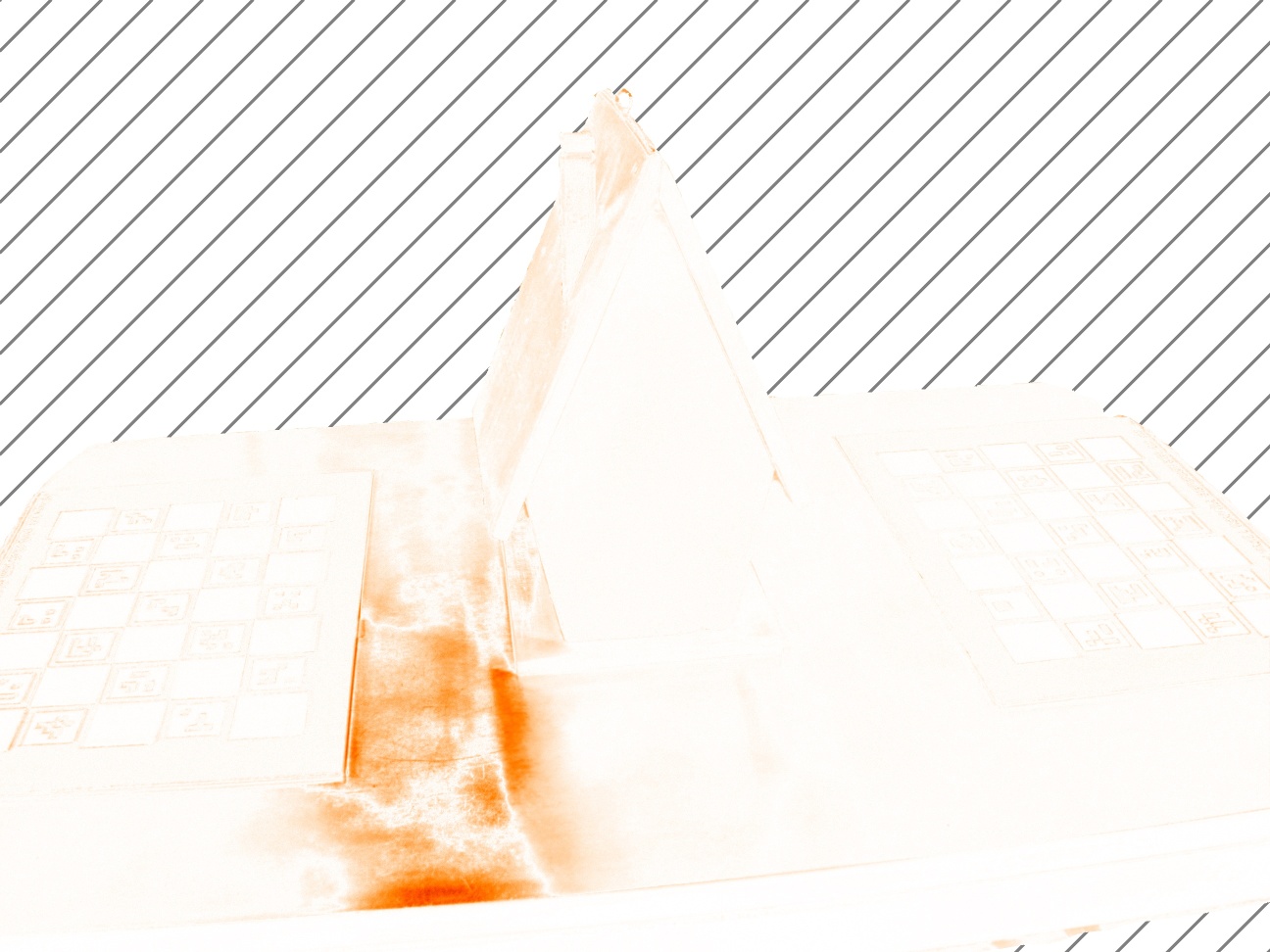} \\
        49 & \includegraphics[width=0.3\linewidth, valign=m]{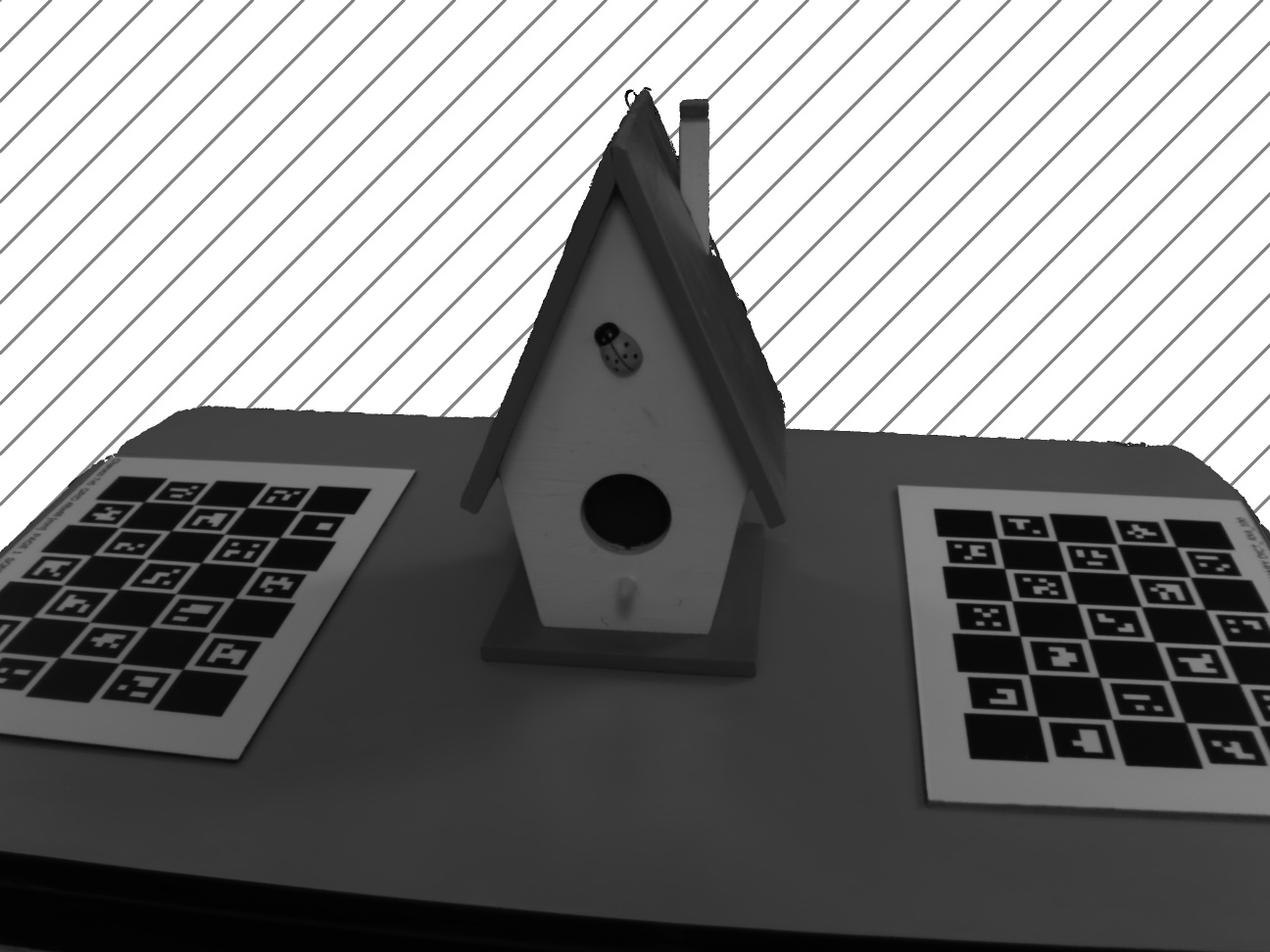} & \includegraphics[width=0.3\linewidth, valign=m]{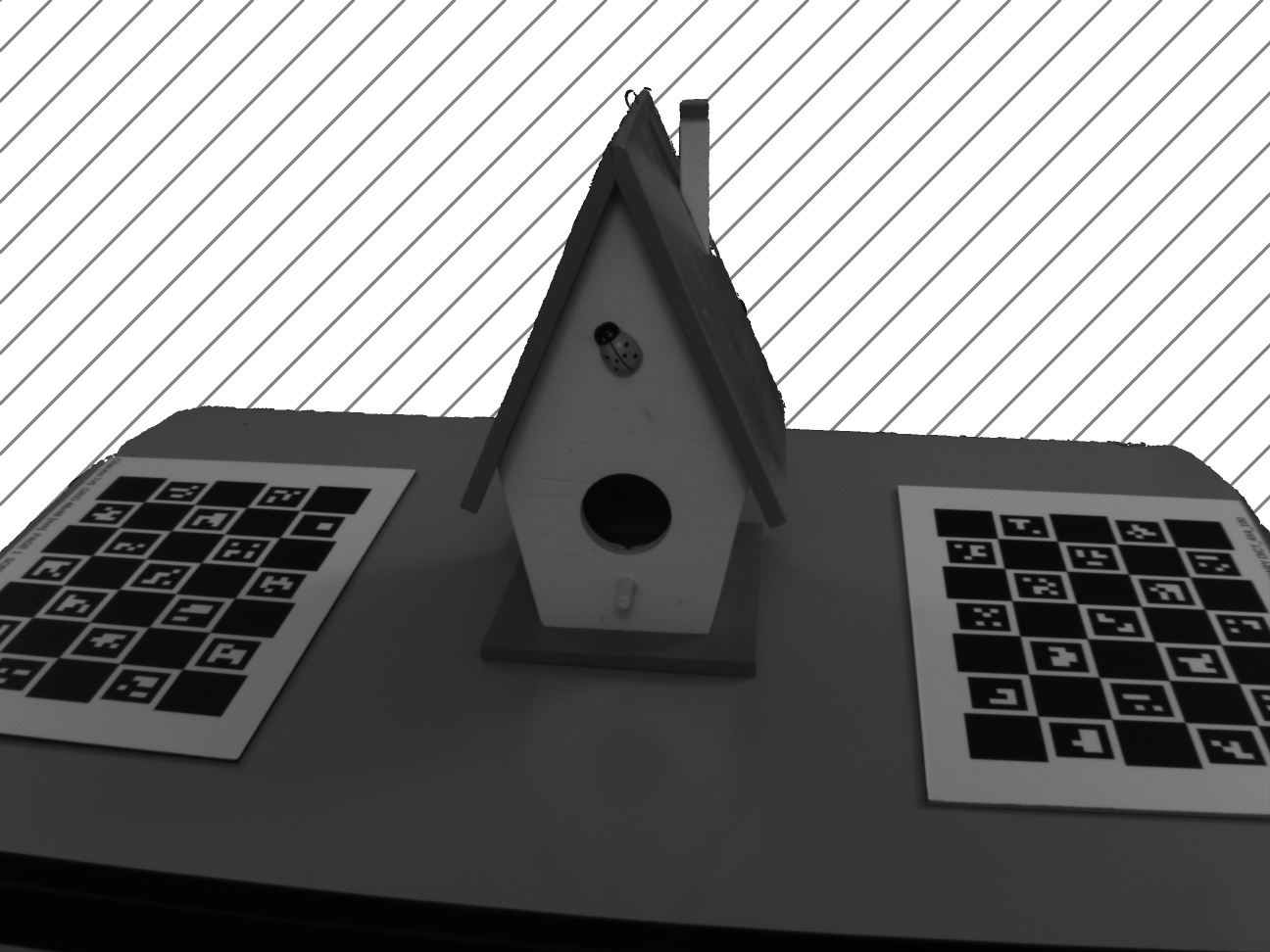} & \includegraphics[width=0.3\linewidth, valign=m]{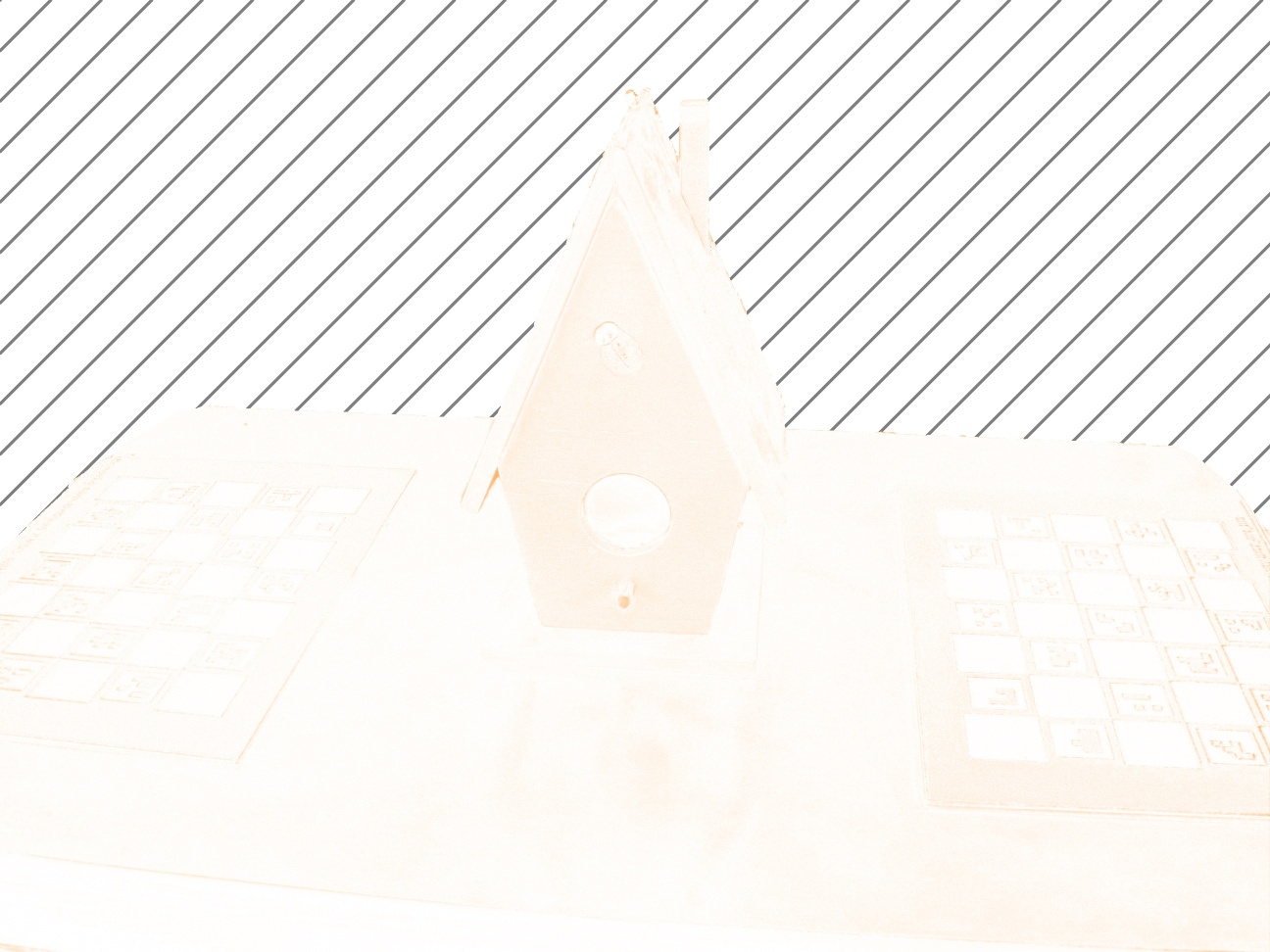} \\
        \midrule
        View & Rendering & Ground Truth & Error \\
    \end{tabular}
    \caption{Mono renderings, ground truth and error maps of the Bird House scene from the five different test views.}
    \label{sup_fig:mono_raw_renedrings}
\end{figure*}

\begin{figure*}
    \centering
    \begin{tabular}{@{\extracolsep{-6pt}}cccc}
        \multicolumn{2}{l}{\hspace{-6pt}Near-Infrared (NIR)} \\
        & & & \\
        View & Rendering & Ground Truth & Error \\
        \midrule
        9 & \includegraphics[width=0.28\linewidth, valign=m]{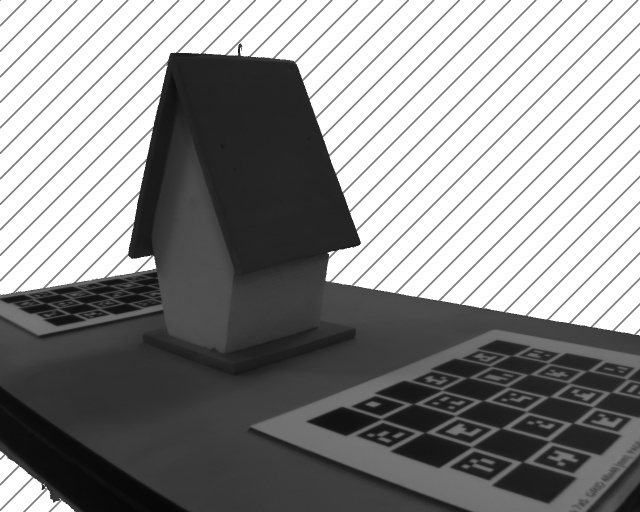} & \includegraphics[width=0.28\linewidth, valign=m]{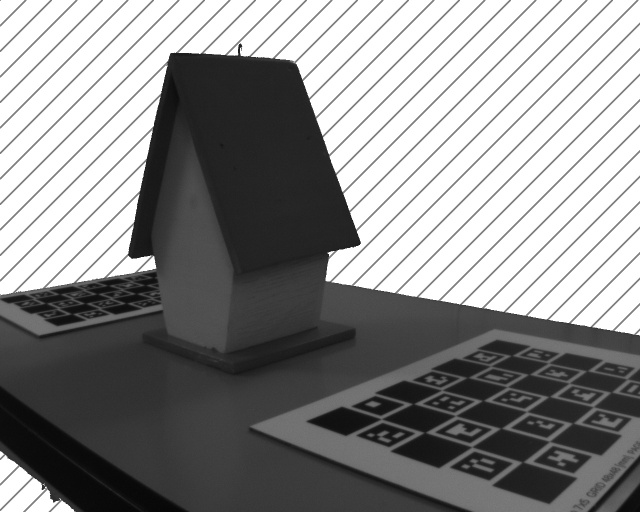} & \includegraphics[width=0.28\linewidth, valign=m]{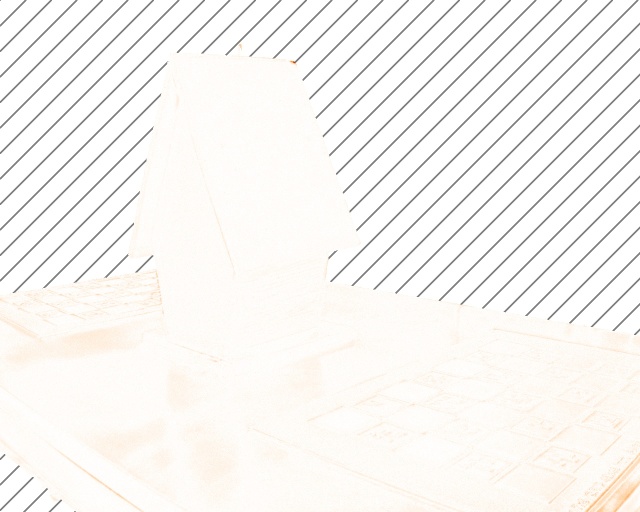} \\
        19 & \includegraphics[width=0.28\linewidth, valign=m]{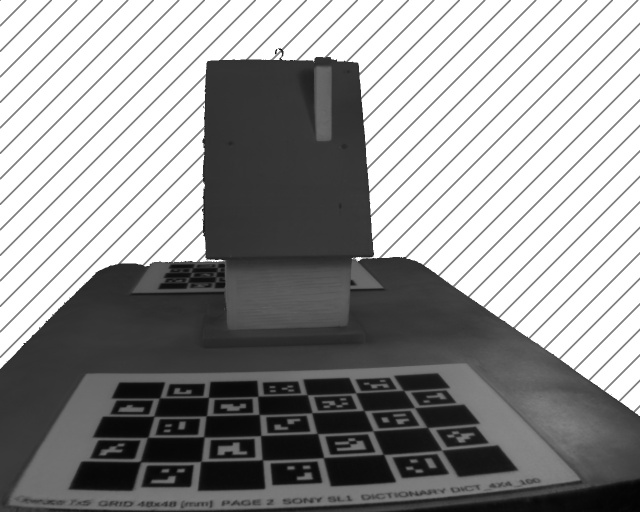} & \includegraphics[width=0.28\linewidth, valign=m]{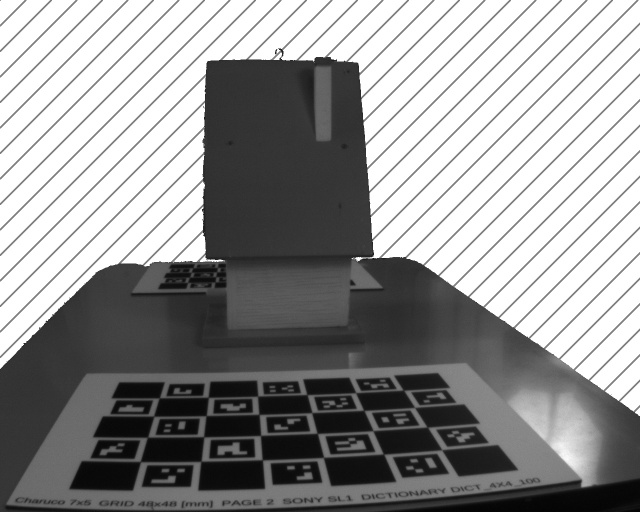} & \includegraphics[width=0.28\linewidth, valign=m]{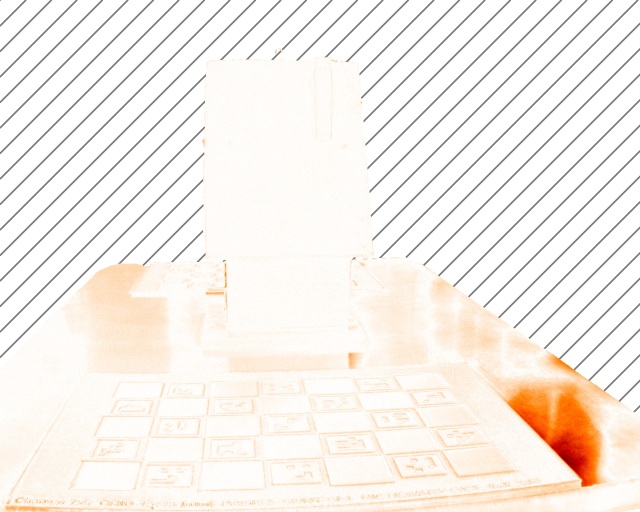} \\
        29 & \includegraphics[width=0.28\linewidth, valign=m]{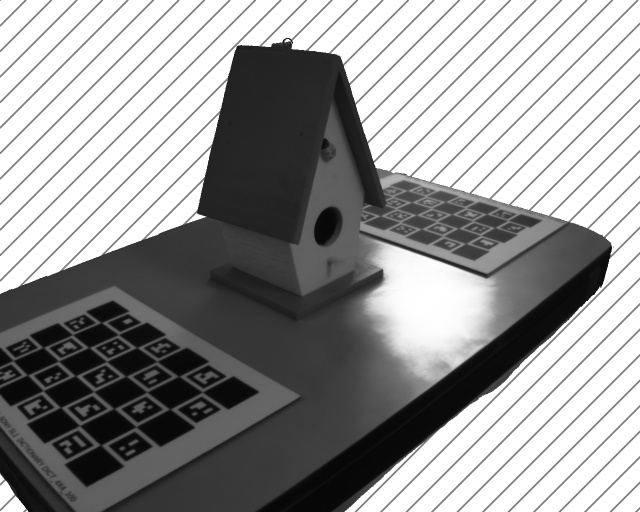} & \includegraphics[width=0.28\linewidth, valign=m]{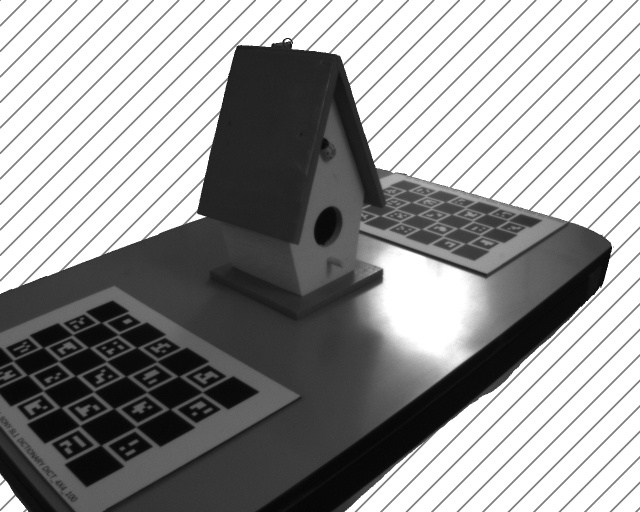} & \includegraphics[width=0.28\linewidth, valign=m]{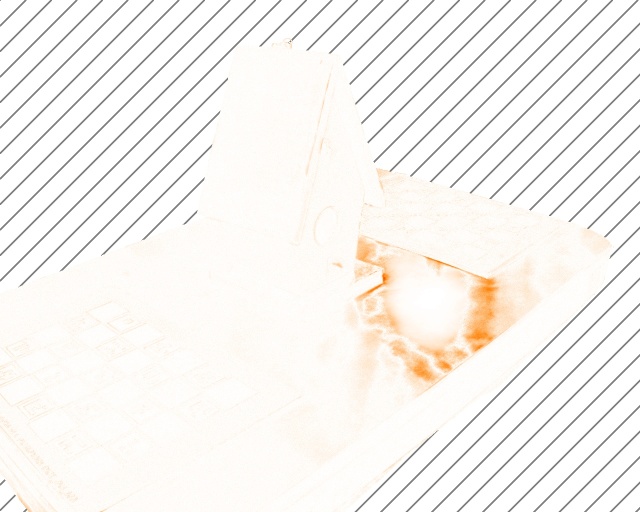} \\
        39 & \includegraphics[width=0.28\linewidth, valign=m]{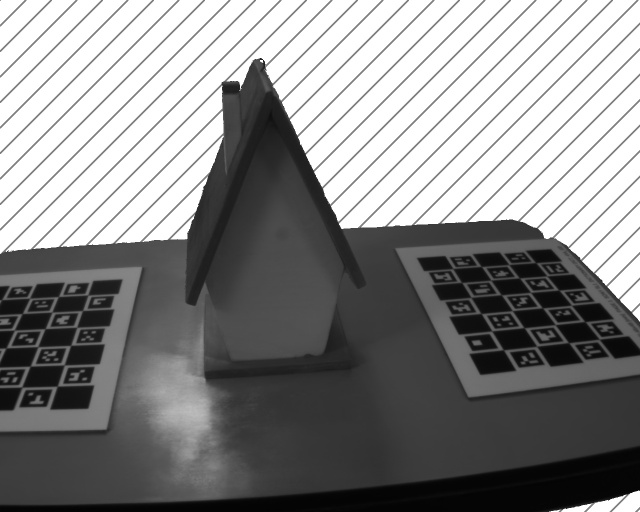} & \includegraphics[width=0.28\linewidth, valign=m]{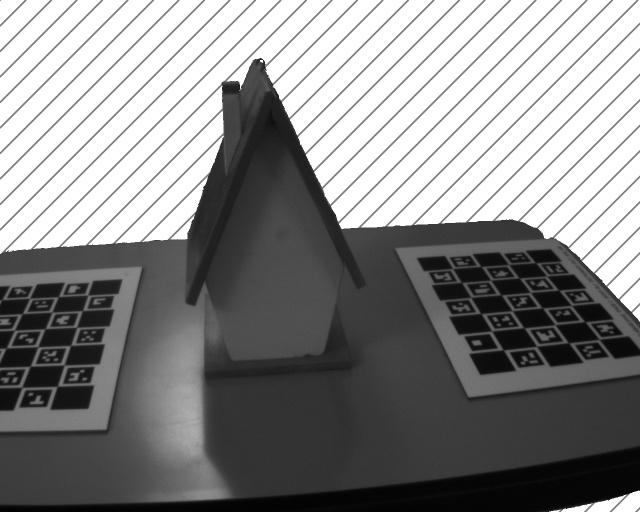} & \includegraphics[width=0.28\linewidth, valign=m]{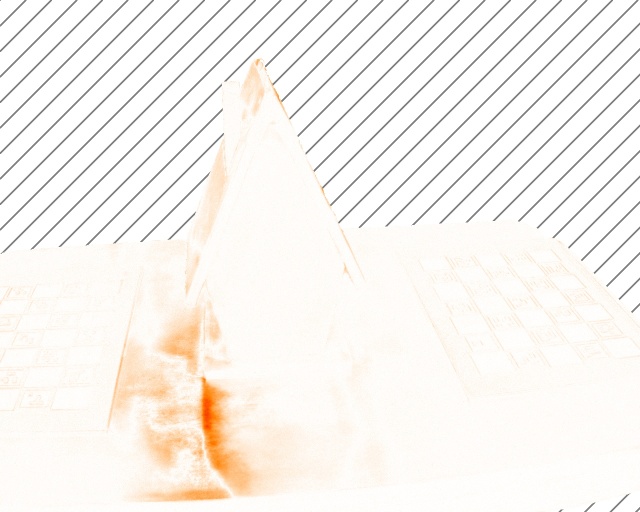} \\
        49 & \includegraphics[width=0.28\linewidth, valign=m]{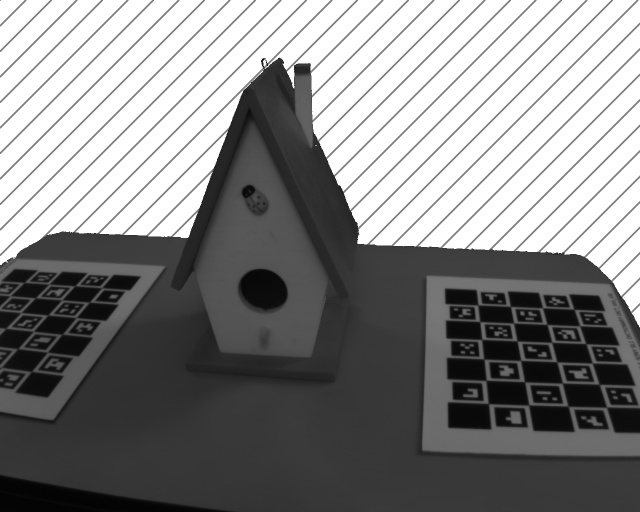} & \includegraphics[width=0.28\linewidth, valign=m]{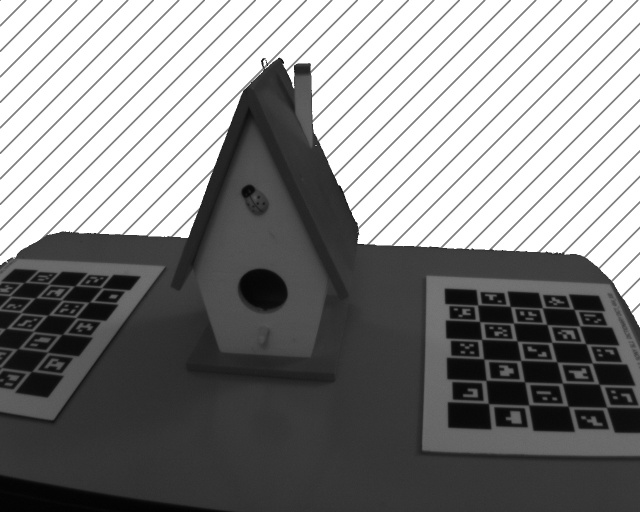} & \includegraphics[width=0.28\linewidth, valign=m]{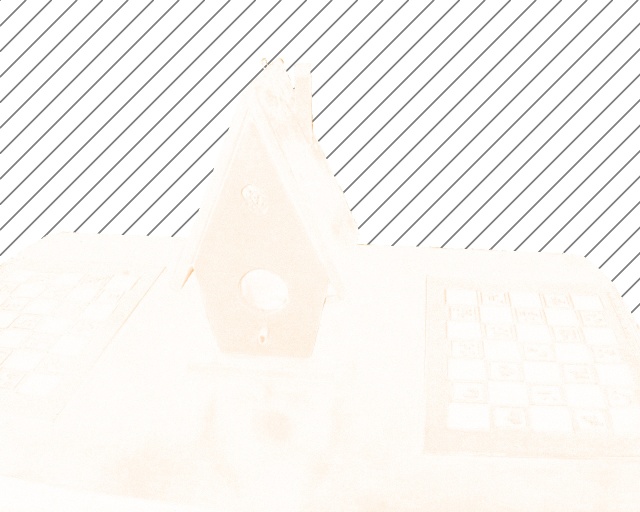} \\
        \midrule
        View & Rendering & Ground Truth & Error \\
    \end{tabular}
    \caption{Near-Infrared renderings, ground truth and error maps of the Bird House scene from the five different test views.}
    \label{sup_fig:infrared_raw_renedrings}
\end{figure*}

\begin{figure*}
    \centering
    \begin{tabular}{@{\extracolsep{-6pt}}cccc}
        \multicolumn{2}{l}{\hspace{-6pt}Polarization (Pol)} \\
        & & & \\
        View & Rendering & Ground Truth & Error \\
        \midrule
        9 & \includegraphics[width=0.27\linewidth, valign=m]{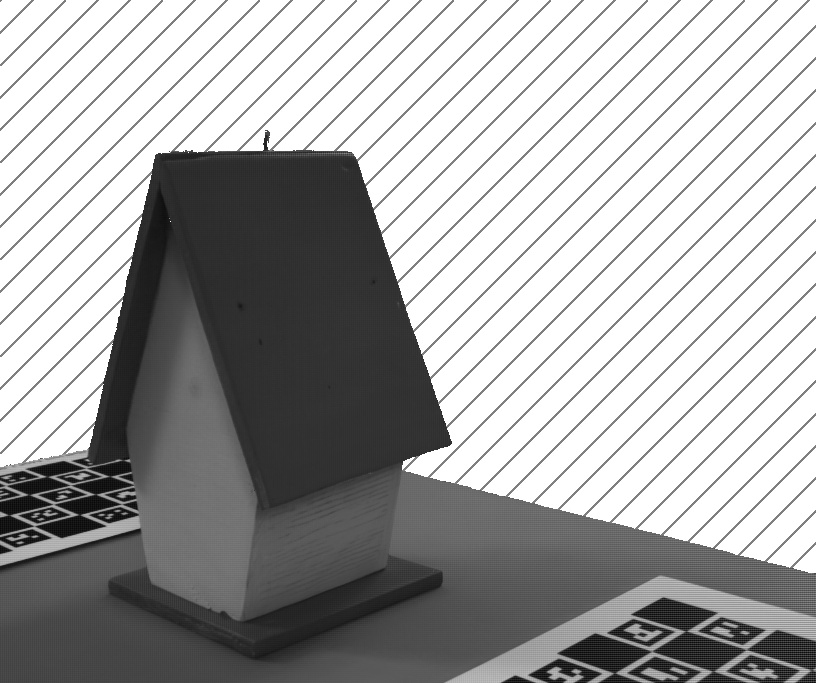} & \includegraphics[width=0.27\linewidth, valign=m]{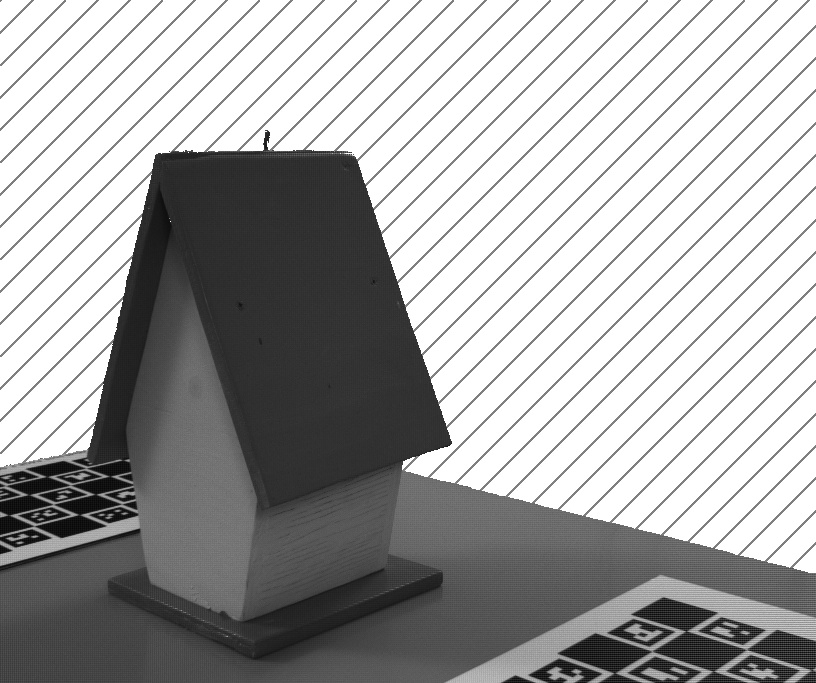} & \includegraphics[width=0.27\linewidth, valign=m]{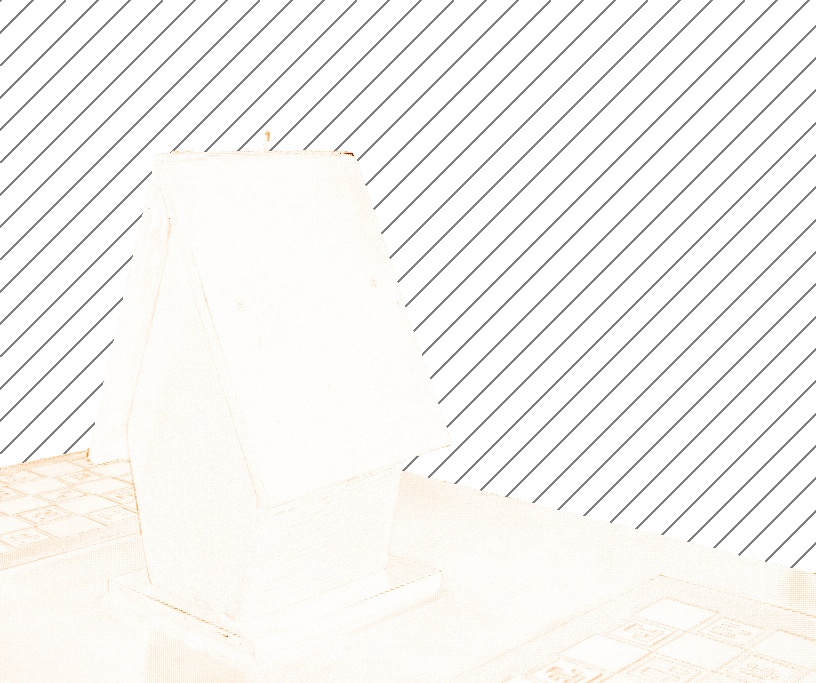} \\
        19 & \includegraphics[width=0.27\linewidth, valign=m]{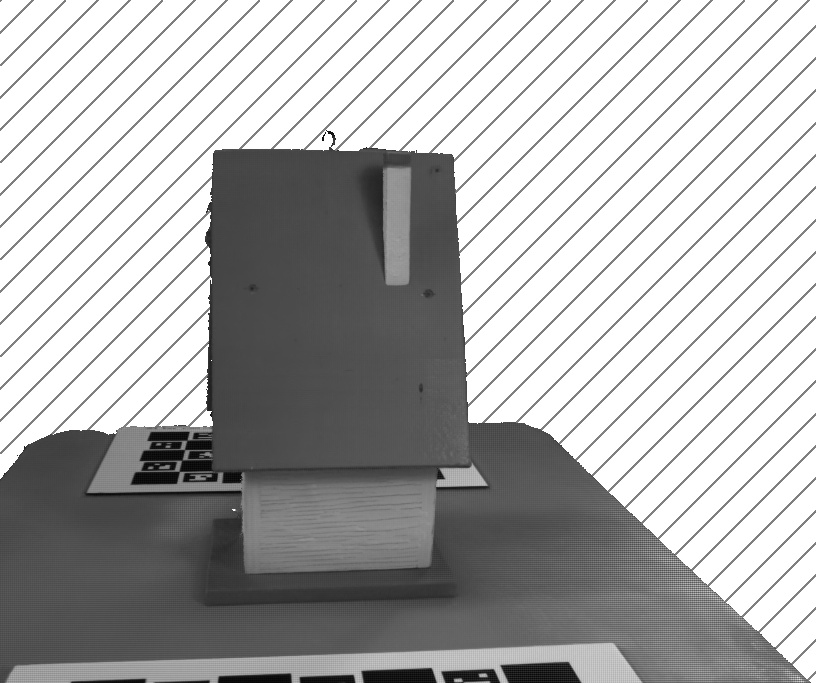} & \includegraphics[width=0.27\linewidth, valign=m]{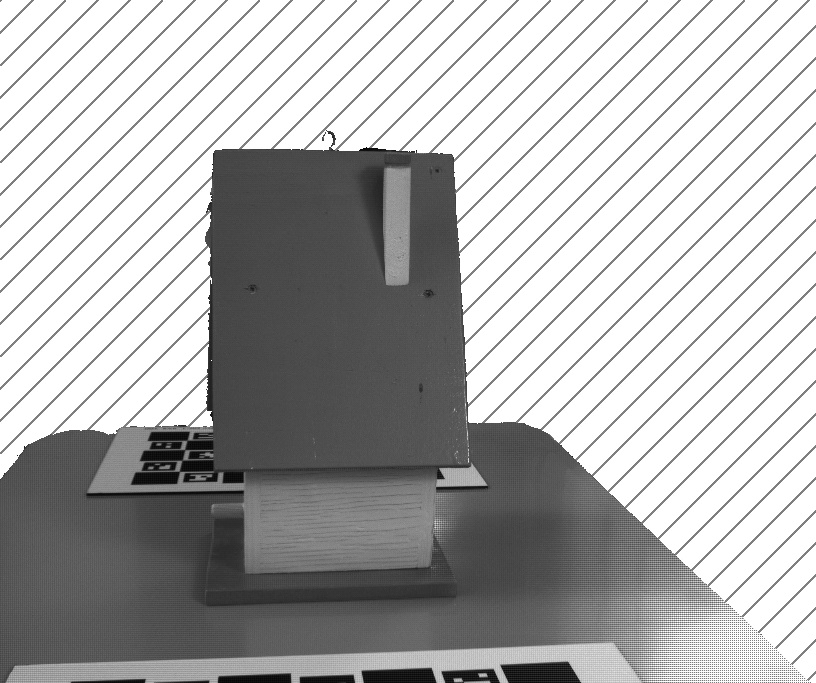} & \includegraphics[width=0.27\linewidth, valign=m]{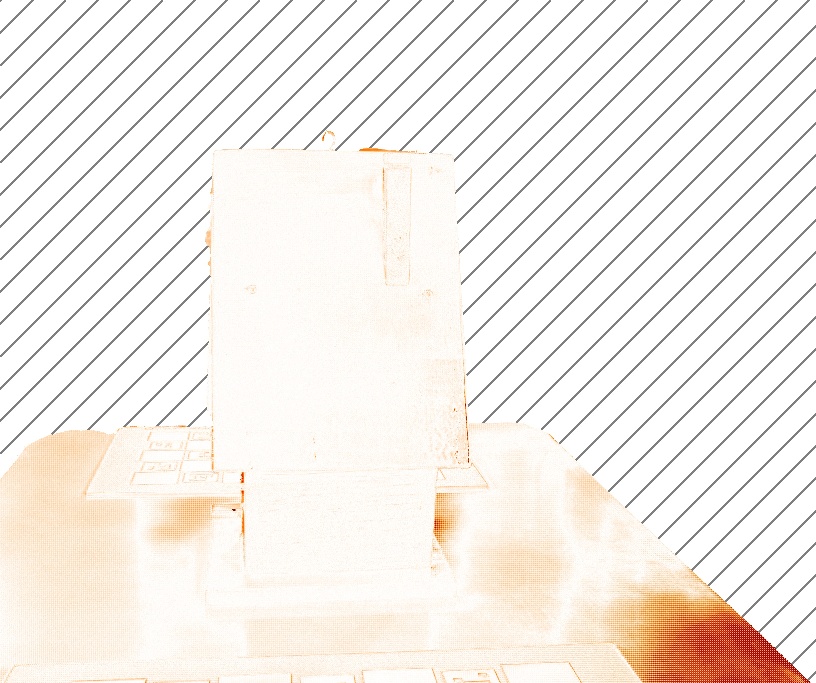} \\
        29 & \includegraphics[width=0.27\linewidth, valign=m]{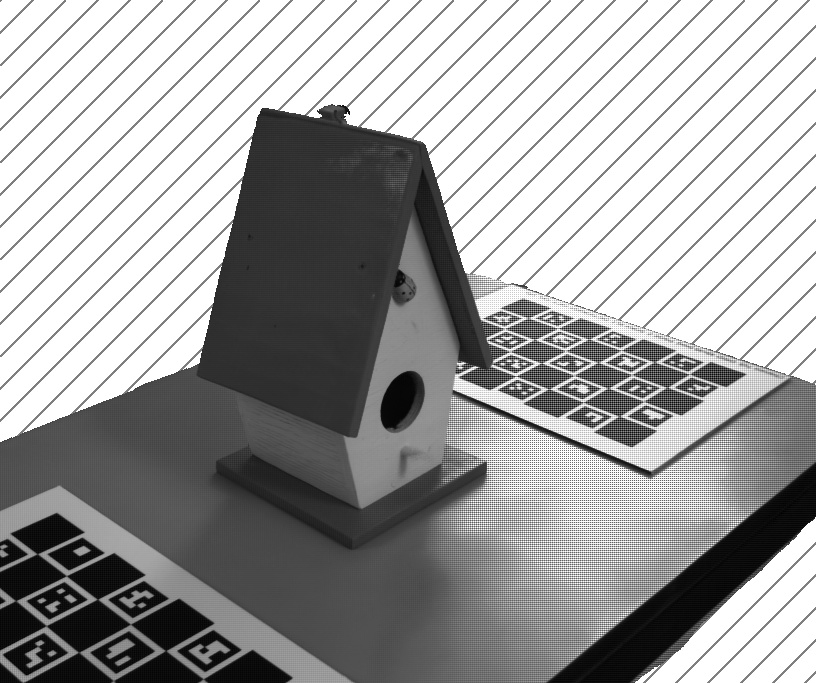} & \includegraphics[width=0.27\linewidth, valign=m]{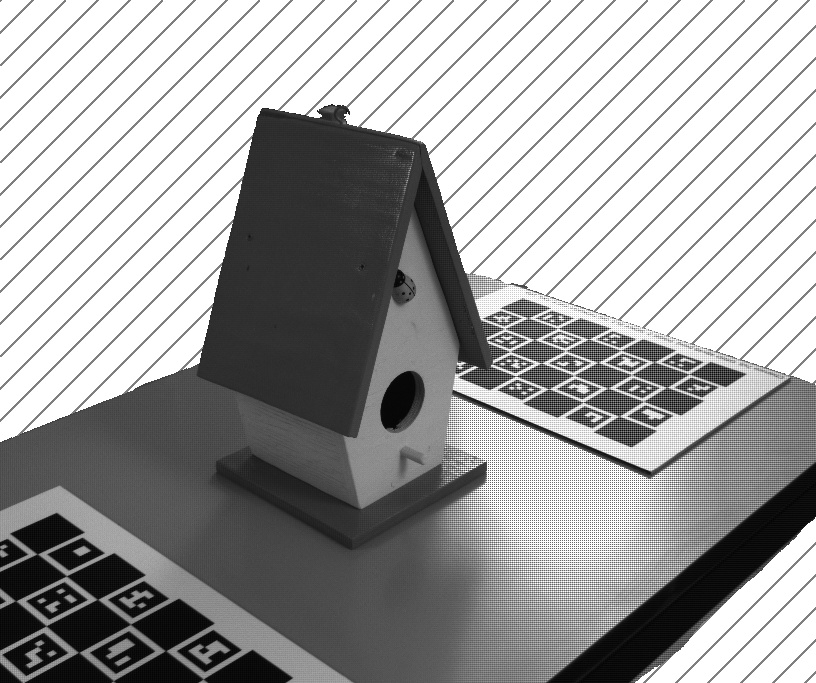} & \includegraphics[width=0.27\linewidth, valign=m]{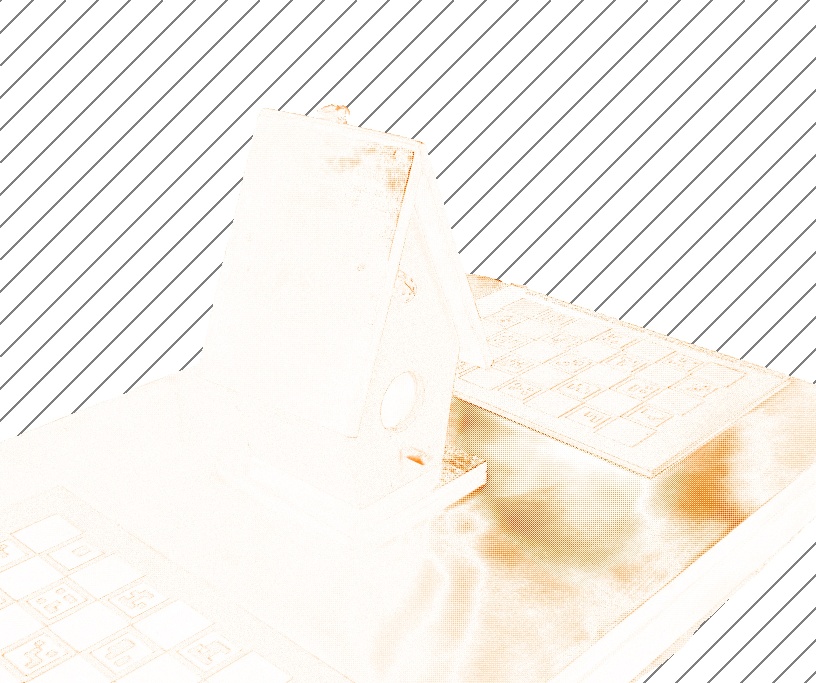} \\
        39 & \includegraphics[width=0.27\linewidth, valign=m]{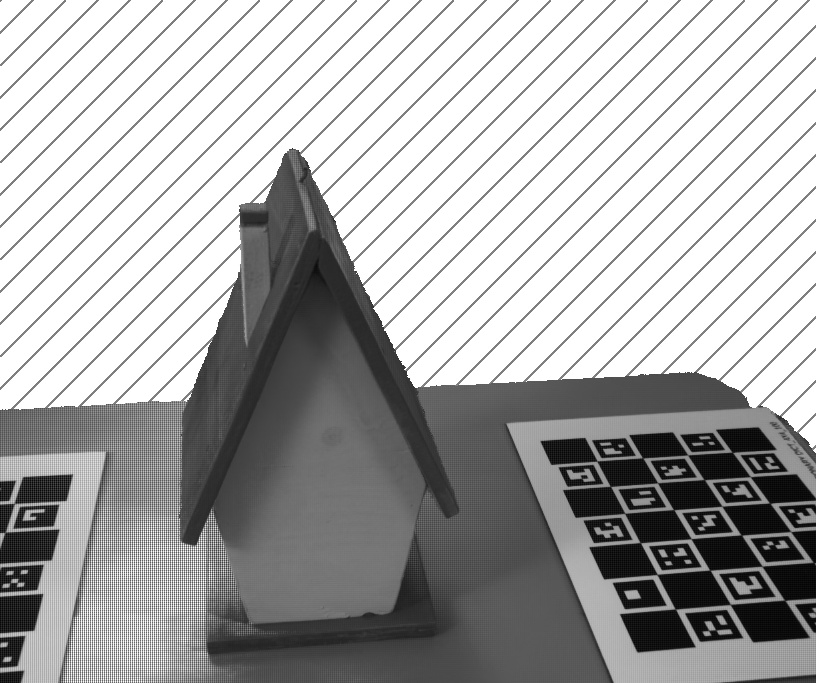} & \includegraphics[width=0.27\linewidth, valign=m]{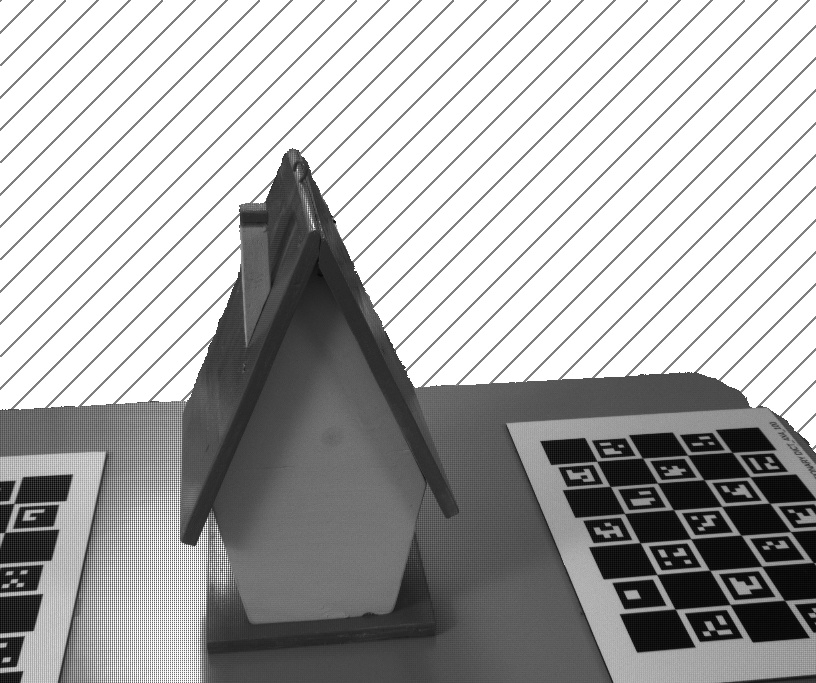} & \includegraphics[width=0.27\linewidth, valign=m]{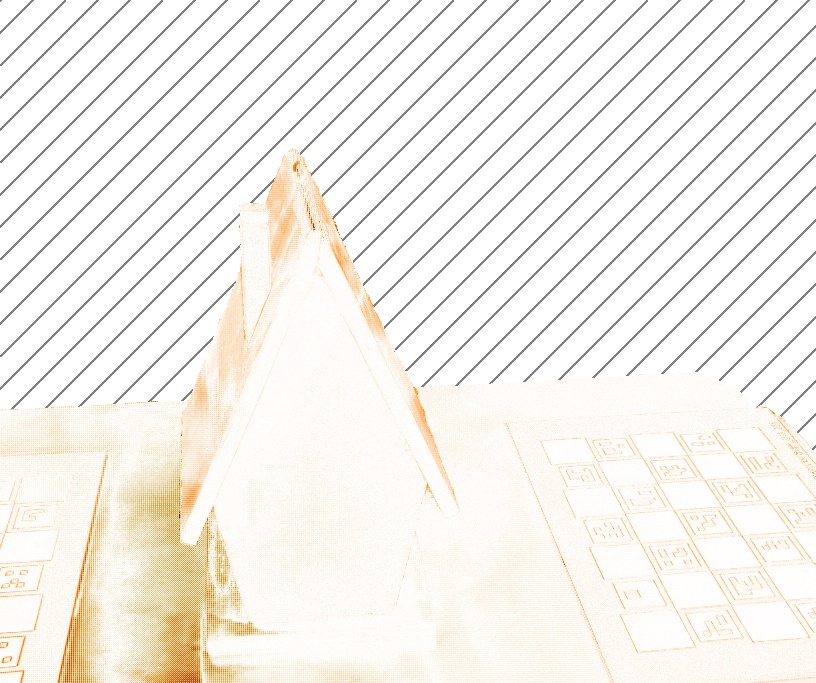} \\
        49 & \includegraphics[width=0.27\linewidth, valign=m]{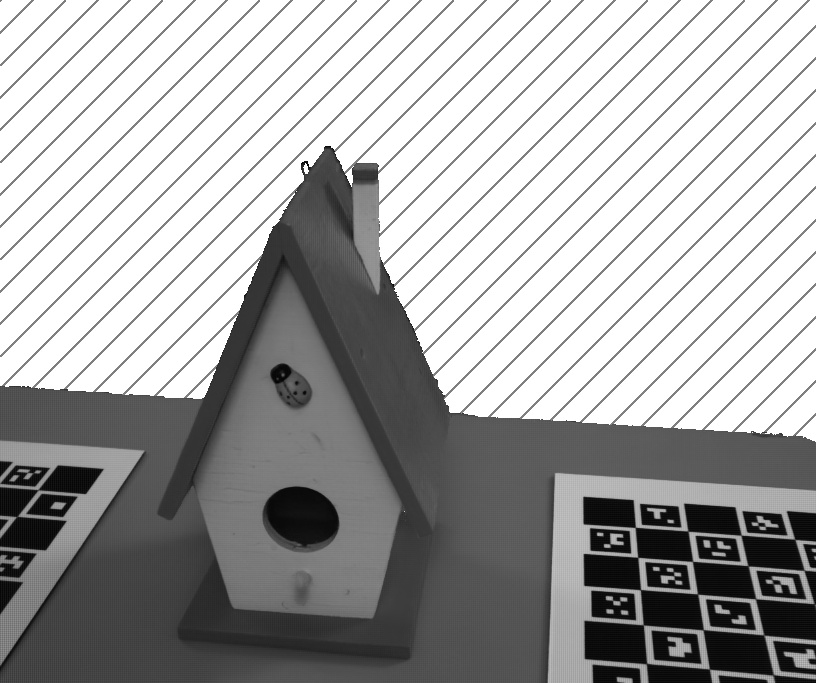} & \includegraphics[width=0.27\linewidth, valign=m]{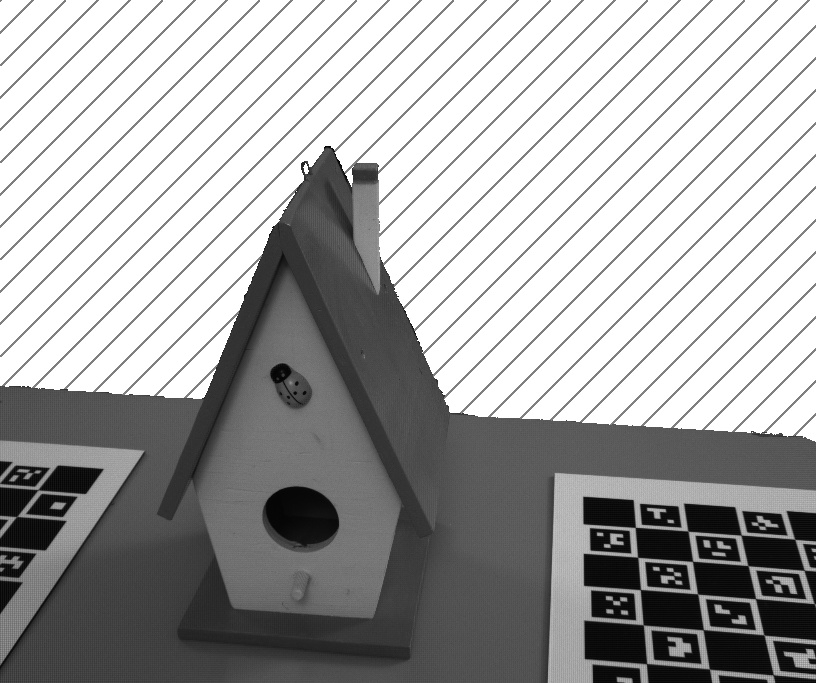} & \includegraphics[width=0.27\linewidth, valign=m]{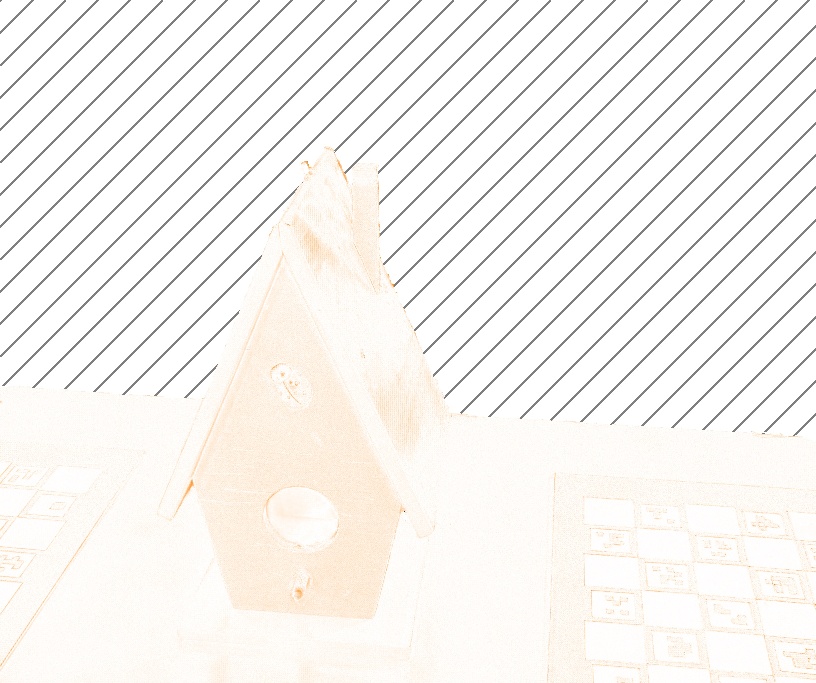} \\
        \midrule
        View & Rendering & Ground Truth & Error \\
    \end{tabular}
    \caption{Mosaicked Polarization renderings, ground truth and error maps of the Bird House scene from the five different test views.}
    \label{sup_fig:polarization_raw_renedrings}
\end{figure*}

\begin{figure*}
    \centering
    \begin{tabular}{@{\extracolsep{-6pt}}cccc}
        \multicolumn{2}{l}{\hspace{-6pt}Multispectral (MS)} \\
        & & & \\
        View & Rendering & Ground Truth & Error \\
        \midrule
        9 & \includegraphics[width=0.29\linewidth, valign=m]{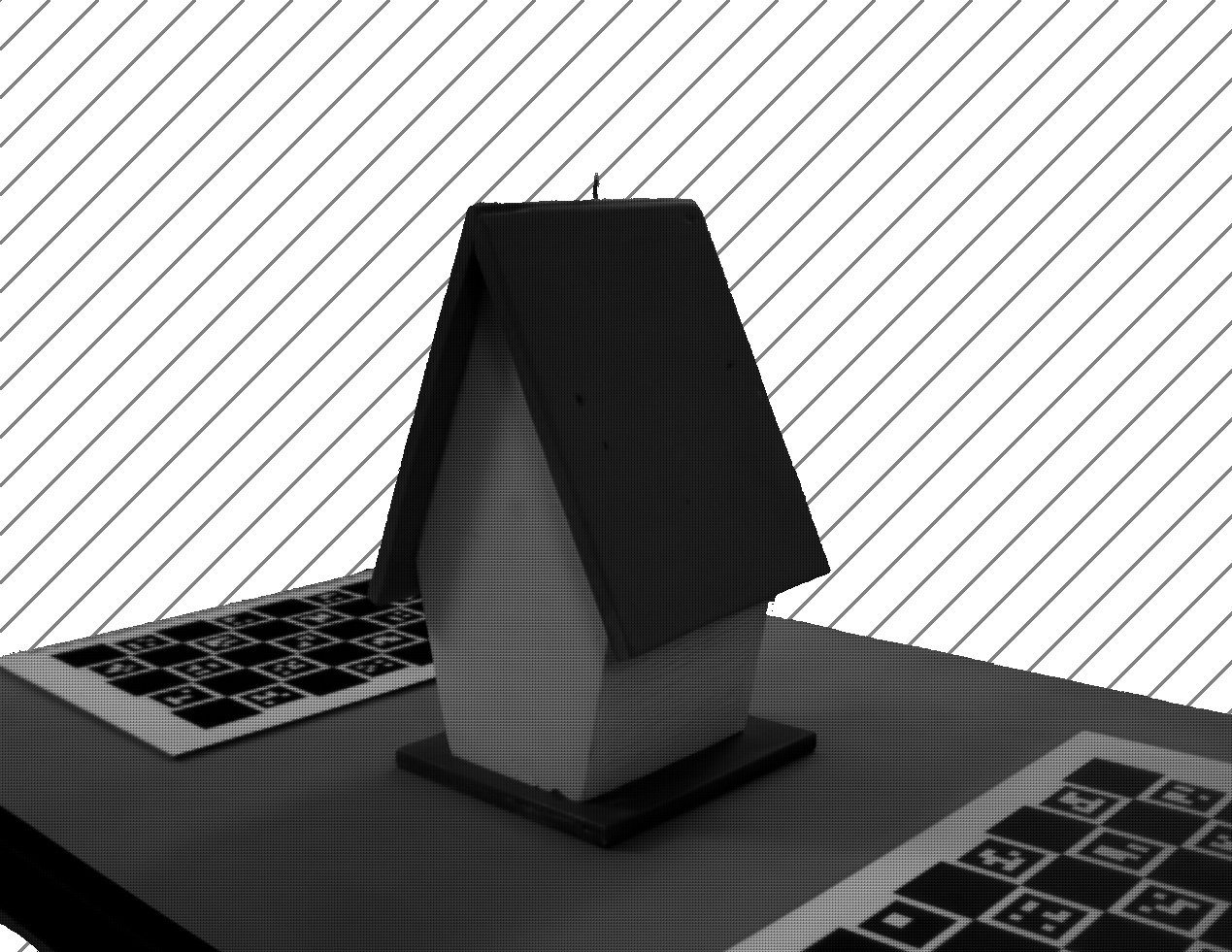} & \includegraphics[width=0.29\linewidth, valign=m]{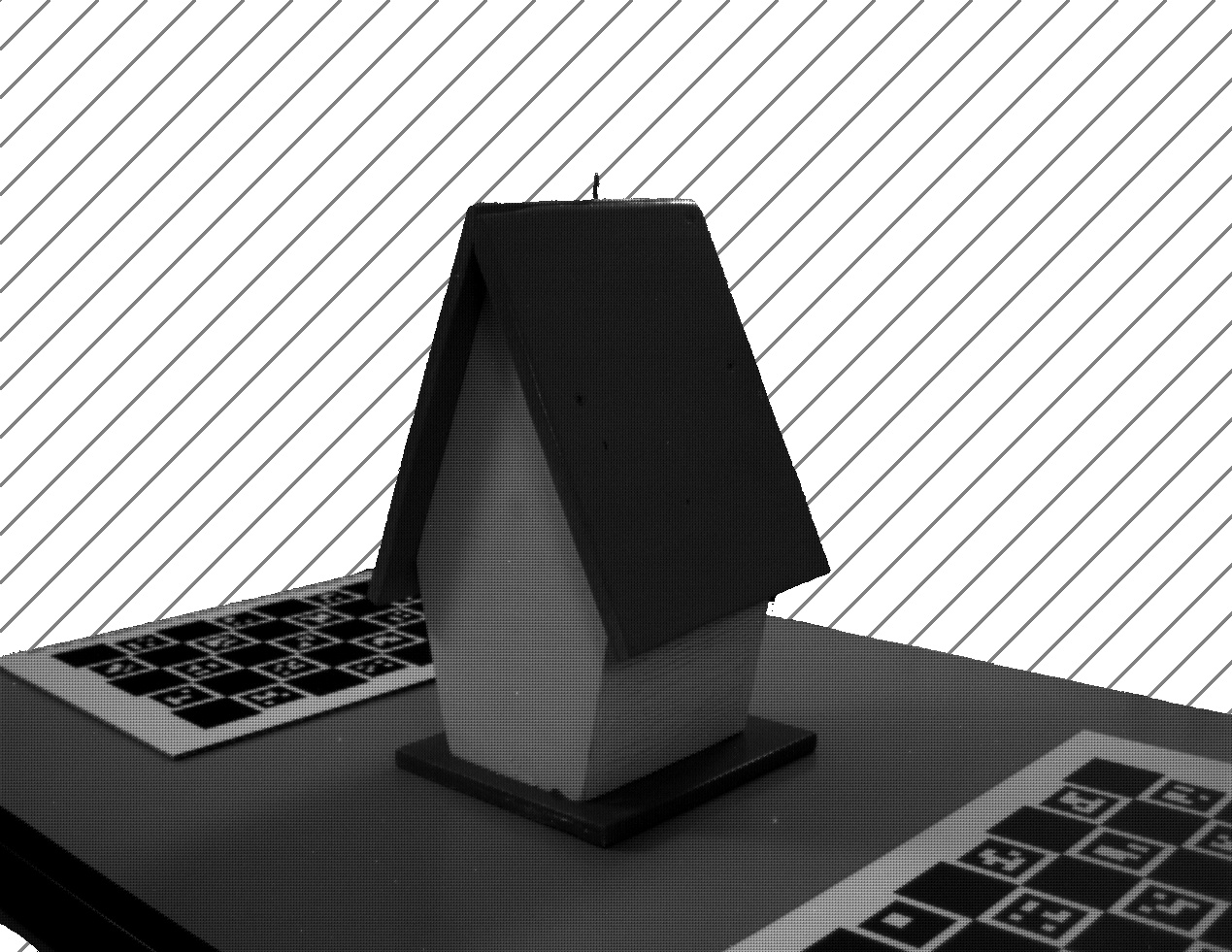} & \includegraphics[width=0.29\linewidth, valign=m]{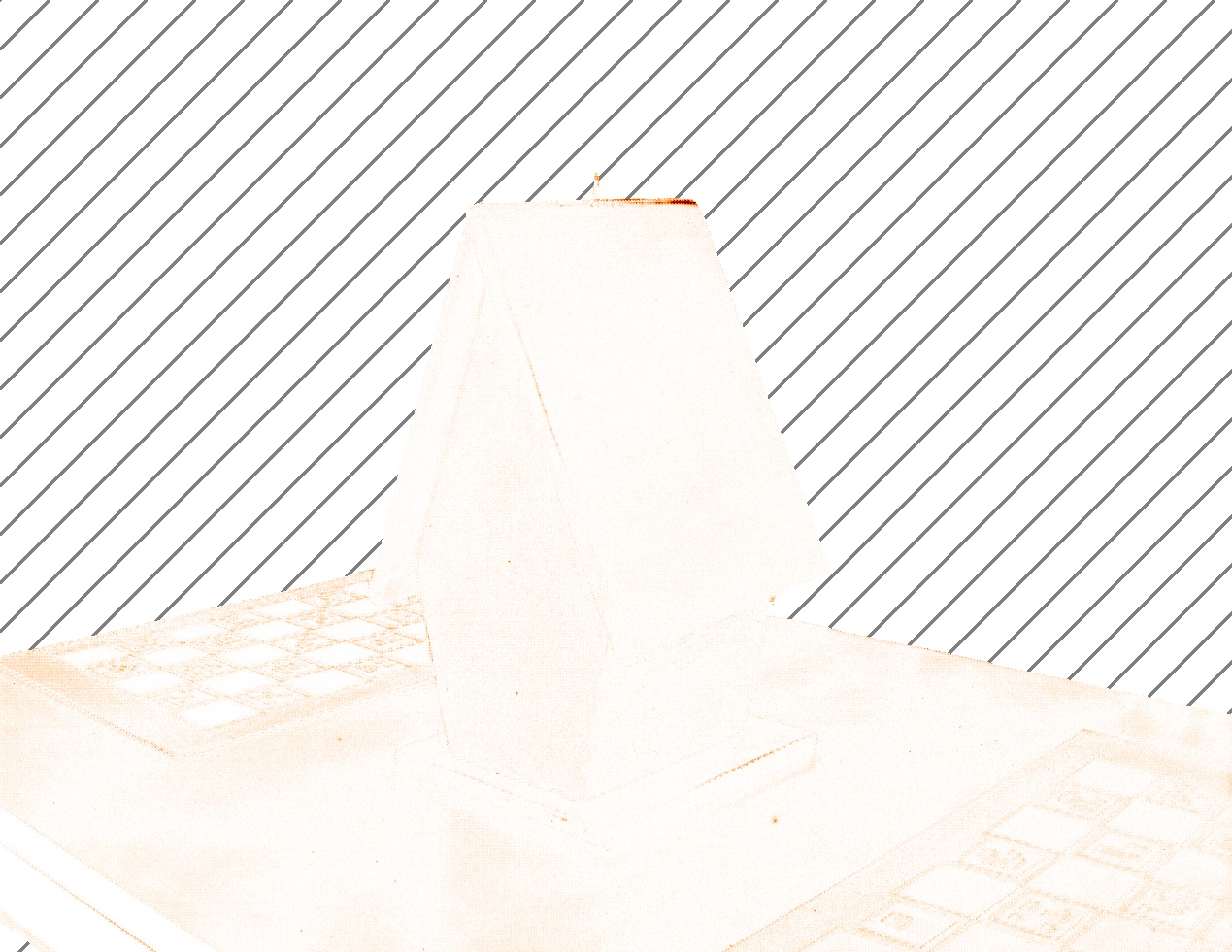} \\
        19 & \includegraphics[width=0.29\linewidth, valign=m]{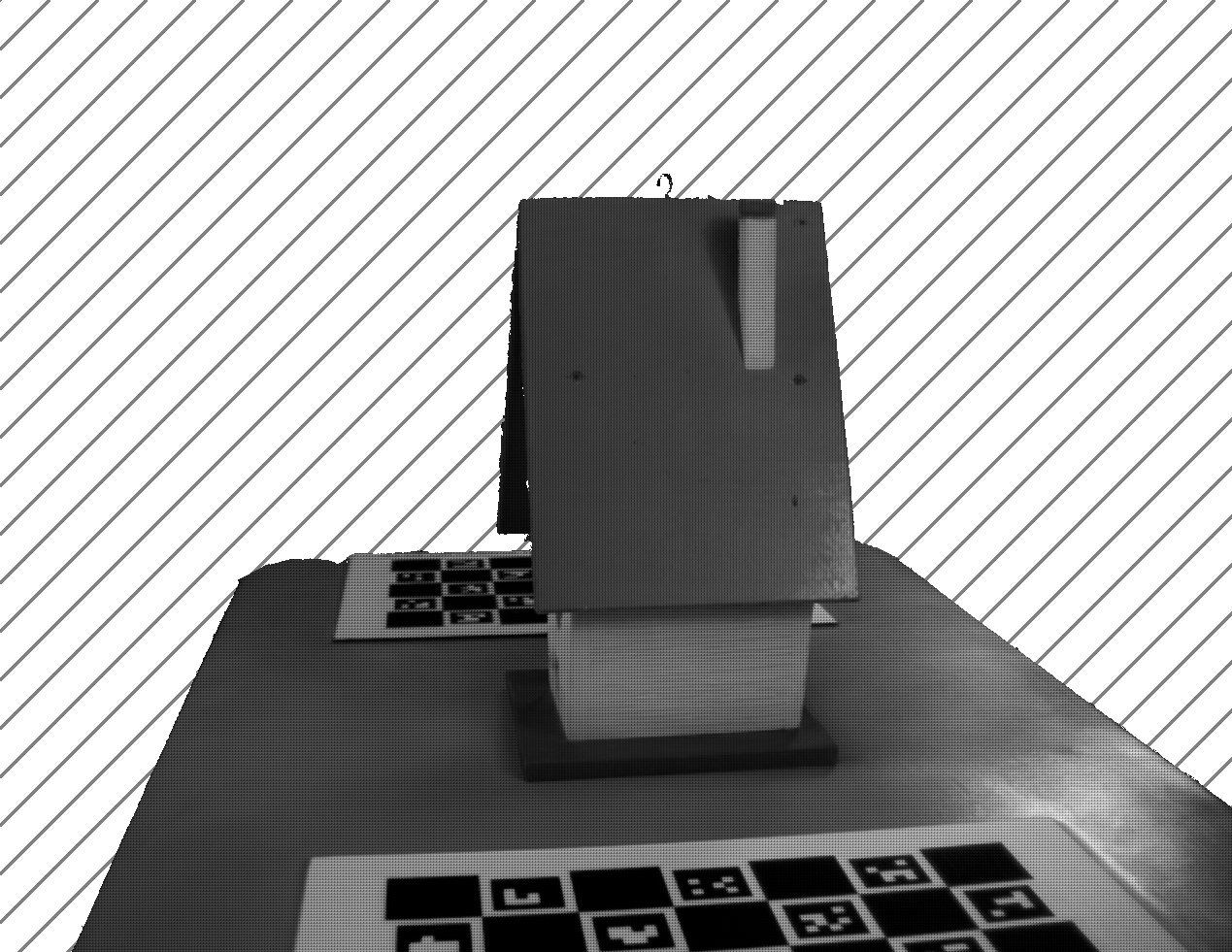} & \includegraphics[width=0.29\linewidth, valign=m]{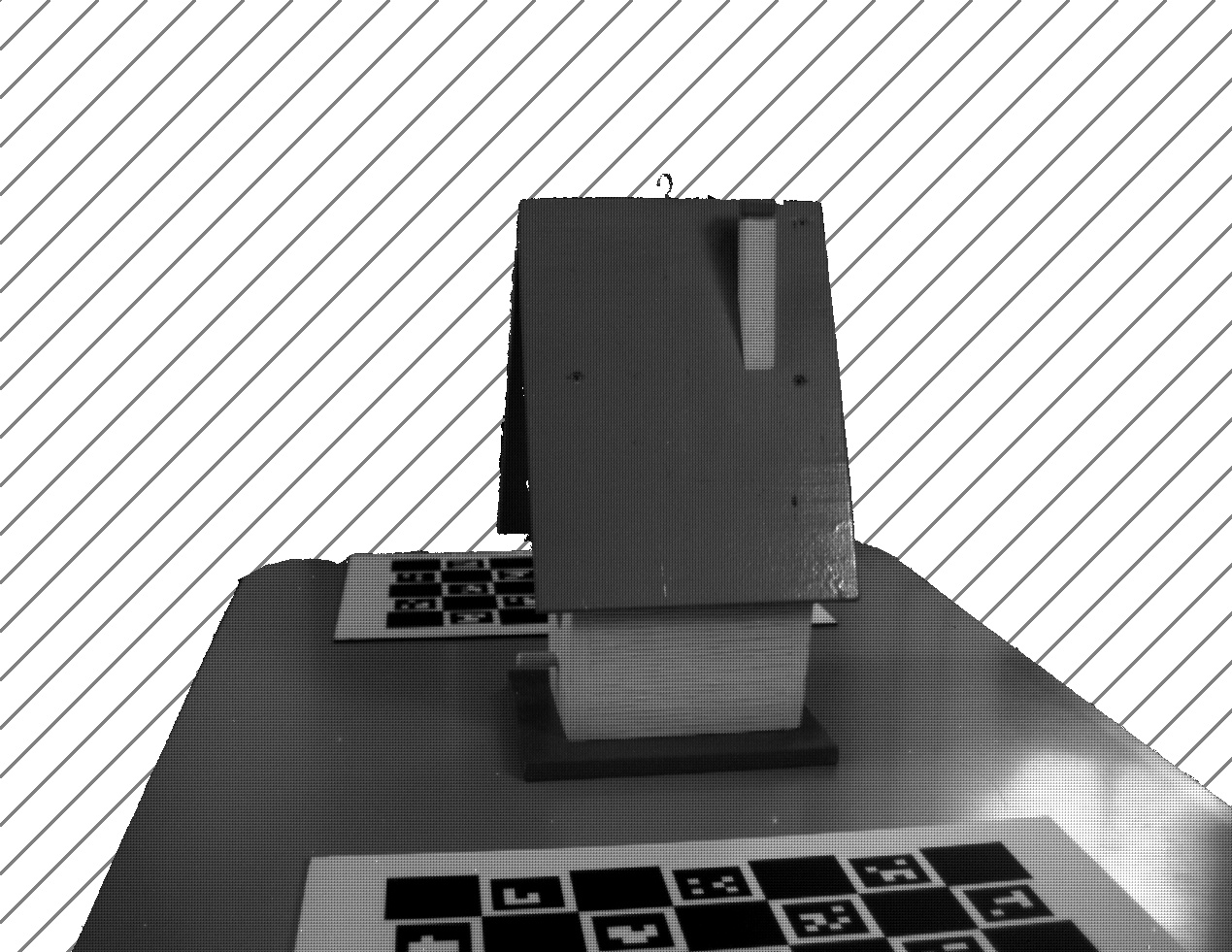} & \includegraphics[width=0.29\linewidth, valign=m]{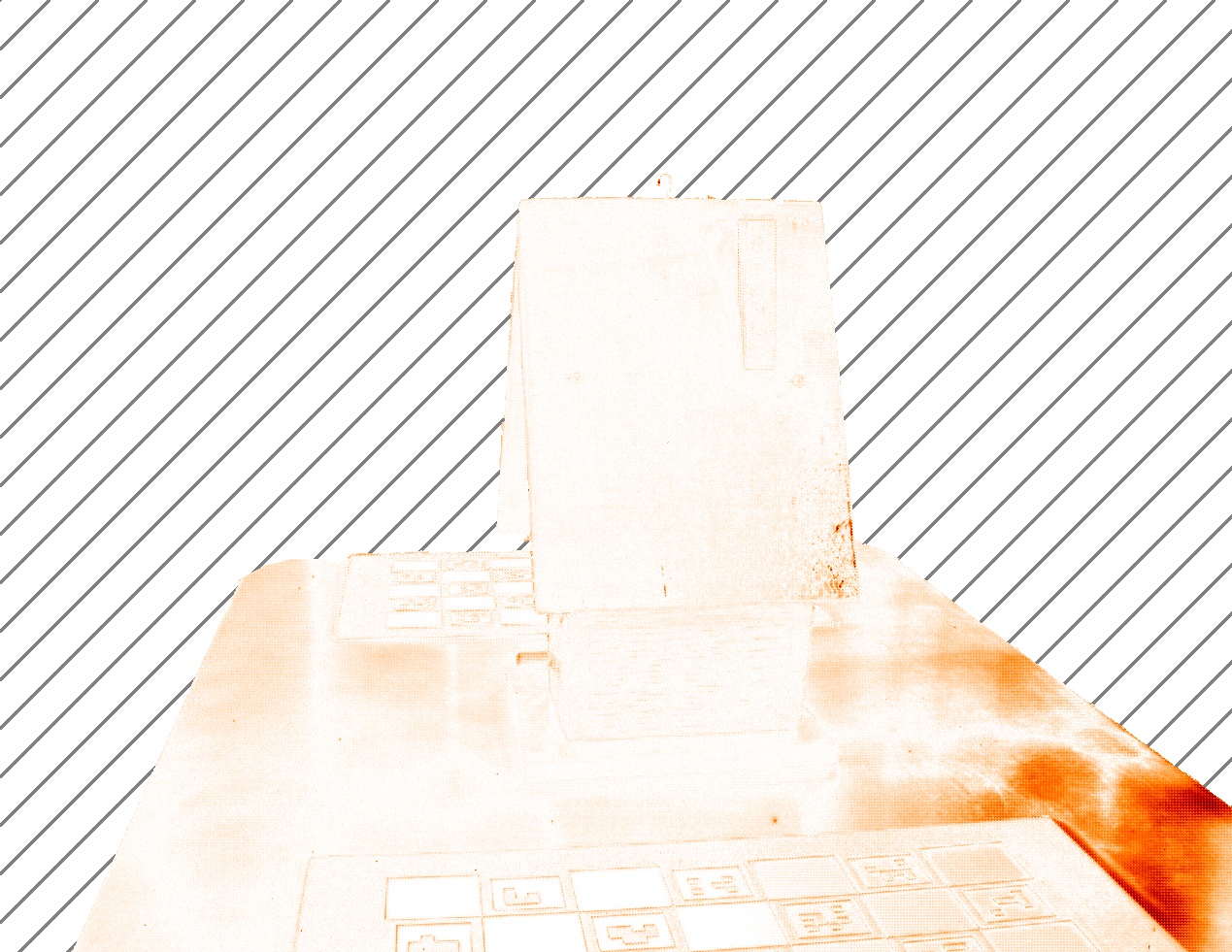} \\
        29 & \includegraphics[width=0.29\linewidth, valign=m]{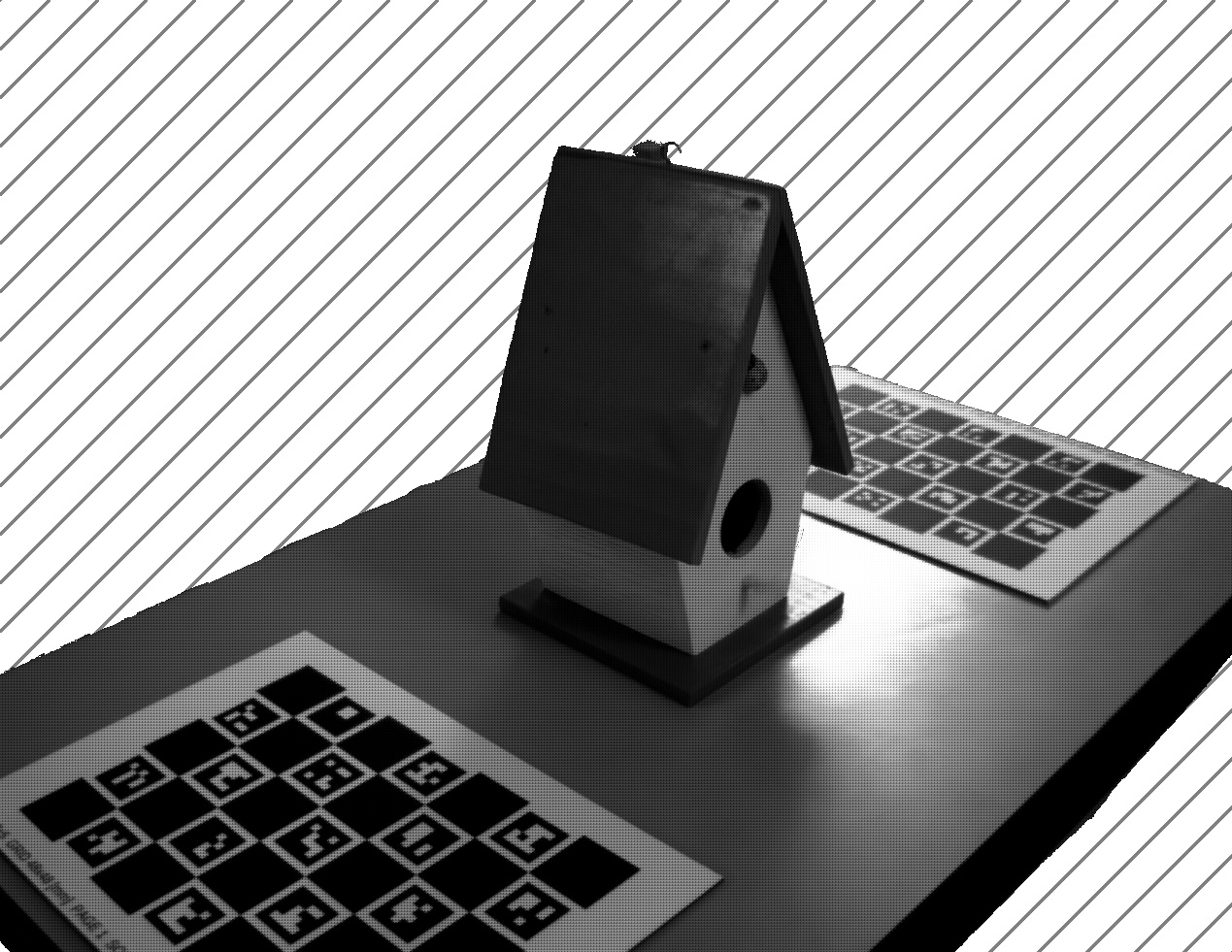} & \includegraphics[width=0.29\linewidth, valign=m]{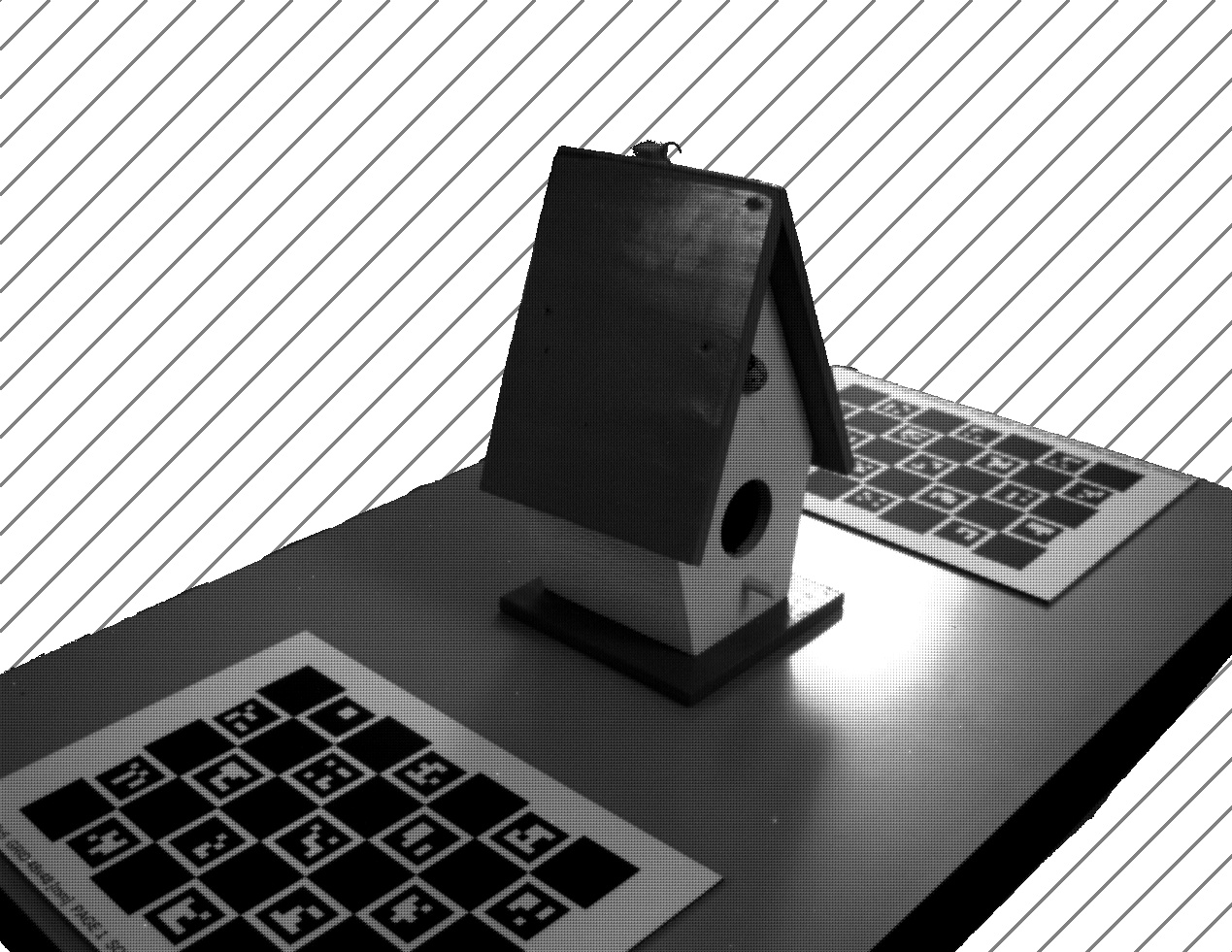} & \includegraphics[width=0.29\linewidth, valign=m]{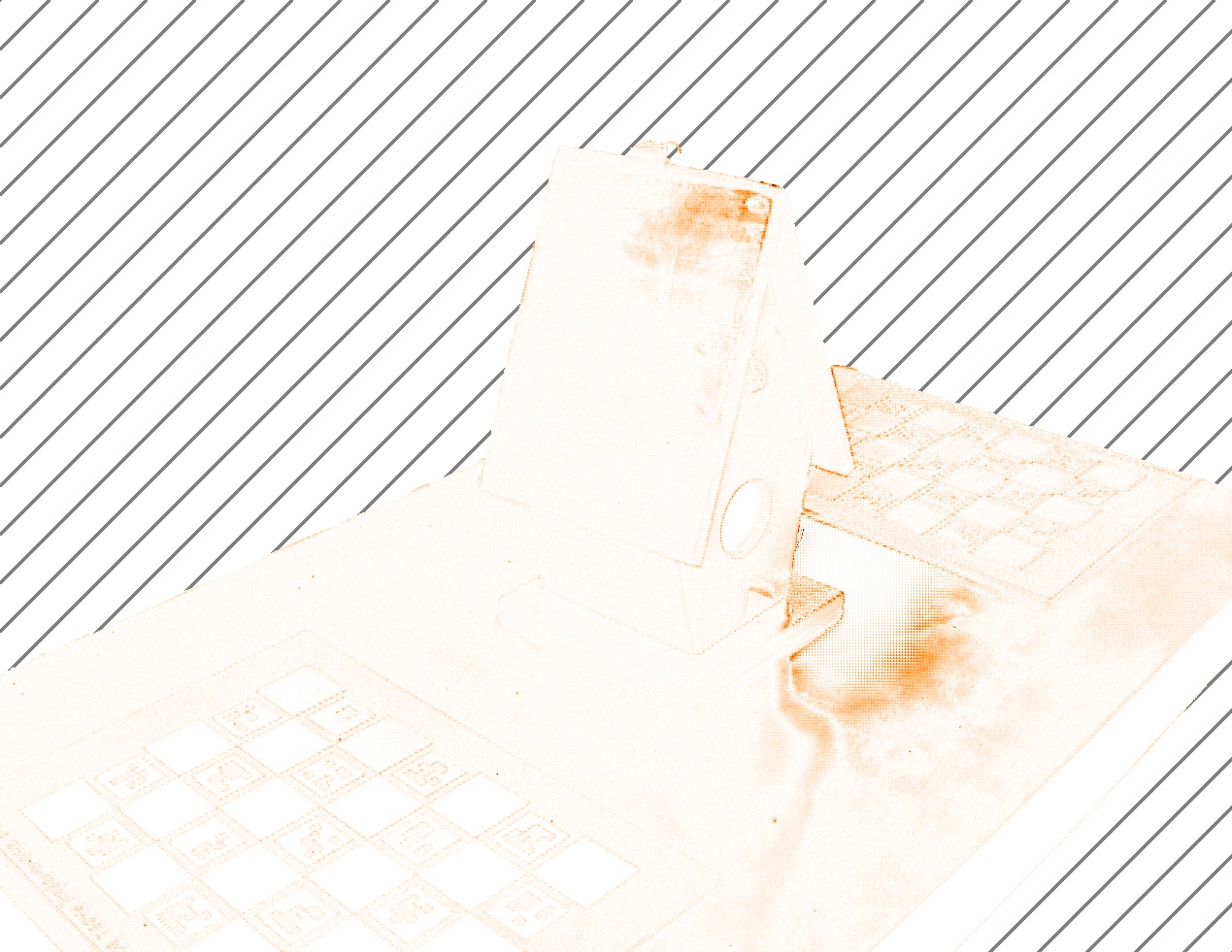} \\
        39 & \includegraphics[width=0.29\linewidth, valign=m]{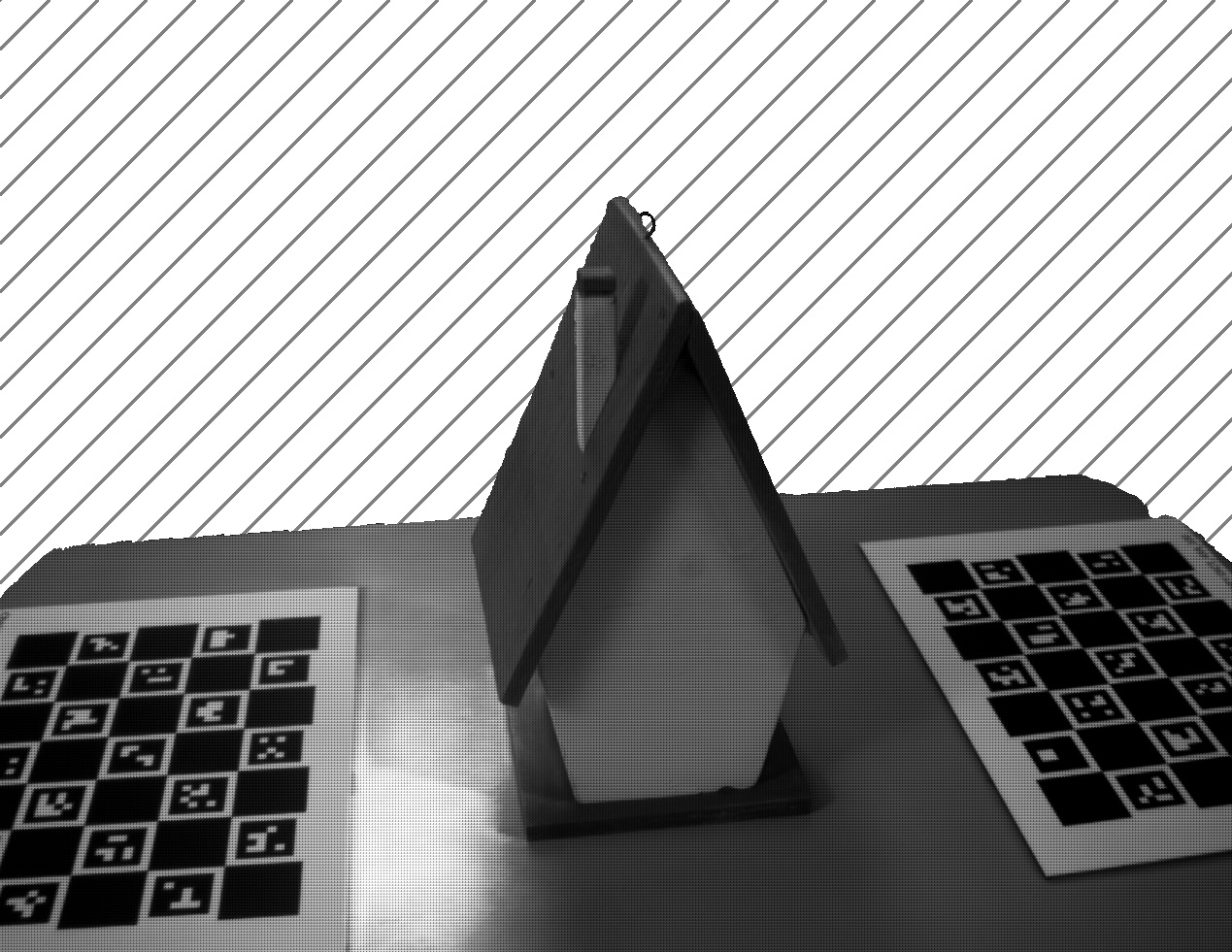} & \includegraphics[width=0.29\linewidth, valign=m]{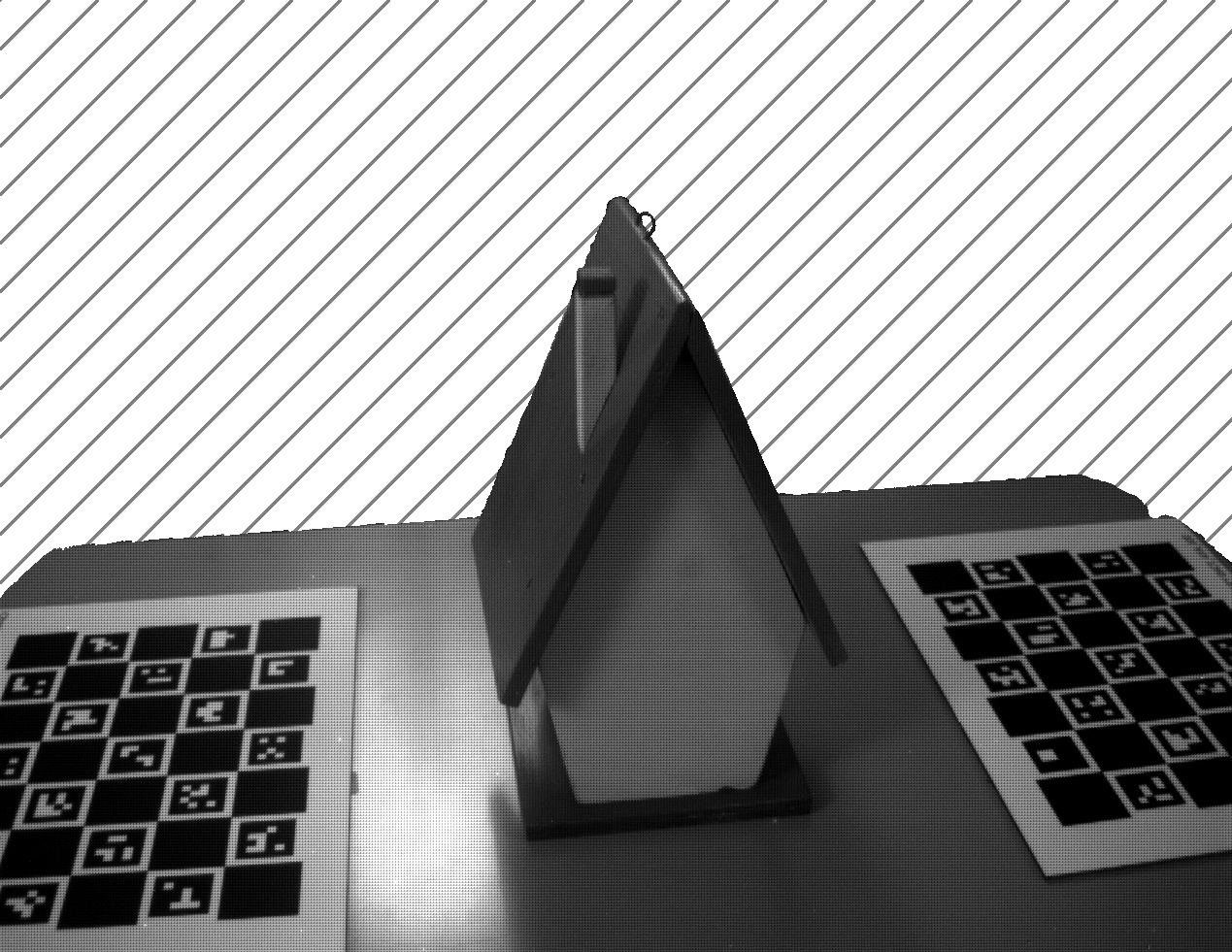} & \includegraphics[width=0.29\linewidth, valign=m]{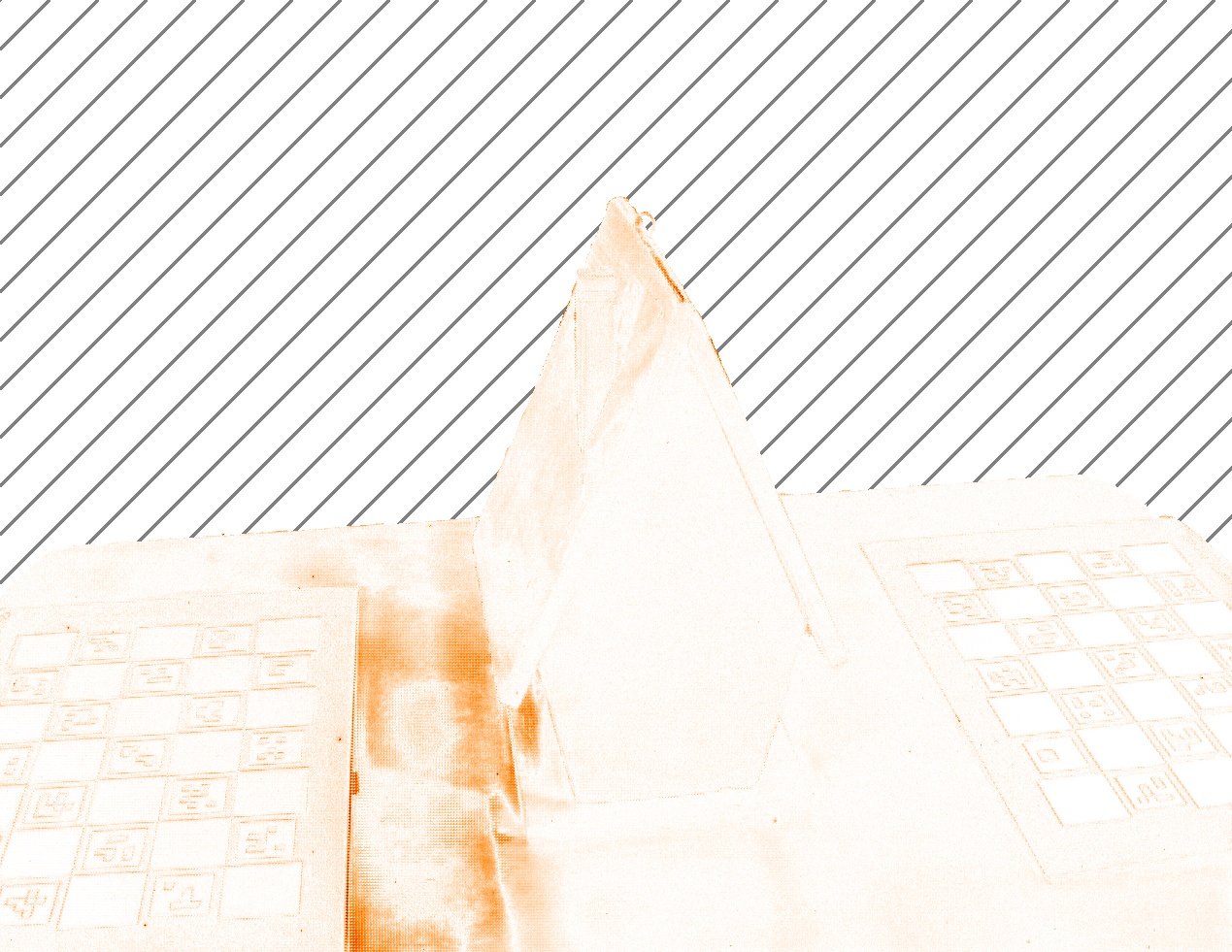} \\
        49 & \includegraphics[width=0.29\linewidth, valign=m]{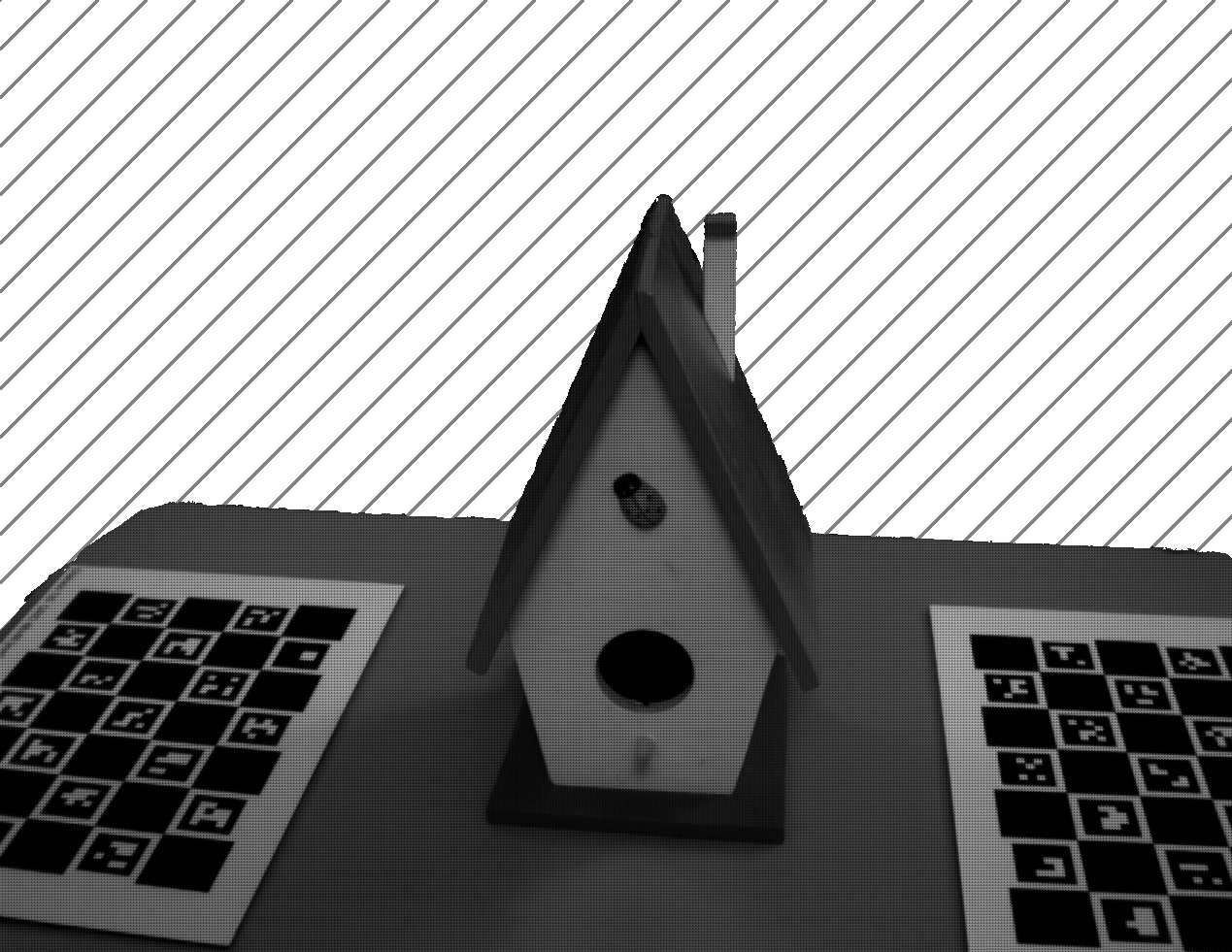} & \includegraphics[width=0.29\linewidth, valign=m]{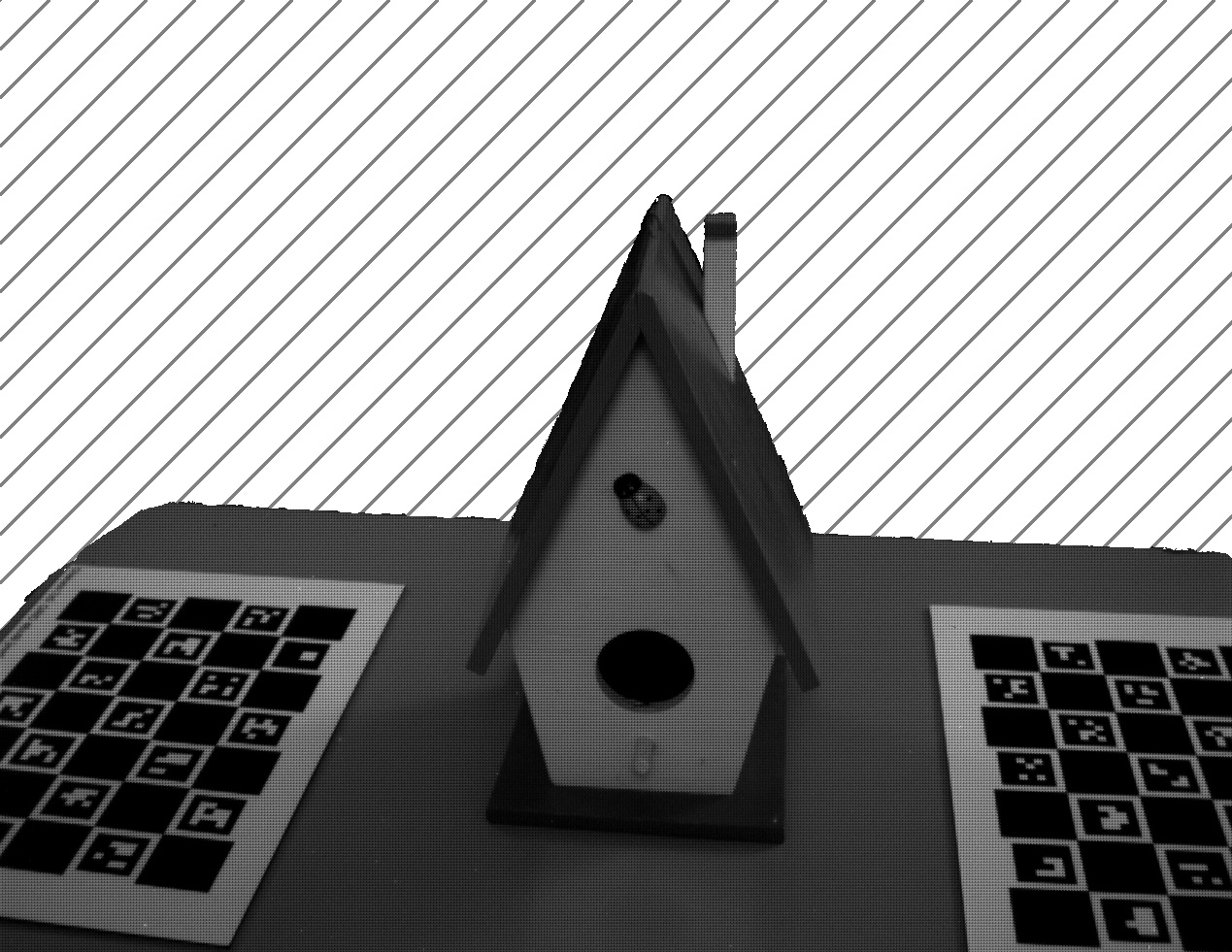} & \includegraphics[width=0.29\linewidth, valign=m]{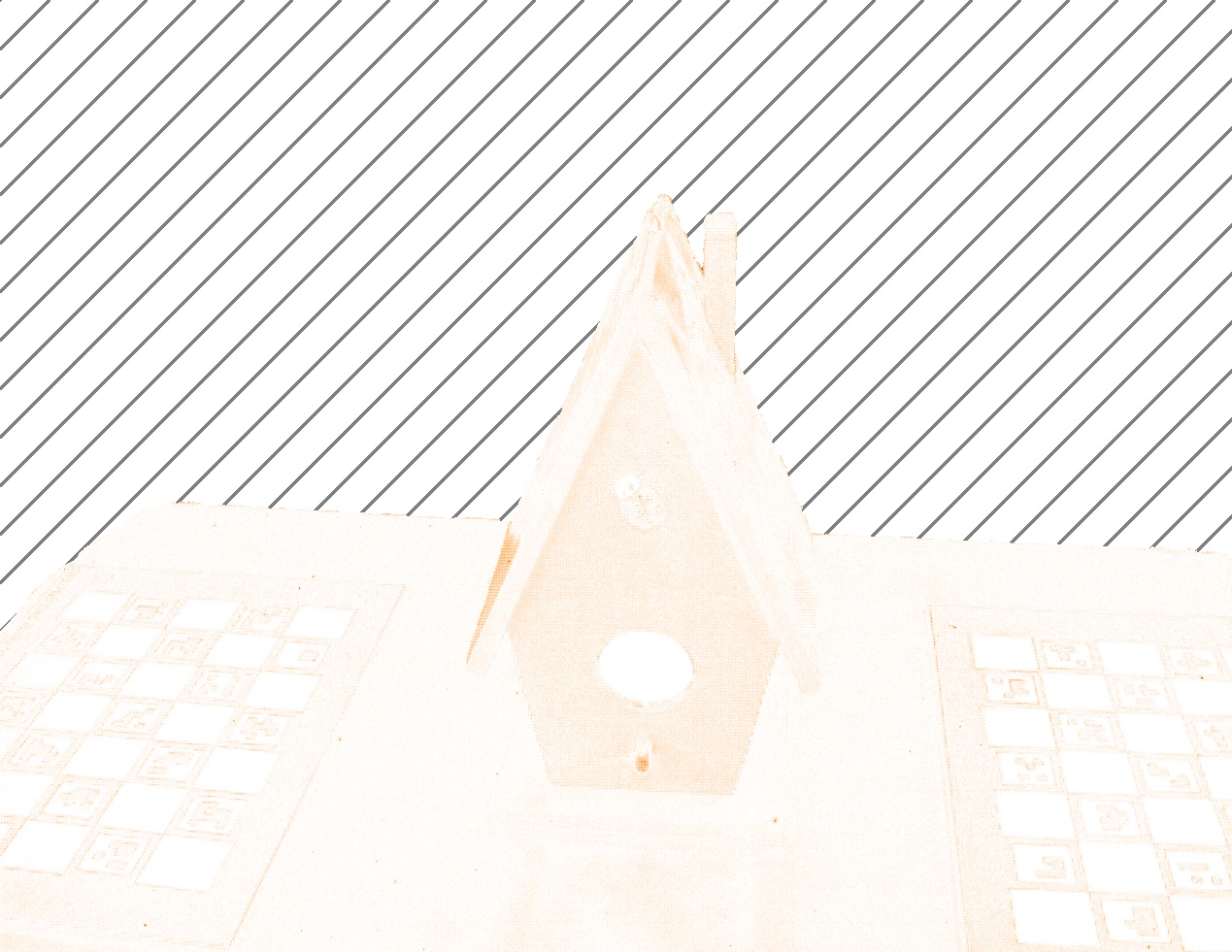} \\
        \midrule
        View & Rendering & Ground Truth & Error \\
    \end{tabular}
    \caption{Mosaicked Multispectral renderings, ground truth and error maps of the Bird House scene from the five different test views.}
    \label{sup_fig:multispectral_raw_renedrings}
\end{figure*}

\begin{figure*}
    \centering
    \begin{tabular}{@{\extracolsep{-6pt}}ccc}
        View & Normals & Depth Maps \\
        \midrule
        9 & \includegraphics[width=0.3\linewidth, valign=m]{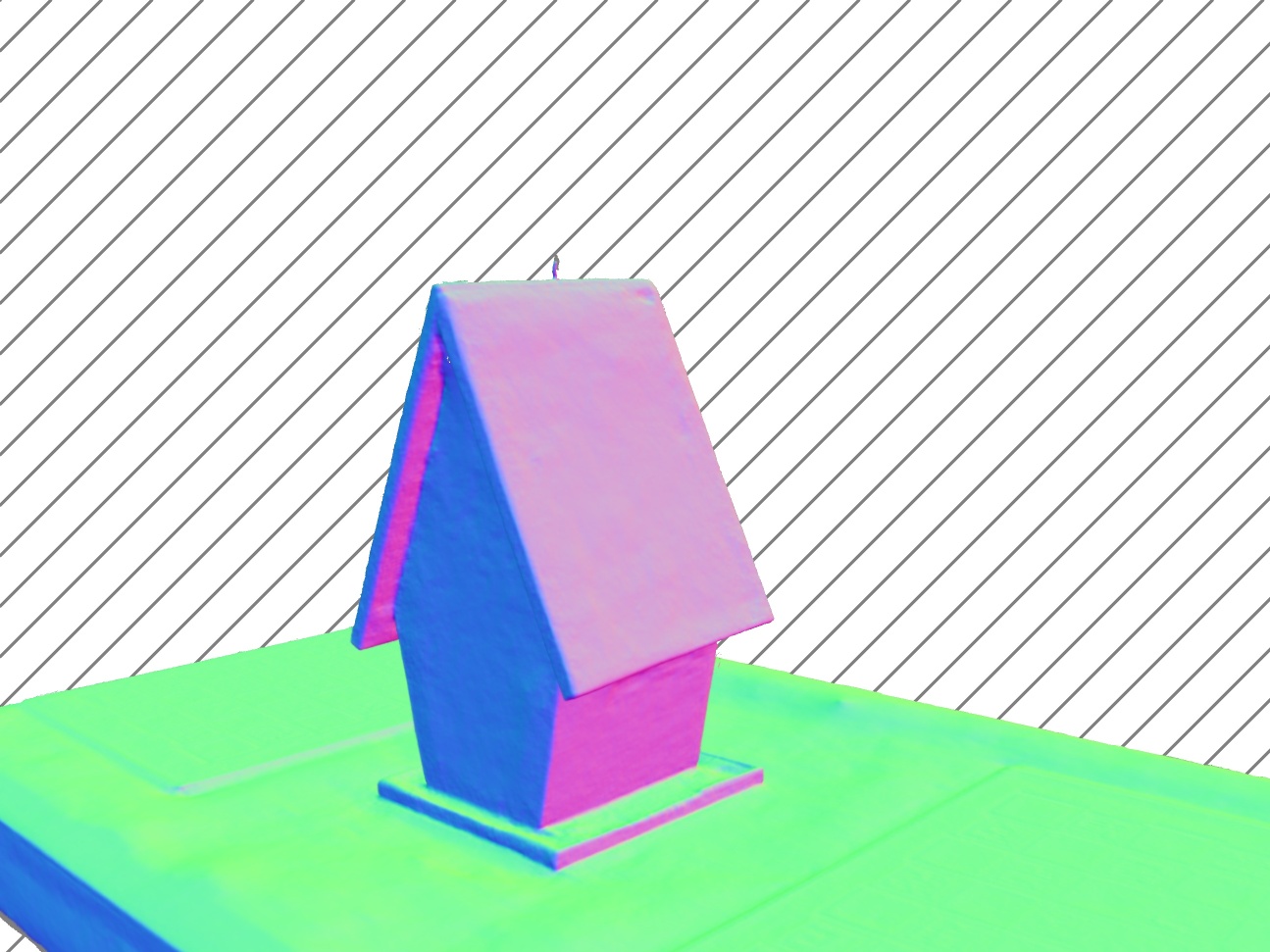} & \includegraphics[width=0.3\linewidth, valign=m]{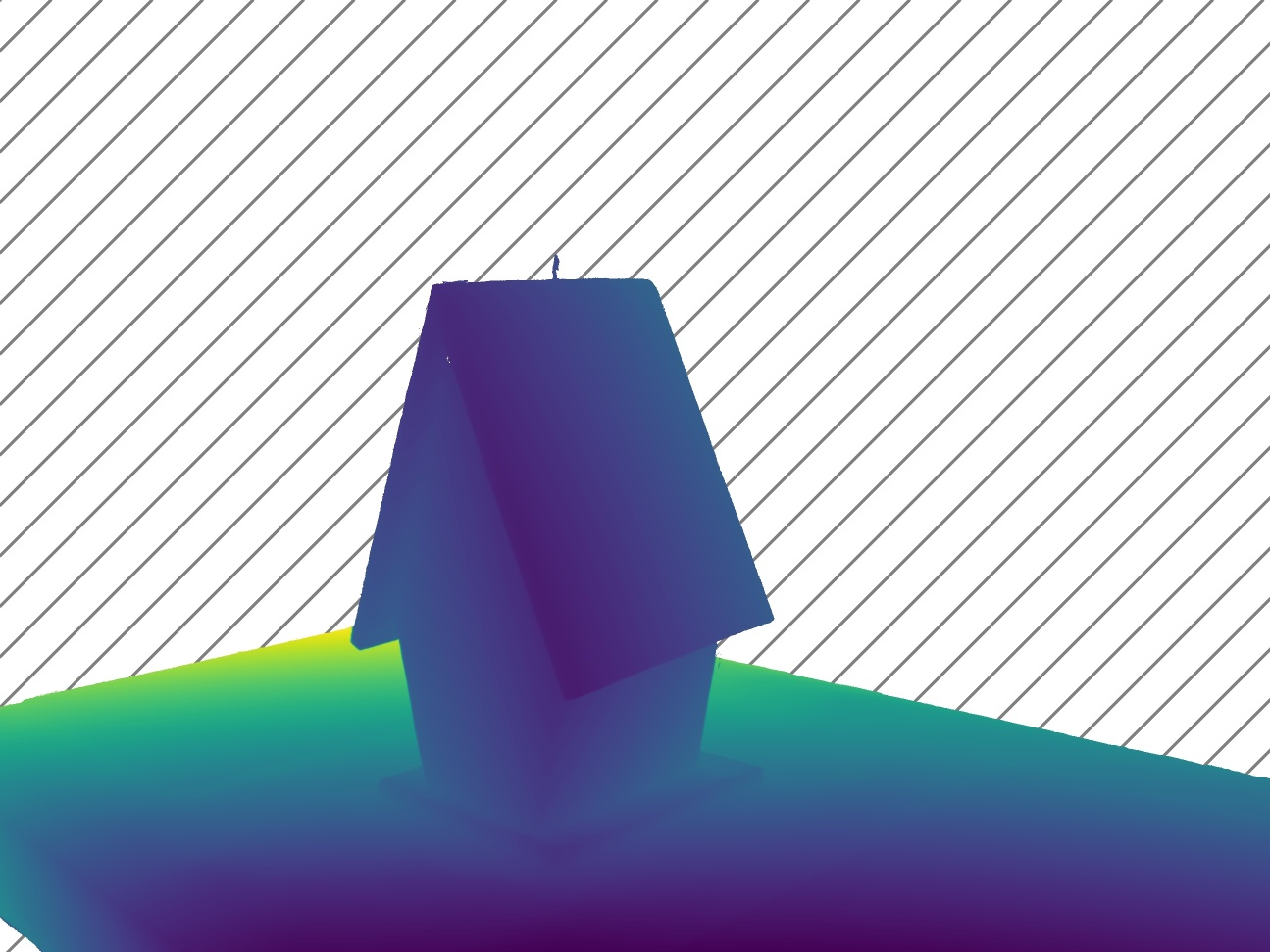} \\
        19 & \includegraphics[width=0.3\linewidth, valign=m]{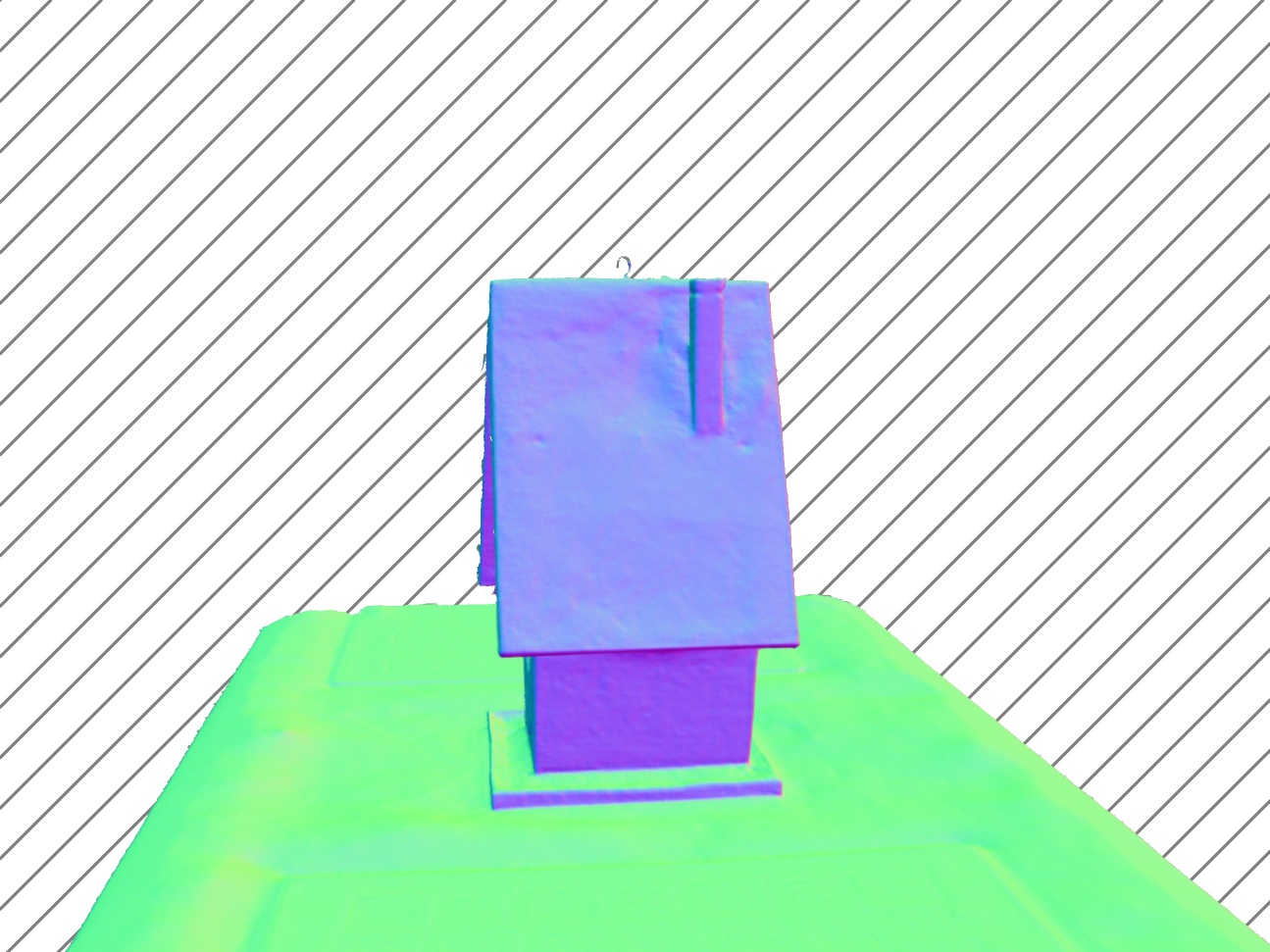} & \includegraphics[width=0.3\linewidth, valign=m]{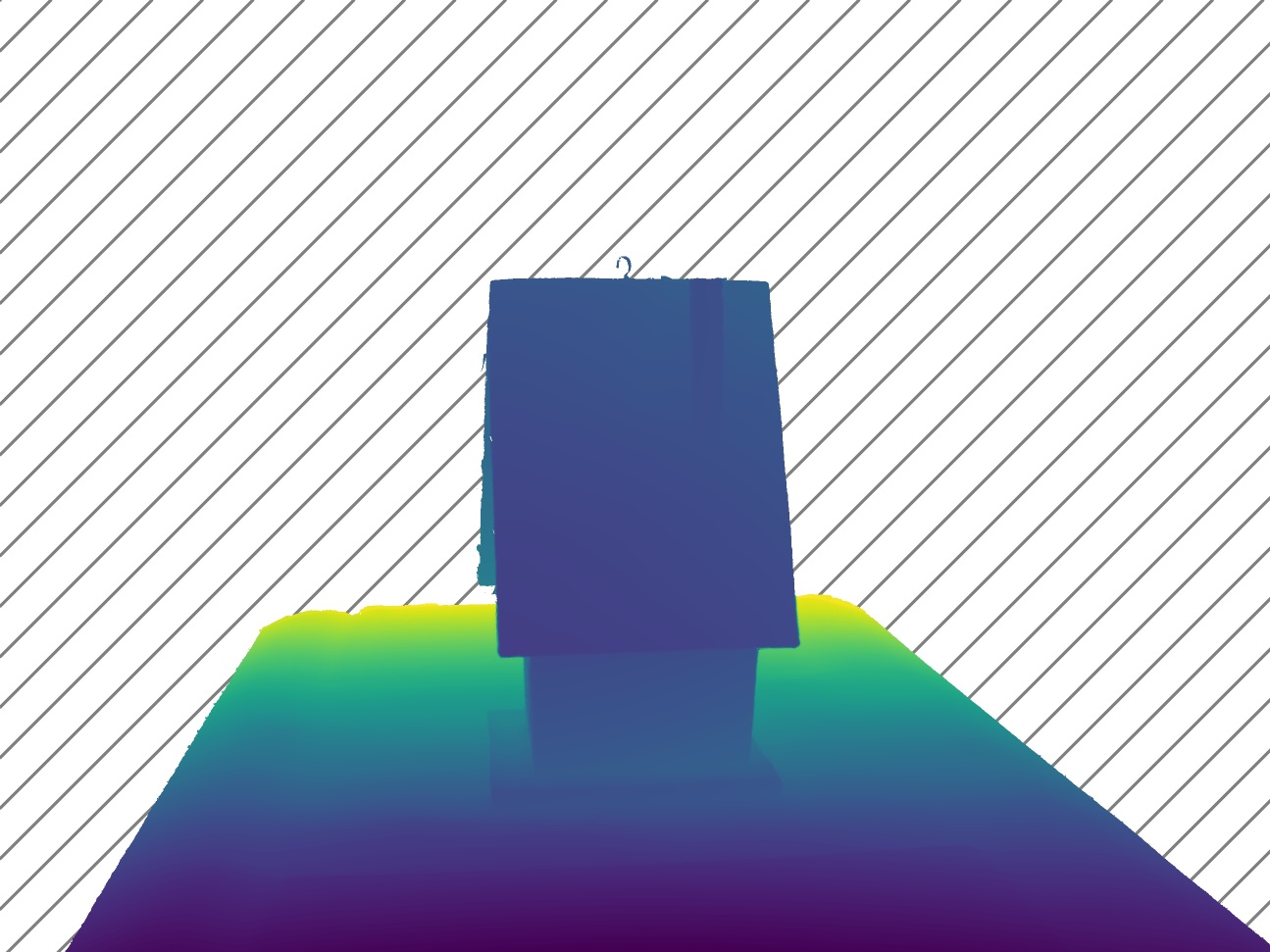} \\
        29 & \includegraphics[width=0.3\linewidth, valign=m]{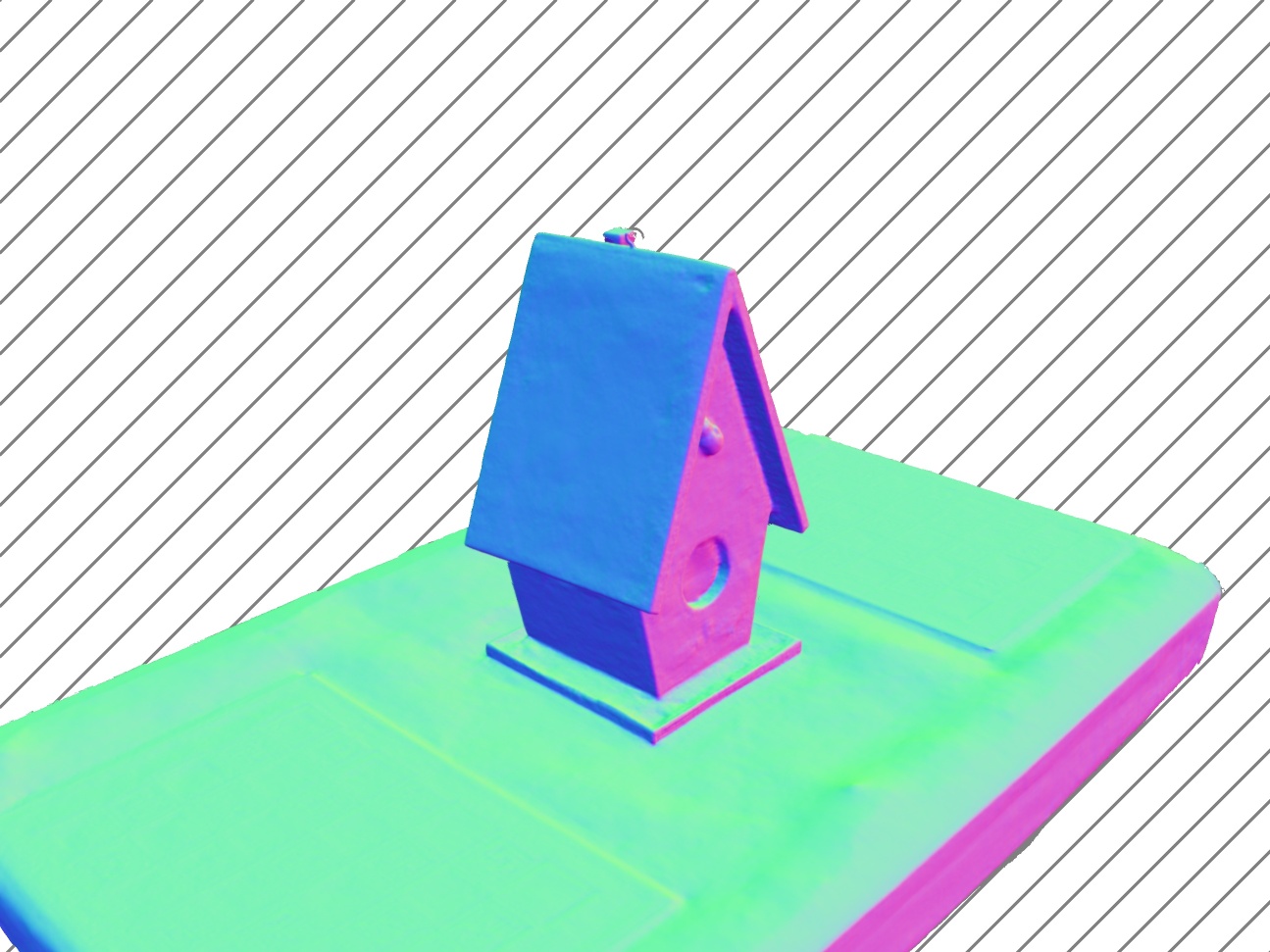} & \includegraphics[width=0.3\linewidth, valign=m]{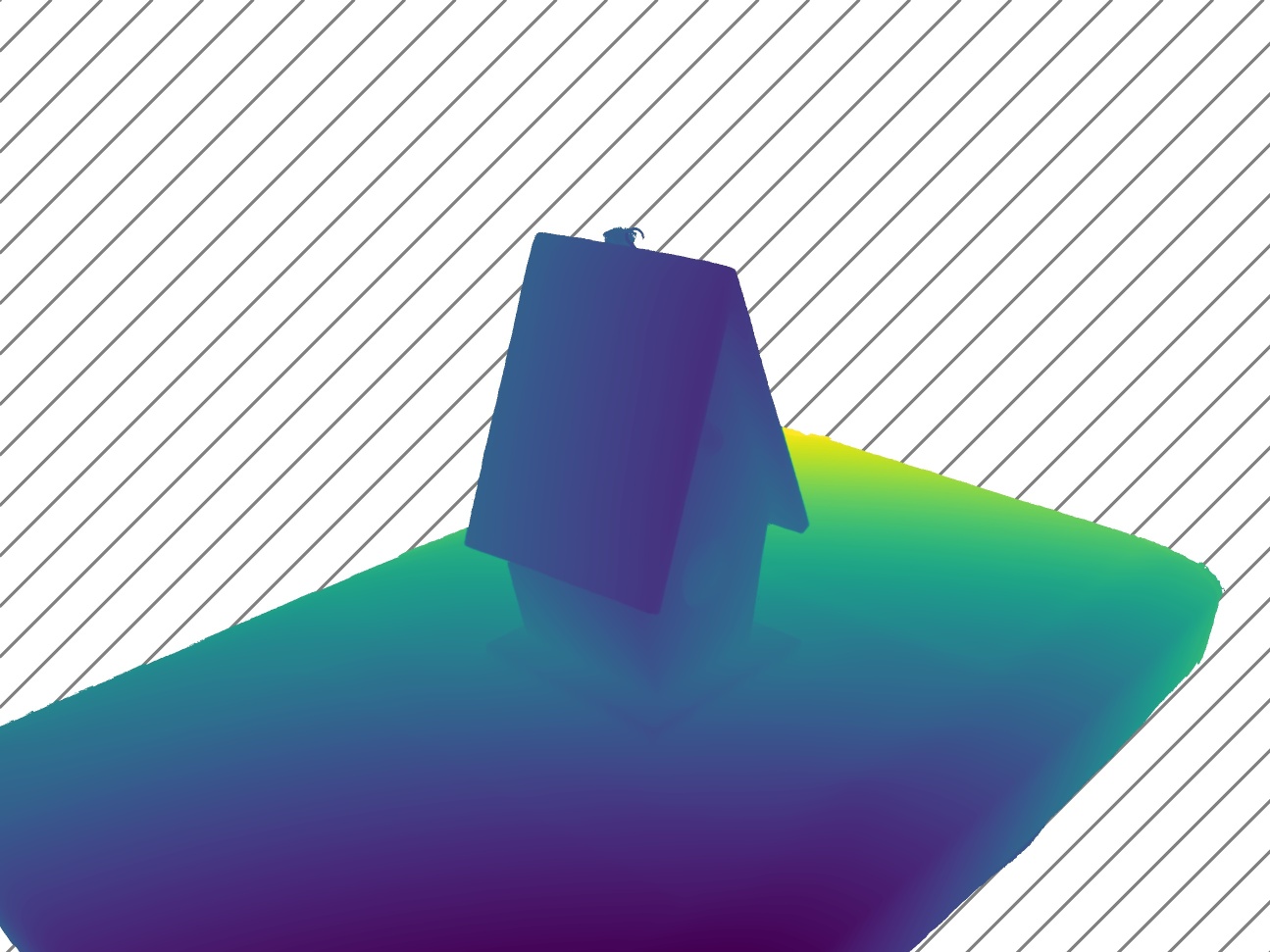} \\
        39 & \includegraphics[width=0.3\linewidth, valign=m]{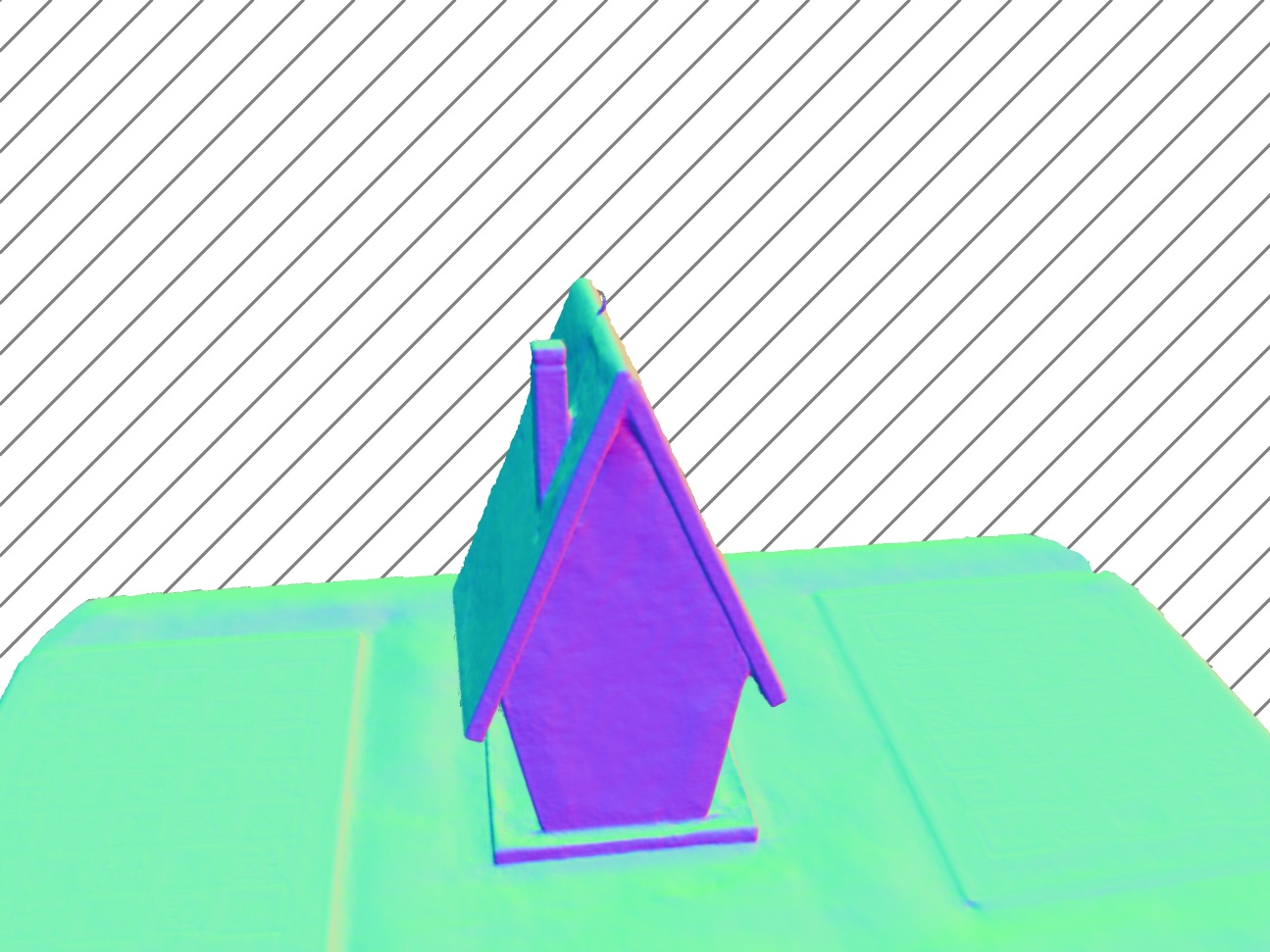} & \includegraphics[width=0.3\linewidth, valign=m]{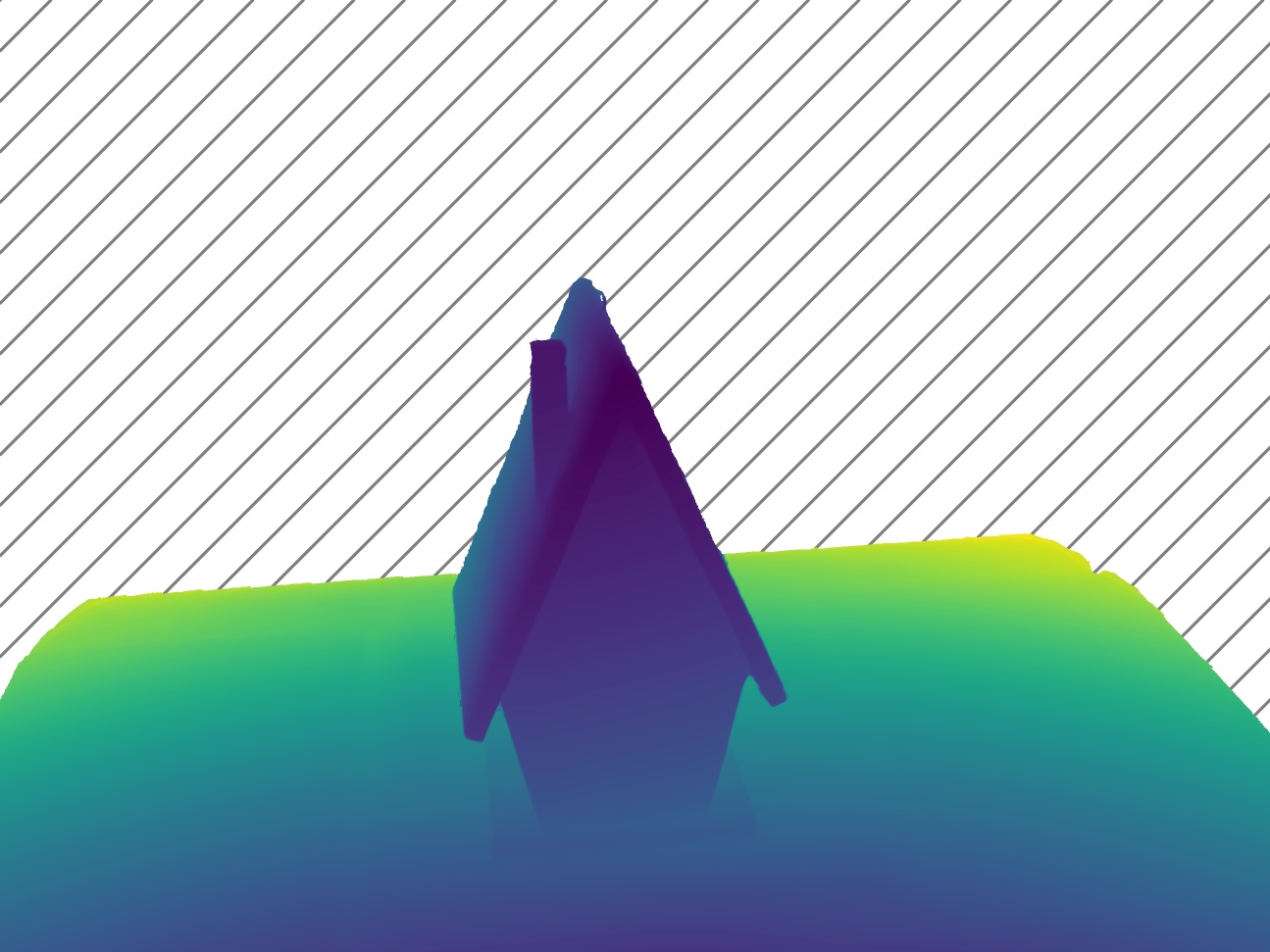} \\
        49 & \includegraphics[width=0.3\linewidth, valign=m]{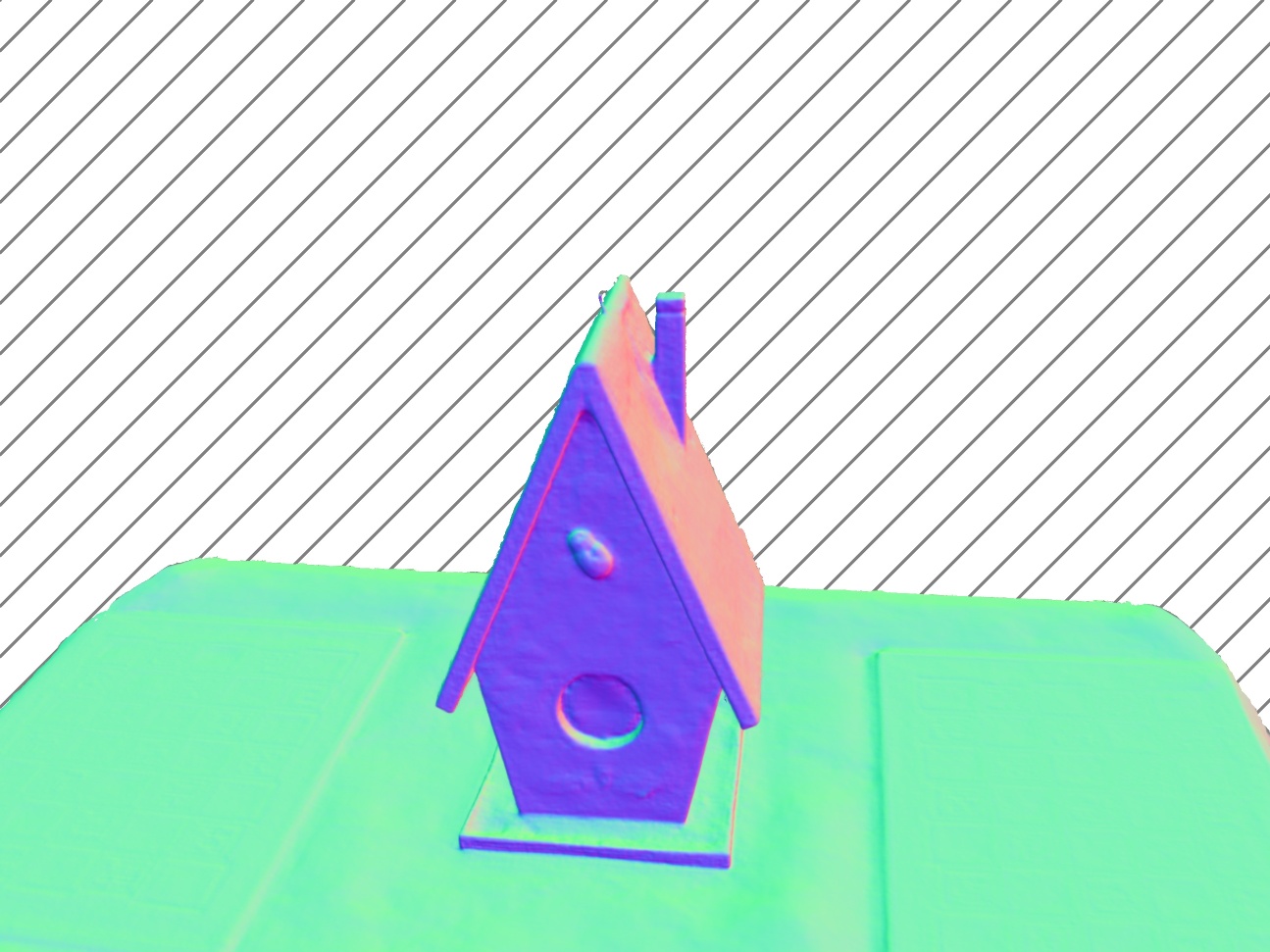} & \includegraphics[width=0.3\linewidth, valign=m]{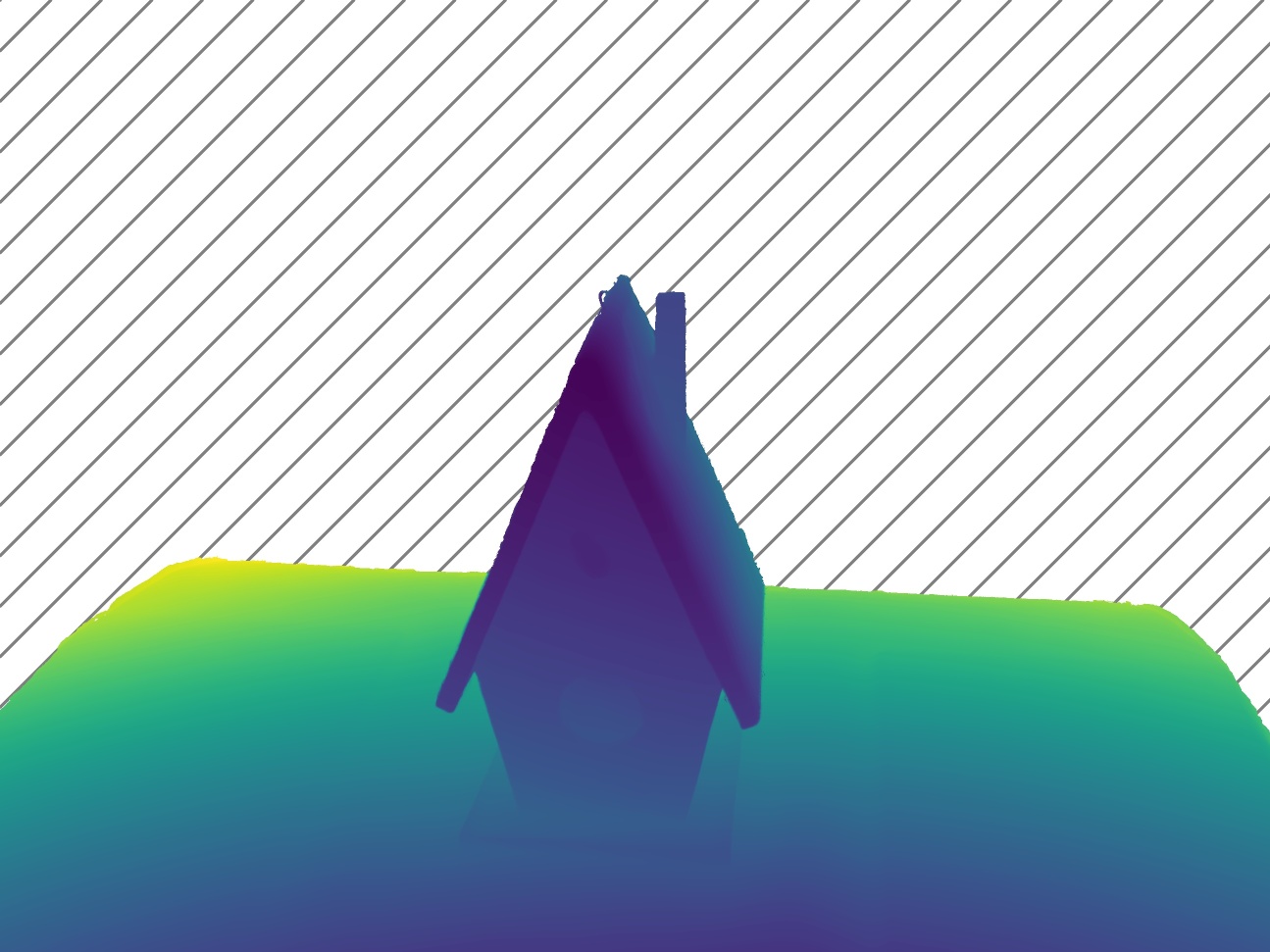} \\
        \midrule
        View & Normals & Depth Maps \\
    \end{tabular}
    \caption{Normal and depth maps of the Bird House scene for the five different test views.}
    \label{sup_fig:geometry}
\end{figure*}

\begin{figure*}
    \centering
    \begin{tabular}{@{\extracolsep{-6pt}}cccc}
        Mod. & Fruits & Aloe & Laurel Wreath \\
        \midrule
        \multirow{1}{*}[5pt]{\rotatebox{90}{RGB}} & \includegraphics[width=0.3\linewidth, valign=m]{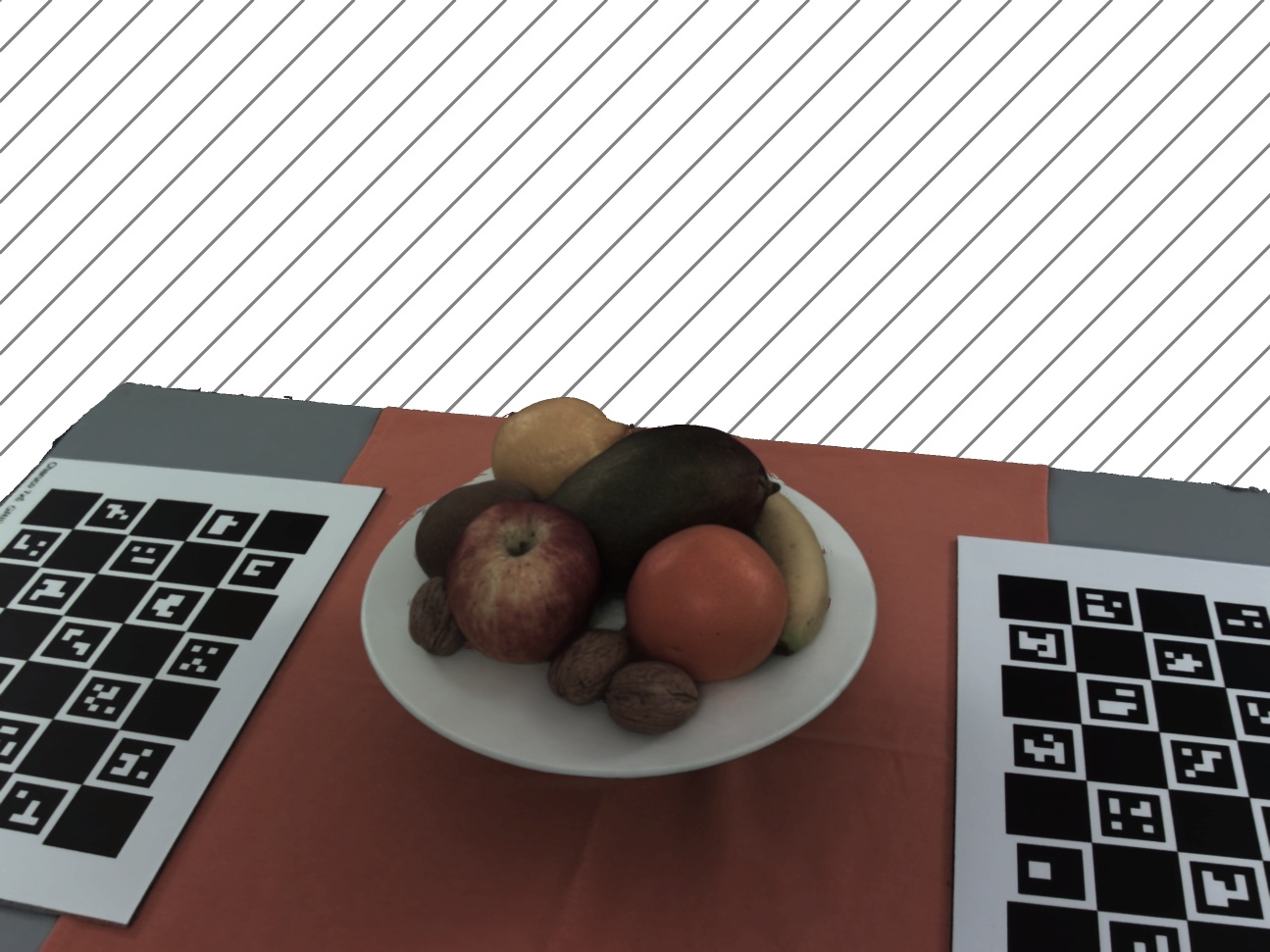} & \includegraphics[width=0.3\linewidth, valign=m]{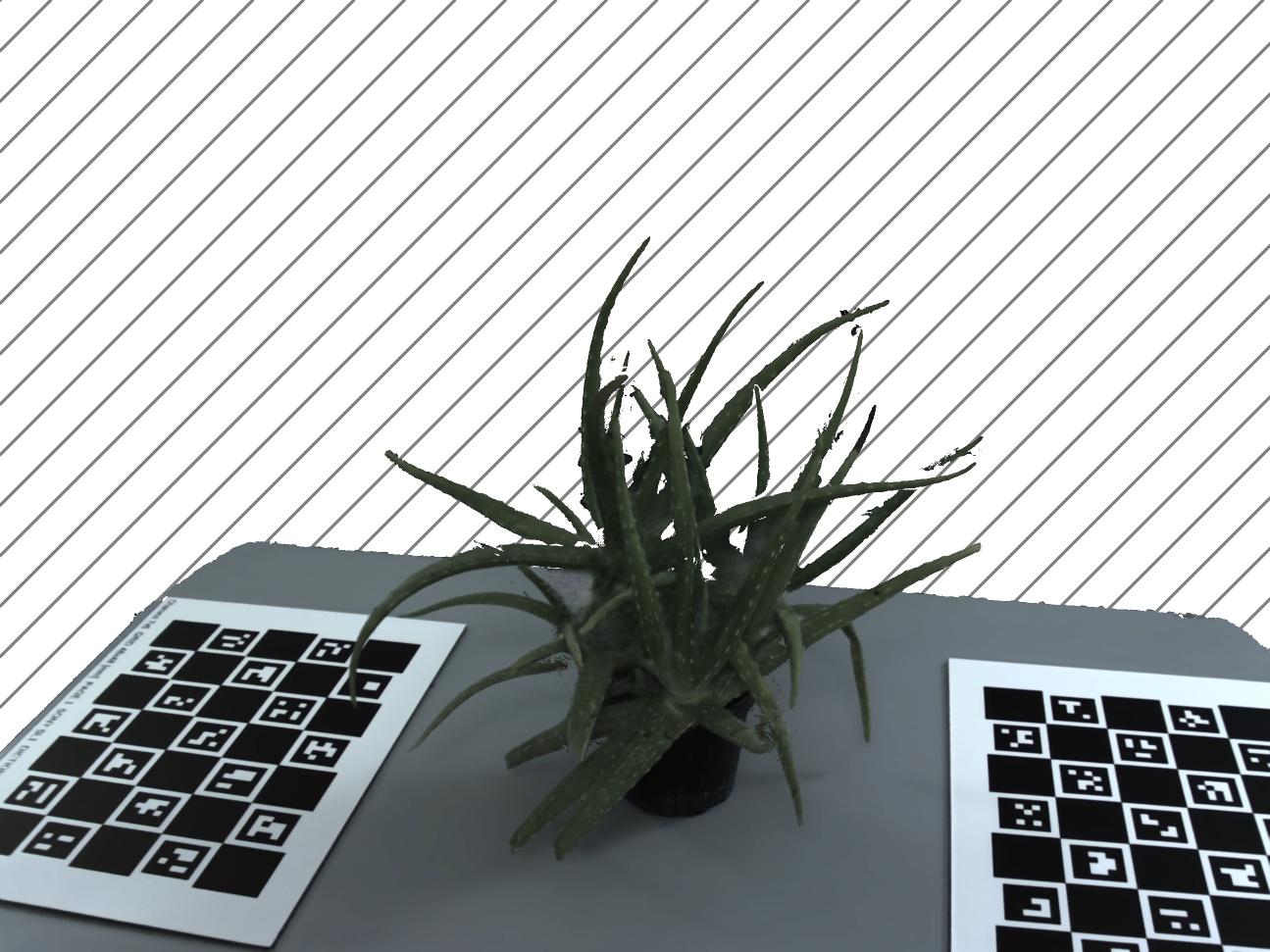} & \includegraphics[width=0.3\linewidth, valign=m]{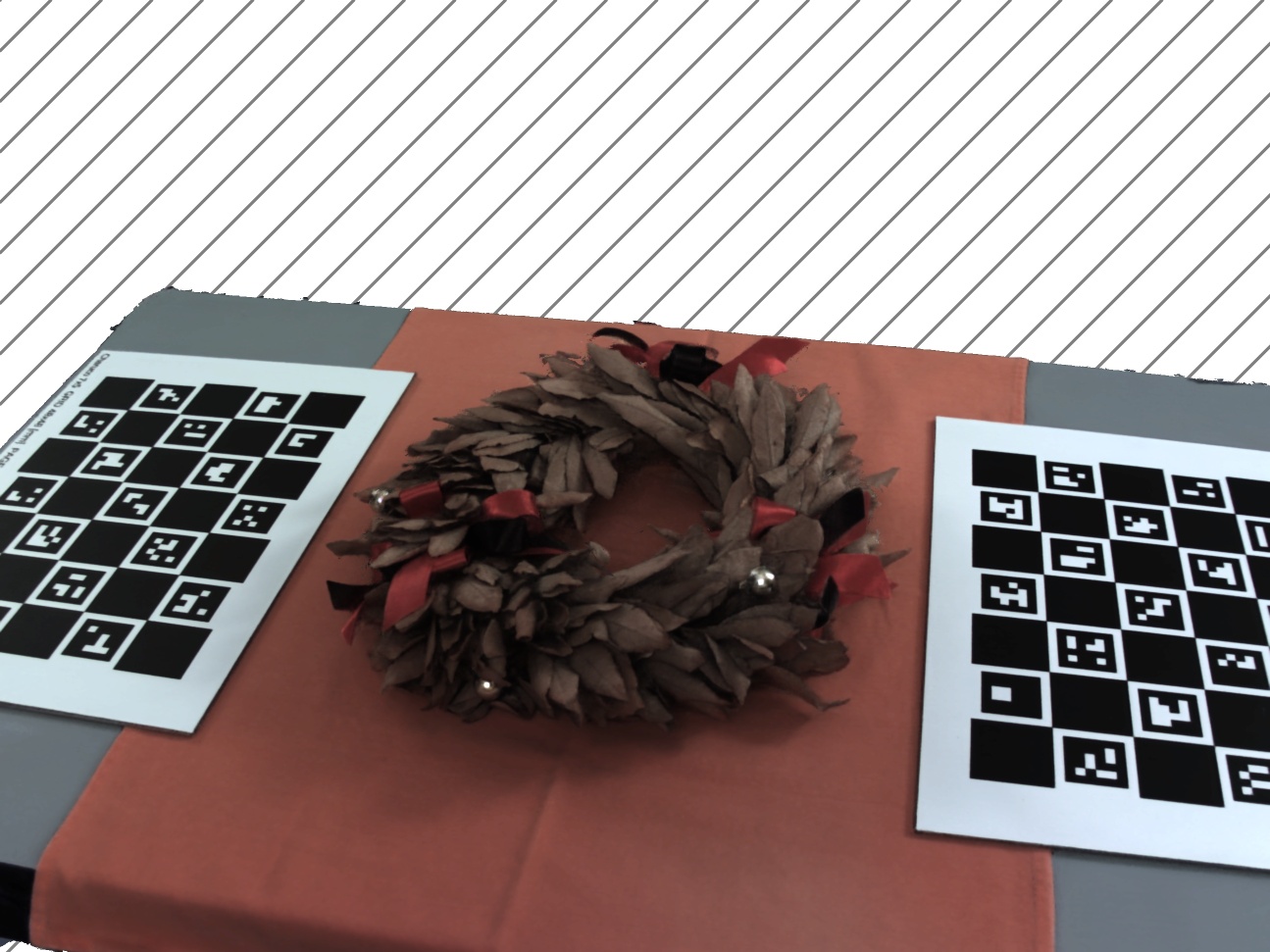} \\
        \multirow{1}{*}[10pt]{\rotatebox{90}{Mono}} & \includegraphics[width=0.3\linewidth, valign=m]{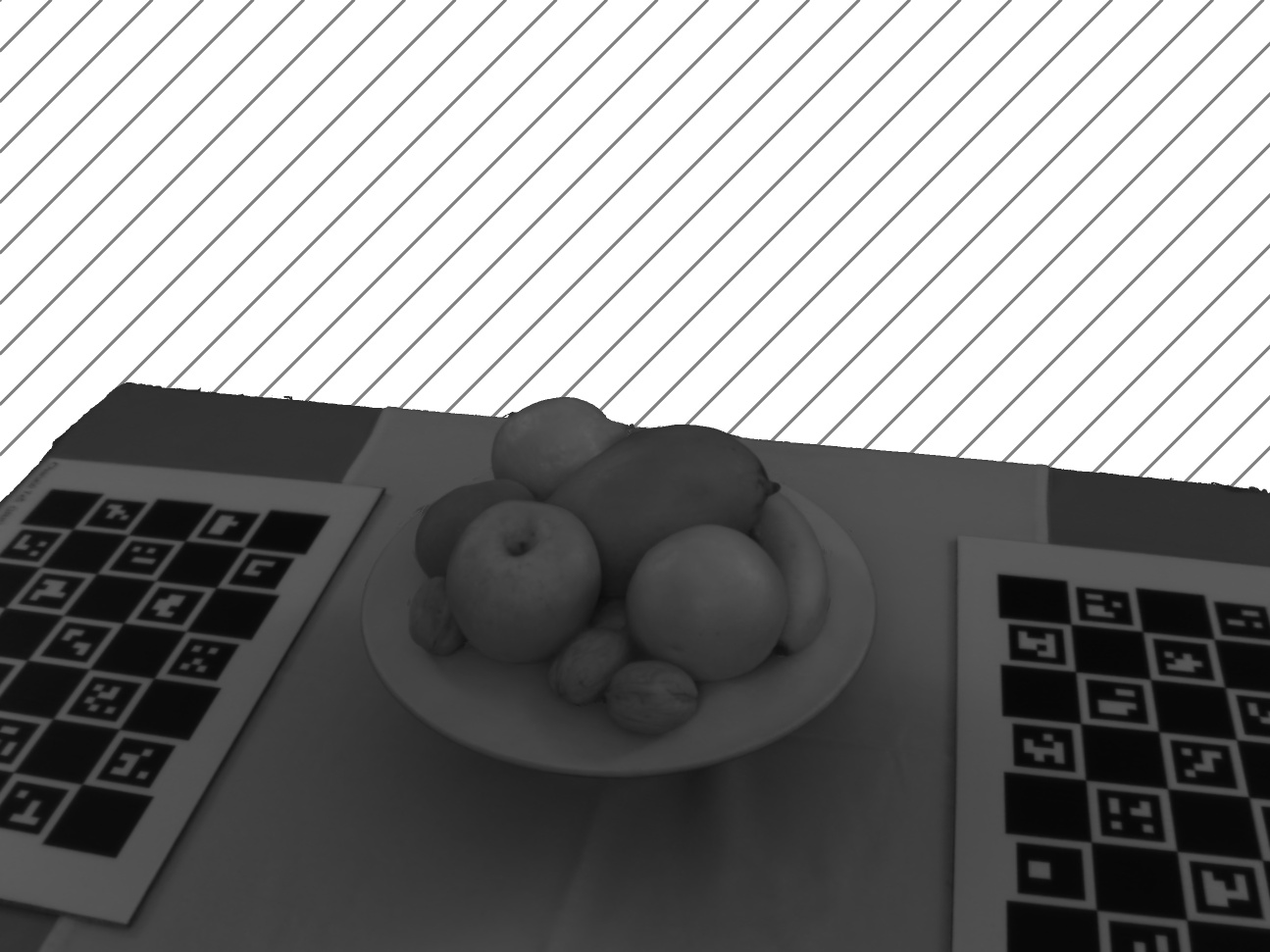} & \includegraphics[width=0.3\linewidth, valign=m]{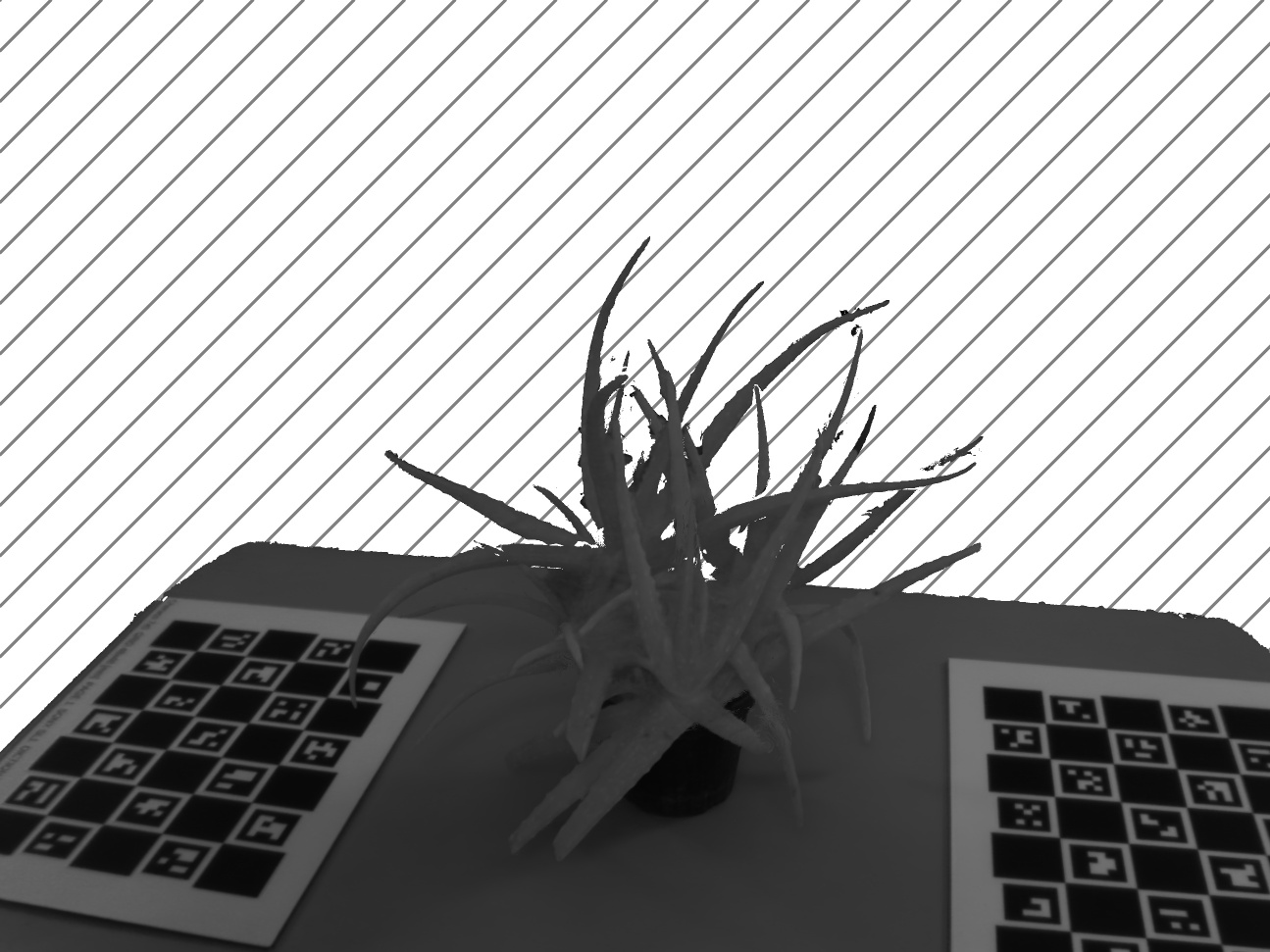} & \includegraphics[width=0.3\linewidth, valign=m]{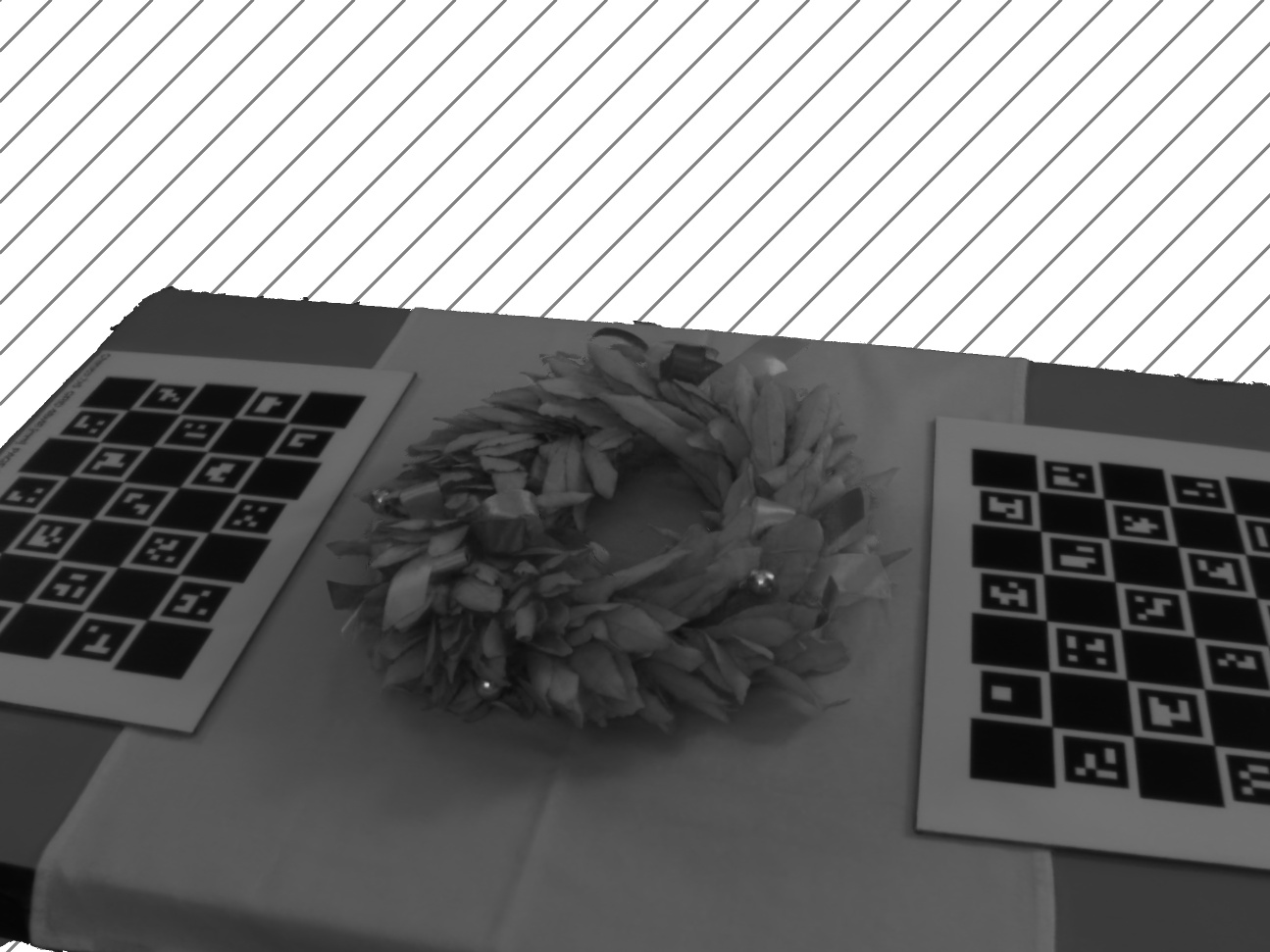} \\
        \multirow{1}{*}[27pt]{\rotatebox{90}{Near-Infrared}} & \includegraphics[width=0.3\linewidth, valign=m]{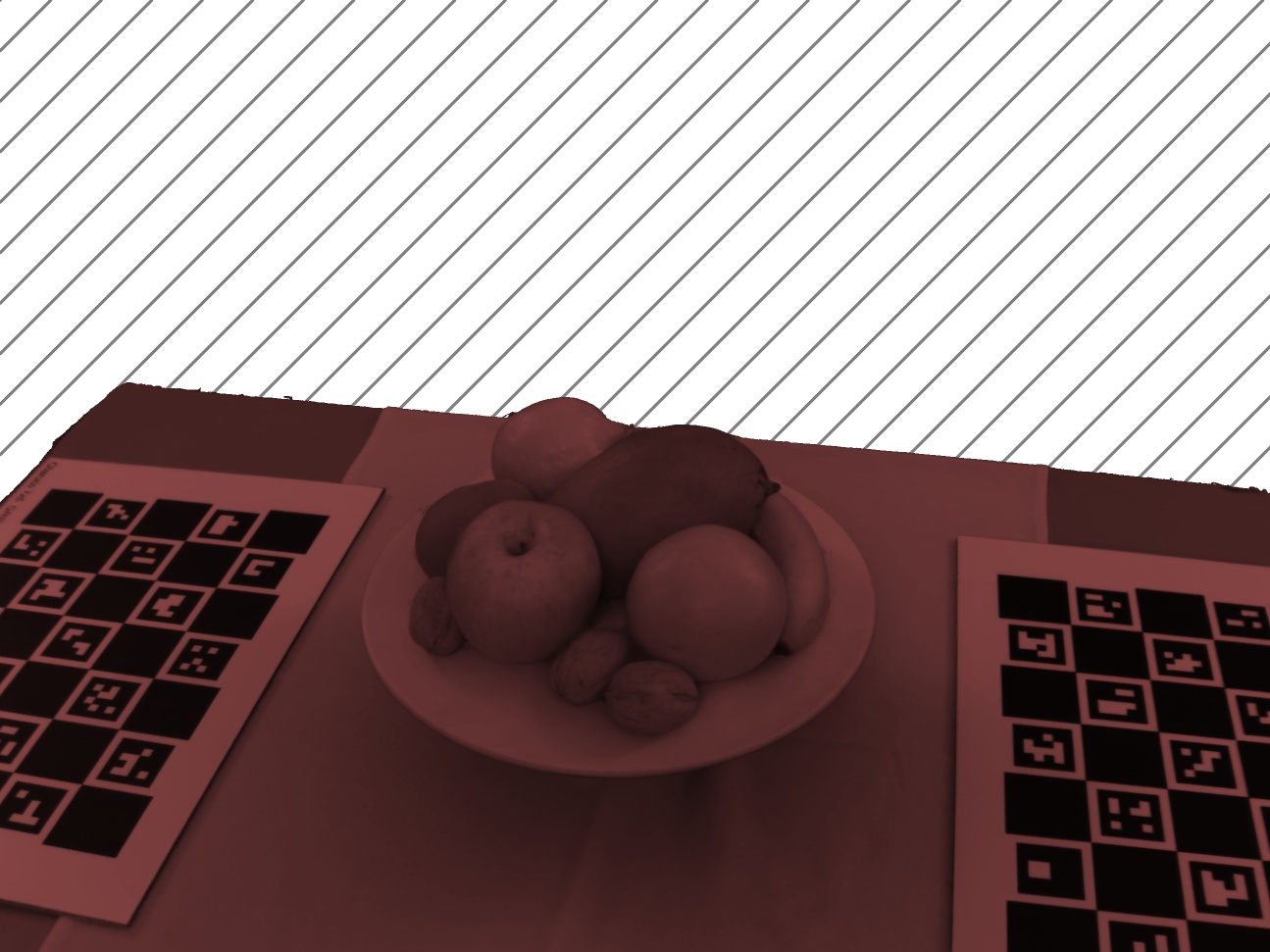} & \includegraphics[width=0.3\linewidth, valign=m]{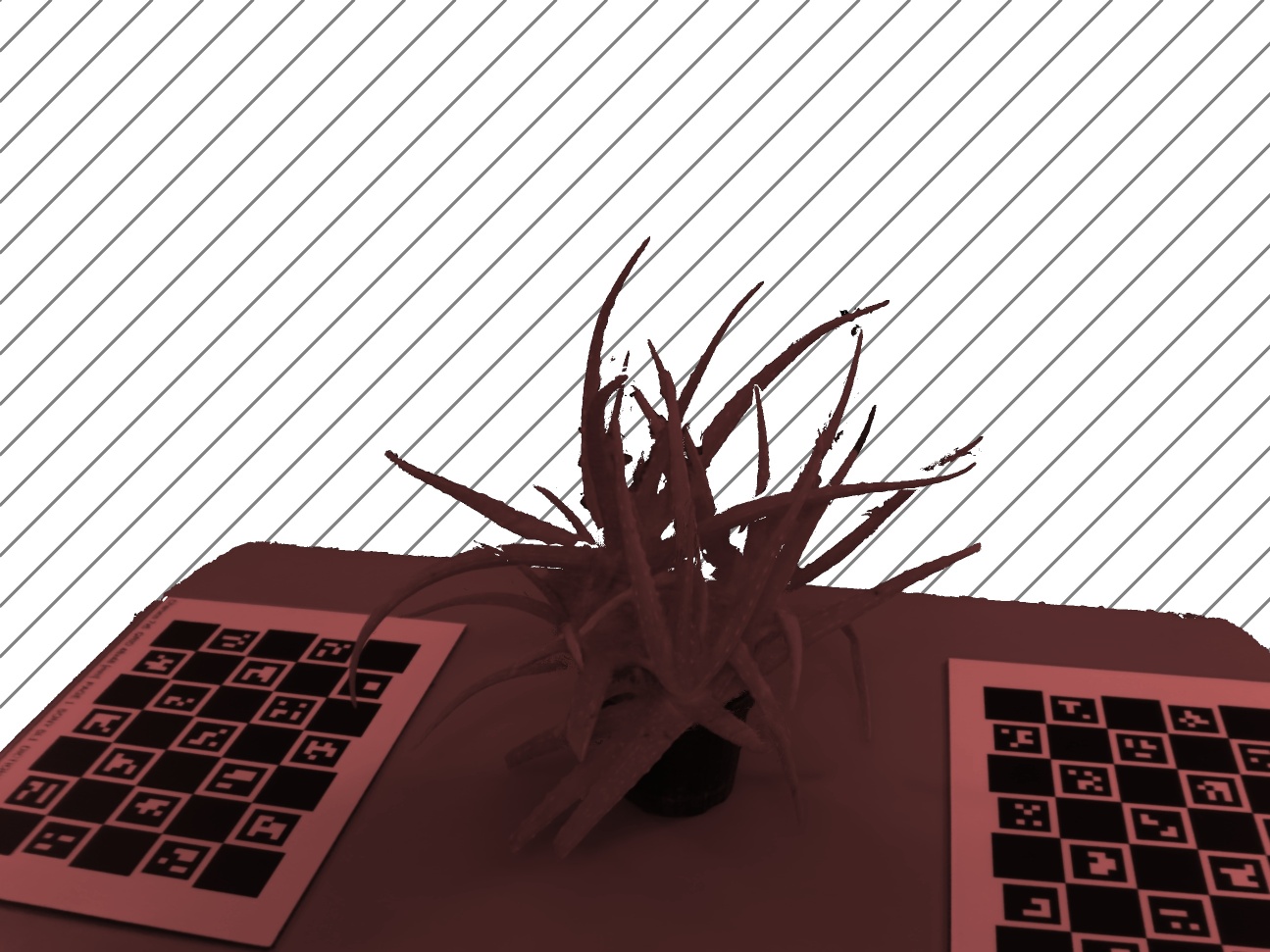} & \includegraphics[width=0.3\linewidth, valign=m]{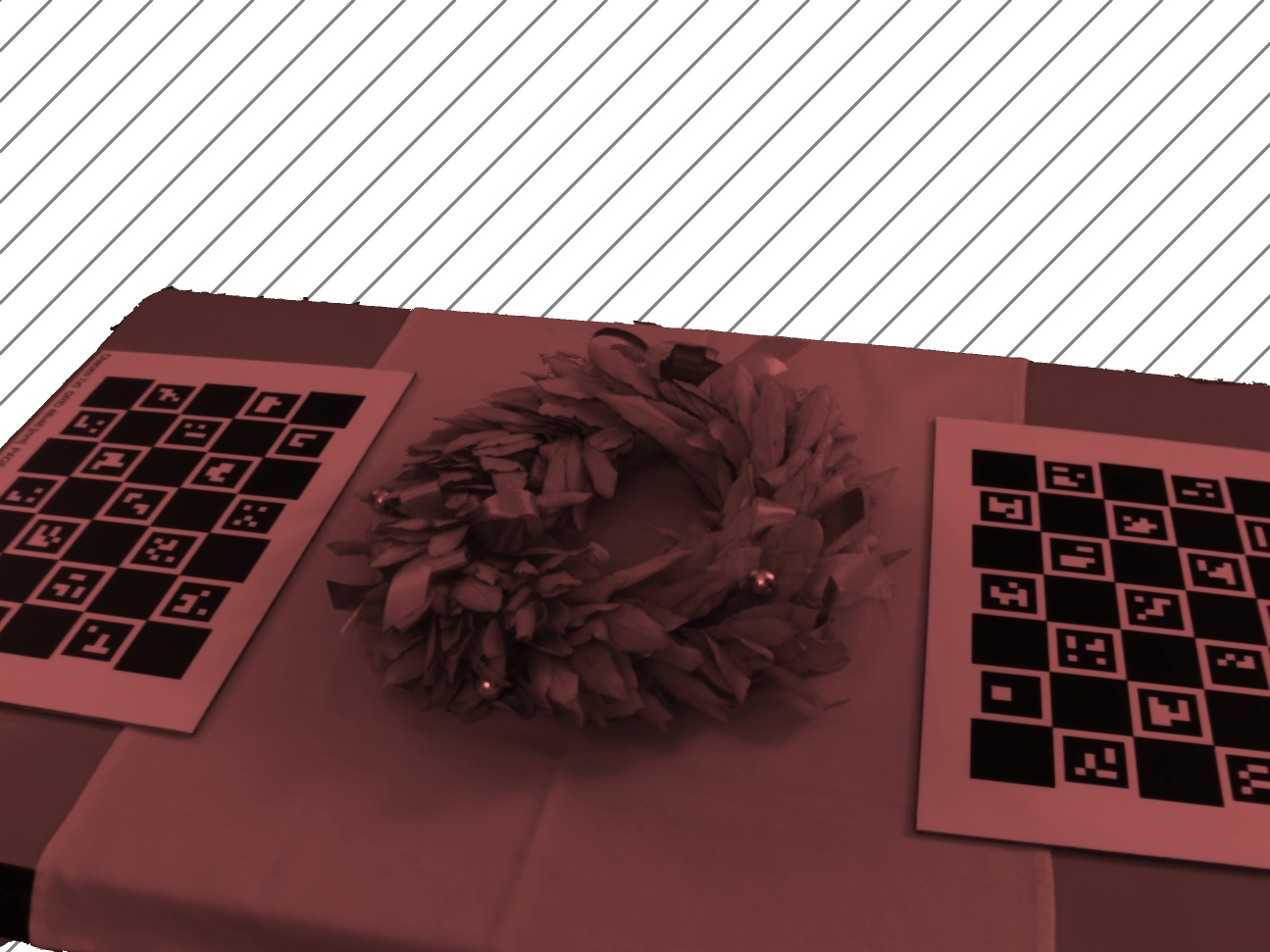} \\
        \multirow{1}{*}[22pt]{\rotatebox{90}{Polarization}} & \includegraphics[width=0.3\linewidth, valign=m]{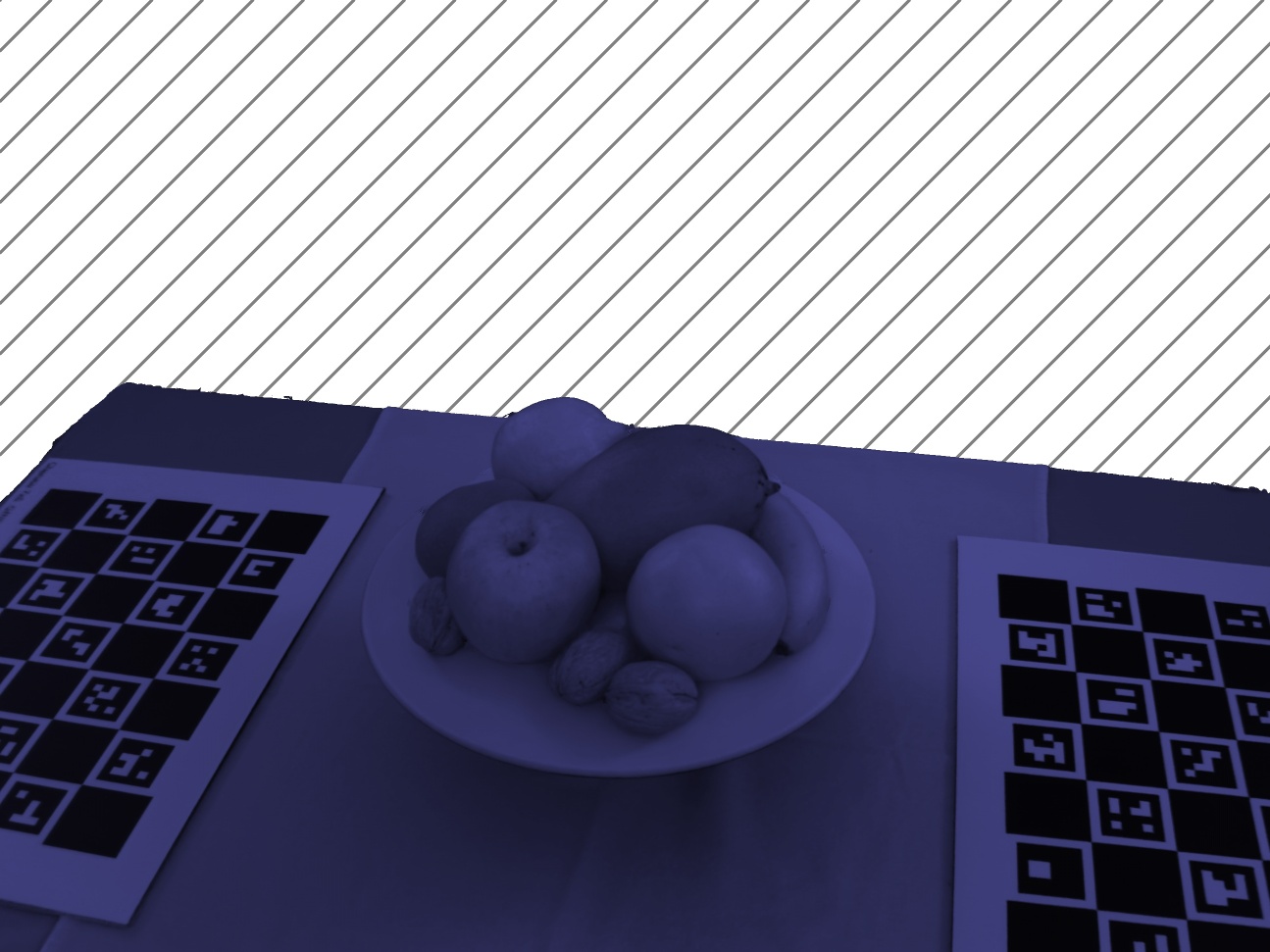} & \includegraphics[width=0.3\linewidth, valign=m]{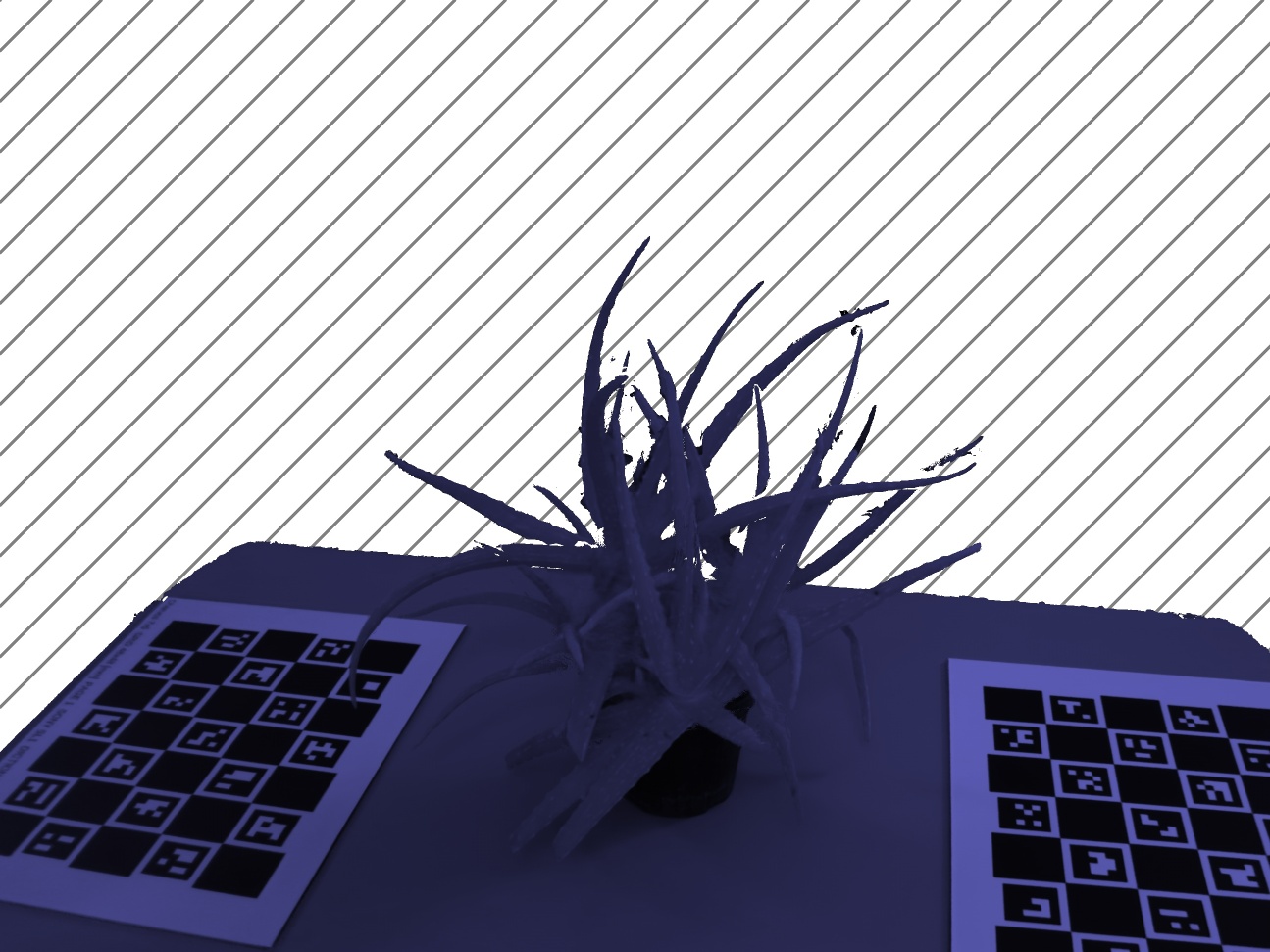} & \includegraphics[width=0.3\linewidth, valign=m]{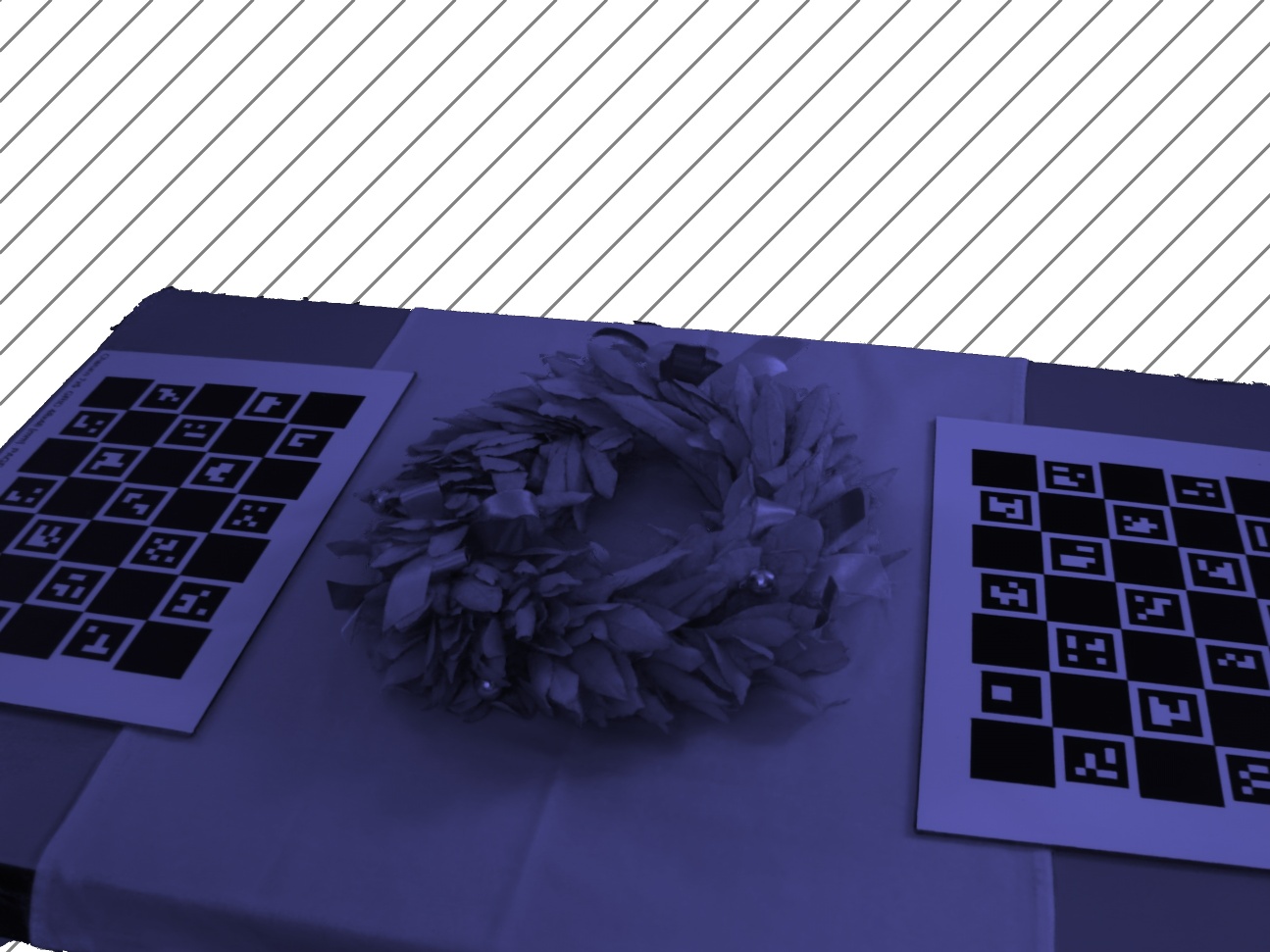} \\
        \multirow{1}{*}[24pt]{\rotatebox{90}{Multispectral}} & \includegraphics[width=0.3\linewidth, valign=m]{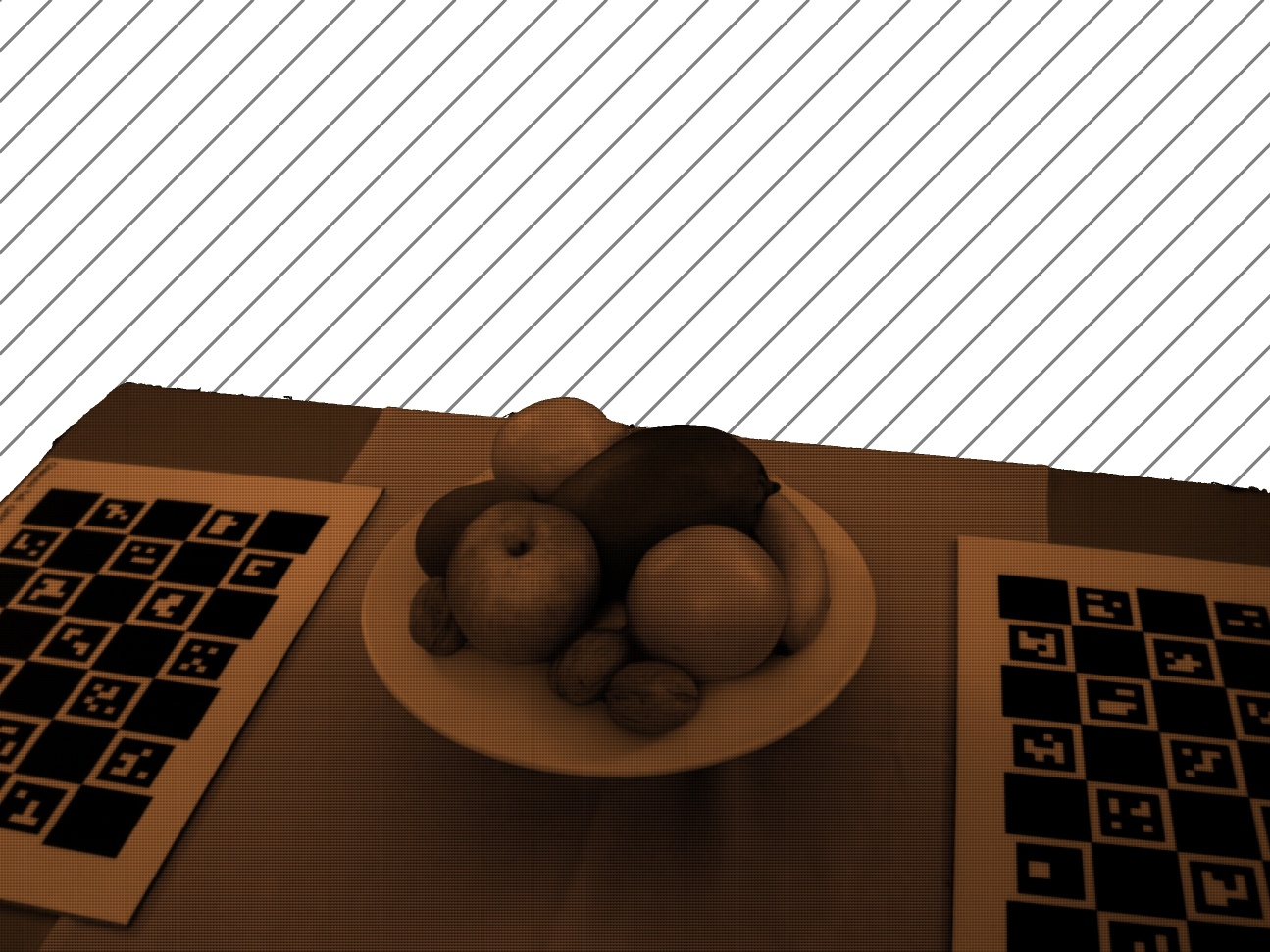} & \includegraphics[width=0.3\linewidth, valign=m]{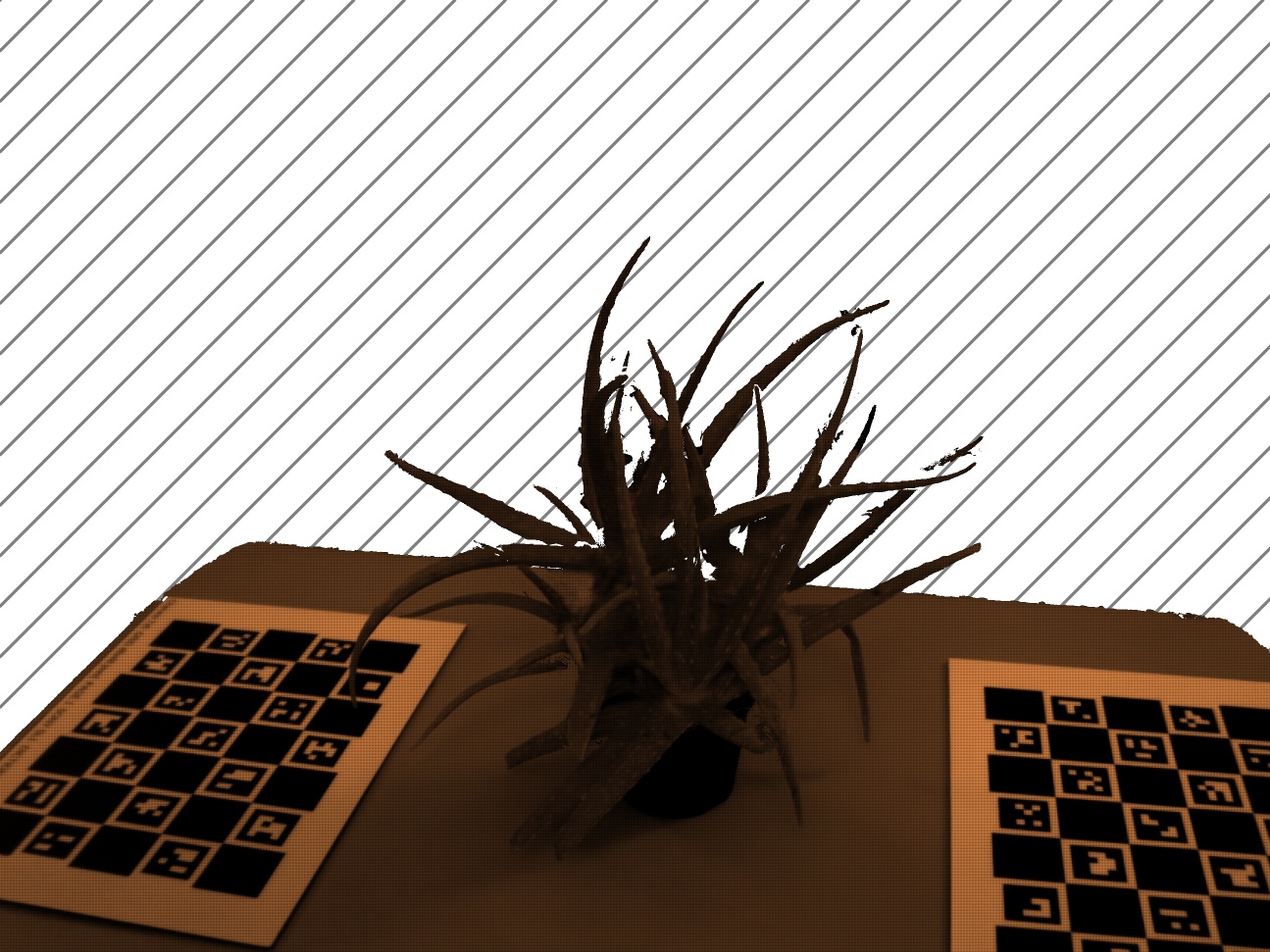} & \includegraphics[width=0.3\linewidth, valign=m]{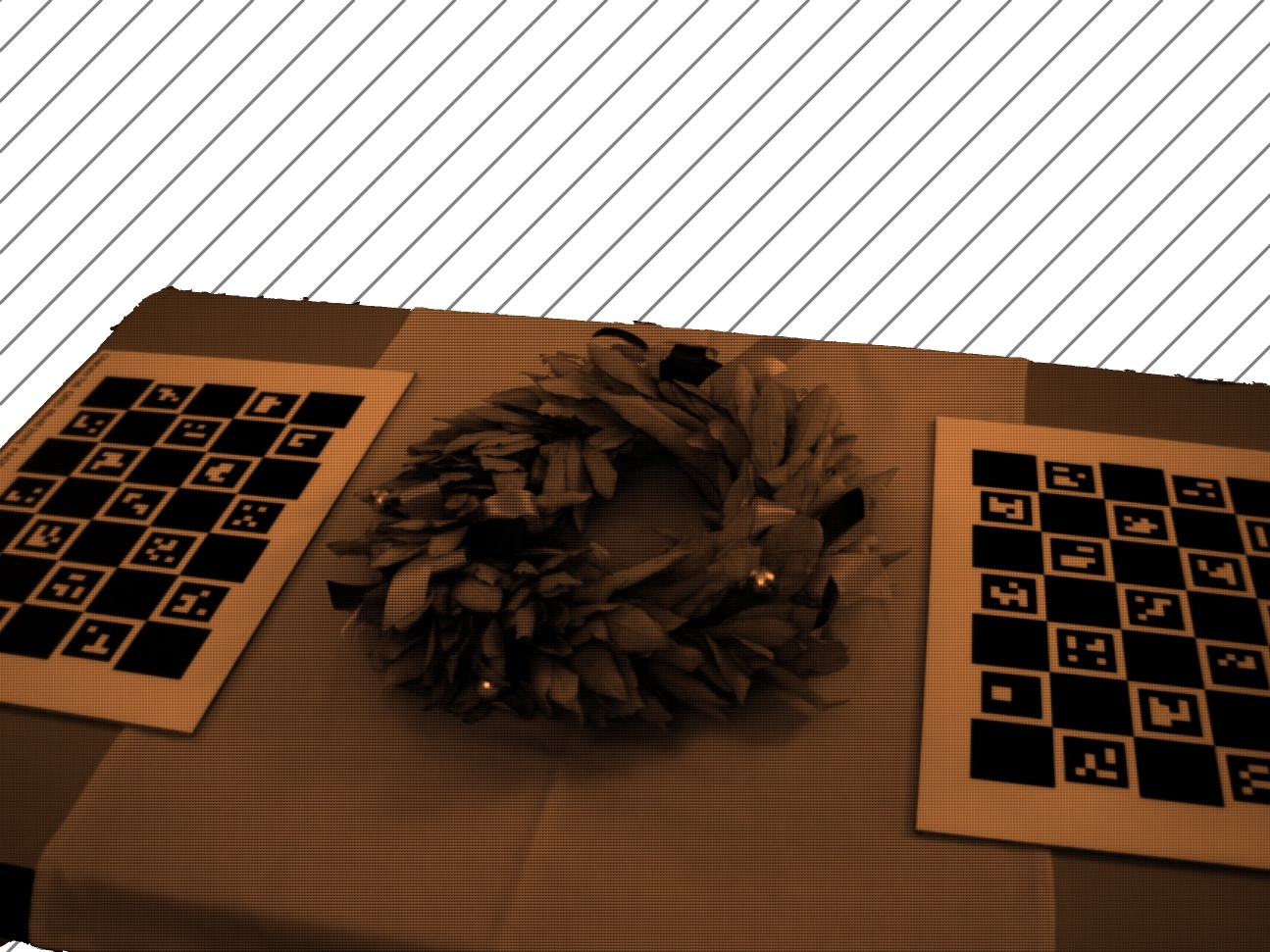} \\
        \midrule
        Mod. & Fruits & Aloe & Laurel Wreath \\
    \end{tabular}
    \caption{Aligned renderings of the Fruits, Aloe and Laurel Wreath scenes from the viewpoint number 49. Colors are for visualization purposes only.}
    \label{sup_fig:aligned}
\end{figure*}

\begin{figure*}
    \centering
    \begin{tabular}{@{\extracolsep{-10pt}}cccc}
        GT Raw Frame & DoLP & AoLP & \\
        \midrule
        \multirow{2}{*}{\includegraphics[width=0.3\linewidth, valign=m]{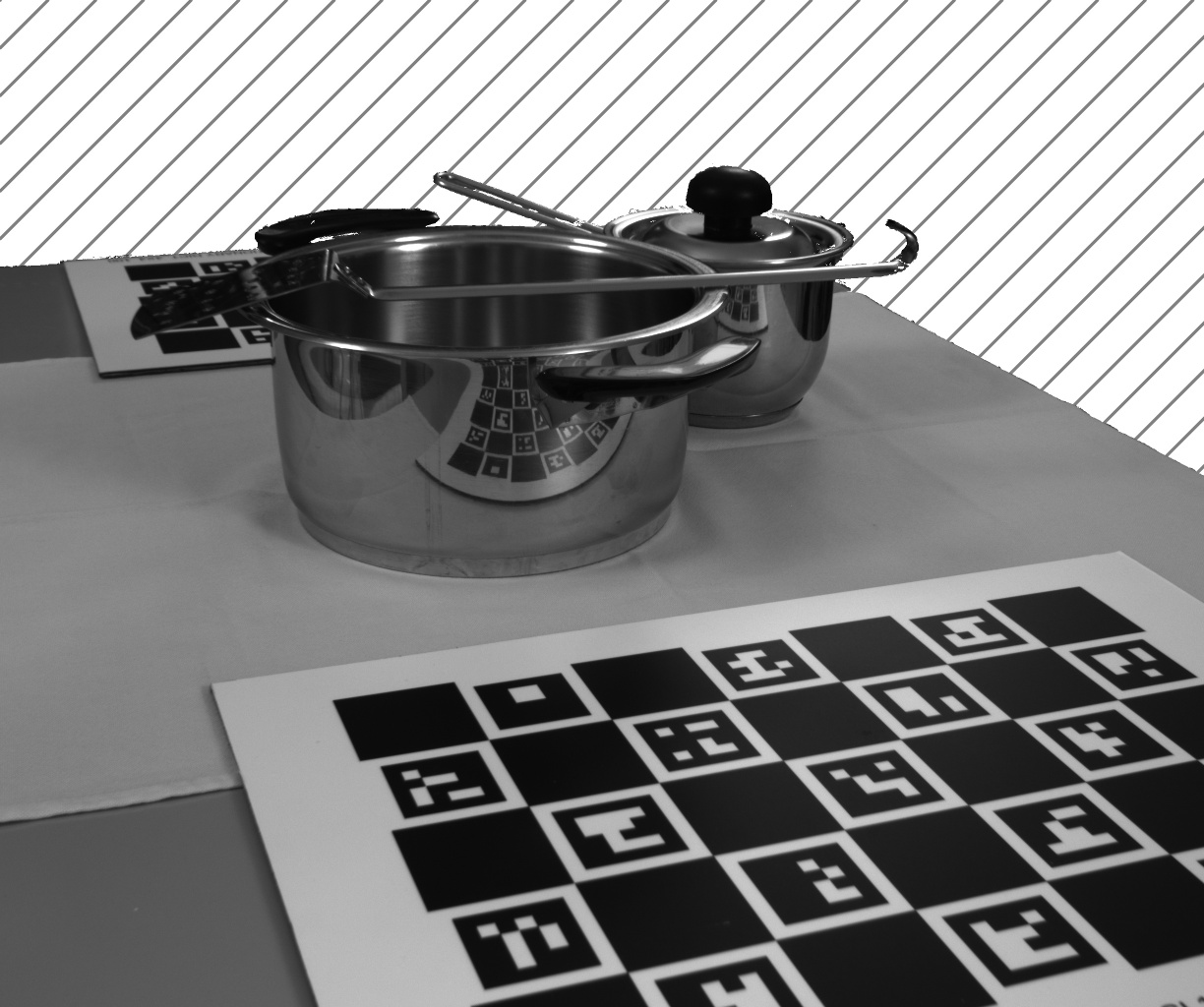}} & \includegraphics[width=0.3\linewidth, valign=m]{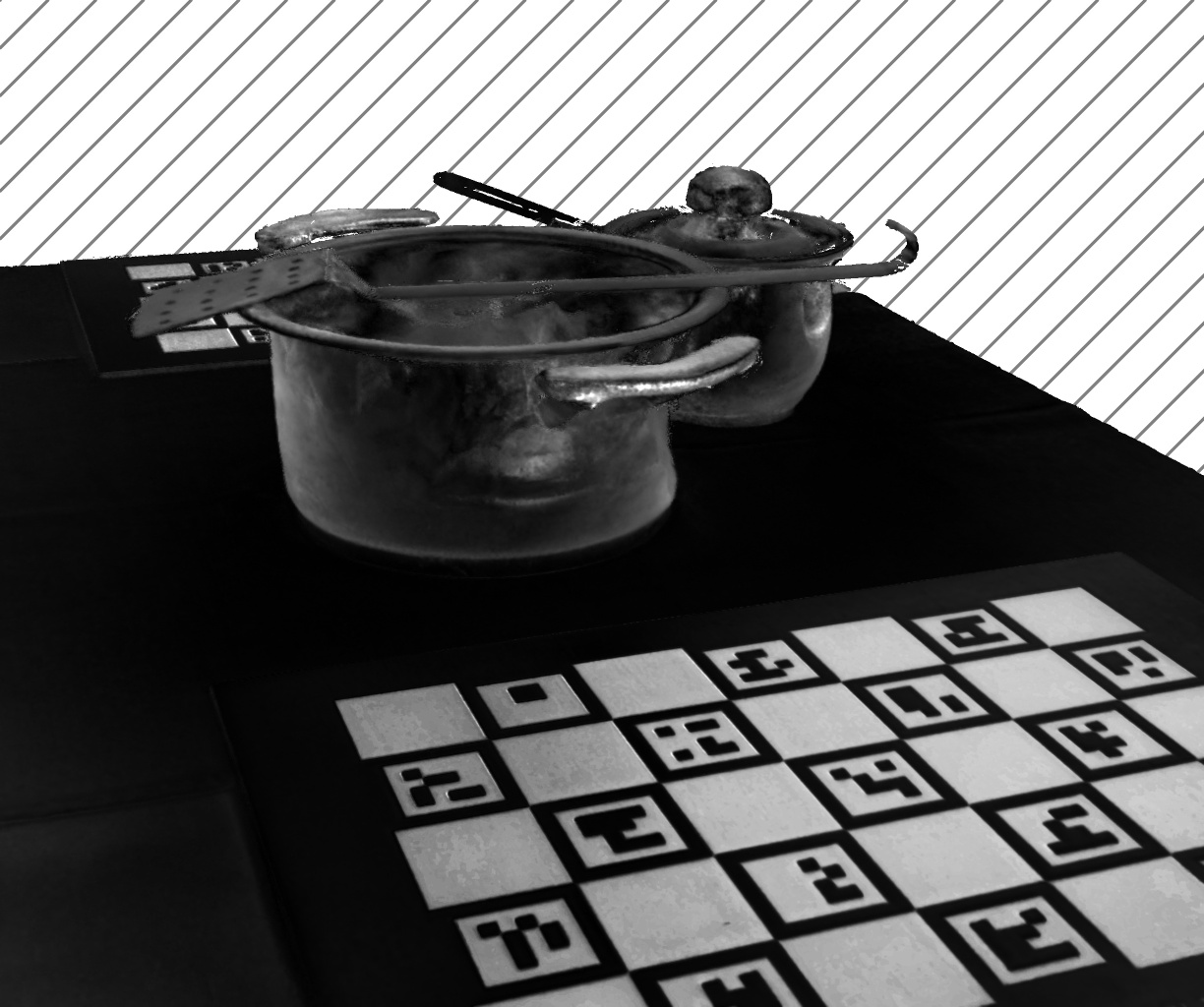} & \includegraphics[width=0.3\linewidth, valign=m]{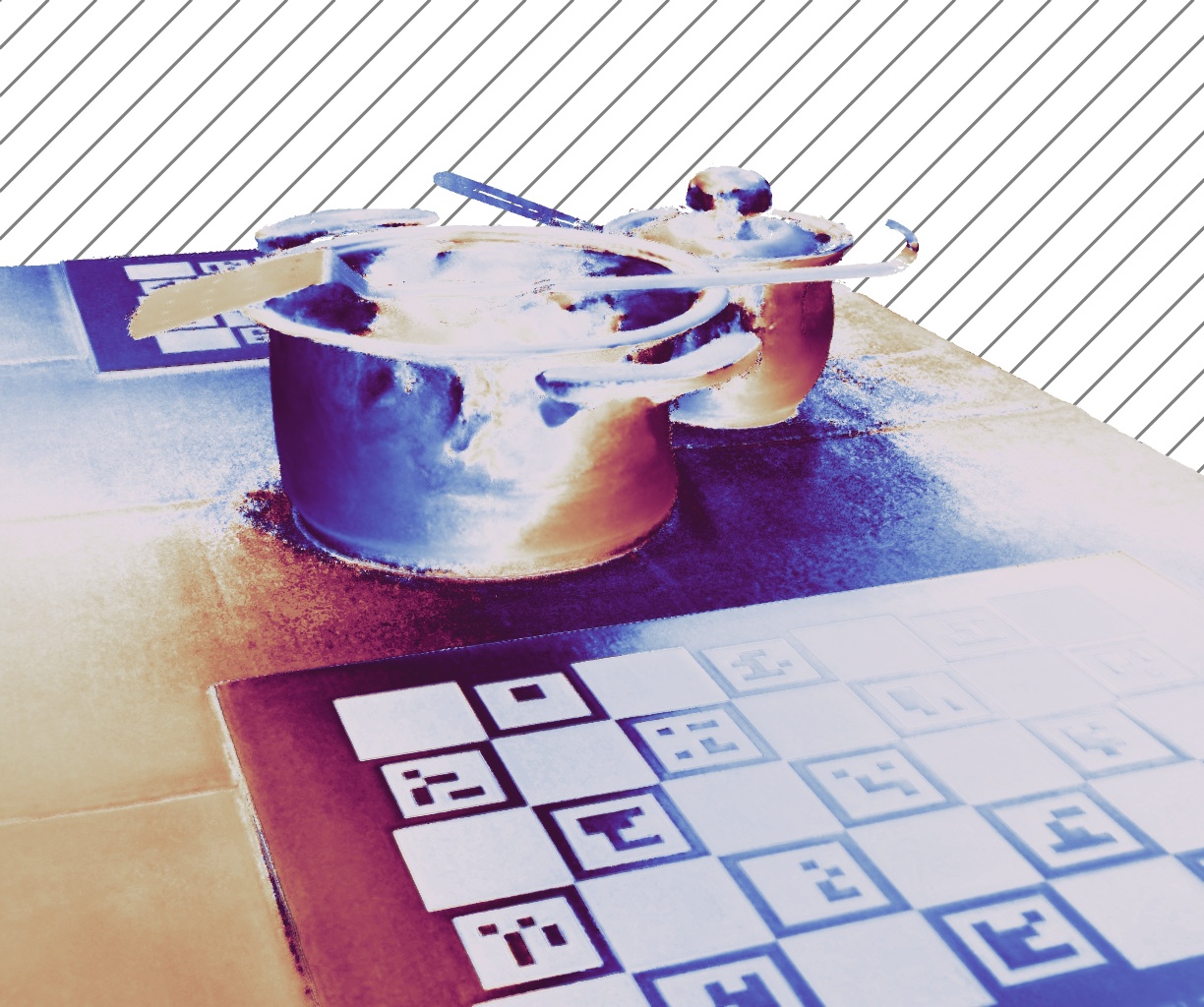} & \multirow{1}{*}[15pt]{\rotatebox{90}{Rendering}} \\
         & \includegraphics[width=0.3\linewidth, valign=m]{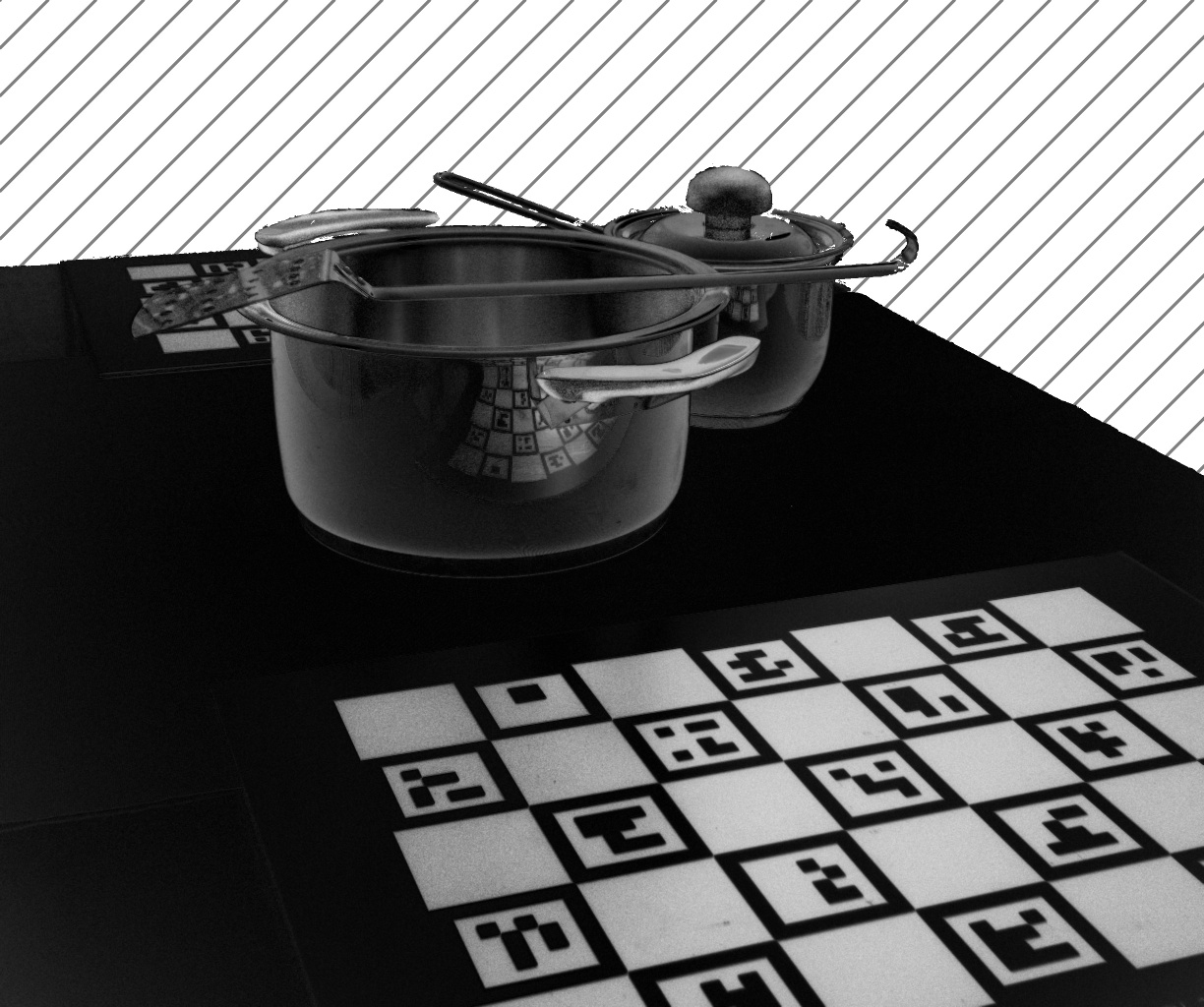} & \includegraphics[width=0.3\linewidth, valign=m]{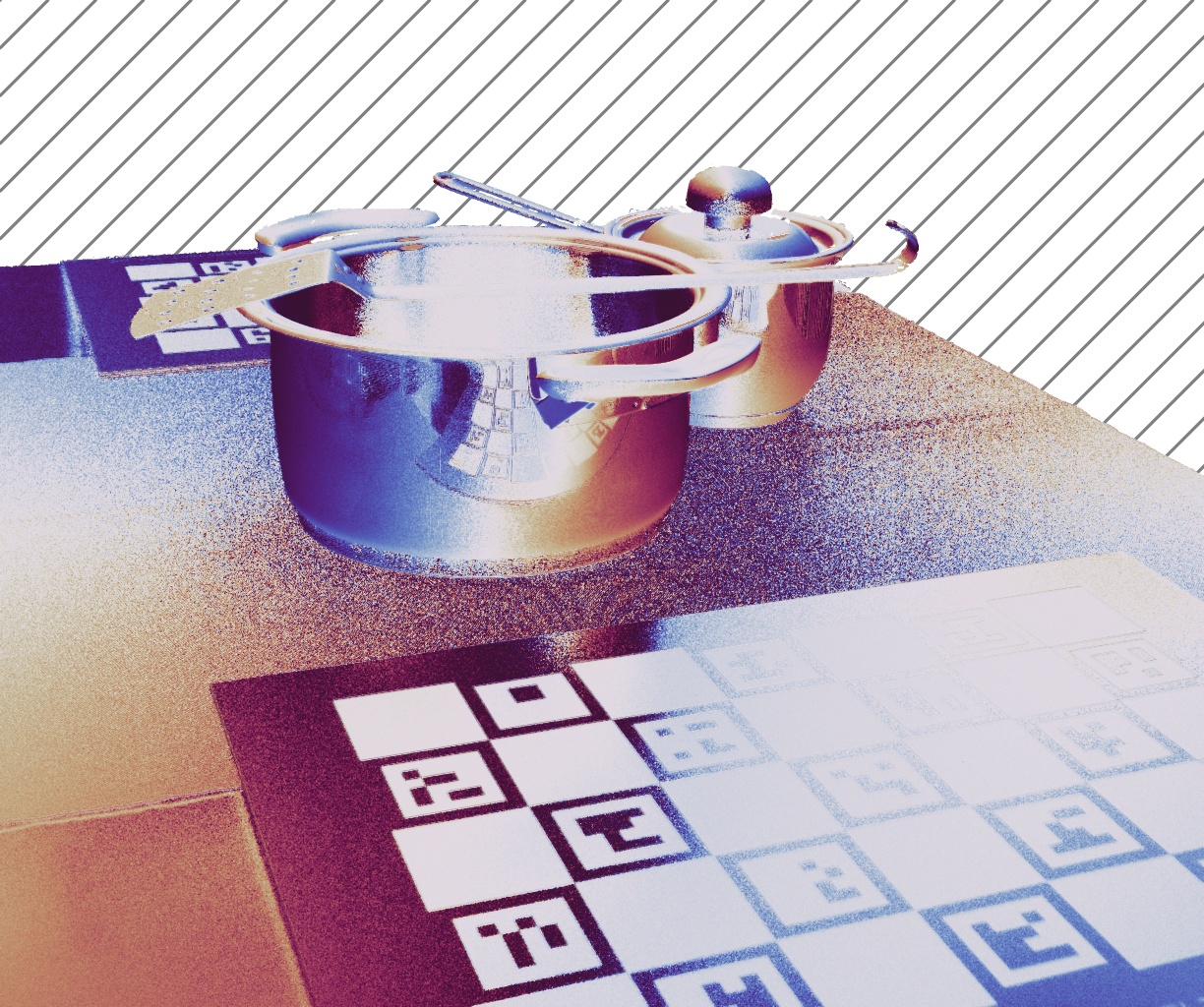} & \multirow{1}{*}[20pt]{\rotatebox{90}{Ground Truth}} \\
        \midrule
        \multirow{2}{*}{\includegraphics[width=0.3\linewidth, valign=m]{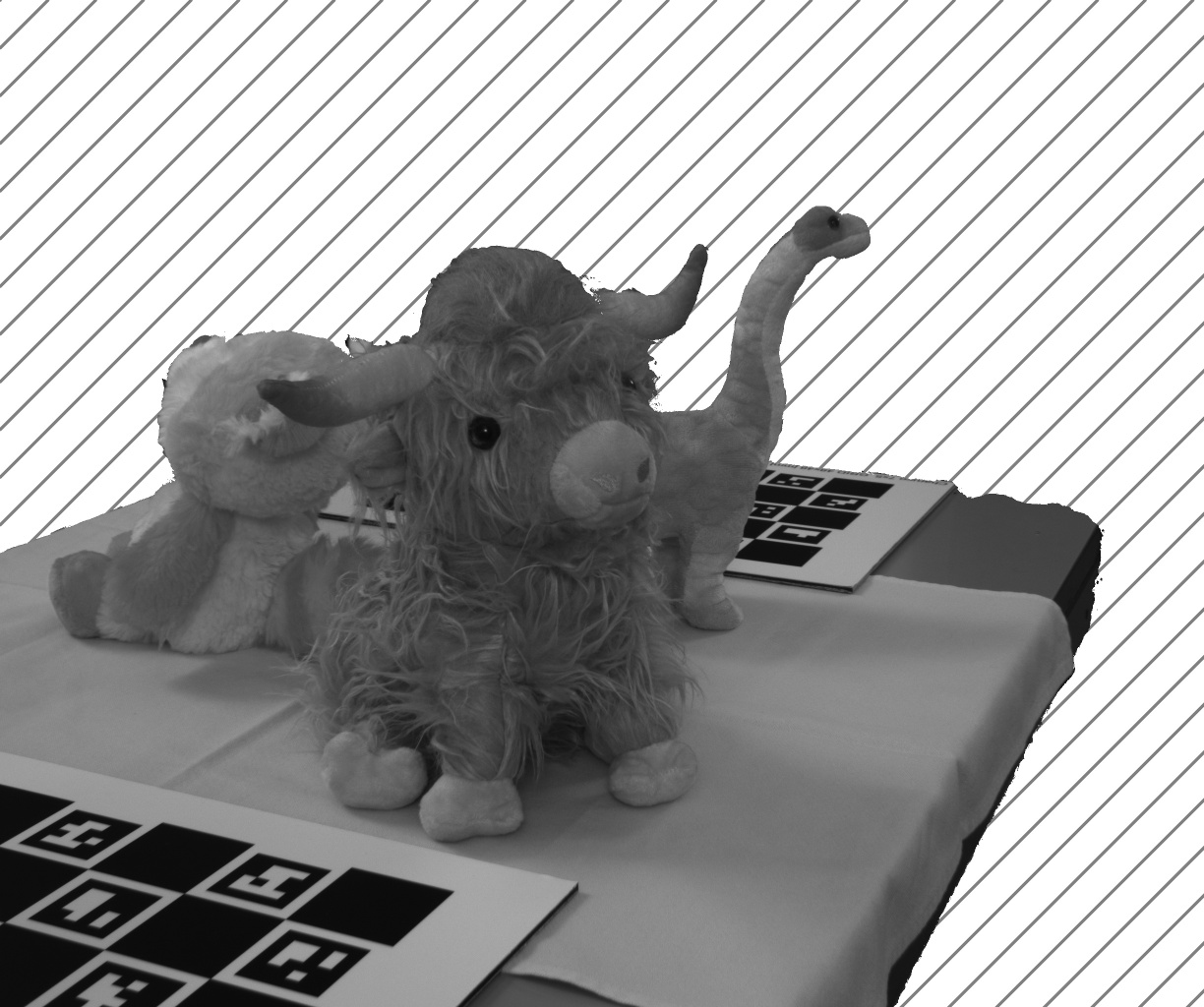}} & \includegraphics[width=0.3\linewidth, valign=m]{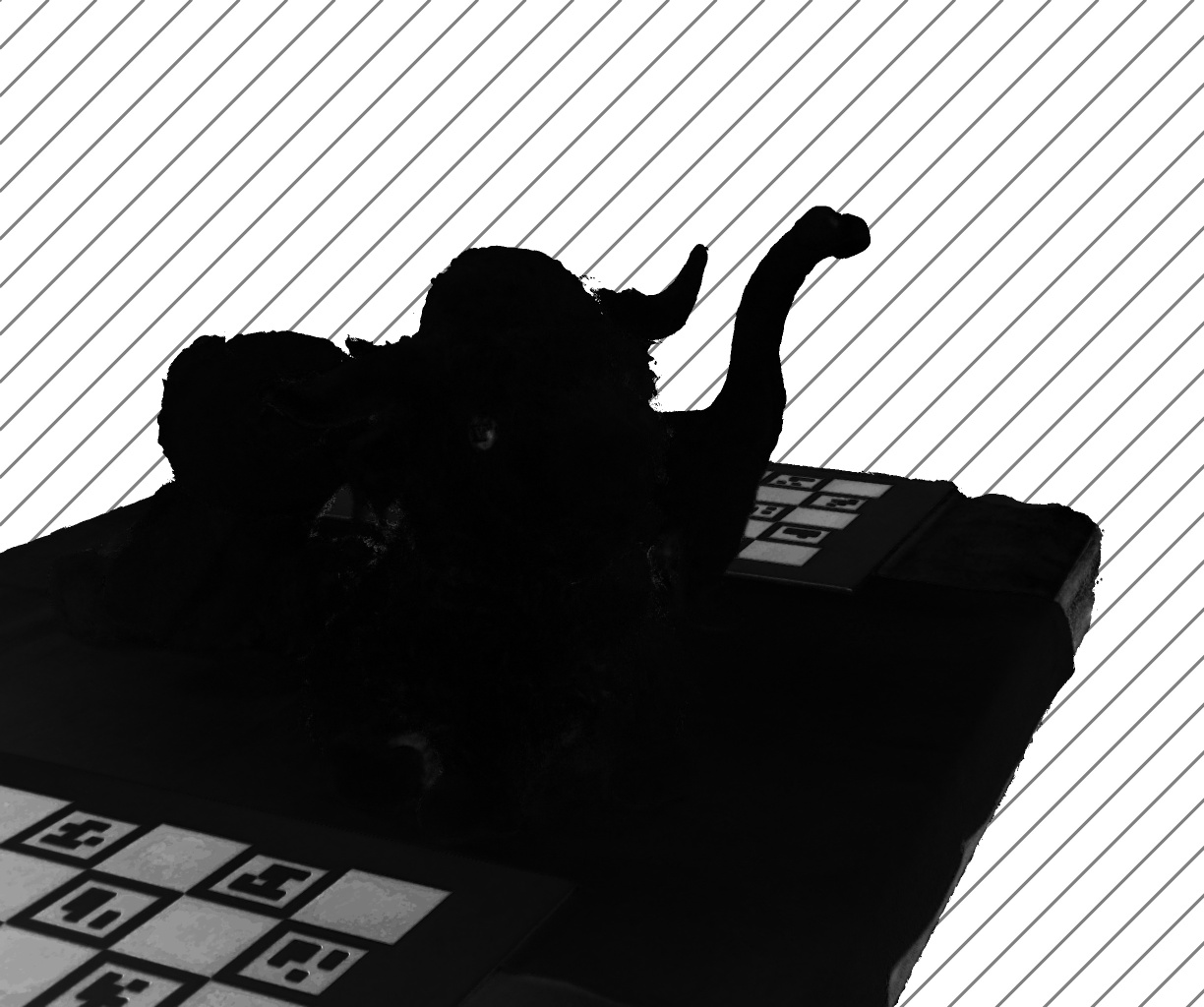} & \includegraphics[width=0.3\linewidth, valign=m]{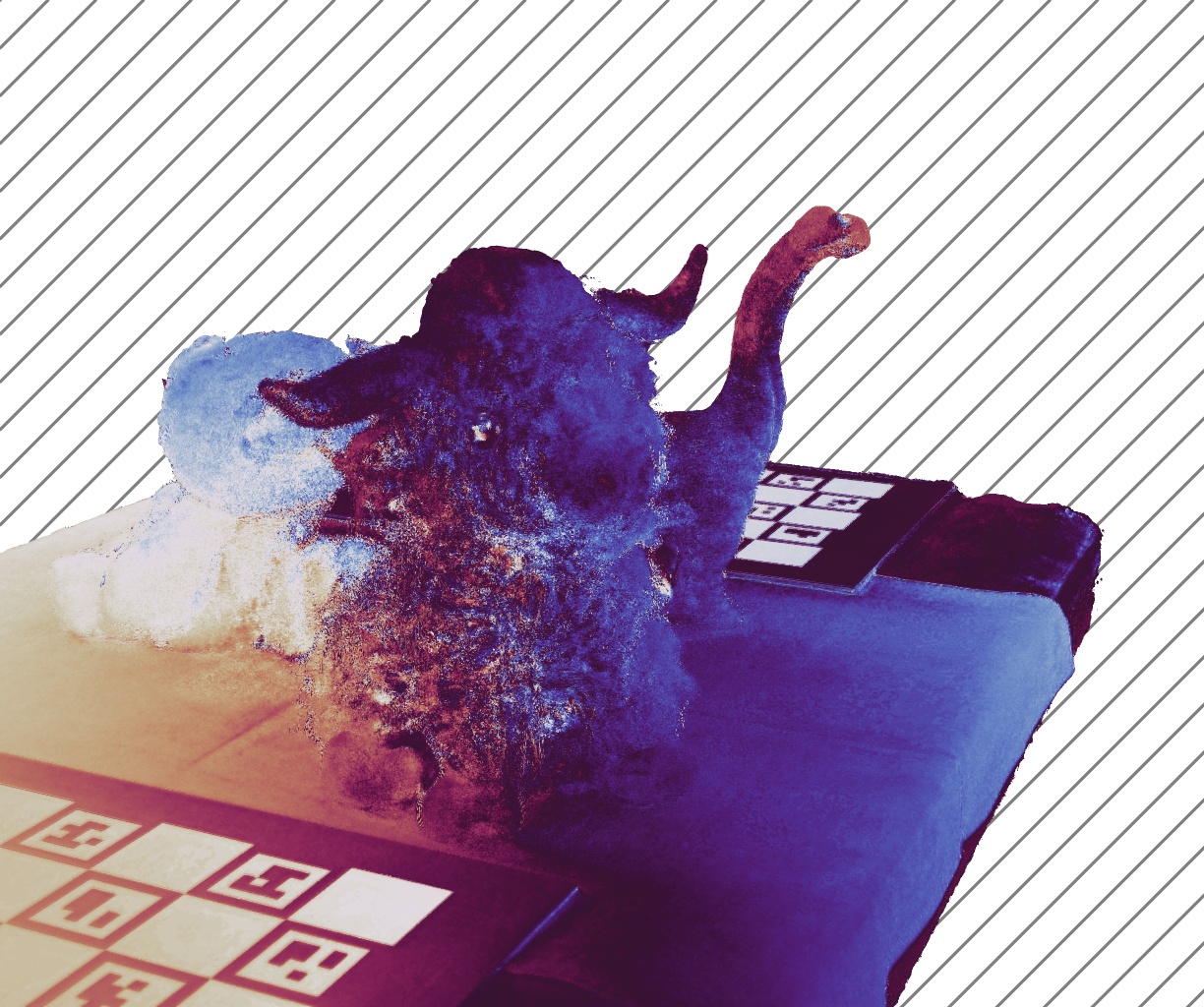} & \multirow{1}{*}[15pt]{\rotatebox{90}{Rendering}} \\
         & \includegraphics[width=0.3\linewidth, valign=m]{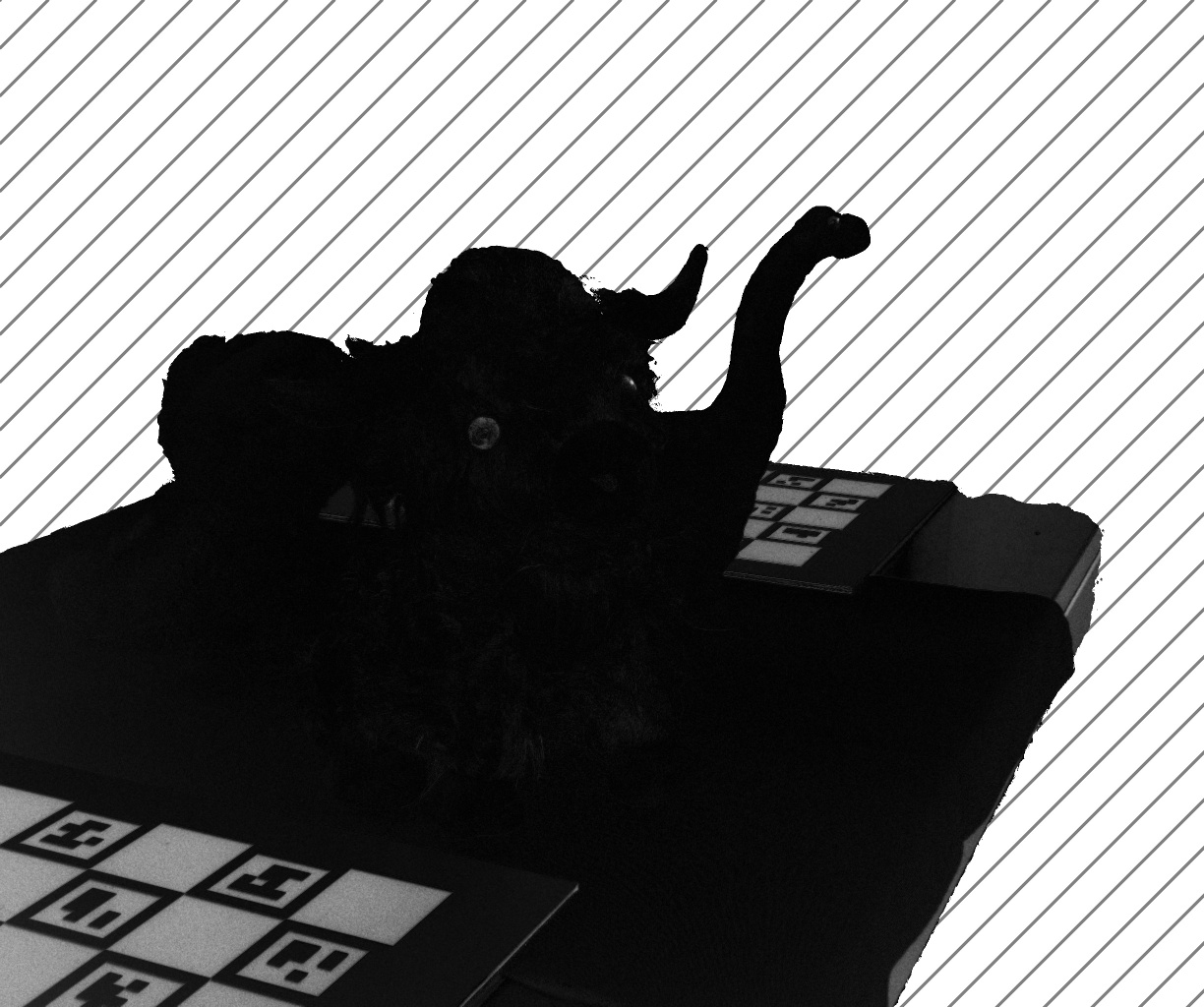} & \includegraphics[width=0.3\linewidth, valign=m]{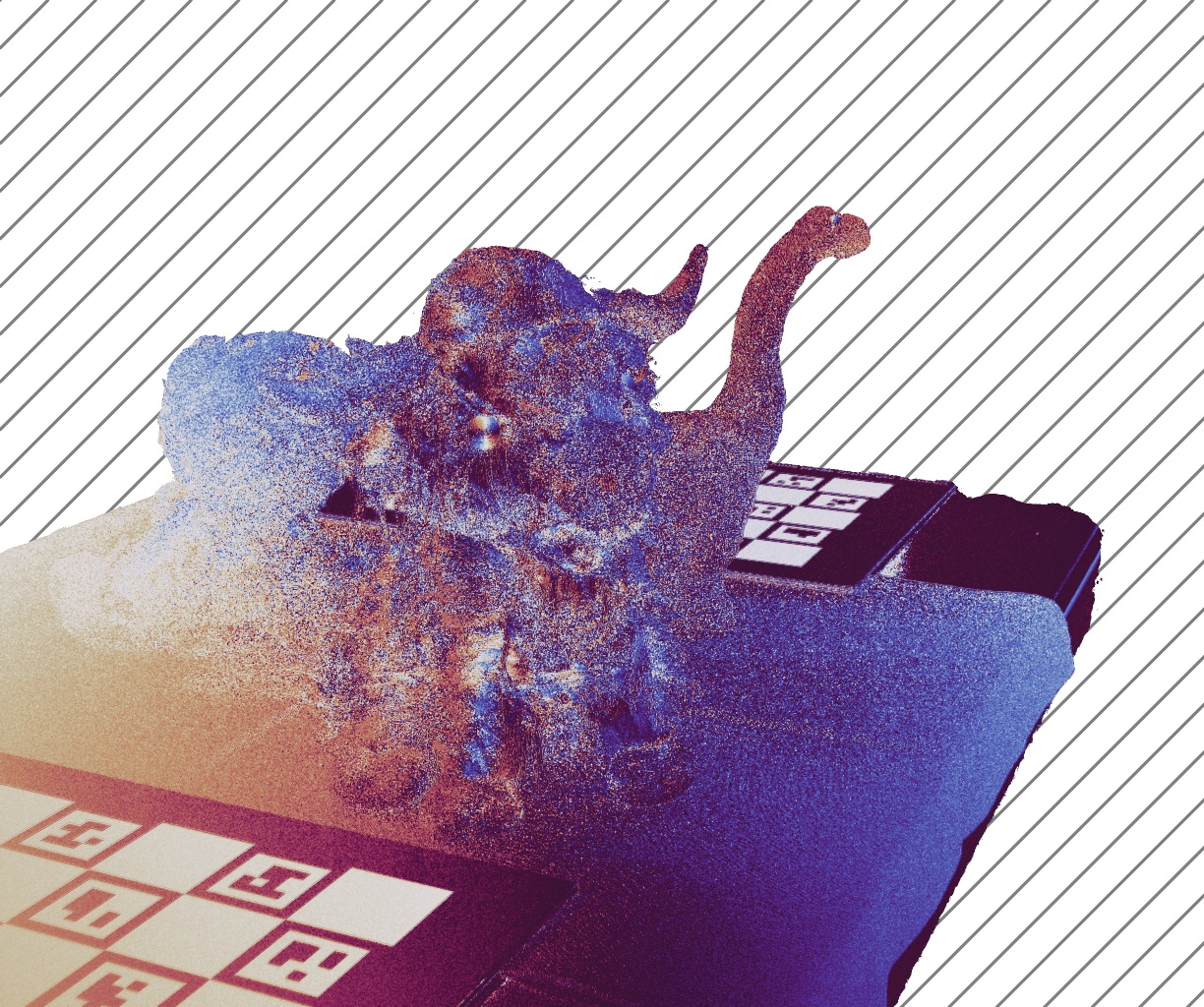} & \multirow{1}{*}[20pt]{\rotatebox{90}{Ground Truth}} \\
         \midrule
        GT Raw Frame & DoLP & AoLP & \\
    \end{tabular}
    \caption{Degree of Linear Polarization (DoLP) and Angle of Linear Polarization (AoLP) of two scenes with different properties: on the upper side, Steel Pot, with mainly reflective materials; on the lower side, Forest Gang 1, with mainly diffusive materials.}
    \label{sup_fig:aop_dop}
\end{figure*}

\begin{figure*}
    \centering\begin{tabular}{c}
        RGB Ground Truth \\
        \includegraphics[width=0.5\linewidth, valign=m]{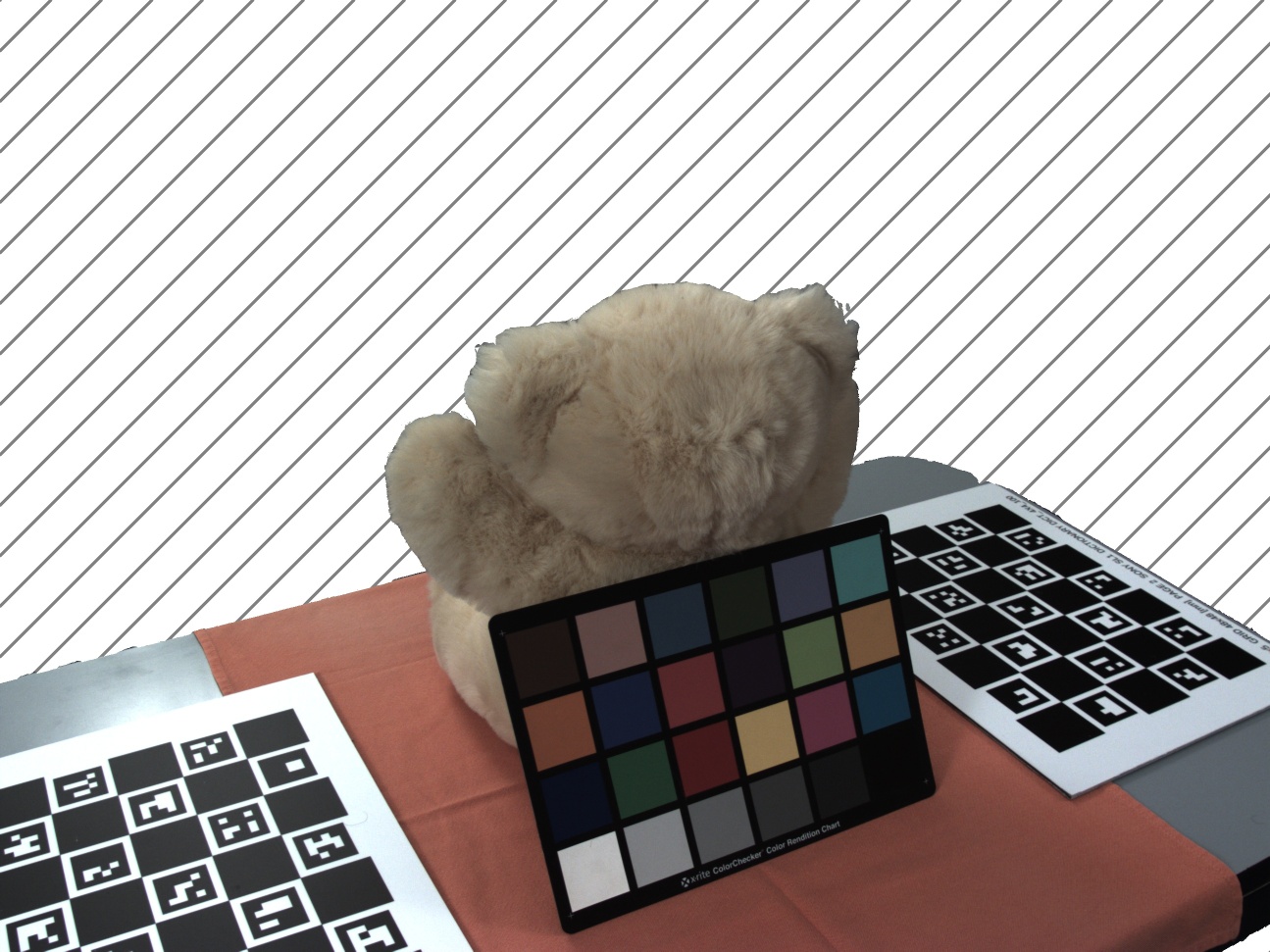}
    \end{tabular} \\
    \vspace{20pt}
    \setlength{\tabcolsep}{4.5pt}
    \begin{tabular}{cccc|cccc}
        Ch. & Band & Rendering & Ground Truth & Ch. & Band & Rendering & Ground Truth \\
        \midrule
        0 & 692~nm & \includegraphics[width=0.17\linewidth, valign=m]{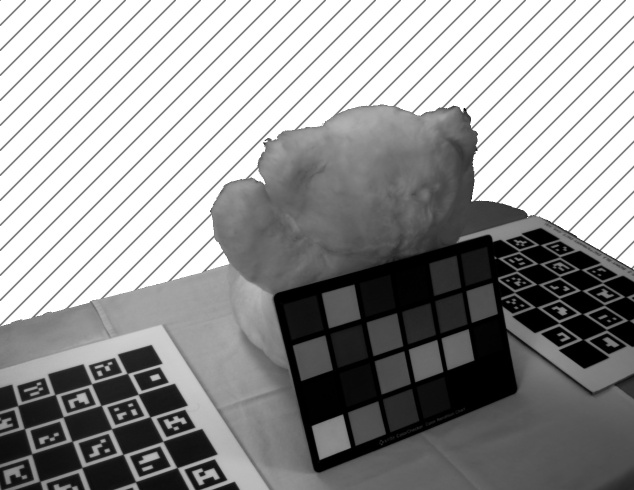} & \includegraphics[width=0.17\linewidth, valign=m]{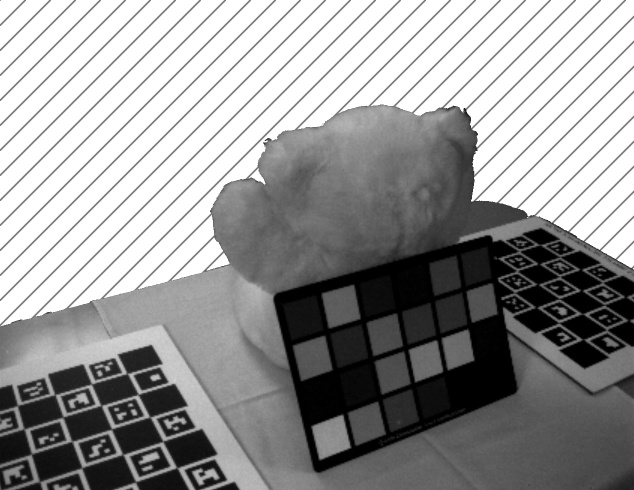} & 1 & 653~nm & \includegraphics[width=0.17\linewidth, valign=m]{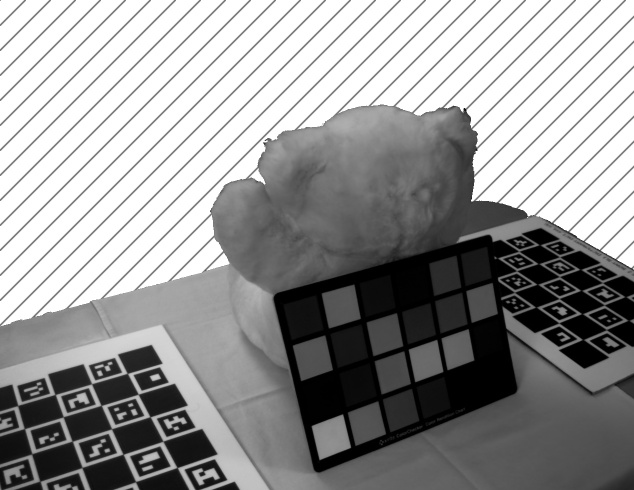} & \includegraphics[width=0.17\linewidth, valign=m]{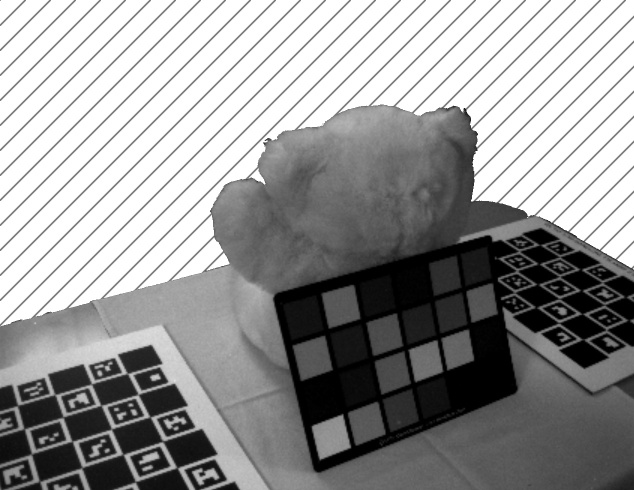} \\
        2 & 611~nm & \includegraphics[width=0.17\linewidth, valign=m]{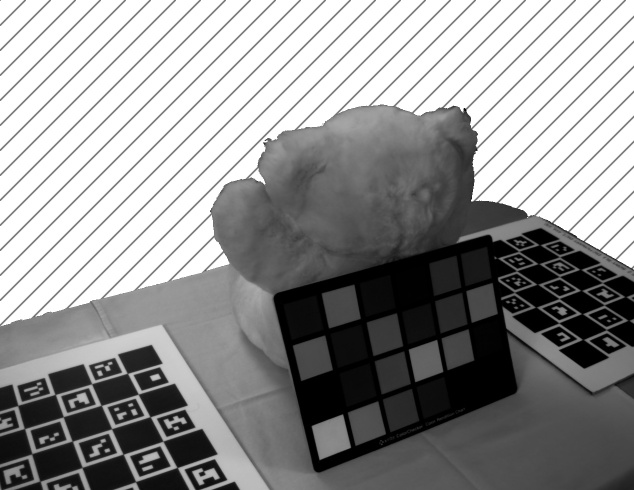} & \includegraphics[width=0.17\linewidth, valign=m]{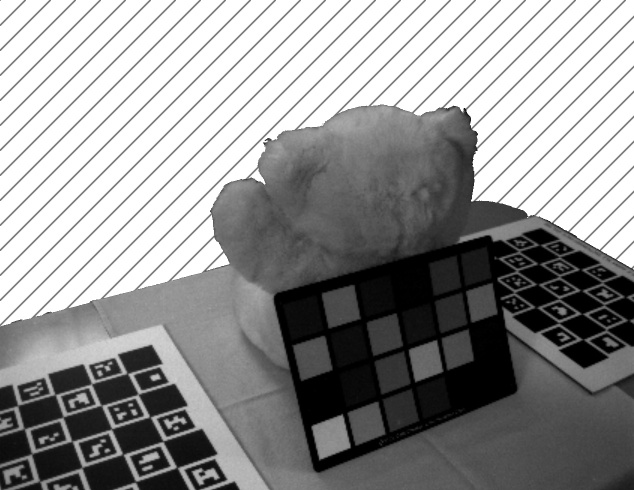} & 3 & 572~nm & \includegraphics[width=0.17\linewidth, valign=m]{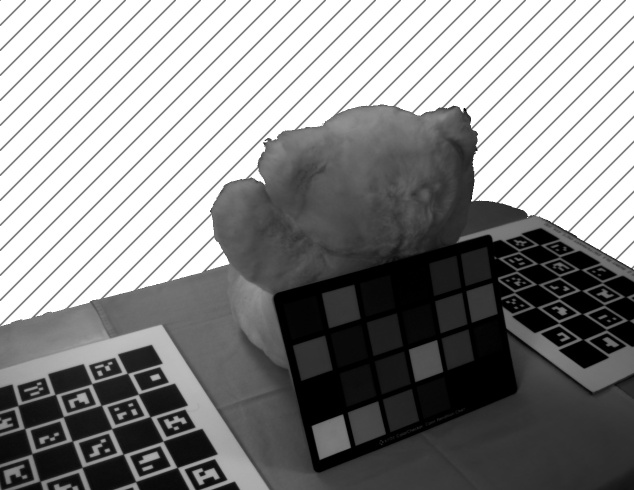} & \includegraphics[width=0.17\linewidth, valign=m]{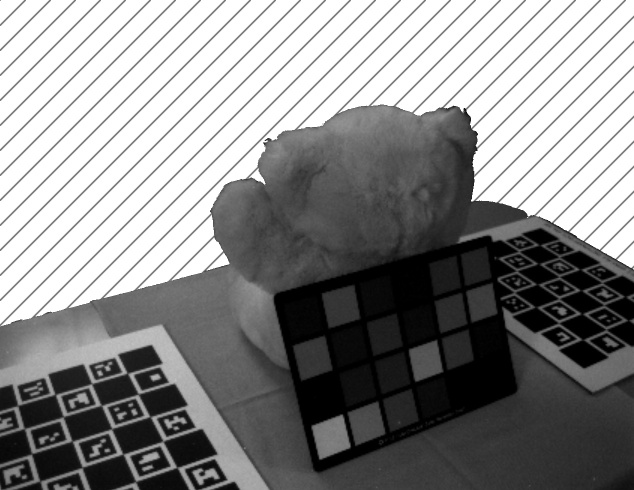} \\
        4 & 541~nm & \includegraphics[width=0.17\linewidth, valign=m]{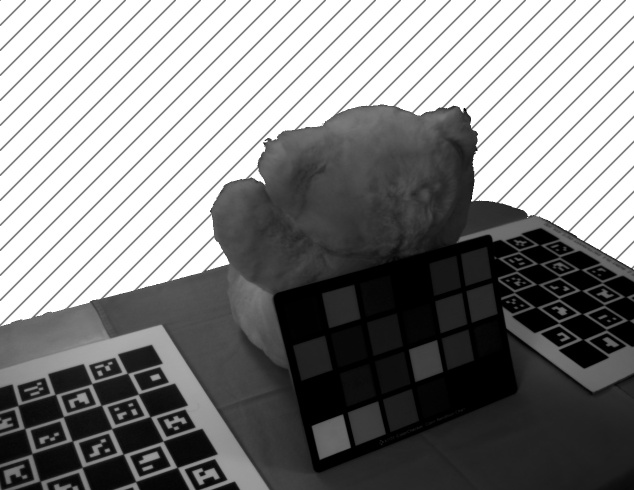} & \includegraphics[width=0.17\linewidth, valign=m]{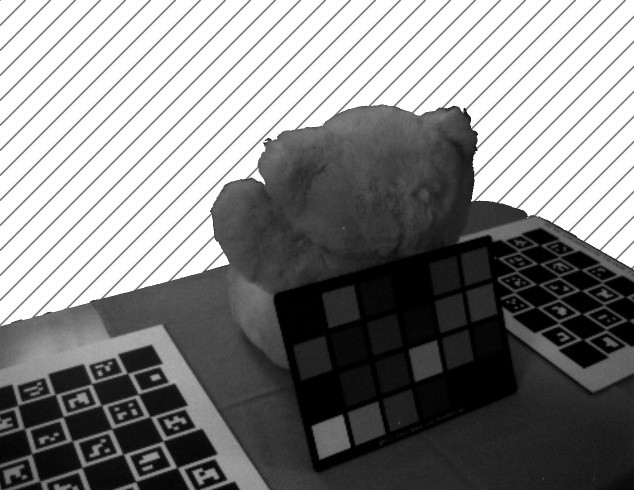} & 5 & 503~nm & \includegraphics[width=0.17\linewidth, valign=m]{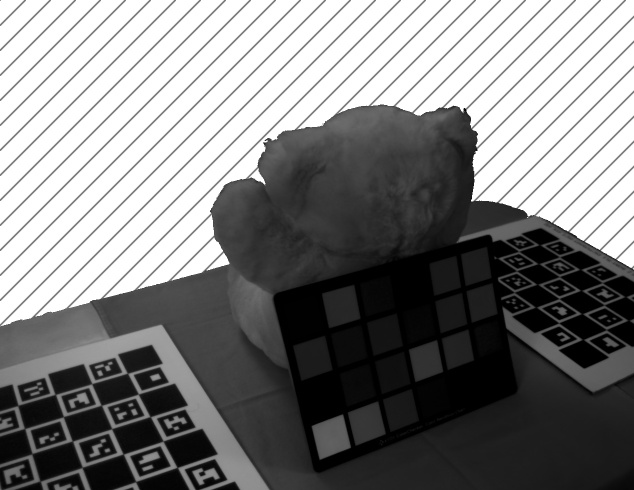} & \includegraphics[width=0.17\linewidth, valign=m]{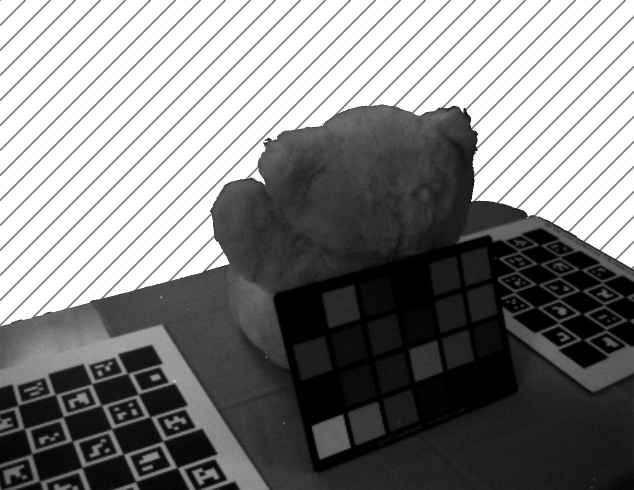} \\
        6 & 464~nm & \includegraphics[width=0.17\linewidth, valign=m]{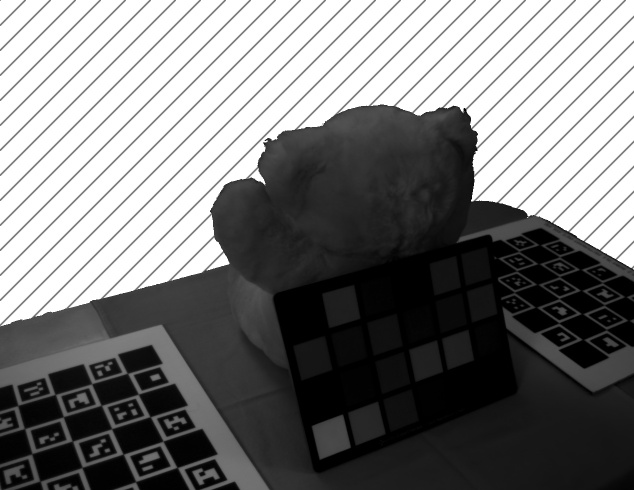} & \includegraphics[width=0.17\linewidth, valign=m]{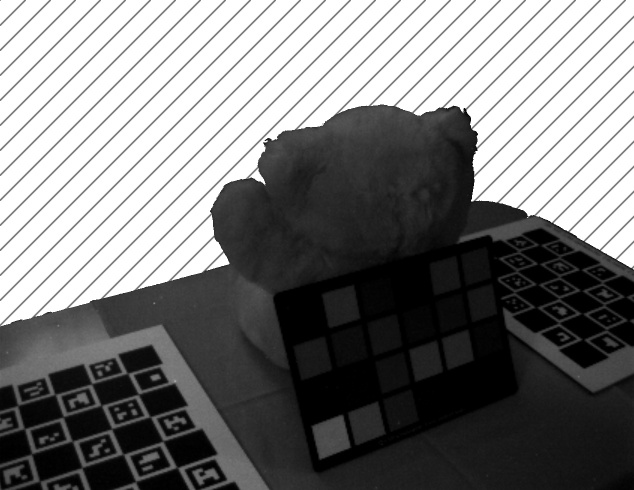} & 7 & 431~nm & \includegraphics[width=0.17\linewidth, valign=m]{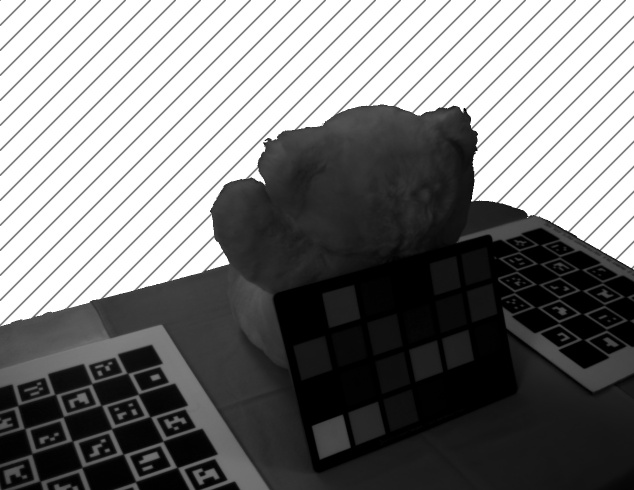} & \includegraphics[width=0.17\linewidth, valign=m]{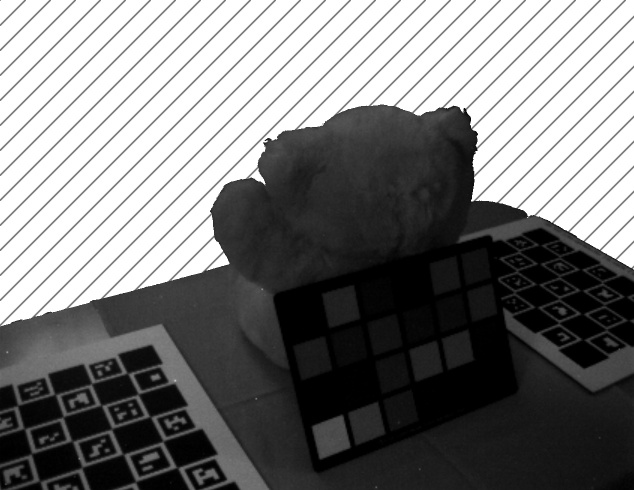} \\
    \end{tabular}
    \caption{Teddy Bear scene. On the top, the RGB ground truth view of the scene, for color reference purposes. On the bottom, the Multispectral individual channel renderings and ground truths.}
    \label{sup_fig:multispectral_single_channels_teddybear}
\end{figure*}

\begin{figure*}
    \centering
    \begin{tabular}{c}
        RGB Ground Truth \\
        \includegraphics[width=0.5\linewidth, valign=m]{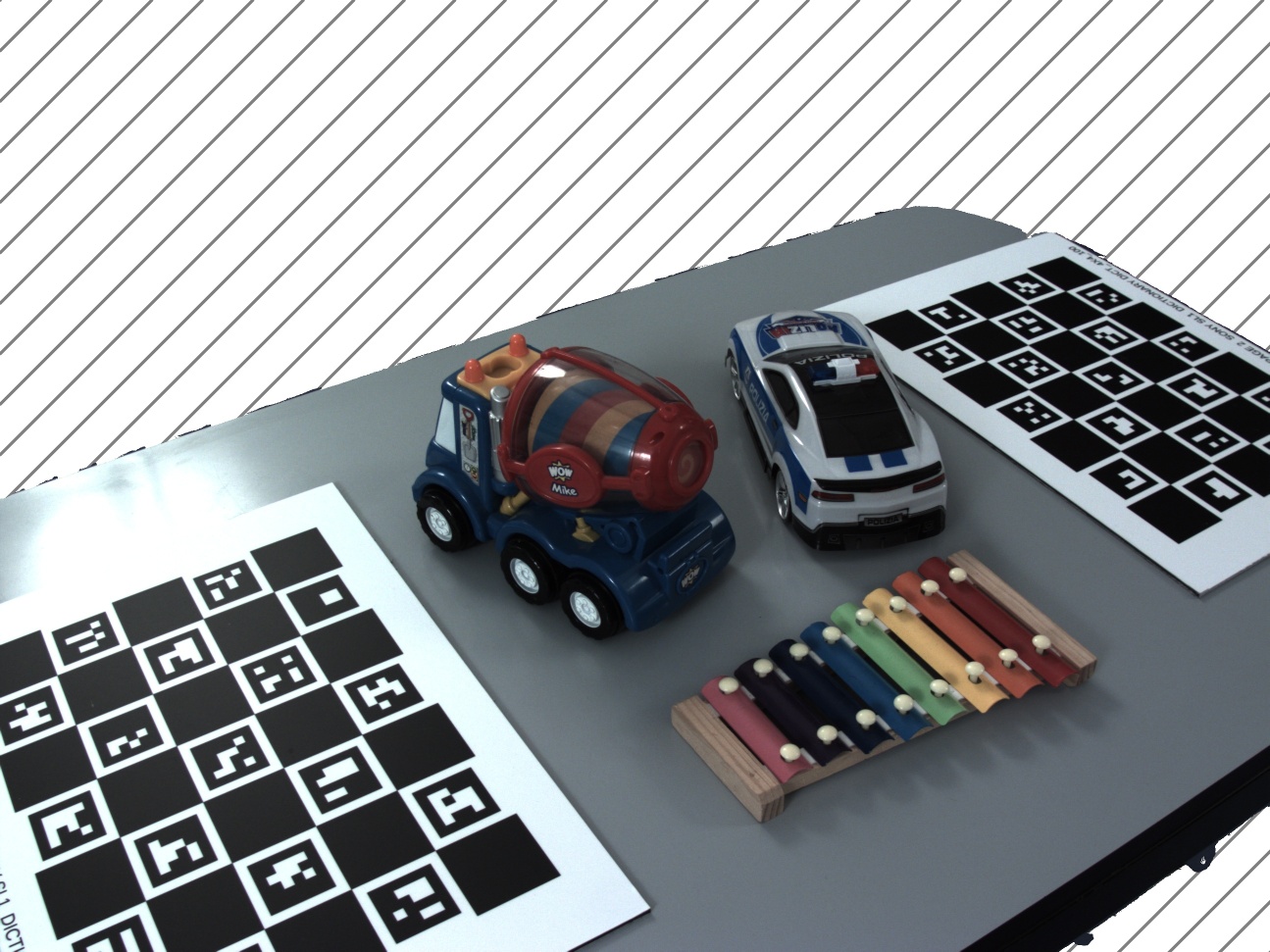}
    \end{tabular} \\
    \vspace{20pt}
    \setlength{\tabcolsep}{4.5pt}
    \begin{tabular}{cccc|cccc}
        Ch. & Band & Rendering & Ground Truth & Ch. & Band & Rendering & Ground Truth \\
        \midrule
        0 & 692~nm &\includegraphics[width=0.17\linewidth, valign=m]{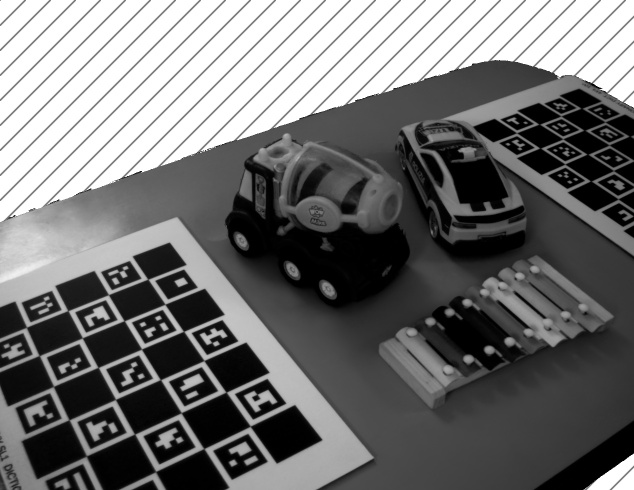} & \includegraphics[width=0.17\linewidth, valign=m]{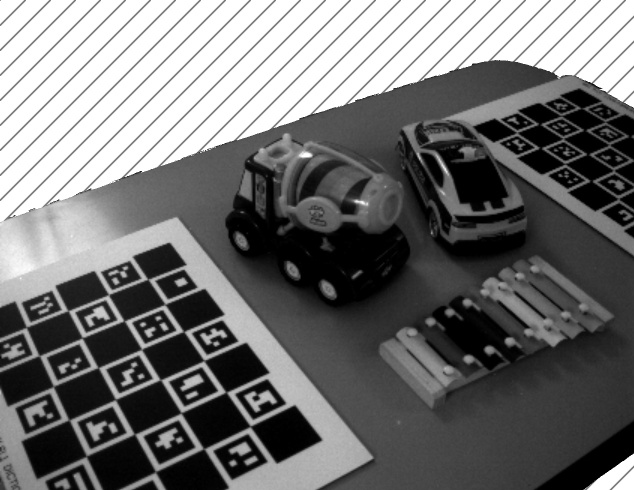} & 1 & 653~nm & \includegraphics[width=0.17\linewidth, valign=m]{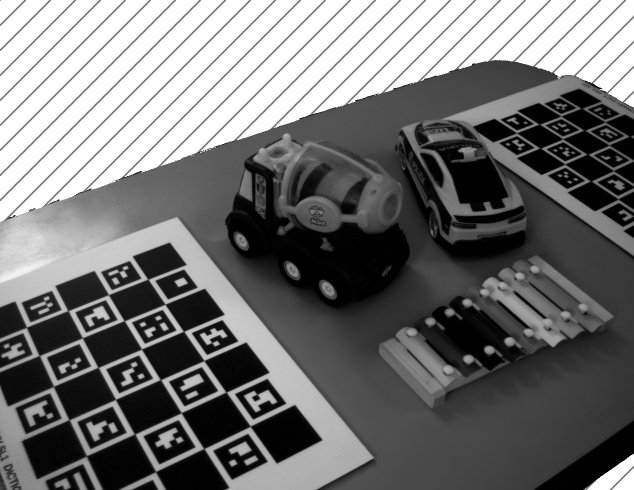} & \includegraphics[width=0.17\linewidth, valign=m]{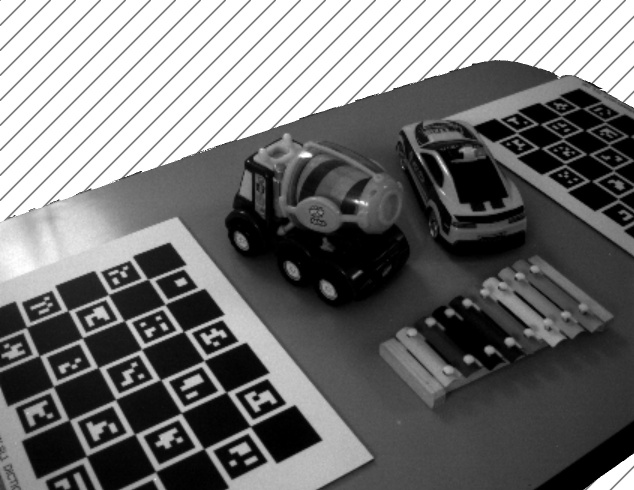} \\
        2 & 611~nm & \includegraphics[width=0.17\linewidth, valign=m]{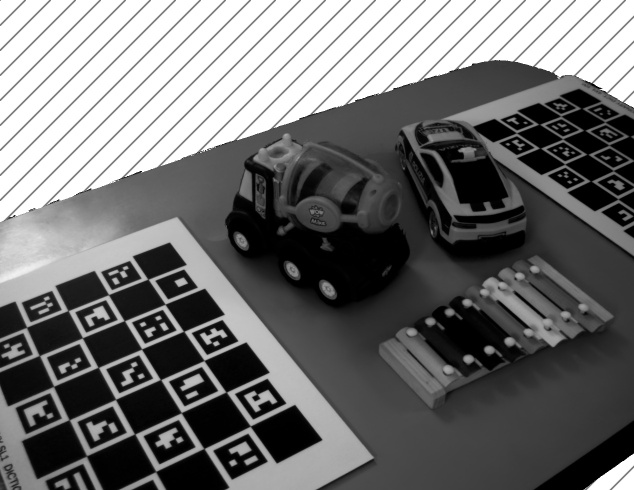} & \includegraphics[width=0.17\linewidth, valign=m]{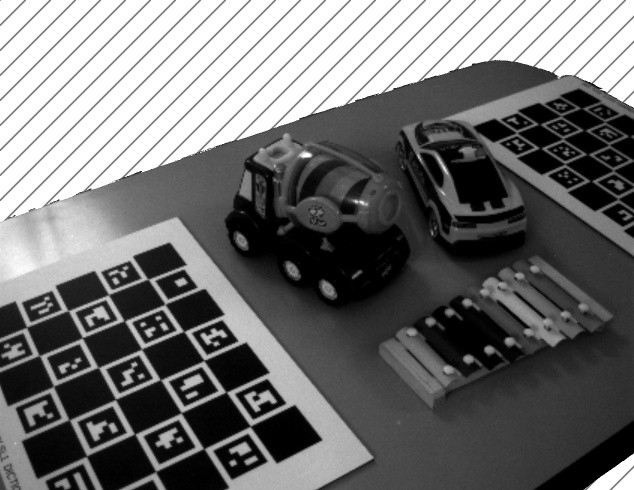} & 3 & 572~nm & \includegraphics[width=0.17\linewidth, valign=m]{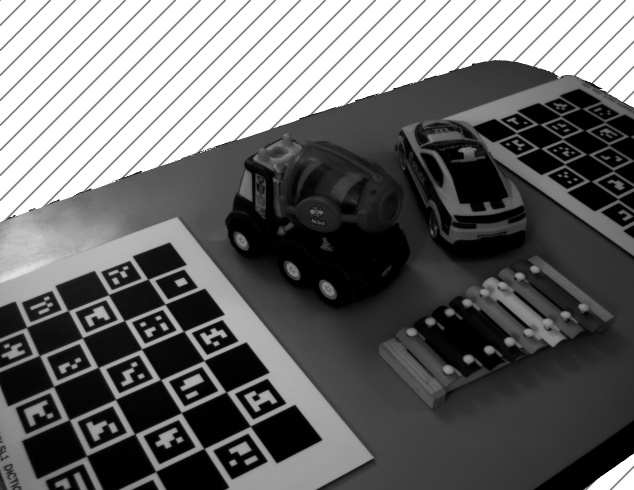} & \includegraphics[width=0.17\linewidth, valign=m]{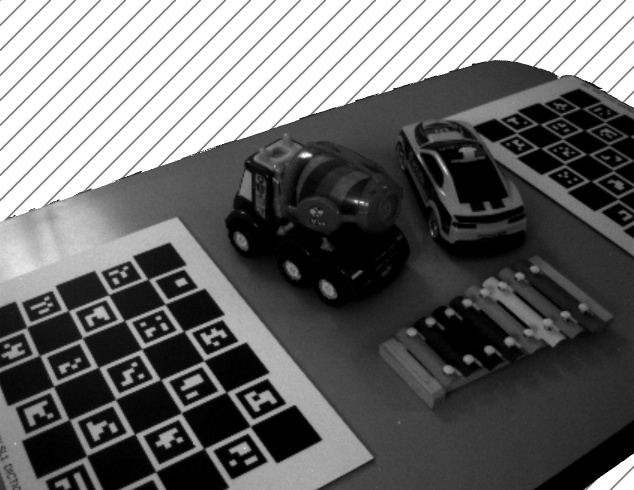} \\
        4 & 541~nm & \includegraphics[width=0.17\linewidth, valign=m]{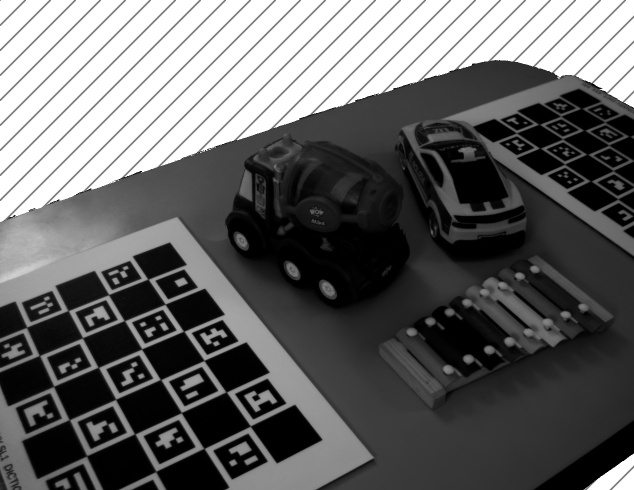} & \includegraphics[width=0.17\linewidth, valign=m]{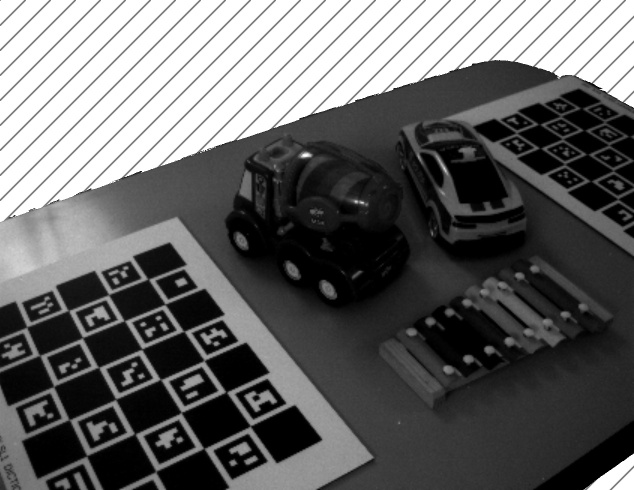} & 5 & 503~nm & \includegraphics[width=0.17\linewidth, valign=m]{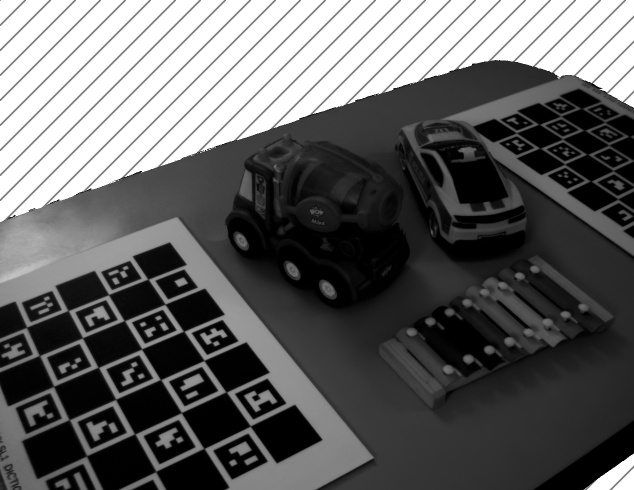} & \includegraphics[width=0.17\linewidth, valign=m]{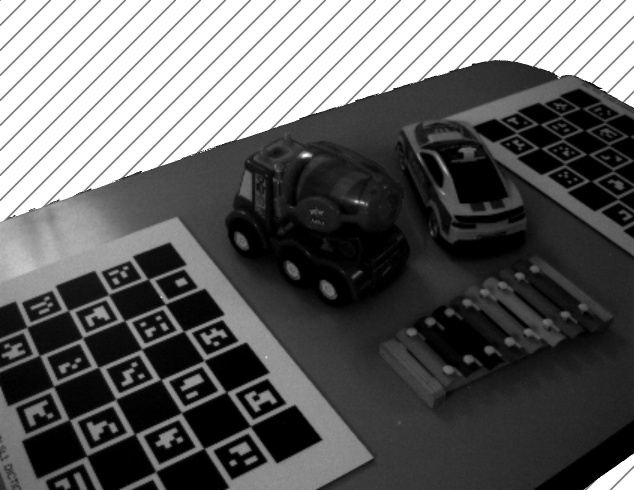} \\
        6 & 464~nm & \includegraphics[width=0.17\linewidth, valign=m]{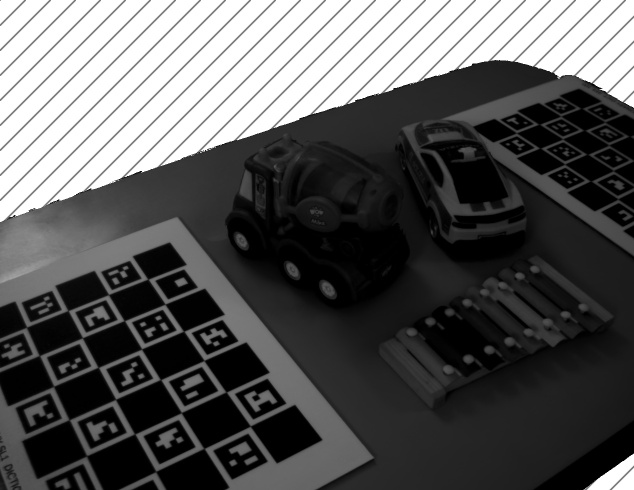} & \includegraphics[width=0.17\linewidth, valign=m]{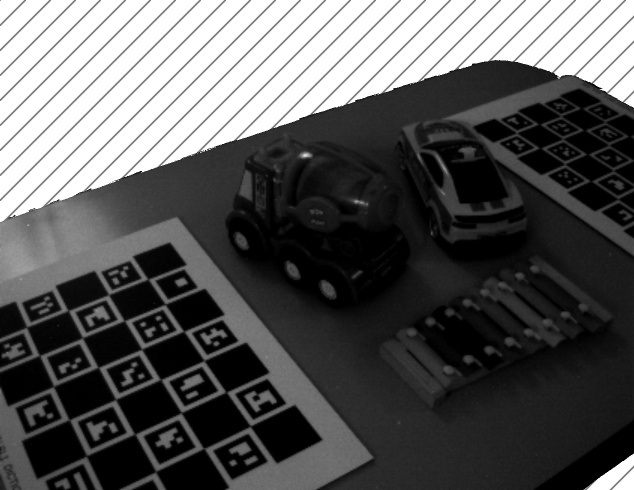} & 7 & 431~nm & \includegraphics[width=0.17\linewidth, valign=m]{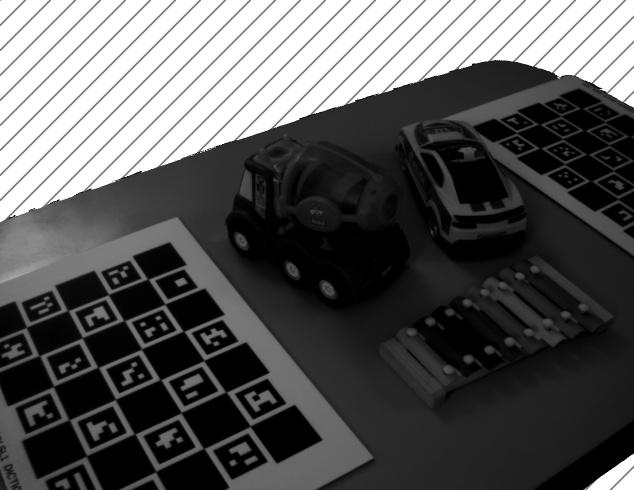} & \includegraphics[width=0.17\linewidth, valign=m]{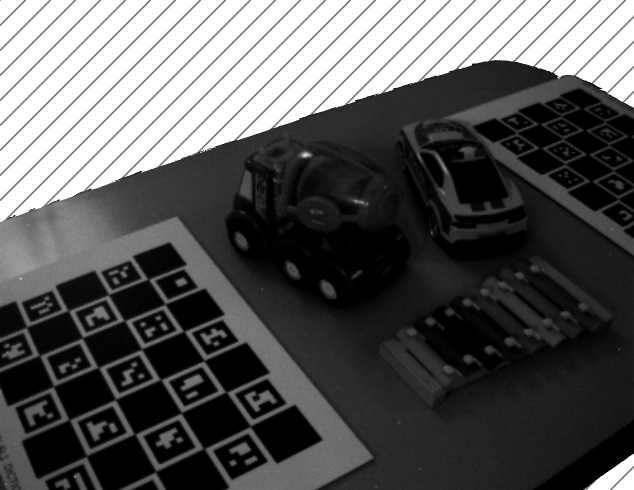} \\
    \end{tabular}
    \caption{Toys scene. On the top, the RGB ground truth view of the scene, for color reference purposes. On the bottom, the Multispectral individual channel renderings and ground truths.}
    \label{sup_fig:multispectral_single_channels_toys}
\end{figure*}

In this section, we provide an extensive overview of the rendering results obtained by the five-modality training described in Sec. 5.2
and in Tab. 4
of the main paper. In \cref{sup_tab:all_scenes} it is possible to see the PSNR achieved for each scene, averaged on the five test views. Analyzing the results, we observe that the model struggles with scenes mostly containing reflective or transparent materials. For instance, these include the Pillow, Steel Pot, Tin Box 1, Tin Box 2, Trophies, and Glass Clock scenes. This mainly happens because of specular reflections, as confirmed by \cref{sup_fig:rgb_raw_renedrings,sup_fig:mono_raw_renedrings,sup_fig:infrared_raw_renedrings,sup_fig:polarization_raw_renedrings,sup_fig:multispectral_raw_renedrings} where the reflection on the desk is challenging to estimate for all the modalities. 
The reflected radiance is a high frequency function because it is highly dependent on the viewpoint, thus it is difficult to predict for arbitrary viewpoints far from the training views. Even if our model employs \gls{sh} encoding~\cite{verbin2022ref} to ease the estimation of view-dependent high-frequency details, it is still not enough to always accurately estimate reflections from novel viewpoints. One possible solution is to introduce the estimation of the \gls{brdf}, as proposed in~\cite{verbin2022ref,boss2021nerd}, which is a parametric model that describes how light is reflected according to the specific properties of the surface materials; we leave it for future work.\\
\indent In \cref{sup_fig:geometry,sup_fig:aligned}, we show some qualitative renderings of the normal and depth maps estimated by our model and some examples of perfectly aligned renderings of different modalities. Finally, in \cref{sup_fig:aop_dop,sup_fig:multispectral_single_channels_teddybear,sup_fig:multispectral_single_channels_toys} we present some Polarization and Multispectral renderings, respectively. For the \gls{pol} renderings, we show both the \gls{aop} and the \gls{dop}: it is possible to observe the limited difference in the measured and estimated polarization for mostly reflective or diffusive scenes. The \gls{ms} renderings show how the different multispectral channels capture different bands of the visible spectrum.


\begin{table*}
    \centering
    \renewcommand{\arraystretch}{0.7}
    \begin{tabular}{@{}lcccccc}
        \toprule
        Training Mod. & Test Mod. & \textbf{African Art} & \textbf{Aloe} & \textbf{Bird House} & \textbf{Book} & \textbf{Bouquet} \\
        \midrule
         & \green{RGB} & 34.87 & 31.47 & 30.40 & 34.84 & 29.73\\
        \arrayrulecolor{black!30}\cmidrule{2-7}
        \multirow{2}{2.15cm}{\centering \green{RGB} - \purple{Mono}} & \purple{Mono} & 35.19 & 32.00 & 31.83 & 35.15 & 30.16\\
        \arrayrulecolor{black!30}\cmidrule{2-7}
        \multirow{2}{*}{\red{NIR} - \blue{Pol} - \orange{MS}} & \red{NIR} & 35.94 & 33.78 & 32.18 & 37.27 & 33.24\\
        \arrayrulecolor{black!30}\cmidrule{2-7}
        & \blue{Pol} & 31.85 & 28.31 & 29.20 & 32.50 & 27.06\\
        \arrayrulecolor{black!30}\cmidrule{2-7}
        & \orange{MS} & 32.91 & 28.94 & 30.38 & 33.96 & 28.07\\
        \arrayrulecolor{black}\midrule
        \midrule
        Training Mod. & Test Mod. & \textbf{Chess} & \textbf{Clock} & \textbf{Easter Egg} & \textbf{Fan} & \textbf{Forest Gang 1} \\
        \midrule
         & \green{RGB} & 33.59 & 30.90 & 29.24 & 31.08 & 33.56\\
        \arrayrulecolor{black!30}\cmidrule{2-7}
        \multirow{2}{2.15cm}{\centering \green{RGB} - \purple{Mono}} & \purple{Mono} & 33.62 & 32.22 & 30.75 & 31.88 & 34.23\\
        \arrayrulecolor{black!30}\cmidrule{2-7}
        \multirow{2}{*}{\red{NIR} - \blue{Pol} - \orange{MS}} & \red{NIR} & 35.84 & 33.68 & 31.14 & 32.70 & 34.90\\
        \arrayrulecolor{black!30}\cmidrule{2-7}
        & \blue{Pol} & 31.53 & 29.04 & 26.71 & 28.04 & 33.09\\
        \arrayrulecolor{black!30}\cmidrule{2-7}
        & \orange{MS} & 31.46 & 30.16 & 28.56 & 30.59 & 32.77\\
        \arrayrulecolor{black}\midrule
        \midrule
        Training Mod. & Test Mod. & \textbf{Forest Gang 2} & \textbf{Fruits} & \textbf{Gamepads} & \textbf{Glass Clock} & \textbf{Globe} \\
        \midrule
         & \green{RGB} & 32.39 & 34.55 & 36.19 & 28.05 & 35.95\\
        \arrayrulecolor{black!30}\cmidrule{2-7}
        \multirow{2}{2.15cm}{\centering \green{RGB} - \purple{Mono}} & \purple{Mono} & 32.82 & 34.56 & 34.56 & 29.75 & 35.29\\
        \arrayrulecolor{black!30}\cmidrule{2-7}
        \multirow{2}{*}{\red{NIR} - \blue{Pol} - \orange{MS}} & \red{NIR} & 33.22 & 34.19 & 36.76 & 31.34 & 37.52\\
        \arrayrulecolor{black!30}\cmidrule{2-7}
        & \blue{Pol} & 32.49 & 32.99 & 33.85 & 25.94 & 32.70\\
        \arrayrulecolor{black!30}\cmidrule{2-7}
        & \orange{MS} & 31.21 & 33.37 & 34.90 & 27.10 & 34.04\\
        \arrayrulecolor{black}\midrule
        \midrule
        Training Mod. & Test Mod. & \textbf{Laptop} & \textbf{Laurel Wreath} & \textbf{Lego Ship} & \textbf{Makeup} & \textbf{Orchid} \\
        \midrule
         & \green{RGB} & 36.88 & 33.37 & 33.51 & 33.68 & 30.50\\
        \arrayrulecolor{black!30}\cmidrule{2-7}
        \multirow{2}{2.15cm}{\centering \green{RGB} - \purple{Mono}} & \purple{Mono} & 36.02 & 34.28 & 34.46 & 31.78 & 31.03\\
        \arrayrulecolor{black!30}\cmidrule{2-7}
        \multirow{2}{*}{\red{NIR} - \blue{Pol} - \orange{MS}} & \red{NIR} & 37.32 & 35.07 & 36.36 & 34.68 & 33.34\\
        \arrayrulecolor{black!30}\cmidrule{2-7}
        & \blue{Pol} & 34.04 & 30.61 & 31.80 & 31.28 & 28.21\\
        \arrayrulecolor{black!30}\cmidrule{2-7}
        & \orange{MS} & 34.45 & 30.81 & 32.11 & 33.09 & 29.00\\
        \arrayrulecolor{black}\midrule
        \midrule
        Training Mod. & Test Mod. & \textbf{Pillow} & \textbf{Plant} & \textbf{Steel Pot} & \textbf{Teddy Bear} & \textbf{Tin Box 1} \\
        \midrule
         & \green{RGB} & 27.67 & 29.89 & 27.52 & 35.83 & 28.20\\
        \arrayrulecolor{black!30}\cmidrule{2-7}
        \multirow{2}{2.15cm}{\centering \green{RGB} - \purple{Mono}} & \purple{Mono} & 28.24 & 31.49 & 28.70 & 36.38 & 29.19\\
        \arrayrulecolor{black!30}\cmidrule{2-7}
        \multirow{2}{*}{\red{NIR} - \blue{Pol} - \orange{MS}} & \red{NIR} & 29.18 & 32.97 & 29.10 & 37.16 & 31.34\\
        \arrayrulecolor{black!30}\cmidrule{2-7}
        & \blue{Pol} & 25.66 & 28.47 & 24.89 & 33.72 & 26.11\\
        \arrayrulecolor{black!30}\cmidrule{2-7}
        & \orange{MS} & 26.28 & 29.66 & 26.17 & 33.74 & 27.37\\
        \arrayrulecolor{black}\midrule
        \midrule
        Training Mod. & Test Mod. & \textbf{Tin Box 2} & \textbf{Toys} & \textbf{Trophies} & \textbf{Truck} & \textbf{Vases} \\
        \midrule
         & \green{RGB} & 29.34 & 33.33 & 25.88 & 31.45 & 34.83\\
        \arrayrulecolor{black!30}\cmidrule{2-7}
        \multirow{2}{2.15cm}{\centering \green{RGB} - \purple{Mono}} & \purple{Mono} & 27.96 & 31.96 & 26.84 & 30.76 & 34.11\\
        \arrayrulecolor{black!30}\cmidrule{2-7}
        \multirow{2}{*}{\red{NIR} - \blue{Pol} - \orange{MS}} & \red{NIR} & 28.51 & 34.48 & 27.99 & 31.84 & 34.96\\
        \arrayrulecolor{black!30}\cmidrule{2-7}
        & \blue{Pol} & 27.31 & 30.31 & 24.34 & 29.72 & 32.48\\
        \arrayrulecolor{black!30}\cmidrule{2-7}
        & \orange{MS} & 28.17 & 31.56 & 25.36 & 29.10 & 34.37\\
        \arrayrulecolor{black}\midrule
        \midrule
        Training Mod. & Test Mod. & \textbf{Watering Can 1} & \textbf{Watering Can 2} & & \textbf{\underline{Mean}} & \textbf{\underline{Std.}}\\
        \midrule
         & \green{RGB} & 33.31 & 32.01 & & \textbf{32.00} & \textbf{2.83}\\
        \arrayrulecolor{black!30}\cmidrule{2-7}
        \multirow{2}{2.15cm}{\centering \green{RGB} - \purple{Mono}} & \purple{Mono} & 34.93 & 31.64 & & \textbf{32.31} & \textbf{2.49}\\
        \arrayrulecolor{black!30}\cmidrule{2-7}
        \multirow{2}{*}{\red{NIR} - \blue{Pol} - \orange{MS}} & \red{NIR} & 35.23 & 33.03 & & \textbf{33.63} & \textbf{2.59}\\
        \arrayrulecolor{black!30}\cmidrule{2-7}
        & \blue{Pol} & 29.74 & 29.59 & & \textbf{29.80} & \textbf{2.79}\\
        \arrayrulecolor{black!30}\cmidrule{2-7}
        & \orange{MS} & 31.75 & 30.76 & & \textbf{30.69} & \textbf{2.64}\\
        \arrayrulecolor{black}\bottomrule
    \end{tabular}
    \caption{Five-modality mosaicked training results of all the scenes of \dname\ in terms of PSNR (dB).}
    \label{sup_tab:all_scenes}
\end{table*}
\section{Few-Shot Experiments}
\label{sup_sec:few_shot}

Given the number of advantages introduced by the use of multiple modalities, we also investigated whether their use could help to obtain better results in the few-shot scenario. For this reason, we performed some additional tests to evaluate the impact of additional modalities on the RGB rendering quality. We conducted two few-shot experiments involving the single-, two-, and five-modality training, by training the model with 5 and 10 macroframes, respectively. The results are reported in \cref{sup_tab:few-shot}. Even if we did not explicitly develop our model to address the few-shot task, it is possible to see that introducing additional modalities is beneficial in terms of RGB rendering PSNR. With the introduction of a single additional modality (namely, the NIR in this case), we obtain an increase in PSNR of 1~dB and 5~dB for the 5 and 10 macroframes cases, respectively.  This trend is also confirmed considering the single- and the five-modality training: we achieve a gain up to 3.5~dB and 9~dB, respectively. Analogously, the SSIM and the LPIPS also improve.\\
\indent These preliminary but encouraging results, obtained with a model not specifically developed to handle the few-shot task, open new possibilities of investigation of the few-shot neural rendering and 3D reconstruction tasks in the multimodal scenario.

\end{document}